\documentclass{article}

\usepackage{arxiv}

\usepackage[utf8]{inputenc} 
\usepackage[T1]{fontenc} 
\usepackage[hidelinks]{hyperref} 
\usepackage{url} 
\usepackage{booktabs} 
\usepackage{amsfonts} 
\usepackage{stmaryrd}
\usepackage{nicefrac} 
\usepackage{microtype} 
\usepackage{lipsum}		
\usepackage{graphicx}
\usepackage{pdflscape}
\usepackage{afterpage}
\usepackage{tabularx}
\usepackage{multirow}
\usepackage{longtable}
\usepackage{stmaryrd} 
\usepackage[table,xcdraw]{xcolor}
\usepackage{geometry} 
\usepackage{makecell}
\usepackage[toc,page,title]{appendix}
\usepackage{array, tabularx, afterpage, lscape}
\usepackage{adjustbox}

\newcolumntype{C}{>{\centering\arraybackslash}X} 
\usepackage[
    backend=biber,
    style=numeric,
    sorting=none,  
    giveninits=true,  
    maxnames=1,  
    uniquename=false,
    url=false,
    doi=false,
    isbn=false,
    eprint=false
]{biblatex}

\AtEveryBibitem{%
    \clearfield{day}%
    \clearfield{month}%
    \clearfield{endday}%
    \clearfield{endmonth}%
    \clearfield{note}%
}

\DeclareNameAlias{author}{family-given}
\DeclareNameAlias{editor}{family-given}
\DeclareNameAlias{translator}{family-given}

\renewbibmacro{in:}{}

\DeclareFieldFormat{pages}{#1}

\usepackage{doi}
\usepackage{tikz}
\usetikzlibrary{mindmap}
\usetikzlibrary{bayesnet} 
\usetikzlibrary{arrows}
\usepackage{caption}
\usepackage{subcaption}
\usepackage{dsfont}
\usepackage{amsmath}
\usepackage{comment}
\usepackage{mathpazo} 
\usepackage{amsthm} 
\usepackage{amssymb}
\usepackage{mathrsfs}
\usepackage{stmaryrd}

\usepackage[section]{placeins}
\usepackage{longtable}
\usepackage[linesnumbered,ruled,vlined]{algorithm2e}
  
\usepackage{booktabs}
\usepackage{multirow}  
\usepackage{pdflscape}
\usepackage{subfiles} 
\usepackage{titletoc}
\usepackage{subfiles}
\usepackage{titlesec}
\titlespacing*{\subsection}
{0pt}{3ex}{1.5ex}
\titlespacing*{\subsubsection}
{0pt}{3ex}{2ex}
\titlespacing*{\paragraph}
{0pt}{2ex}{1ex}

\title{Comparison of Bayesian methods for extrapolation of treatment effects: a simulation study}

\author{
    Tristan Fauvel   \thanks{Corresponding author: \texttt{t.fauvel@quinten-health.com}} \\
    Quinten Health, 8 rue Vernier, Paris, France
    \And
    Julien Tanniou \\
    Quinten Health; now with Saryga
    \And
    Pascal Godbillot \\
    Quinten Health
    \And
    Marie Génin \\
    Quinten Health
    \And
    Billy Amzal \\
    Quinten Health
}

\date{}

\renewcommand{\headeright}{}
\renewcommand{\undertitle}{}
\renewcommand{\shorttitle}{Comparison of Bayesian methods for extrapolation of treatment effects}

\hypersetup{
pdftitle={Comparison of Bayesian methods for extrapolation of treatment effects: a simulation study},
pdfsubject={},
pdfauthor={Quinten Health},
pdfkeywords={Health},
}

\begin{document}
\maketitle

\begin{abstract}
Extrapolating treatment effects from related studies is a promising strategy for designing and analyzing clinical trials in situations where achieving an adequate sample size is challenging. Bayesian methods are well-suited for this purpose, as they enable the synthesis of prior information through the use of prior distributions. While the operating characteristics of Bayesian approaches for borrowing data from control arms have been extensively studied \parencite{viele_use_2014}, methods that borrow treatment effects—quantities derived from the comparison between two arms—remain less well understood.  

In this paper, we present the findings of an extensive simulation study designed to address this gap, across six clinically motivated case studies covering continuous, binary, recurrent-event, and time-to-event endpoints. We evaluate the frequentist operating characteristics of these methods, including the probability of success, mean squared error, bias, precision, credible interval coverage, and interval score. We varied target sample size, treatment-effect drift, and endpoint-specific design assumptions. 
Power was compared with a frequentist analysis of the target data alone calibrated to the same actual type I error rate, and estimation performance was compared across configurations producing similar type I error rates.

Our results provide insights into the strengths and limitations of existing methods in the context of confirmatory trials: borrowing generally increased both power and type I error. At matched actual type I error rates, power differences were usually small, with occasional gains over the reference analysis and larger losses for some adaptive procedures. Borrowing reduced MSE when source and target effects were consistent, but increasing drift eroded this benefit and could substantially reduce interval coverage. Commensurate priors with appropriately chosen hyperpriors offered the most favorable balance between estimation gains and robustness to conflict. The conditional power prior was competitive for estimation at comparable type I error rates, despite its lack of adaptation. Test-then-pool procedures often performed poorly. No method dominated across all criteria. Method selection should be supported by external evidence of treatment-effect consistency and design-specific evaluation of type I error, estimation accuracy, and coverage.

\textbf{Disclaimer:} This document expresses the opinion of the authors of the paper, and may not be understood or quoted as being made on behalf of or reflecting the position of Quinten Health or the European Medicines Agency or one of its committees or working parties.
\end{abstract}

\section{Introduction}

Borrowing information from historical or concurrent studies may support the evaluation of medicines in populations in which conventional randomized controlled trials are difficult to perform, including children and patients with rare diseases. Regulatory guidance recognizes the potential role of external information in trial design and analysis, provided its relevance to the target population and the assumptions supporting its use  are justified. ICH E11(R1) emphasizes the need to avoid unnecessary enrollment of children and discusses the use of existing evidence in pediatric development. The subsequent ICH E11A guideline on pediatric extrapolation (EMA/CHMP/ICH/205218/2022) includes recommendations on Bayesian approaches to trial design and analysis. These guidelines provide a framework for the use of extrapolation in drug development. 

 The frequentist operating characteristics of Bayesian borrowing methods, including type I error rates, have been studied extensively when borrowing is restricted to the control arm \parencite{viele_use_2014, lesaffre_review_2024, su_comparative_2022}. Many methods were developed for this setting \parencite{hobbs_hierarchical_2011, hobbs_commensurate_2012, neuenschwander_note_2009, schmidli_robust_2014, banbeta_modified_2019, jiang_optimal_2021, hupf_bayesian_2021, yang_sam_2023}.
Other approaches borrow information from both arms of the source trial \parencite{rietbergen_incorporation_2011, kaizer_bayesian_2018, feisst_incorporating_2020}. Borrowing directly on a treatment effect contrast is a distinct formulation.

When borrowing treatment effects, a central consideration is the discrepancy between the true treatment effect in the target population and the observed estimate from the source study. We refer to this discrepancy as drift \parencite{viele_phase_2018, lim_reducing_2020, best_classical_2025}. 
Although frequentist operating characteristics have been evaluated for some of the individual methods in this setting \parencite{nikolakopoulos_dynamic_2018, psioda_bayesian_2018, quan_bayesian_2019}, comparisons across methods and scenarios within a common framework remain limited.

We conducted a simulation study to compare Bayesian and frequentist approaches to borrowing treatment effects across a range of clinically motivated scenarios.  We varied the target trial sample size, the true target treatment effect and the parameters governing borrowing. We assessed the probability of meeting the trial success criterion and measures of estimation performance. Borrowing trades bias for variance: it adds bias when the source and target treatment effects differ, and it reduces the variance of the estimate. A primary objective was therefore to quantify this trade-off with the mean squared error (MSE) of the estimate, and to find how much drift each method tolerates before its MSE exceeds that of an analysis of the target data alone. Because the amount of borrowing depends on tuning parameters that have no common scale across methods, we compared the power and the MSE of the methods at a matched type I error rate.

We focused on methods that incorporate source evidence through aggregate study results \parencite{holzhauer_evidence_2018}. This scope was consistent with the project's objective of informing regulatory assessment of borrowing methods, including settings in which the available source evidence consists of study summaries. These methods are also applicable when the summaries are derived from analyses of individual patient data (IPD). Methods requiring specifically IPD for borrowing, and approaches that explicitly model heterogeneity across multiple source studies, were outside the scope of this manuscript. Within this framework, we compared the operating characteristics of the selected methods to inform their use in the design and analysis of trials involving extrapolation of treatment effects.
 
\section{Methods}
  
\subsection{Scenarios considered}

The scenarios were motivated by pediatric extrapolation, in which evidence on treatment effects in adults may inform the design and analysis of trials in children. We considered borrowing from source populations for which aggregate results were available.

Adjustment for prognostic factors and treatment effect modifiers may improve comparability between source and target populations. However, our evaluation did not include covariates or borrowing methods requiring IPD, such as the propensity score-integrated power prior \parencite{lin_incorporating_2022, wang_propensity_2022}. Indeed, published aggregate results do not generally allow for such adjustment. In addition, adjustment for age in adult-to-pediatric extrapolation requires structural assumptions. Age also reflects developmental, physiological, pharmacokinetic, pharmacodynamic, and disease-related differences that may affect treatment response and may not be fully measured.
Thus, assumptions about the relevance of source treatment effects remain necessary even when adjustment for measured covariates is possible.  

\subsubsection{Selected case studies}
\label{sec:case_studies}

We selected published adult and pediatric studies evaluating treatment efficacy in comparable clinical settings. The case studies covered different endpoint types, treatment effect measures, therapeutic areas, and sample sizes, as summarized in Supplementary Table \ref{tab:RCTs_selection}.  
 
\paragraph{Botox for the treatment of lower limb spasticity (continuous endpoint)}
 
 We considered a case study on botox introduced in \textcite{wang_bayesian_2022}, based on a published phase 3 placebo-controlled RCT evaluating a single treatment with botox (4 U/kg or 8 U/kg) combined with standardized physical therapy in pediatric patients with lower limb spasticity. In this trial, 412 patients aged 2 to 16 years were randomized in a 1:1:1 ratio to botox 8 U/kg, botox 4 U/kg or control. Following \textcite{wang_bayesian_2022}, we considered only the botox 4 U/kg arm ($N = 126$) and the control arm ($N = 130$). The primary endpoint was change in a clinical score. The estimated treatment difference between botox 4 U/kg and control was 0.10 (SD 0.10), and superiority was not established. Botox was previously approved in adults for a similar indication, based on a single phase 3 placebo-controlled trial (233 patients receiving botox and 235 controls), which we used as the source study. The estimated treatment difference was 0.20 (SD 0.10). Both treatment differences are the posterior means and standard deviations obtained by \textcite{wang_bayesian_2022} from a Bayesian re-analysis of the original frequentist analyses.

\paragraph{Dapagliflozin for the management of type 2 diabetes  (continuous endpoint)}
For the second case study with a continuous endpoint, we considered the randomized controlled trial reported by \textcite{shehadeh_dapagliflozin_2023}, evaluating the efficacy of dapagliflozin in uncontrolled type 2 diabetes in children (N = 81 in the dapagliflozin group, N = 76 in the placebo group). The primary endpoint was change in HbA1c at week 26. The trial demonstrated efficacy of dapagliflozin. As a source study, we considered a phase 3 trial in adults with type 2 diabetes  and inadequate glycemic control despite daily metformin \parencite{bailey_effect_2010}. To align the adult and pediatric comparisons, we focused on the arm receiving 5 mg daily dapagliflozin (N = 133) and the placebo arm (N = 134). The estimated treatment effect, expressed as the between-group difference in mean reduction in HbA1c from baseline to week 24, was 0.36 (95\% CI 0.16 to 0.56) \parencite{bailey_dapagliflozin_2013}. 

\paragraph{Belimumab for the treatment of seropositive systemic lupus erythematosus (binary endpoint)}
\label{par:binary_endpoint}
For a binary endpoint, we considered intravenous belimumab in children aged 5-17 years with active, seropositive systemic lupus erythematosus (SLE) \parencite{psioda_bayesian_2020, best_classical_2025}. 
A post-marketing pediatric randomized controlled trial enrolled 92 participants \parencite{brunner_safety_2020}, and a post hoc Bayesian analysis borrowed information on the treatment effect from phase 3 adult evidence \parencite{pottackal_application_2025}. Following the published case study \parencite{best_classical_2025}, we pooled the two adult trials and treated them as a single source of historical information. The pooled odds ratio based on a total of $N_S = $ 1125 subjects from these studies was 1.62 (95\% CI, 1.27 - 2.05). 

\paragraph{Aprepitant for the prevention of postoperative nausea and vomiting (binary endpoint).}
\label{par:aprepitant}
For the second binary endpoint, we considered aprepitant for the prevention of postoperative nausea and vomiting in children \cite{jin_bayesian_2021}.  An adult trial with sample sizes 293 and 280 in the treatment and control groups reported response proportions of 62.8\% and 55.0\% in the treatment and control groups, respectively \cite{diemunsch_singledose_2007}. A similar randomized phase 2b study was completed in children \cite{salman_pharmacokinetics_2019}. Response was defined as no vomiting and no use of rescue therapy during the first 24 hours after surgery.  In the pediatric trial, the difference in response proportions between the treatment and control groups was 3.4 percentage points. 

In this case study (in which the treatment effect is a difference in proportions), we followed an approach initially described in \textcite{jin_bayesian_2021}, in which a prior is specified for the target study control rate (such as a beta prior or a uniform prior in the [0,1] range), and a prior is specified for the target study treatment effect (such as a truncated normal). 

\paragraph{Teriflunomide for the treatment of multiple sclerosis (time-to-event endpoint)} 
For the time-to-event endpoint, we considered the trial of teriflunomide versus placebo in pediatric multiple sclerosis (TERIKIDS) \parencite{chitnis_safety_2021}, which assessed teriflunomide in children (57 placebo vs 109 teriflunomide). \textcite{bovis_reinterpreting_2022} applied a Bayesian approach for estimating the effect of teriflunomide in children in the TERIKIDS study, by integrating the available knowledge on teriflunomide in adults. As source studies, they used published data from 2 randomized clinical trials testing teriflunomide in adult patients with MS (TEMSO: 363 placebo vs 359 teriflunomide \parencite{oconnor_randomized_2011}, and   
TOWER: 389 placebo vs 372 teriflunomide \parencite{confavreux_oral_2014}). TEMSO and TOWER had annualized relapse rate as their primary endpoint. In TERIKIDS, the primary endpoint was the time to first relapse, and the treatment effect summary measure was the log hazard ratio for active treatment compared to placebo (assumed to be normally distributed). \textcite{bovis_reinterpreting_2022} combined the log hazard ratios for time to first relapse using inverse-variance weighting and derived the pooled hazard ratios (HRs) and 95\% confidence interval (CI). The estimated HRs for time to first relapse with teriflunomide versus placebo in TEMSO, TOWER, and in TERIKIDS were 0.72 (95\% CI, 0.58-0.90), 0.63 (95\% CI, 0.50-0.79), and 0.66 (95\% CI, 0.39-1.11), respectively. 
  
\paragraph{Mepolizumab for the management of severe asthma (recurrent event endpoint)} 
For the recurrent event endpoint, we considered the case study described by \textcite{best_assessing_2021}, based on a post hoc analysis of the MENSA trial of mepolizumab in severe asthma \parencite{ortega_mepolizumab_2014} by  \textcite{keene_use_2020}. In the MENSA trial, the primary endpoint was the rate of clinically significant exacerbations per year. The treatment effect was expressed as the log rate ratio for active treatment versus placebo, estimated using negative binomial regression of exacerbation counts and analyzed using a normal approximation. The trial randomized patients to placebo, 75 mg intravenous mepolizumab or 100 mg subcutaneous mepolizumab, and, following \textcite{best_assessing_2021}, we combined the two active arms. The trial included 25 adolescents (9 control and 16 treated patients) and 551 adults (182 control and 369 treated patients). The estimated log rate ratio in adolescents was -0.40 with standard error 0.703, whereas the corresponding estimate in adults was -0.69, with a standard error of 0.13 \parencite{best_assessing_2021}. To determine the rate in the adult control group, we used the data from \textcite{ortega_mepolizumab_2014}. We assumed that the contribution of the pediatric subgroup to the overall rate was negligible, and therefore set the adult control rate equal to the overall control rate, 1.74. The corresponding treatment group rate, consistent with this control rate and the estimated log rate ratio, was 0.87.
   
\subsubsection{Sample sizes}
Within each case study, the source sample size $N_S$ was fixed. We considered target sample sizes $N_T = N_S$, $N_S/2$, $N_S/4$ and $N_S/6$, going from trials of the same size as the source study to smaller trials relevant to pediatric development. The corresponding sample sizes are provided in Supplementary Table \ref{tab:drift_ranges_and_sample_sizes}.  Allocation in the target trial was balanced between the treatment and control arms: each arm had $n_T = N_T/2$ participants, and we write $N_T^{(c)} = N_T^{(t)} = n_T$ when the arm matters.  
 
\subsubsection{Drift in treatment effect}
\label{ref:drift_treatment_effect}
 
The drift in treatment effect is defined as the difference between the true treatment effect in the target population and the observed source treatment effect estimate, $\delta = \theta_T - \hat{\theta}_S$ \parencite{viele_use_2014, lim_reducing_2020, best_classical_2025}. 
Drift quantifies disagreement between the target treatment effect and the fixed source estimate. We are particularly interested in drift values corresponding to a target treatment effect  $\theta_T$ between $\theta_0$ and $\hat{\theta}_S$, where $\theta_0$ is the boundary of the null hypothesis space $\Theta_0$. This corresponds to a drift $\delta \in \left[\mathrm{min}(0, \theta_0 - \hat{\theta}_S), \mathrm{max}(0, \theta_0 - \hat{\theta}_S) \right]$. We focused in particular on the three following scenarios:
\begin{enumerate}
    \item the true treatment effect in the target population is the same as the estimated treatment effect in the source population ("consistent treatment effect"),
    \item the true treatment effect in the target population is half the estimated treatment effect in the source population ("partially consistent treatment effect"),
    \item there is no treatment effect in the target population.
\end{enumerate}
Where needed, the interval $\theta_T \in [\theta_0, \hat{\theta}_S]$ was extended to characterize the operating characteristics over a wider range of discrepancies. Details on the approach used to calculate extended limits and the specific ranges considered for each use case are provided in the supplementary section \ref{supp:definition_drift_ranges} of the supplemental material.

\subsubsection{Drift in the control arm event rate}
\label{sec:control_drift}

In the primary simulation scenarios, the underlying response or event rate in the control arm was held constant between the source and target studies. This choice was made to isolate the impact of the treatment effect drift while limiting the number of simulation dimensions. It should therefore be interpreted as a reference scenario rather than as an assumption that the source and target control populations are exchangeable.

To investigate departures from this reference scenario, we additionally considered control arm drift in the teriflunomide case study. Let $\lambda_S^{(c)}$ and $\lambda_T^{(c)}$ denote the source and target
control arm event rates, respectively. We define $ \kappa =
    \log\left\{
    \frac{\lambda_T^{(c)}}{\lambda_S^{(c)}}
    \right\}$, so that
$ \lambda_T^{(c)}  =  \lambda_S^{(c)}\exp(\kappa)$.  The treatment effect drift was varied independently according to $\delta = \theta_T-\hat{\theta}_S$, where $\theta_T$ is the target log hazard ratio. Thus, $\kappa$ modifies the underlying event rate in the target population without, by itself, altering the target treatment effect.

\subsubsection{Sensitivity to control response and event rates}
\label{sec:control_rate_sensitivity}

We examined whether the control response or event rate affected the operating characteristics of methods borrowing treatment effects expressed as odds ratios, rate ratios or hazard ratios. For the belimumab, mepolizumab and teriflunomide case studies, we multiplied the reference control-arm odds of response (belimumab) or event rate (mepolizumab, teriflunomide) by 0.5 and 1.5, while keeping the source treatment effect estimate and the source and target sample sizes fixed. We refer to this multiplier as the source denominator factor; a factor of 0.5 gives a lower source control rate. In the aprepitant case study, analyzed with the binomial likelihood, the factor was kept at 1.

For the belimumab and mepolizumab case studies, the source treatment-arm response probability or event rate was adjusted to preserve the source odds ratio or rate ratio, respectively. The standard error of the source treatment effect was then recalculated from the modified arm-level quantities. The target control response probability or event rate was set equal to the modified source value, and the target treatment-arm value was determined by the treatment effect specified for each drift scenario. These scenarios therefore changed both the precision of the source estimate and the sampling distribution of the target estimate.

For the teriflunomide case study, the published pooled source log hazard ratio and its standard error were held fixed. The factors of 0.5 and 1.5 were applied to the reference placebo event rate used to generate target trial data. These scenarios therefore affected the number of observed target events and the precision of the target estimate, without changing the source information used for borrowing.

\subsection{Data generation and sampling approximations}
\subsubsection{Data generation for continuous endpoints}

For continuous endpoints, the target data were summarized by the mean of $n_T$ paired treatment differences, each distributed as $\mathcal{N}\left(\hat{\theta}_S + \delta, \sigma_T^2\right)$. Rather than generating the individual differences, we drew their sufficient statistics directly from their exact sampling distributions: the sample mean from $\mathcal{N}\left(\hat{\theta}_S + \delta, \sigma_T^2 / n_T\right)$ and, independently, the sample variance from $\sigma_T^2 \chi^2_{n_T - 1} / (n_T - 1)$, which is equivalent in distribution. The sampling standard deviation $\sigma_T$ was specified for each scenario relative to that of the source study, through the ratio $\sigma_T / \sigma_S$, where $\sigma_S = \hat{\sigma}_{\theta_S} \sqrt{n_S}$, $\hat{\sigma}_{\theta_S}$ is the standard error of the source estimate and $n_S$ is the equivalent source sample size per arm. For analysis, the variance estimated from each simulated target dataset, $\hat{\sigma}_T^2$, was treated as fixed.

\subsubsection{Data generation for binary endpoints}
\label{sec:binary_generation}
  
For the log odds ratio analysis, we generated arm-specific response counts from binomial distributions: $y_T^{(c)} \sim \operatorname{Bin}(N_T^{(c)}, p_T^{(c)})$ and $y_T^{(t)} \sim \operatorname{Bin}(N_T^{(t)}, p_T^{(t)})$, where:
\begin{itemize}
\item $y_T^{(a)}$: number of responders in arm $a$ ($c$: control, $t$: treatment) of the target trial.
\item $N_T^{(a)}$: number of subjects in arm $a$ of the target trial.
\item $p_T^{(a)}$: response rate in arm $a$ of the target trial.
\end{itemize}
We calculated the observed response proportions, $\hat{p}_T^{(a)} = y_T^{(a)}/N_T^{(a)}$, and the estimated log odds ratio, $\hat{\theta}_T = \log \left( \frac{\hat{p}_T^{(t)}/(1-\hat{p}_T^{(t)})}{\hat{p}_T^{(c)}/(1-\hat{p}_T^{(c)})}\right)$.  
To keep the estimate and its standard error finite when an arm has no responder or only responders, 0.5 was added to each of the four cells of the $2 \times 2$ table before computing them, in every trial. With the binomial likelihood of the aprepitant case study, the analysis is defined for every pair of responder counts, including 0 and $N_T^{(a)}$. In both binary case studies, the operating characteristics were computed by enumerating the pairs of responder counts. 
We assumed that the response rates are the same in the source and target studies control arms. So, for drift $\delta$,  the response rate in the target study treatment arm is: $p_T^{(t)} = \frac{e^\delta}{e^\delta +1/\text{odds}^{(t)}_S}$, where $\text{odds}^{(t)}_S$ is the observed odds in the source study treatment arm.

\subsubsection{Data generation for time-to-event endpoints}

The simulated target trial was inspired by the TERIKIDS design. Let $\hat{\theta}_S=\log(\widehat{HR}_S)$ denote the pooled treatment effect estimate from the adult TEMSO and TOWER trials, obtained by inverse-variance weighting of the two study-specific log hazard ratios (Section~\ref{sec:case_studies}). We used $\hat{\theta}_S=-0.393$ and $\hat{\sigma}_{\theta_S}=0.081$, and defined the target treatment effect as $\theta_T=\hat{\theta}_S+\delta$, where $\delta$ is the treatment effect drift.

\paragraph{Placebo event rate}
Neither adult trial reported a placebo hazard of first relapse, since both had the annualized relapse rate (ARR) as primary endpoint. We therefore anchored the source placebo rate on the ARR, $\lambda_S^{(c)}=0.534$ relapses per patient-year, and used it as the hazard of first relapse. 
Under between-patient heterogeneity in relapse risk, the population hazard of first relapse decreases over time and is lower on average than the ARR, so this choice tends to overstate the number of first relapses. With the control-arm drift $\kappa$ of Section~\ref{sec:control_drift}, $\lambda_T^{(c)}=\lambda_S^{(c)}\exp(\kappa)$ and $\lambda_T^{(t)}=\lambda_T^{(c)}\exp(\theta_T)$. In the primary design, $\kappa=0$. For this case study, the change in the denominator of the source summary measure (Section~\ref{sec:control_rate_sensitivity}) multiplies $\lambda_S^{(c)}$, and therefore the target placebo rate, while the source estimate and its standard error are kept at their published values.
We also studied a target placebo relapse rate equal to two thirds of, or 1.5 times, that of the source trials ($\kappa = -0.405$ or $+0.405$ on the log hazard scale). At a fixed treatment-effect drift, this control-arm drift does not alter the true target hazard ratio, because both target-arm hazards are shifted alike, but it changes the number of observed events and therefore the precision of the target treatment-effect estimate.

\paragraph{Accrual and administrative censoring}
In the original TERIKIDS design, each participant could be followed for up to 96 weeks after randomization.
Under this design, accrual would not affect the individual follow-up distribution if all participants were allowed to complete 96 weeks of follow-up. To investigate the effect of staggered accrual, we therefore considered a modified fixed-calendar design that retained the TERIKIDS maximum follow-up of $L=96$ weeks but closed the database $F=48$ weeks after the end of recruitment.
Patients entered the study over an accrual period of $t_{\mathrm{acc}}=3.41$ years, with calendar-time entry generated as $R_i\sim\mathcal U(0,t_{\mathrm{acc}})$. The database was closed at calendar time $t_{\mathrm{close}}=t_{\mathrm{acc}}+F$. The administrative follow-up available to patient $i$ was $t_{\mathrm{close}}-R_i$, so the administrative censoring time was $C_i=\min\{L,t_{\mathrm{close}}-R_i\}$.  

\paragraph{Event times}
In the primary design, times to first relapse were exponential, $T_i\mid A_i\sim\operatorname{Exp}\{\lambda_T^{(c)}\exp(\theta_T A_i)\}$, where $A_i = 1$ if patient $i$ is in the treatment arm and $A_i = 0$ otherwise, so that the treatment-arm hazard is $\lambda_T^{(t)}$. As a sensitivity analysis, event times were generated from a Weibull proportional-hazards model with survival function $S(t\mid A_i)=\exp\{-(t/b_c)^q\exp(\theta_T A_i)\}$, where $q$ is the common shape parameter and $b_c$ the control-arm scale. This implies a hazard ratio of $\exp(\theta_T)$ and a treatment-arm scale $b_t=b_c\exp(-\theta_T/q)$. We used $q=0.72$, estimated from the aggregate TERIKIDS relapse-free probabilities, so that the hazard of relapse decreases over follow-up. The control-arm scale was calibrated to the 96-week relapse-free probability observed in the TERIKIDS placebo arm, $S_c(L)=0.442$, giving $b_c=L/[-\log\{S_c(L)\}]^{1/q}=2.45$ years. The exponential placebo arm, anchored on the adult ARR, implies $S_c(L)=\exp(-\lambda_S^{(c)}L)\approx0.37$, whereas the Weibull placebo arm reproduces the pediatric value of $0.44$. The Weibull design thus changes both the shape of the hazard and the placebo event rate, and yields fewer events than the primary design.

\paragraph{Delayed treatment effect.}
To examine a departure from proportional hazards, we also simulated a treatment effect that starts $d$ weeks after randomization, with $d\in\{12,24\}$. The treatment-arm hazard equals the control-arm hazard until $d$, and is multiplied by $\exp(\beta_d)$ afterwards. The Cox model is then misspecified, and its estimate
converges to a weighted average of the time-varying log hazard ratio \parencite{struthers_misspecified_1986}. We defined the treatment effect of the scenario as this limit, computed under the censoring of the trial, and solved for the
post-delay log hazard ratio $\beta_d$ that makes it equal to $\theta_T$. For a
consistent treatment effect ($\theta_T=\hat{\theta}_S$), this gives
$\beta_{12}=-0.51$ and $\beta_{24}=-0.69$. Drift, bias and the null hypothesis therefore keep the same meaning as under proportional hazards.

\paragraph{Loss to follow-up}
Independent loss to follow-up was generated as $D_i\sim\operatorname{Exp}(\lambda_D)$, with $\lambda_D=-\log(1-p_D)/L$, where $p_D$ is the probability of loss to follow-up over 96 weeks. We considered $p_D\in\{0,0.05,0.10\}$.
The observed follow-up time was $X_i=\min(T_i,D_i,C_i)$, with event indicator $\Delta_i=\mathbb{1}\{T_i\leq D_i,\ T_i\leq C_i\}$.

\paragraph{Analysis}
Each simulated dataset was analyzed with the Cox proportional-hazards model
$h_i(t)=h_0(t)\exp(\beta A_i)$. The estimate $\hat{\theta}_T=\hat{\beta}$ and its model-based standard error $\hat{\sigma}_{\theta_T}=\operatorname{SE}(\hat{\beta})$ were passed to the Bayesian borrowing procedures, as in the other case studies. Simulated trials in which the maximum partial-likelihood estimate did not exist were excluded.
 
The sensitivity designs (loss to follow-up, Weibull event times, control-arm drift and a delayed treatment effect) were analyzed in the same way. They were simulated with $n_T = 123$ participants per arm, the smallest target sample size of this case study, and a source denominator factor of 1, changing one factor at a time from the primary design.

\subsubsection{Data generation for recurrent event endpoints}

For the recurrent event endpoint, we simulated individual patients' event counts and estimated the treatment effect from them. In each arm, the counts were drawn from a negative binomial distribution with mean $\mu^{(a)}$, the event rate of arm $a$, and size (dispersion) parameter $\phi$, so that the variance is $\mu^{(a)} + (\mu^{(a)})^2 / \phi$. Following the original case study, we set $\phi = 0.8$ in both arms.

In each simulated trial, an intercept-only negative binomial model was fitted by maximum likelihood to the counts of each arm. The maximum likelihood estimate of the rate, $\hat{\mu}^{(a)}$, is the arm's sample mean, and the size
parameter $\hat{\phi}^{(a)}$ was estimated by maximizing the profile likelihood. The treatment effect was estimated by the log rate ratio $\hat{\theta}_T = \log(\hat{\mu}^{(t)} / \hat{\mu}^{(c)})$, with the standard error obtained from the Fisher information of the two models,
\[
\hat{\sigma}_{\theta_T}
= \sqrt{\frac{1}{N_T^{(t)}}\left(\frac{1}{\hat{\mu}^{(t)}} + \frac{1}{\hat{\phi}^{(t)}}\right)
      + \frac{1}{N_T^{(c)}}\left(\frac{1}{\hat{\mu}^{(c)}} + \frac{1}{\hat{\phi}^{(c)}}\right)},
\]
which coincides with the delta-method approximation evaluated at the estimates. 
The analyses then used a normal likelihood for the estimated log rate ratio, with this standard error treated as known. A trial in which an arm recorded no event, so that the log rate ratio is not defined, was excluded from the analysis.

\paragraph{Excluded trials}
\label{sec:data_generation_exclusions}
In the mepolizumab and teriflunomide case studies, the operating characteristics are therefore conditional on the treatment effect being estimable. In the mepolizumab case study, the probability that an arm records no event is $\{\phi/(\phi+\mu^{(a)})\}^{\phi n_T}$, so the share of excluded trials can be computed exactly for every design. It was below $10^{-11}$ under the null hypothesis, and below $10^{-3}$ in every design except the smallest target samples with the lower source control rate ($n_T = 45$ and 68, source denominator factor 0.5), where it reached 1.8\% and 0.2\% at the most extreme treatment effects, which have the lowest treatment-arm event rates. In the teriflunomide case study, we simulated 10{,}000 further trials for each of the 627 combinations of design and treatment effect, including the sensitivity designs, and none had a non-estimable Cox estimate. In the binary case studies, the probability of the excluded outcomes was below $10^{-10}$ (Section~\ref{sec:binary_generation}). Exclusions therefore had no material effect on the results.

\subsection{Analysis of the target trial}
\subsubsection{Decision criterion}

We considered a one-sided null hypothesis $\theta_T \leq \theta_0$ for all case studies except the teriflunomide and mepolizumab case studies, for which the null hypothesis was $\theta_T \geq \theta_0$. For all scenarios, we set $\theta_0 = 0$.

We denote by $\Theta_0$ the null hypothesis space. Given observed data $d_S$ and $d_T$ in the source and target studies respectively, we concluded that $\theta_T \notin \Theta_0$ if the posterior probability $\operatorname{Pr}(\theta_T \notin \Theta_0 \mid \mathbf{D}_T = d_T, \mathbf{D}_S = d_S) > \rho$, with $\rho = 0.975$. This critical value is equivalent to requiring the 95\% equal-tailed posterior credible interval for the treatment effect (with limits at the 2.5\% and 97.5\% posterior quantiles) to lie entirely outside $\Theta_0$. Its lower limit must therefore exceed $\theta_0$ when $\Theta_0 = \{\theta_T \leq \theta_0\}$, and its upper limit must fall below $\theta_0$ when $\Theta_0 = \{\theta_T \geq \theta_0\}$. The type I error rate (TIE) of a method is the probability that this rule concludes $\theta_T \notin \Theta_0$ when $\theta_T \in \Theta_0$. Its estimation is described in Section~\ref{sec:frequentist_ocs}.

\subsubsection{Likelihood}
\label{sec:hypotheses_TE_distribution}

For all case studies except the aprepitant case study, we assumed during analysis that the summary measure of the target study is normally distributed. Therefore, in these cases, the likelihood is given by $p(\hat{\theta}_T \mid \theta_T) = \mathcal{N}(\hat{\theta}_T \mid \theta_T, \hat{\sigma}^2_{\theta_T})$, where $\hat{\sigma}_{\theta_T}$ is the standard error of the target estimate, which, as explained above, is estimated from the target data and treated as known.  
 For log odds ratios, log hazard ratios, and log rate ratios, this is a large-sample approximation to the sampling distribution of $\hat{\theta}_T$.  

For binary endpoints, we included one case study in which the summary measure was the log odds ratio, modeled on the log scale using a normal distribution (belimumab, Section~\ref{par:binary_endpoint}), and one case study (aprepitant, Section~\ref{par:aprepitant}) in which, by contrast, the source data consist of $N^{(c)}_S$ (resp. $N^{(t)}_S$ ) Bernoulli trials with $y_S^{(c)}$ (resp. $y_S^{(t)}$) successes in the control arm (resp. the treatment arm). The target data follow:
\begin{equation}
\begin{aligned}
y_T^{(c)} \mid p_T^{(c)} & \sim \operatorname{Bin}\left(N_T^{(c)}, p_T^{(c)}\right)  \\
y_T^{(t)} \mid p_T^{(t)} & \sim \operatorname{Bin}\left(N_T^{(t)}, p_T^{(t)}\right)
\end{aligned}
\end{equation}

The corresponding model structure is described in Figure  \ref{fig:graphical_model_binomial_likelihood}.

Given $\theta_T$, the treatment-arm rate is determined by the control-arm rate, $p_T^{(t)} = p_T^{(c)} + \theta_T$. Both rates must lie in $[0,1]$, so the control-arm rate is restricted to the set of values compatible with $\theta_T$:
\begin{equation}
\mathcal{P}(\theta_T) = \left[\max(0, -\theta_T),\ \min(1, 1-\theta_T)\right],
\qquad \theta_T \in [-1, 1].
\end{equation}
The likelihood $\mathscr{L}(\theta_T \mid \mathbf{D}_T)$ is obtained by integrating out the nuisance parameter $p_T^{(c)}$ over this set:
\begin{equation}
\label{eq:binomial_integrated_likelihood}
\begin{aligned}
p(\mathbf{D}_T \mid \theta_T)
&= \int_{\mathcal{P}(\theta_T)} p\left(\mathbf{D}_T \mid p_T^{(t)} = p_T^{(c)} + \theta_T,\ p_T^{(c)}\right)
   p\left(p_T^{(c)} \mid \theta_T\right) dp_T^{(c)} \\
&= \int_{\max(0,-\theta_T)}^{\min(1,1-\theta_T)}
   \operatorname{Bin}\left(y_T^{(c)} \mid N_T^{(c)}, p_T^{(c)}\right)
   \operatorname{Bin}\left(y_T^{(t)} \mid N_T^{(t)}, p_T^{(c)} + \theta_T\right)
   p\left(p_T^{(c)} \mid \theta_T\right) dp_T^{(c)}.
\end{aligned}
\end{equation}
The conditional prior $p\left(p_T^{(c)} \mid \theta_T\right)$ depends on the joint prior of $(p_T^{(c)}, \theta_T)$, and therefore on the borrowing method. Every method was specified by a joint prior $\pi\left(p_T^{(c)}, \theta_T\right)$, and its posterior was computed as $\pi\left(p_T^{(c)}, \theta_T \mid \mathbf{D}_T\right) \propto \operatorname{Bin}\left(y_T^{(c)} \mid N_T^{(c)}, p_T^{(c)}\right) \operatorname{Bin}\left(y_T^{(t)} \mid N_T^{(t)}, p_T^{(c)} + \theta_T\right) \pi\left(p_T^{(c)}, \theta_T\right)$, then integrated over $p_T^{(c)}$ (Supplementary Section~\ref{supp:binomial_lattice}). For the separate and pooled analyses, both response rates have independent uniform priors: $p_T^{(c)} \sim \mathcal{U}(0,1)$ and $\theta_T \mid p_T^{(c)} \sim \mathcal{U}(-p_T^{(c)}, 1-p_T^{(c)})$. This joint prior is uniform on the set of admissible pairs. Its marginal for $\theta_T$ is triangular, with density $1-|\theta_T|$, and its conditional for the control rate is uniform on $\mathcal{P}(\theta_T)$, $p\left(p_T^{(c)} \mid \theta_T\right) = 1/(1-|\theta_T|)$ for $p_T^{(c)} \in \mathcal{P}(\theta_T)$, and $0$ otherwise. Equation~\eqref{eq:binomial_integrated_likelihood} with this conditional gives the integrated likelihood of the separate analysis. The methods that borrow from the source study have other joint priors, because the source control rate, the power parameter or the commensurability parameter are integrated out (Section~\ref{sec:binomial_adaptation} and Supplementary Sections~\ref{supp:normalized_power_prior_binomial}--\ref{supp:rmp_binomial}). Their conditional prior of $p_T^{(c)}$ given $\theta_T$ is then not uniform, and Equation~\eqref{eq:binomial_integrated_likelihood} holds with that conditional.

For all case studies, we considered, for simplicity and because this is the most standard setting,  that the source and target data likelihoods belong to the same family of distributions.

When assuming a Gaussian likelihood, we considered that the standard error of the summary measure in the target population is known, and we set the standard deviation to the sample standard deviation in the target study, as is often done in meta-analytic approaches and in Bayesian borrowing \parencite{weber_how_2018,best_assessing_2021}. However, in practice, the variance of the individual outcome may be substantially larger in the target study. For example, pediatric populations tend to be less homogeneous compared to adults because, for instance, of change in weight with age, organ maturation, and body composition differences \parencite{kern_challenges_2009}. Therefore, we included an additional simulation scenario for the case studies with continuous endpoints (botox and dapagliflozin, see section \ref{sec:case_studies}) where the simulated standard deviation in the pediatric data is twice as large as the standard deviation observed in adults. 

\subsubsection{Prior on the source study treatment effect}

When multiple source studies were available for a given target study, we followed the aggregation used in the corresponding published case study. For the belimumab example, \textcite{best_classical_2025} pooled the response data from the two adult trials and treated them as a single historical source.  For teriflunomide, \textcite{bovis_reinterpreting_2022} combined the study-specific log hazard ratios using inverse-variance weighting.
With two source studies, the between-study heterogeneity parameter would not be estimable from the data without a very strong prior \parencite{friede_metaanalysis_2017, seide_likelihoodbased_2019}.

Even if the source data are kept fixed, Bayesian borrowing methods need an initial prior $\pi_0$ on the treatment effect, i.e. the prior before any source information is used. With a normal summary likelihood, we used a flat initial prior for every method, so that, for example, the pooled analysis uses the source likelihood $\mathcal{N}(\hat{\theta}_S, \hat{\sigma}_{\theta_S}^2)$ as its prior and the conditional power prior uses $\mathcal{N}(\hat{\theta}_S, \hat{\sigma}_{\theta_S}^2/\gamma)$. The separate analysis, and the power priors with $\gamma = 0$, used a vague $\mathcal{N}(\hat{\theta}_S, 1000)$ prior, whose influence on the posterior is negligible. With the binomial likelihood of the aprepitant case study, the response rates in each arm have uniform priors. For the separate and pooled analyses, this means $p_T^{(c)} \sim \mathcal{U}(0,1)$ and $\theta_T \mid p_T^{(c)} \sim \mathcal{U}(-p_T^{(c)}, 1-p_T^{(c)})$. For the power priors, which share the treatment effect between the source and target studies, the source and target control rates $p_S^{(c)}$ and $p_T^{(c)}$ have independent uniform priors and the treatment effect is uniform on $(-\min(p_S^{(c)}, p_T^{(c)}), 1 - \max(p_S^{(c)}, p_T^{(c)}))$, the values that keep all four response rates in $[0, 1]$. For the commensurate priors, each study has its own treatment effect, with a uniform initial prior given its control rate (Supplementary Section~\ref{supp:commensurate_binomial}).

\subsection{Statistical methods for information borrowing}
\label{sec:selected_methods}
 The choice of statistical methods considered for the simulation study is based on an extensive literature review. 
However, our evaluation did not include covariates or borrowing methods requiring IPD, such as the latent exchangeability prior (LEAP) \parencite{alt_leap_2024}, which requires IPD to identify latent subgroups of exchangeable and non-exchangeable source observations. Our selection was intended to represent a range of approaches used in clinical applications, rather than to provide an exhaustive comparison of all available methods.
For each method, we varied the parameters that affect the amount of borrowing. These parameters are summarized in Table \ref{tab:study_protocol_methods_range}.  The configurations used are summarized in Table \ref{tab:simulation_configuration}. 
Below we introduce each method in the case of a normal likelihood, and later detail how each was adapted to the aprepitant case study which relied on a binomial likelihood.

Table~\ref{tab:abbreviations} lists the abbreviations of the methods and their parameters, as used in the figures.

\begin{table}[ht]
\centering
\small
\renewcommand{\arraystretch}{1.1}
\begin{tabularx}{\textwidth}{@{}>{\raggedright\arraybackslash}p{2.3cm}>{\raggedright\arraybackslash}X>{\raggedright\arraybackslash}X@{}}
\hline
Abbreviation & Method & Parameters \\
\hline
Separate & Separate analysis of the target data alone & -- \\
Pooling & Pooled analysis of the source and target data & -- \\
CPP & Conditional power prior & $\gamma$: power parameter \\
TtP (diff) & Test-then-pool, test for difference & $\eta$: significance level of the test \\
TtP (eq) & Test-then-pool, test for equivalence & $\eta$: significance level; $\lambda$: equivalence margin \\
NPP & Normalized power prior & $\xi_\gamma$, $\sigma_\gamma$: mean and SD of the beta prior on $\gamma$ \\
NPP (KL) & KL-calibrated normalized power prior & $\lambda_{\mathrm{KL}}$: weight of the no-drift scenario; $c$: shape of the reference beta distributions; $k_\text{MTD}$: multiplier of the maximum tolerable discrepancy \\
EBPP & Empirical Bayes power prior & -- \\
PDCCPP & Prior-data conflict calibrated power prior & Target TIE (0.065) \\
p-PP & $p$-value-based power prior & $k$: shape parameter; $\lambda$: equivalence margin \\
Com.\ PP & Commensurate power prior & Prior on the commensurability parameter $\tau$ \\
Com.\ prior & Commensurate prior & Prior on the commensurability parameter $\tau$ \\
RMP & Robust mixture prior & $w$: weight of the informative component \\
Empirical RMP & Robust mixture prior with an empirical weight & $\alpha_{\mathrm{PC}}$: conflict threshold (0.05) \\
\hline
\end{tabularx}
\caption{Abbreviations of the borrowing methods and of their parameters. In the figures, the CPP is labeled ``Conditional PP''. The values of the parameters are given in Supplementary Table~\ref{tab:study_protocol_methods_range}.}
\label{tab:abbreviations}
\end{table}

\subsubsection{Frequentist reference}

For each borrowing method, power was compared with that of a frequentist test of the target data alone, both at the nominal one-sided level of 2.5\% and at the type I error rate that the method actually attains. The reference test was calibrated on its actual, rather than nominal, type I error rate, so that a borrowing method can show a power gain only by being more powerful than this test at the same type I error rate (Section~\ref{sec:methods_reference}).

\subsubsection{Conditional power prior (CPP)} 

As a Bayesian baseline, and to study borrowing that does not adapt to prior-data conflict, we first examined fixed borrowing, in which the source posterior is discounted by a fixed amount before being used as the prior.
In order to incorporate a fixed amount of information from source studies into the prior for $\theta_T$, \textcite{ibrahim_power_2000} introduced the power prior (also referred to as the conditional power prior, CPP \parencite{neelon_bayesian_2010}): 
\begin{equation}
\label{eq:conditional_power_prior}
\pi(\theta_T \mid \mathbf{D}_S, \gamma) \propto
\mathscr{L}(\theta_T \mid \mathbf{D}_S)^{\gamma}\,\pi_0(\theta_T),
\end{equation}
where $\gamma \in [0,1]$ is the power parameter and $\pi_0(\theta_T)$ is the initial prior for $\theta_T$. The prior is called conditional because it conditions on both the source data and a fixed value of the power parameter.
The power parameter controls how much the source data influence the posterior, which makes it a simple way of discounting prior information. In all case studies, the initial prior was noninformative. With a normal likelihood and a flat initial prior, $\gamma = 1$ gives the pooled analysis, $\gamma = 0$ gives the separate analysis, and intermediate values move smoothly from no borrowing to pooling. With the binomial likelihood of the aprepitant case study, these two identities hold only approximately (Section~\ref{sec:binomial_adaptation}). Because the source likelihood is evaluated at $\theta_T$, the method assumes that the parameter of interest is the same in the source and target studies. 

For normal likelihoods, raising the source likelihood to the power $\gamma$ inflates its variance by a factor $1/\gamma$. The prior is then $\mathcal{N}(\hat{\theta}_S, \hat{\sigma}_{\theta_S}^2/\gamma)$, and the posterior is available in closed form. The binomial version used in the aprepitant case study is described in Section~\ref{sec:binomial_adaptation}.

\subsubsection{Frequentist test-then-pool (TtP)}

With the frequentist test-then-pool method \parencite{viele_use_2014}, the difference between the source and target data is assessed before deciding whether to pool them. The hypothesis $H_0: \theta_T = \theta_S$ is tested at significance level $\eta$. If $H_0$ is rejected, the data are not pooled and the target data are analyzed on their own. \textcite{liu_dynamic_2018} argued that testing the difference between $\theta_S$ and $\theta_T$ may not be the best approach, and proposed testing an equivalence hypothesis instead: $H_0: |\theta_S - \theta_T| \geq \lambda$ versus $H_1: |\theta_S - \theta_T| < \lambda$, where $\lambda > 0$ is a predetermined equivalence margin. The \textit{p}-value is the larger of the \textit{p}-values of the two one-sided tests of $H_{0a} : \theta_S - \theta_T \geq \lambda$ and $H_{0b} : \theta_S - \theta_T \leq -\lambda$ \parencite{schuirmann_comparison_1987}. A \textit{p}-value below $\eta$ rejects non-equivalence, and the data are then pooled.

We implemented both approaches with Welch two-sample $t$-tests on summary statistics. 
The source study enters through its estimate $\hat{\theta}_S$ and standard error, as a sample of its equivalent sample size per arm $n_S$ with standard deviation $\hat{\sigma}_{\theta_S} \sqrt{n_S}$. The target study enters through its estimate, standard deviation and sample size per arm. Once the test has decided whether to pool, the target data were analyzed with the Bayesian pooled analysis or the Bayesian separate analysis, both with a noninformative initial prior. The amount of borrowing is set by the significance level $\eta$ and, for the equivalence test, by the margin $\lambda$. For the difference test we considered $\eta \in \{0.01, 0.1, 0.4, 0.8\}$. For the equivalence test, $\eta \in \{0.1, 0.5\}$ and $\lambda \in \{0.1, 0.5, 0.8\}$. The margin is expressed on the scale of each case study's treatment effect (risk difference, log rate ratio, log hazard ratio or mean difference).

\subsubsection{Normalized power prior (NPP)}
\label{subsec:NPP}

In the power prior approach, the power parameter $\gamma$ can be treated as a random variable and inferred, by giving it a prior $\pi(\gamma)$ in a hierarchical model. This gives the normalized power prior (NPP, \textcite{duan_evaluating_2006}, \textcite{neuenschwander_note_2009}):
\begin{equation}
\label{eq:normalized_power_prior}
\pi(\theta_T, \gamma \mid \mathbf{D}_S) = Z(\gamma)\,\mathscr{L}(\theta_T \mid \mathbf{D}_S)^{\gamma}\,\pi_0(\theta_T)\,\pi(\gamma),
\end{equation}
where $Z(\gamma)$ is the normalizing constant
\begin{equation}
\label{eq:power_prior_normalization}
Z(\gamma) = 1\bigg/\int \mathscr{L}(\theta_T \mid \mathbf{D}_S)^{\gamma}\,\pi_0(\theta_T)\, d\theta_T, \qquad \gamma \in (0, 1].
\end{equation}
We used a beta prior on the power parameter, $\gamma \sim \operatorname{Beta}(a_\gamma, b_\gamma)$, a common choice \parencite{gravestock_adaptive_2017, shi_novel_2023}, with a noninformative initial prior $\pi_0$. The beta prior was specified by its mean $\xi_\gamma = 0.5$ and its standard deviation $\sigma_\gamma \in \{0.1, 0.2, 0.4\}$, from which $a_\gamma$ and $b_\gamma$ follow. For a normal likelihood with known standard deviation, where the standard errors of the source and target estimates are plugged in, the prior and posterior distributions are derived analytically in the supplementary material (Section~\ref{supp:normalized_power_prior}). Conditional on $\gamma$, the posterior of $\theta_T$ is normal, and we integrated over $\gamma$ numerically.

We also included the KL-calibrated NPP variant proposed by \textcite{shen_optimal_2024}. 
In this variant, the shape parameters $a_\gamma$ and $b_\gamma$ of the beta prior on $\gamma$ are chosen at the design stage, so that the marginal posterior of $\gamma$ concentrates near 1 when the target estimate equals the source estimate, and near 0 at a maximum tolerable discrepancy $d_{\mathrm{MTD}} = k_{\mathrm{MTD}}\,|\hat{\theta}_S - \theta_0|$ between them. We used the default values of \textcite{shen_optimal_2024} for the weight of the two criteria ($\lambda_{\mathrm{KL}} = 0.5$) and for the shape of the reference beta distributions ($c = 10$), and $k_{\mathrm{MTD}} \in \{0.5, 1\}$. With $k_{\mathrm{MTD}} = 1$, there is almost no borrowing when there is no treatment effect in the target population, and $k_{\mathrm{MTD}} = 0.5$ borrows more conservatively. The beta parameters were calibrated once per design, with the design-expected target standard error, and then held fixed across simulated trials. The criterion and its computation are given in Supplementary Section~\ref{supp:npp_kl}. 

\subsubsection{Empirical Bayes power prior (EBPP) and prior-data conflict calibrated power prior (PDCCPP)}
\textcite{gravestock_adaptive_2017} proposed an empirical Bayes adaptation of the normalized power prior. 
When using a power prior defined as:
\begin{equation}
 \pi(\theta_T \mid \gamma, \mathbf{D}_S) = Z(\gamma)\,\pi_0(\theta_T)\,\mathscr{L}(\theta_T \mid \mathbf{D}_S)^\gamma,
\end{equation}
the empirical Bayes estimate of $\gamma$ is: 
\begin{equation}
\label{eq:marginal_likelihood PP}
\widehat{\gamma} = \underset{\gamma \in [0,1]}{\operatorname{argmax}} p(\mathbf{D}_T| \gamma, \mathbf{D}_S).
\end{equation}

 For a normal likelihood with a flat initial prior, $\widehat{\gamma}$ and the resulting posterior are available in closed form (Supplementary Section~\ref{supp:ebpp_closed_form}).

However, the estimate $\widehat{\gamma}$ suggested by \textcite{gravestock_adaptive_2017} is not calibrated to control the type I error rate, which would be appealing from a regulatory standpoint. For this reason, \textcite{nikolakopoulos_dynamic_2018} suggested using a point estimate of $\gamma$ that controls type I error. Their approach, developed for normal likelihoods, is based on the Box p-value (also called the prior-predictive p-value, PPP). The two-sided PPP for a test statistic $T(\mathbf{D}_T)$ is defined as: 
\begin{equation}
ppp(\mathbf{D}_S) = 2 \min \left[\operatorname{Pr}\{T(\mathbf{D}_T) \geq T(d_T) \mid \mathbf{D}_S\}, \operatorname{Pr}\{T(\mathbf{D}_T) \leq T(d_T) \mid \mathbf{D}_S\}\right],
\end{equation}
where $d_T$ denotes the observed target trial data.
They suggest choosing a fixed value for $\gamma$ using: 
\begin{equation}
\widehat{\gamma} = \min \left[ \underset{\gamma \in [0,1]}{\max}\{ \gamma: ppp(\mathbf{D}_S, \gamma) \geq c_{\mathrm{PPP}} \}, 1\right],
\label{eq:PDCCPP}
\end{equation}
where $ppp(\mathbf{D}_S, \gamma)$ indicates that the PPP is evaluated conditionally on $\mathbf{D}_S$ and $\gamma$, and the threshold $c_{\mathrm{PPP}}$ is chosen so that the TIE equals a prespecified value, here 0.065.
\textcite{nikolakopoulos_dynamic_2018} call this variant the prior-data conflict calibrated power prior (PDCCPP). 

\subsubsection{\texorpdfstring{$p$}{p}-value-based power prior (p-PP)}

In a generalization of the test-then-pool approach, \textcite{liu_dynamic_2018} proposed selecting the power parameter $\gamma$ of the conditional power prior from the \textit{p}-value of an equivalence test between the source and target data, $H_0: |\theta_S - \theta_T| \geq \lambda$ versus $H_1: |\theta_S - \theta_T| < \lambda$. $p$ is the larger of the \textit{p}-values of the two one-sided tests. The power parameter is:
\begin{equation}
\label{eq:pvalue_based_power_prior}
\gamma = \exp\left[\frac{k}{1-p}\ln(1-p) \right] = (1-p)^{k/(1-p)},
\end{equation}
where $k > 0$ is a shape parameter. More source data are borrowed when the \textit{p}-value is close to 0, i.e., when non-equivalence is strongly rejected ($\gamma = 1$ at $p = 0$), and $\gamma$ decreases to 0 as $p$ approaches 1. Larger values of $k$ discount the source data more for a given \textit{p}-value. The method can thus be viewed as an extension of test-then-pool, in which the power parameter moves smoothly from no borrowing to pooling. We considered $k \in \{0.01, 0.1, 1, 10, 20\}$ and $\lambda \in \{0.1, 0.5\}$. As for test-then-pool, in the normal-likelihood case studies the two one-sided tests were Welch $t$-tests on summary statistics.  

\subsubsection{Commensurate prior (Com.\ prior) and commensurate power prior (Com.\ PP)}
 
The commensurate power prior of \textcite{hobbs_hierarchical_2011} is
  \begin{equation}
   \pi(\theta_T, \gamma, \tau \mid \mathbf{D}_S)
   = \int \pi(\theta_T \mid \theta_S, \tau)\,
     \frac{\mathscr{L}(\theta_S \mid \mathbf{D}_S)^\gamma \pi_0(\theta_S)}
          {\int \mathscr{L}(\theta_S \mid \mathbf{D}_S)^\gamma \pi_0(\theta_S)\, d\theta_S}\, d\theta_S
     \times p(\gamma \mid \tau)\, p(\tau),
  \end{equation}
  where $\pi_0(\theta_S)$ is an initial prior for $\theta_S$. \textcite{hobbs_hierarchical_2011} considered Gaussian likelihoods and chose:
  \begin{equation*}
  \theta_T \mid \theta_S, \tau \sim \mathcal{N}\left(\theta_S, \frac{1}{\tau}\right)
  \quad \text{and} \quad
  \gamma \mid \tau \sim \operatorname{Beta}(g(\tau), 1), \quad g(\tau) = \max(\log\tau, 1),
  \end{equation*}
  with a heavy-tailed $\operatorname{Cauchy}(0, 30)$ prior on $\log\tau$. The commensurability parameter $\tau$ is the precision of the target effect around the source effect. When the target data provide little evidence of commensurability, the posterior of $\tau$ concentrates near zero, which inflates the variance of the prior on $\theta_T$. For $\tau \leq e$, $g(\tau) = 1$ and $\gamma$ is uniform on $[0, 1]$, whereas for larger $\tau$ the prior on $\gamma$ concentrates near 1. The amount of borrowing therefore adapts in two ways: through the power parameter and through the commensurability parameter.

  We used a flat initial prior $\pi_0(\theta_S)$, so that the source posterior is $\mathcal{N}(\hat{\theta}_S, \hat{\sigma}^2_{\theta_S})$, and considered five priors on $\tau$: the $\operatorname{Cauchy}(0, 30)$ prior on $\log\tau$ of \textcite{hobbs_hierarchical_2011}, two inverse-gamma priors on $\tau^2$, $\tau^2 \sim \operatorname{IG}(\alpha, 1)$ with $\alpha \in \{1/3, 1/7\}$ (written IG(0.33, 1) and IG(0.14, 1) in the results), and two half-normal priors on $\tau$, $\tau \sim \operatorname{HN}(\sigma)$ with $\sigma \in \{1, 5\}$. The Cauchy prior is centered on $\log\tau = 0$ and places half its mass on $\tau < 1$. On the $\log\tau$ scale, $\operatorname{IG}(\alpha, 1)$ on $\tau^2$ vanishes for small $\tau$ and decays at rate $2\alpha$ for large $\tau$, so that it is nearly flat above $\tau \approx 1$. Smaller values of $\alpha$ therefore put more prior mass on large precisions, and hence on stronger borrowing. Conversely, the $\operatorname{HN}(1)$ prior concentrates $\tau$ near zero and favors weak borrowing. The posterior was computed by numerical integration over $\tau$ and, for the commensurate power prior, over $\gamma$ given $\tau$.

As a frequently used variant, we also evaluated the Gaussian commensurate prior \parencite{hobbs_hierarchical_2011}, which keeps the relationship $\theta_T \mid \theta_S, \tau \sim \mathcal{N}(\theta_S, 1/\tau)$ but omits the power parameter. Integrating out $\theta_S$ gives
\begin{equation}
\pi(\theta_T \mid \mathbf{D}_S, \tau) = \int \mathcal{N}\left(\theta_T \mid \theta_S, \frac{1}{\tau}\right) p(\theta_S \mid \mathbf{D}_S)\, d\theta_S
= \mathcal{N}\left(\theta_T \mid \hat{\theta}_S, \hat{\sigma}^2_{\theta_S} + \frac{1}{\tau}\right).
\end{equation}
We used the same five priors on $\tau$ as for the commensurate power prior.

\subsubsection{Robust mixture prior (RMP)}

\textcite{schmidli_robust_2014}, building on earlier work by \textcite{greenhouse_robust_1995}, introduced the concept of robust mixture prior for combining an informative component with a weak or vague component to guard against potential conflict between prior information and current data. The informative component may come from any prior source, whether based on historical data or other external information. In the context of treatment effect borrowing, the general formulation is:

\begin{equation}
\begin{aligned}
\pi(\theta_T|\mathbf{D}_S= d_S) &= w\pi(\theta_T | M_{\text{source}},\mathbf{D}_S= d_S) + (1-w)\pi(\theta_T | M_{\text{weak}},\mathbf{D}_S= d_S),
\end{aligned}
\end{equation}

where $M_{\text{source}}$ represents a model reflecting some form of consistency or exchangeability across studies, and $M_{\text{weak}}$ represents a model corresponding to unrelated treatment effects. The weight $w=\operatorname{Pr}(M_{\text{source}}|\mathbf{D}_S)$ expresses the prior belief in the source model. $\pi(\theta_T | M_{\text{source}},\mathbf{D}_S)$ corresponds to an informative component, which in our case is based on the assumption that studies are related, whereas $\pi(\theta_T | M_{\text{weak}},\mathbf{D}_S)$ is typically a vague component. In our simulation study, the posterior distribution from the source study was used as the informative component $\pi(\theta_T | M_{\text{source}},\mathbf{D}_S)$.

The posterior distribution of the target study treatment effect $\theta_T$ is a weighted average of the posterior distributions under each model, weighted by their respective posterior model probabilities:
\begin{equation}
\begin{aligned}
\pi(\theta_T \mid \mathbf{D}_T = d_T, \mathbf{D}_S= d_S)&= \tilde{w} \pi\left(\theta_T \mid M_{\text {source }}, \mathbf{D}_T= d_T, \mathbf{D}_S = d_S\right) \\ &+(1-\tilde{w}) \pi\left(\theta_T \mid M_{\text {weak }}, \mathbf{D}_T = d_T, \mathbf{D}_S = d_S\right),
\end{aligned}
\end{equation}
where the updated weight $\tilde{w}$ corresponds to the posterior $\operatorname{Pr}\left(M_{\text {source }} \mid \mathbf{D}_T = d_T, \mathbf{D}_S = d_S\right)$, that is:
\begin{equation}
\tilde{w}  
= \frac{w \, p(\mathbf{D}_T \mid M_{\text{source}}, \mathbf{D}_S)}
{w \, p(\mathbf{D}_T \mid M_{\text{source}}, \mathbf{D}_S) + (1 - w) \, p(\mathbf{D}_T \mid M_{\text{weak}}, \mathbf{D}_S)}.
\label{eq:posterior_weight}
\end{equation}

So the mixture introduces robustness by allowing the vague prior to dominate if the heterogeneity between source and target trials is large compared to within-trial variance.

As recommended by \textcite{schmidli_robust_2014}, the vague component was a unit-information prior: a normal distribution centered on the null value $\theta_0$, whose variance is the variance of a treatment effect estimate based on one subject per arm in the target study, $n_T\,\hat{\sigma}^2_{\theta_T}$, where $\hat{\sigma}_{\theta_T}$ is the estimated standard error of the target treatment effect and $n_T$ the target sample size per arm. Because $\hat{\sigma}_{\theta_T}$ is estimated from the target data, the vague component was recomputed for each simulated trial. The informative component was the source posterior, $\mathcal{N}(\hat{\theta}_S, \hat{\sigma}^2_{\theta_S})$. Under a normal likelihood, both components are conjugate, so the posterior is a two-component normal mixture whose component parameters and posterior weight $\tilde{w}$ (Equation~\ref{eq:posterior_weight}) are available in closed form.

Following the approach proposed by \textcite{egidi_avoiding_2022}, we also included a variant of the RMP where the weights are empirically determined. The prior is written as:

$$ \pi_{\psi}(\theta_T)
= (1-\psi)\pi_{\mathrm{S}}(\theta_T)
+ \psi\pi_{\mathrm{W}}(\theta_T),
\qquad 0\leq\psi\leq1,
$$

where $\psi$ is the weight assigned to the weakly informative component. 
The weight $\widehat{\psi}$ is the smallest weight for which the observed target data no longer conflict with the mixture prior, according to a prior-predictive $p$-value at the level $\alpha_{\mathrm{PC}} = 0.05$, as in \textcite{egidi_avoiding_2022}. When no $\psi < 1$ satisfies this criterion, $\widehat{\psi} = 1$. The method therefore keeps the source-based prior in the absence of prior-data conflict, and adds only the smallest amount of the weakly informative component needed to remove the conflict. Because $\widehat{\psi}$ depends on the target data, it was computed in every simulated trial (Supplementary Section~\ref{supp:egidi}). We later refer to this empirical variant of the RMP as the empirical RMP.

\subsubsection{Adaptation of existing methods to the settings of interest}
\label{sec:binomial_adaptation}

In the aprepitant case study, the analyses used the arm-level responder counts, with the binomial model shown in Figure~\ref{fig:graphical_model_binomial_likelihood}, following \textcite{jin_bayesian_2021}. The treatment effect is the difference in response probabilities, $\theta_T = p_T^{(t)} - p_T^{(c)}$, and inference accounts jointly for $\theta_T$ and the control probability $p_T^{(c)}$, subject to $0 \leq p_T^{(c)} + \theta_T \leq 1$. Every method was used with this binomial likelihood, and every prior that borrows from the source study was derived from the binomial likelihoods of its two arms (rather than from a normal approximation of its risk difference). The derivations are given in Supplementary Sections~\ref{supp:normalized_power_prior_binomial}--\ref{supp:rmp_binomial}.

The separate and pooled analyses use independent uniform priors on the response probabilities of each arm, which are conjugate. In the pooled analysis, the source and target data of each arm are combined, so the two studies share their control rate as well as their treatment effect. 

For the conditional power prior, the source study has its own control probability $p_S^{(c)}$, the two control probabilities have independent uniform priors, and $\theta_T$, shared by the two studies, has a uniform initial prior on $\left(-\min(p_T^{(c)}, p_S^{(c)}),\, 1 - \max(p_T^{(c)}, p_S^{(c)})\right)$, so that the treatment-arm probabilities of both studies remain in $[0, 1]$. The power parameter discounts the source likelihood of both arms. With this model, the limits of the conditional power prior differ from the separate and pooled analyses. With $\gamma = 0$, the initial prior of $\theta_T$ is restricted to the values admissible for both studies, which, after integration over $p_S^{(c)}$, puts more prior mass near $\theta_T = 0$ than the separate analysis does. The type I error rate was then 0.018 to 0.024, against 0.024 to 0.026 for the separate analysis, and the probability of success differed by at most 0.026. With $\gamma = 1$, the source and target studies share the treatment effect but not the control rate, whereas pooling shares both: the probability of success differed from that of pooling by at most 0.021. In the aprepitant case study, the conditional power prior with $\gamma = 0$ and $\gamma = 1$ should therefore be read as close to, but not identical with, the separate and pooled analyses (see also Supplementary Section~\ref{supp:binomial_lattice}).

The normalized power prior, its KL-calibrated variant and the empirical Bayes power prior are built on the same model: for the normalized power prior, the power parameter is integrated out against its beta prior with the normalizing constant computed for the binomial model. The KL calibration evaluates its criterion with the binomial marginal likelihood of the design's expected responder counts, and the empirical Bayes estimate of the power parameter maximizes the binomial marginal likelihood of the target data. 

For the commensurate priors, the source and target studies have their own risk differences, linked by $\theta_T \mid \theta_S, \tau \sim \mathcal{N}(\theta_S, 1/\tau)$ truncated to the admissible values given $p_T^{(c)}$, and the source information enters through the binomial likelihoods of the source arms. 

For the robust mixture prior and its empirical variant, the informative component is the source posterior of the risk difference under the binomial model, and the weak component gives both target response probabilities independent uniform priors. For the empirical variant, the prior-predictive conflict $p$-value was computed exactly, by summing the prior-predictive probabilities of all pairs of responder counts at most as probable as the observed pair.

Three methods set the amount of borrowing with a frequentist rule applied to the estimated risk differences and their standard errors, as in their original formulations, and then analyze the data with the binomial likelihood: test-then-pool (Welch two-sample $t$-tests, followed by the binomial separate or pooled analysis), the $p$-value-based power prior ($z$-tests on the estimated risk differences, followed by the binomial conditional power prior with the resulting power parameter, and the PDCCPP (the rule of \textcite{nikolakopoulos_dynamic_2018}, followed by the binomial conditional power prior).  
For the PDCCPP, whose calibration relies on closed-form expressions under a normal likelihood, the calibration parameter was instead chosen on the exact type I error rate of the binomial analysis, computed by enumerating all pairs of responder counts under the null hypothesis. Because this type I error rate is a step function of the calibration parameter, we took the largest step not exceeding 0.065.

\subsection{Estimation of posterior distributions}  
\label{sec:posterior_computation}

In the Bayesian framework, all information about the target treatment effect is summarized in the posterior distribution $p(\theta_T\mid\mathbf{D}_S = d_S,\mathbf{D}_T= d_T)$. In our study, every posterior was computed either in closed form or by deterministic numerical integration, which is exact up to a quadrature error, and much faster than Markov chain Monte Carlo (MCMC). Because the computation is deterministic, simulated trials that yield identical data also yield identical analyses, which were therefore computed once and reused.

 With a normal prior or a mixture of normal priors, the posterior was computed in closed form. The normalized power priors and the commensurate priors were reduced to finite normal mixtures by quadrature over their hyperparameters. The quadrature rules and the computation of credible intervals are given in Supplementary Section~\ref{supp:posterior_computation}.

 In the aprepitant case study, all these methods were instead analyzed with binomial likelihoods, as described in Supplementary Section~\ref{supp:binomial_lattice}.

\section{Operating characteristics}

\subsection{Frequentist operating characteristics}
\label{sec:frequentist_ocs}

The steps of the simulation study are summarized in Supplementary Figure~\ref{fig:simulation_pipeline}.

For each method and each scenario, we evaluated the probability of study success, the MSE and bias of the posterior mean, the precision (the half-width of the 95\% credible interval), the coverage probability of the 95\% credible interval, and the interval score \parencite{gneiting_strictly_2007}.

The type I error rate of the test with borrowing, $\alpha_B$, and its power for $\theta_T \notin \Theta_0$, $1-\beta_B(\theta_T)$, were estimated by Monte Carlo as
\begin{equation}
\begin{aligned}
\hat{\alpha}_B &= \frac{1}{N_{\text{sims}}}\sum_{i=1}^{N_{\text{sims}}}\varphi_B(d^{(i)}_T \mid d_S), \quad d_T^{(i)} \sim p(\mathbf{D}_T \mid \theta_T = \theta_0), \\
1-\hat{\beta}_B(\theta_T) &= \frac{1}{N_{\text{sims}}}\sum_{i=1}^{N_{\text{sims}}} \varphi_B(d^{(i)}_T \mid d_S), \quad d_T^{(i)} \sim p(\mathbf{D}_T \mid \theta_T),
\end{aligned}
\end{equation}
where $N_{\text{sims}}$ is the number of datasets drawn from $p(\mathbf{D}_T \mid \theta_T)$, and $\varphi_B(d^{(i)}_T \mid d_S)$ indicates whether the success criterion with borrowing is met for dataset $d^{(i)}_T$. In the aprepitant and belimumab case studies, whose binary outcomes take finitely many values, we replaced these Monte Carlo averages by exact sums over all possible pairs of responder counts, each weighted by its probability under $p(\mathbf{D}_T \mid \theta_T)$. The same exact computation was used for all other operating characteristics in these two case studies, so they carry no Monte Carlo error. These rates are conditional: the source data $d_S$ are fixed at their published values, and the nuisance parameters (control rate, standard deviation, dispersion or control hazard) at the values of the design. They are therefore not unconditional error rates over repeated source and target studies. We evaluated the type I error rate at the null boundary, $\theta_T = \theta_0$. This is the largest rejection rate over the composite null only if the rejection rate increases with $\theta_T$ (decreases, for the case studies whose null is $\theta_T \geq \theta_0$), which we did not prove for the adaptive methods. We checked it on the drift grid instead: over the 3,528 method-setting-design combinations and their null-interior treatment effects (27,440 pairs in total), the rejection rate inside the null was never larger than at the boundary. The type I error rate reported is thus the rejection rate at the null boundary, which was the largest rejection rate over the null treatment effects evaluated. Its dependence on the nuisance parameters is examined in Section~\ref{sec:methods_reference}.

To compare the power of the tests with and without borrowing fairly, we followed \textcite{kopp-schneider_simulating_2024}. We evaluated the type I error rate $\alpha_B$ of the test with borrowing, and compared its power $1-\beta_B(\theta_T)$ with the power $1-\beta(\theta_T)$ of a test without borrowing whose type I error rate is also $\alpha_B$. How this reference test was calibrated to $\alpha_B$ is described in Section~\ref{sec:methods_reference}.

Because the strength of borrowing, and consequently the bias-variance trade-off, depends on method-specific tuning parameters, comparisons of estimation performance at arbitrarily selected parameter values may confound differences between borrowing methods with differences in borrowing strength. We therefore used the frequentist type I error rate as a common reference scale for comparing estimation performance across methods. Specifically, each method-parameter combination was characterized by its type I error rate, and the operating characteristic of interest was compared with that of other method-parameter combinations at similar type I error rates. This follows the same general principle as comparing power at a common type I error rate \parencite{kopp-schneider_power_2020}.

For each operating characteristic estimated by simulation, we reported the uncertainty due to the finite number of simulated trials through 95\% Monte Carlo confidence intervals. For the probability of success and the coverage, which are binomial proportions, we used exact (Clopper--Pearson) intervals. For the other metrics, we used normal-approximation intervals for the mean, since every scenario was simulated with $N_{\text{sims}} = 10\,000$ trials. Operating characteristics computed exactly by enumeration have no Monte Carlo error. For the difference between the power of a borrowing method and that of the reference test at the same type I error rate (Section~\ref{sec:methods_reference}), the interval combines two sources of Monte Carlo error. First, the reference power depends on the method's estimated type I error rate $\hat{\alpha}_B$, which sets the reference's rejection threshold. We drew $\alpha_B$ from the Jeffreys posterior of a binomial proportion, $\operatorname{Beta}(x + 0.5, N_{\text{sims}} - x + 0.5)$, where $x$ is the number of null trials the method rejected, recomputed the reference threshold and power for each draw (adding the Monte Carlo error of the simulated power, where the power was simulated), and took the 2.5\% and 97.5\% quantiles. The interval of the reference power therefore does not condition on the estimated threshold. Second, the interval of the difference combines this interval with the Clopper--Pearson interval of the method's power by the method of variance estimates recovery \parencite{newcombe_interval_1998}, assuming the two estimates independent. Both are computed on the same simulated trials and are positively correlated, so this choice gives wider intervals than a paired analysis would. In the recurrent-event and time-to-event case studies, the reference threshold is the empirical quantile of the null trials on which $\hat{\alpha}_B$ was estimated: the reference then rejects the same proportion of these trials as the method, but its population type I error rate equals that of the method only up to Monte Carlo error. In the binary case studies, both type I error rates are exact and the power difference has no Monte Carlo error.


\subsection{Prior effective sample size}

The amount of borrowing is most easily measured using the concept of prior effective sample size (ESS). Prior ESS corresponds to the number of pseudo-observations required to update a vague conjugate prior to the prior of interest (viewed as the posterior from previous analysis). It is a measure of the informativeness of the prior distribution in terms of number of samples.
For instance, in a beta-binomial model, the parameters of the $\operatorname{Beta}(a, b)$ prior can be interpreted as the posterior obtained after observing $a$ successes and $b$ failures, starting from a vague Beta prior (with $a$ and $b$ arbitrarily small). Similarly, if the data of one subject per arm give a treatment effect estimate with variance $\sigma^2$, a normal prior with variance $\sigma^2/n$ carries the information of $n$ subjects per arm: starting from a flat prior, it is the posterior after $n$ such subjects, and its prior ESS is $n$.
However, the prior ESS is not clearly defined for non-conjugate priors. We used several prior ESS measures: the moment-based prior ESS, the precision-based prior ESS, and the ELIR prior ESS.

 \paragraph{Reference scale}
All three effective sample sizes compare a distribution of the treatment effect with the information in the target data, measured against a normal reference: a treatment effect estimate based on one subject per arm, with standard deviation $\sigma$. We set $\sigma$ to the sampling standard deviation of one subject per arm estimated in each simulated target trial, $\sigma = \sqrt{n_T}\,\hat{\sigma}_{\theta_T}$ (in the continuous case studies, $\sigma = \hat{\sigma}_T$), so that effective sample sizes are expressed in subjects per arm. The vague component of the RMP has variance $\sigma^2$.  

\paragraph{Moment-based prior ESS}
Following the moment-matching approach of RBesT \parencite{ag_rbest_2023}, a normal distribution with variance $v$ corresponds to $\sigma^2 / v$ subjects. We matched the posterior of the treatment effect by its variance, which is available exactly: the posterior ESS is $\sigma^2 / \operatorname{Var}(\theta_T \mid \mathbf{D}_T, \mathbf{D}_S)$. The prior ESS is the posterior ESS minus the target sample size per arm, $n_T$. Because it is derived from the posterior, this quantity measures the information that the method borrowed for the observed target data. For a adaptive-borrowing methods,  method, it changes from one simulated trial to the next. We report its mean over the simulated trials. The same applies to the precision-based ESS. Only the ELIR is computed from the prior alone.

\paragraph{Precision-based prior ESS}
The precision-based ESS matches the width of the 95\% credible interval rather than the variance. The normal distribution with the same 95\% interval width has standard deviation $\sigma_{\mathrm{CI}} = (u - l) / (2 z_{0.975})$, where $l$ and $u$ are the interval bounds and $z_{0.975}$ is the standard normal quantile, which gives a posterior ESS of $\sigma^2 / \sigma^2_{\mathrm{CI}}$. The prior ESS again subtracts $n_T$. The two ESS agree for a normal posterior, and differ when the posterior has heavier tails than a normal of the same variance, as a mixture posterior may.

\paragraph{ELIR method for prior ESS}
\textcite{neuenschwander_predictively_2020} introduced an information-based ESS, the expected local-information-ratio (ELIR), which has the property of being ``predictively consistent'', meaning that the expected posterior ESS for a sample of $n_T$ subjects per arm is equal to the sum of the prior ESS and $n_T$.
The ELIR is defined as follows:
\begin{equation}
ELIR = \mathbb{E}_\pi \left[\frac{\mathcal{I}_\pi(\theta)}{\mathcal{I}_1(\theta)}\right],
\end{equation}
where $\mathcal{I}_\pi(\theta) = -\partial^2_\theta \log \pi(\theta)$ is the information of the prior $\pi$, and $\mathcal{I}_1(\theta)$ is the expected Fisher information for one information unit,
\begin{equation}
\mathcal{I}_1(\theta) = - \mathbb{E}\left[\frac{\partial^2\log \mathscr{L}(\theta \mid \mathbf{D}_1)}{\partial \theta^2} \bigg|\theta \right],
\end{equation}
with $\mathbf{D}_1$ a dataset with one subject per arm. Against the normal reference, $\mathcal{I}_1 = 1/\sigma^2$, so that $ELIR = \sigma^2\, \mathbb{E}_\pi\!\left[-\partial^2_\theta \log \pi(\theta)\right]$. With a Gaussian summary likelihood, the prior is a normal mixture, and we evaluated this expectation by numerical integration under the exact prior. In the aprepitant case study, the prior of the treatment effect has no closed form for any method: it was computed on the lattice of Supplementary Section~\ref{supp:binomial_lattice}. We approximated it by a normal mixture fitted with RBesT \parencite{ag_rbest_2023} to 1,000 draws from it, computed the ELIR of the fitted mixture with RBesT, and averaged the result over 10 fits made under a fixed seed. This ELIR is therefore an approximation, subject to the random draws and to the mixture fit, unlike the moment-based and precision-based ESS, which were computed from the posterior without random draws.

\subsection{Comparison with a frequentist test of the target data alone}
\label{sec:methods_reference}

\textcite{kopp-schneider_power_2020} showed that, when a uniformly most powerful (UMP) test exists, borrowing information cannot provide more power than this test at the same TIE: the increase in power is then simply bought at the expense of TIE inflation. 

 To assess this, we compared the power of each method with that of the standard frequentist analysis of the target data alone, the one-sided Wald test, performed at the TIE the method actually incurs, at every treatment effect in the alternative hypothesis. This reference is the analysis that would be run without borrowing, but not necessarily the most powerful test. For the botox and dapagliflozin case studies (normally distributed endpoints), it corresponds to the $t$-test, which is uniformly most powerful among unbiased tests. For the binary, recurrent-event and time-to-event case studies, where the Wald test is only approximately normal and not optimal, we also computed the power of the most powerful test with the nuisance parameters known (the control rate, and the dispersion or the control hazard), which bounds the power of any test at the same TIE in that design. The derivation of these tests, for each endpoint, is given in Supplementary Section~\ref{supp:most_powerful_test}. How the reference was matched to the actual, rather than the nominal, TIE is described below.

\paragraph{Matching on the actual TIE.} The reference test must spend the same type I error rate $\alpha_B$ as the borrowing method it is compared with. Its calibration depended on how the data of each case study were generated.

In the botox and dapagliflozin case studies, the data were Gaussian, so the variance estimate follows a scaled $\chi^2$ distribution and the one-sample $t$-test is exact. Its power was computed in closed form, at the level $\alpha_B$ estimated for the borrowing method.

In the mepolizumab and teriflunomide case studies, the data of each patient were simulated (negative binomial event counts and censored event times), and each simulated trial was summarized by its estimate and standard error (log rate ratio from a negative binomial regression, log hazard ratio from a Cox model). The reference was the one-sided test based on the Wald statistic, the estimate divided by its standard error. This statistic is only approximately normal, so a test at the nominal level $\alpha_B$ does not spend a type I error rate of $\alpha_B$. Its rejection threshold was instead set to the empirical quantile of the statistic over the trials simulated under $\theta_T = \theta_0$, the same trials on which $\alpha_B$ was estimated, so that the reference rejects the same proportion of these trials as the method. Its power was then estimated on the trials simulated under each alternative. Trials in which the treatment effect could not be estimated were excluded, as for the borrowing methods.

In the aprepitant and belimumab case studies, the reference was the test based on the Wald statistic of the difference in proportions and of the log odds ratio, respectively, and its size and power were computed exactly by enumerating the pairs of responder counts. Because the outcome is discrete, a test that is not randomized attains only a few type I error rates. We therefore used the randomized test that rejects every outcome whose statistic is beyond a threshold and a fixed fraction of the outcomes at the threshold, chosen so that its type I error rate is exactly $\alpha_B$.

These calibrations make sure that the comparison is not distorted by the asymptotic approximation of the test statistic. As a check, the separate Bayesian analysis should be as powerful as the reference at its own type I error rate (Section~\ref{sec:results_power_matched}).



\paragraph{TIE across control rates.} As is usual in simulation studies of borrowing methods, the TIE of each design was evaluated at a single value of the nuisance parameters. For the configurations with a power gain over the reference, we also evaluated the TIE when the true target control rate differed from that of the design. In the binary case studies, the TIE was computed exactly, for each method as specified for the design, on a grid of true target control response rates with a step of 0.01, within 0.05 and within 0.10 of the design rate (additive changes on the probability scale). In the recurrent-event and time-to-event case studies, it was estimated by simulation, for every design and setting with a gain larger than 0.02, with the true control event rate or hazard multiplied by $\exp(\kappa)$ for $\kappa \in \{0, \pm 0.05, \pm 0.10, \pm 0.20\}$. A change of $\pm 0.10$ on the log scale multiplies the rate by 0.905 or 1.105, that is, it changes it by about 10\%. The power comparison was then repeated with the reference allowed the method's largest TIE over the evaluated range. This largest TIE is not a worst case over all control rates: it holds only for the range evaluated.

\subsection{Comparison between borrowing methods at a matched TIE}
\label{sec:methods_frontier}

Borrowing methods can be governed by tuning parameters (the power parameter of the conditional power prior, the mixture weight of the robust mixture prior, the hyperprior on the commensurability parameter, and so on), and bias, precision and MSE all vary substantially with these choices. There is generally no direct correspondence between the tuning parameters of different methods: a mixture weight of \(w=0.5\) for a robust mixture prior has no natural equivalence to a power parameter of \(\gamma=0.5\) for a power prior. A comparison between methods at arbitrarily chosen parameter settings would therefore reflect the settings at least as much as the methods themselves. We therefore indexed each method-parameter configuration by the TIE it produces, which is the quantity of primary regulatory interest: each configuration is represented by a pair such as $\bigl\{\alpha_B(\zeta),\,\mathrm{MSE}(\zeta)\bigr\}$, where \(\zeta\) denotes the method-specific borrowing parameter or hyperparameter configuration, and estimation performance is compared among configurations yielding similar values of \(\alpha_B\). This is conceptually analogous to comparisons of power at a common TIE \parencite{kopp-schneider_power_2020}. A related but distinct strategy would be to prospectively calibrate the parameters of every method to attain an exactly prespecified TIE \(\alpha^\star\), and subsequently compare the methods under that calibration. Such an approach would require a nested simulation or numerical calibration for every method and scenario and was not undertaken here. Our approach instead evaluates a broad grid of parameter values and compares configurations producing similar, rather than exactly identical, TIEs. Each design has 56 method-parameter configurations.

\paragraph{Frontiers.} In each design, we define the power-TIE frontier as the highest probability of success reached by any of the 56 method-parameter configurations with a TIE no larger than a given value, and the MSE-TIE frontier the smallest MSE reached by any configuration with a TIE no larger than a given value, interpolated linearly in TIE between Pareto-optimal configurations. Each method-parameter configuration can therefore be compared with the frontier at its own TIE. The frontier is a benchmark interpolated over the grid of configurations we evaluated, not the best performance attainable by any tuning of any method. A point between two Pareto-optimal configurations is not an observed procedure: it is attained only by a randomized procedure that uses one configuration or the other with a fixed probability, whose TIE, power and MSE are the same mixture of those of the two configurations. The frontier also depends on the grid. Methods with more settings, or with settings closer together, are more likely to lie on it. In the case studies analyzed by simulation, the frontier is the extremum of noisy estimates, which makes it slightly optimistic. As an indication of this effect, the separate analysis lay 3\% above the estimated MSE--TIE frontier in median (Section~\ref{sec:results_mse}). In the binary case studies, the operating characteristics are exact and this bias is absent. A configuration close to the frontier is therefore among the best of the configurations evaluated, at its TIE, and the comparison does not show that a method is better than every possible tuning of the others.

\paragraph{Local comparisons under no treatment effect.} Under no treatment effect, MSE, coverage and interval score all worsen with the TIE, so a one-sided frontier would only reproduce the separate analysis. For these metrics, each configuration was compared with the configurations whose TIE lies within a factor 1.25 of its own (a two-sided window): we report the median shortfall from the highest coverage, and the relative excess over the smallest interval score, in that window. For the coverage, we also fitted a common decreasing (isotonic) curve of coverage against TIE to the 56 configurations of each design, and report the share of configurations more than 0.02 below it. The same local window was used as a check of the MSE comparison under a partially consistent effect.

\paragraph{TIE budget.} For TIE budgets of 0.05, 0.10 and 0.20, we computed the smallest MSE under a partially consistent effect that each method reaches with a TIE no larger than the budget, as a relative excess over the smallest MSE reached by any method within the same budget. This view favors methods with many settings, since a finer grid is more likely to have a setting just under the budget.

\paragraph{Matched prior ESS.} In the botox and dapagliflozin case studies, where the prior ESS does not depend on the drift through the target standard error, we also compared the configurations at a similar moment-based prior ESS. For a configuration $j$ with ESS $E_j$, the reference is the smallest MSE, under a partially consistent treatment effect, among the configurations $i$ of the same design with $|E_i - E_j| \leq b$, where $b = 0.10\,E_{\text{pool}}$ and $E_{\text{pool}}$ is the ESS of pooling, that is, the full source information. The window is therefore centered on each configuration, and pooling only sets its half-width. We report the relative excess of the MSE of configuration $j$ over this reference, and repeated the comparison with $b = 0.05\,E_{\text{pool}}$ and $b = 0.20\,E_{\text{pool}}$.

\paragraph{Break-even drift.} For every configuration and design, we define the break-even drift as the smallest drift between the target and the source treatment effect beyond which the MSE of the configuration exceeds that of the separate analysis, by linear interpolation on the drift grid. It is expressed in units of the target standard error. 

 \section{Results}

\subsection{Probability of success and power at a matched TIE}

\paragraph{Impact of the drift on the TIE}
\label{sec:results_tie}

TIE inflation, that is, a TIE above the value $\alpha$ that a Bayesian separate analysis with a critical value $\rho = 1 - \alpha$ would attain, is the main concern when using partial extrapolation in clinical trials. Borrowing from the source study inflated the TIE for most methods and parameter settings.
Methods that borrow fully from the source study  reached a TIE of 0.77 to 1 in the belimumab, dapagliflozin, mepolizumab and teriflunomide case studies. In these four case studies, the source estimate is precise and far from the null relative to the target standard error, so that borrowing it leads to claiming success regardless of the target data (Supplementary Table~\ref{tab:supp_tie_power}).

Most configurations that did not inflate the TIE reduce to the separate analysis under conflict: the RMP with $w = 0$, the CPP with $\gamma = 0$, and settings that stop borrowing at the slightest sign of prior-data conflict, such as the p-PP with an equivalence margin of 0.1 and $k \geq 10$, or TtP with an equivalence test ($\eta = \lambda = 0.1$).

 The others were found in the botox case study when the target standard deviation was twice the source one. In this case study, the source data favor the alternative hypothesis by a small margin only: $\operatorname{Pr}(\theta_T > 0 \mid \mathbf{D}_S) = 0.977$, just above the threshold of 0.975. A discounted prior gives a lower probability, for example 0.92 for the CPP with $\gamma = 0.5$ and 0.84 with $\gamma = 0.25$. In a trial with null treatment effect, borrowing has two opposite effects. It adds evidence for a positive effect, which increases the risk of a false rejection. It also shrinks a high target estimate towards the source estimate, which decreases this risk. When the target data are imprecise, they get little weight, and the second effect is larger. Thus, with a doubled target standard deviation, the TIE of the CPP with 39 participants per arm fell from 0.035 to 0.008 ($\gamma = 0.25$) and from 0.075 to 0.024 ($\gamma = 0.5$), whereas that of the separate analysis stayed at 0.030.

The PDCCPP was the only method whose TIE was controlled by construction, with a target of 0.065. With the normal summary likelihood, its calibration relies on closed-form normal approximations, which explains the slight departures from the target (0.046 to 0.077). With the binomial likelihood of the aprepitant case study, it was calibrated on the exact TIE and attained 0.062 to 0.064.

\begin{figure}[htbp]
\centering
\includegraphics[width=0.95\linewidth, height=0.95\textheight,keepaspectratio]{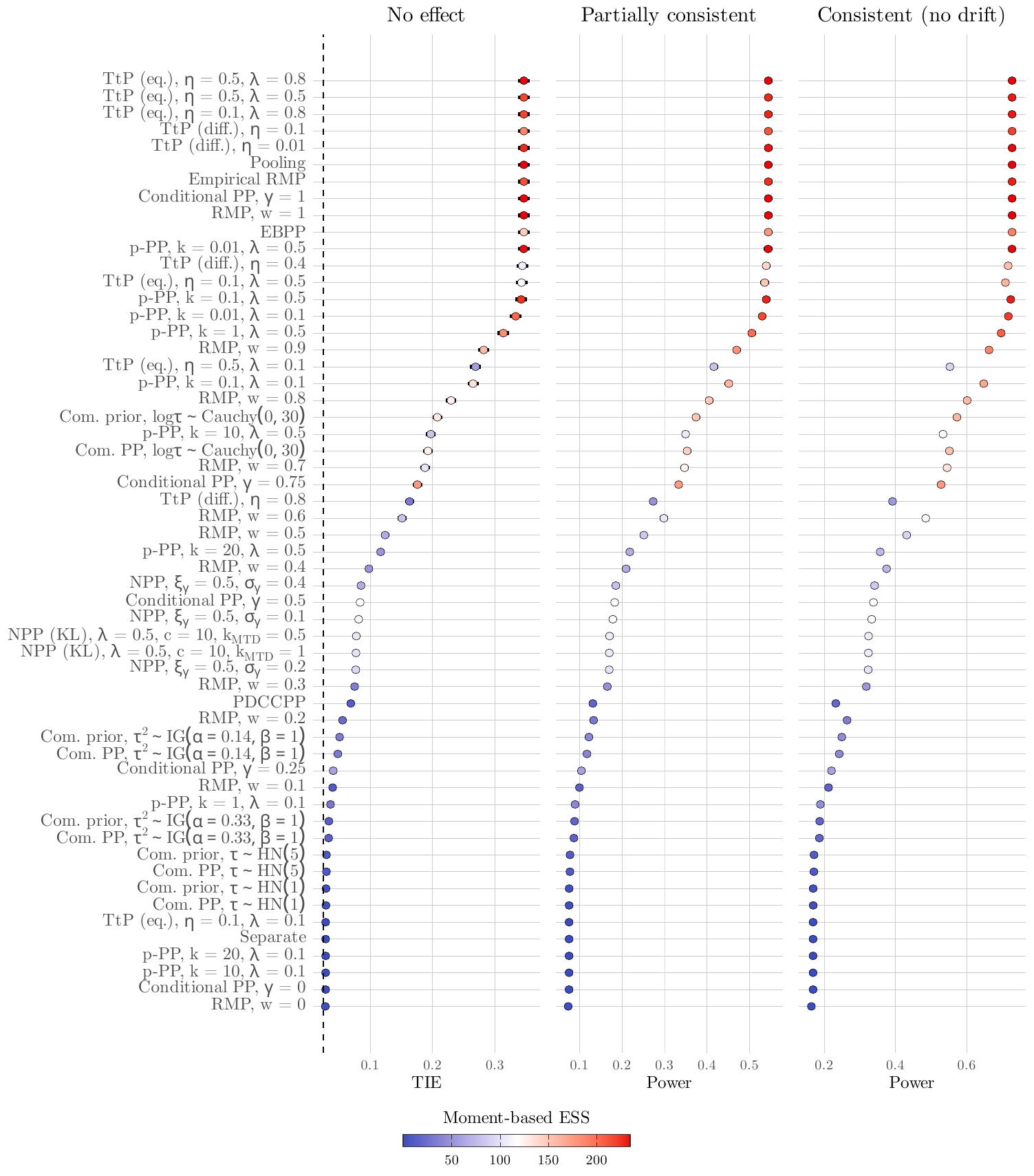}
\caption{Probability of success for the three main treatment effects considered, in the botox case study ($n_T = 58$). Error bars are 95\% CI. The dashed vertical line indicates the nominal TIE. Method abbreviations and parameters as in Table~\ref{tab:abbreviations}.}
\label{fig:botox_success_proba_forest_plot_58}
\end{figure}

The same methods that inflated TIE increased power when the source and target treatment effects were consistent or partially consistent. For example, in the aprepitant case study with 47 participants per arm, power rose from 0.111 with a separate analysis to 0.598 with pooling. In each design, the configurations mostly kept the same order under no effect and under a consistent effect: a configuration with a higher TIE mostly also had a higher power. The exceptions were configurations of TtP and the PDCCPP, which lay below the power--TIE frontier (see below). The amount of borrowing, as measured by the moment-based ESS, explained most of this order. It explained less of it in the mepolizumab and teriflunomide case studies, where the methods that decide from the data how much to borrow did not follow their average ESS. 
 Figure~\ref{fig:botox_success_proba_forest_plot_58} illustrates this in the botox case study (see Figures \ref{fig:belimumab_success_proba_forest_plot_target_sample_size_per_arm_140}, \ref{fig:dapagliflozin_success_proba_forest_plot_target_sample_size_per_arm_66}, \ref{fig:mepolizumab_success_proba_forest_plot_target_sample_size_per_arm_68}, \ref{fig:teriflunomide_success_proba_forest_plot_target_sample_size_per_arm_123} and \ref{fig:aprepitant_success_proba_forest_plot_target_sample_size_per_arm_143} for the other case studies). Figures \ref{fig:botox_relative_success_proba_forest_plot_target_sample_size_per_arm_58}, \ref{fig:belimumab_relative_success_proba_forest_plot_target_sample_size_per_arm_140}, \ref{fig:dapagliflozin_relative_success_proba_forest_plot_target_sample_size_per_arm_66}, \ref{fig:mepolizumab_relative_success_proba_forest_plot_target_sample_size_per_arm_68}, \ref{fig:teriflunomide_relative_success_proba_forest_plot_target_sample_size_per_arm_123} and \ref{fig:aprepitant_relative_success_proba_forest_plot_target_sample_size_per_arm_143} show the probability of study success relative to a separate analysis of the target study data.

\paragraph{Power gains at an equivalent TIE}
\label{sec:results_power_matched}
 
As expected, the separate Bayesian analysis was exactly as powerful as the reference frequentist analysis in the binary case studies, and within Monte Carlo error for all other case studies.  

Over all designs and settings, 93.6\% of the comparisons with the reference test differed by less than 0.01 in power, and 2.0\% a loss larger than 0.02, over all designs including the sensitivity designs. The largest gains and losses of each method are given in Table~\ref{tab:method_comparison}. Gains were small, and confined to a few case studies and to some adaptive borrowing methods. In the primary designs, the largest gain at the consistent treatment effect was obtained by TtP with a difference test ($\eta = 0.4$) in the mepolizumab case study with 45 participants per arm ($+0.044$, 95\% interval 0.025 to 0.059, at a TIE of 0.070; Figure~\ref{fig:mepolizumab_power_gain_vs_tie_sample_size=45_treatment_effect=consistent_source_denominator_change_factor=1}). The largest gains in each case study are given in Supplementary Table~\ref{tab:supp_matched_tie} (Figures~\ref{fig:teriflunomide_power_gain_vs_tie_sample_size=123_treatment_effect=consistent_source_denominator_change_factor=1}, \ref{fig:aprepitant_power_gain_vs_tie_sample_size=143_treatment_effect=consistent}, \ref{fig:aprepitant_power_gain_vs_tie_sample_size=71_treatment_effect=consistent} and \ref{fig:botox_power_gain_vs_tie_sample_size=58_treatment_effect=consistent_target_to_source_std_ratio=1}). In most settings, the power of a borrowing method was therefore that of a frequentist test of the target data alone at the same TIE, as predicted by \textcite{kopp-schneider_power_2020} (see Figure~\ref{fig:power_equivalent_tie} for an example with the CPP).

We observed that gains were largest with a lower source control rate (source denominator factor 0.5), reaching 0.062 in the mepolizumab case study. The single exception to small gains was the p-PP in the belimumab case study with a lower source control rate: with 93 participants per arm, $k = 20$ and $\lambda = 0.5$, it gained up to 0.109 in power at a TIE of 0.031. This gain was computed exactly, and it does not contradict the result of \textcite{kopp-schneider_power_2020}, for two reasons (Section~\ref{sec:methods_reference}). First, the standard test is far from the most powerful test in this design: with the control rate known, the most powerful test reaches a power of 0.854 at the same TIE, against 0.573 for the standard test. Over all designs of the binary case studies, the most powerful test was up to 0.31 more powerful than the standard test, and no method exceeded it. Second, the TIE of 0.031 holds only at the control rate of the design (0.24): it rose to 0.055 when the true control rate was within 0.05 of the design rate, whereas that of the separate analysis stayed between 0.023 and 0.024 (Figure~\ref{fig:belimumab_tie_vs_control_rate_sample_size=93_source_denominator_change_factor=0.5}). When the reference was allowed the method's largest TIE within 0.05 of the design control rate, the gain fell to at most 0.017, and no gain larger than 0.02 remained in any design of the belimumab case study. In the aprepitant case study, the largest gain, 0.038 for the p-PP ($k = 1$, $\lambda = 0.1$) with 47 participants per arm at a TIE of 0.024, depended less on the control rate: the TIE of this configuration rose only to 0.025 when the true control rate was within 0.05 of that of the design (0.55), and to 0.029 within 0.10. With the reference allowed these TIEs, however, the gain fell to 0.022 and 0.011, and no gain larger than 0.02 remained in any design when the control rate could differ by 0.10 (the sensitivity of each method's TIE to the control rate is given in Supplementary Table~\ref{tab:supp_tie_control_rate}). In that scenario, the most powerful test with the control rate known reached a power of 0.990, against 0.840 for the standard test and 0.879 for the p-PP.

In the recurrent-event and time-to-event case studies, gains larger than 0.02 arose for TtP, the p-PP, the NPP (KL), the EBPP and the RMP, mostly with small target samples and a lower source control rate. The largest was 0.062 (mepolizumab, 68 participants per arm, source denominator factor of 0.5, TtP with a difference test, $\eta = 0.4$). We repeated the two checks for every design and setting with such a gain. With the nuisance parameters known, the most powerful test was in median 0.13 to 0.21 more powerful than the standard Wald test, and no method exceeded it. When the reference was allowed the method's largest TIE over true control rates within about 10\% of the design rate ($|\kappa| \leq 0.10$), only 14 of the 180 teriflunomide gains larger than 0.02 remained so, whereas in the mepolizumab case study the largest gain fell only from 0.062 to 0.050 and 95 of 180 remained larger than 0.02.

\begin{table}[ht]
\centering
\small
\begin{tabular}{lccccccc}
\hline
Method & TIE & \multicolumn{2}{c}{Power at matched TIE} & \multicolumn{3}{c}{MSE relative to separate} & Coverage \\
 & & Largest gain & Largest loss & Consistent & Partial & Null & (null) \\
\hline
Com.\ PP        & 0.027--0.189 & $+0.015$ & $-0.020$ & 0.70 & 0.82 & 0.98 & 0.94 \\
Com.\ prior     & 0.028--0.205 & $+0.015$ & $-0.019$ & 0.69 & 0.82 & 0.98 & 0.94 \\
RMP                    & 0.025--0.957 & $+0.019$ & $-0.022$ & 0.27 & 0.76 & 1.35 & 0.88 \\
NPP                    & 0.100--0.361 & $+0.010$ & $-0.013$ & 0.23 & 0.61 & 1.58 & 0.78 \\
NPP (KL)               & 0.141--0.230 & $+0.022$ & $-0.014$ & 0.23 & 0.60 & 1.56 & 0.81 \\
EBPP                   & 0.201 & $+0.017$ & $-0.016$ & 0.36 & 0.86 & 1.22 & 0.78 \\
Empirical RMP  & 0.361 & $+0.014$ & $-0.009$ & 0.09 & 0.75 & 2.21 & 0.62 \\
p-PP                   & 0.025--0.870 & $+0.038$ & $-0.129$ & 0.39 & 1.00 & 1.23 & 0.76 \\
TtP (diff)             & 0.039--0.384 & $+0.046$ & $-0.280$ & 0.59 & 1.03 & 1.29 & 0.78 \\
TtP (eq)               & 0.026--0.797 & $+0.038$ & $-0.235$ & 0.31 & 1.00 & 1.44 & 0.68 \\
CPP         & 0.025--0.957 & $+0.009$ & $-0.015$ & 0.20 & 0.66 & 1.78 & 0.55 \\
Pooling                & 0.957 & $+0.006$ & $-0.005$ & 0.08 & 0.72 & 2.66 & 0.04 \\
PDCCPP                 & 0.065 & $+0.006$ & $-0.223$ & 0.85 & 0.99 & 1.05 & 0.91 \\
Separate               & 0.025 & $+0.004$ & $-0.016$ & 1.00 & 1.00 & 1.00 & 0.95 \\
\hline
\end{tabular}
\caption{Comparison of the methods across case studies and target sample sizes (primary designs, target-to-source standard deviation ratio and source denominator factor of 1). TIE: range, across a method's parameter settings, of the median type I error rate across designs. Largest gain and largest loss: extreme power differences with a frequentist test of the target data alone calibrated to the same actual type I error rate, over all designs, settings and treatment effects in the alternative hypothesis. MSE: median MSE relative to the separate analysis under a consistent, a partially consistent and a null (conflicting) treatment effect. Coverage: median coverage of the 95\% credible interval under a null treatment effect.}
\label{tab:method_comparison}
\end{table}

\paragraph{Power loss due to borrowing}

\textcite{kopp-schneider_simulating_2024} reported that, in some ``extreme borrowing'' cases, Bayesian borrowing methods lead to tests that are not UMP, and therefore to a loss of power compared with a frequentist test at the same TIE (illustrated in Figure~\ref{fig:power_loss}). In our simulations, losses were larger than gains, and concentrated in the methods that decide how much to borrow from a test or a threshold on the observed conflict (Table~\ref{tab:method_comparison}). TtP with a difference test ($\eta = 0.8$) lost up to 0.280 in power in the aprepitant case study with 47 participants per arm, the PDCCPP up to 0.22 in the same design, and the p-PP up to 0.129 in the botox case study with 39 participants per arm. These largest losses occurred at treatment effects 2.2 to 2.4 times the source estimate. At the consistent treatment effect, the losses were smaller, for example 0.138 for TtP and 0.083 for the PDCCPP in the aprepitant case study with 71 participants per arm (Figure~\ref{fig:aprepitant_power_gain_vs_tie_sample_size=71_treatment_effect=consistent} shows TtP; see Figure~\ref{fig:botox_power_gain_vs_tie_sample_size=58_treatment_effect=consistent_target_to_source_std_ratio=1} for the botox case study). These losses vanished as the sample size grew: from the smallest to the largest target sample, the largest loss fell from 0.280 to 0.001 in the aprepitant, from 0.241 to 0.011 in the botox and from 0.085 to 0.009 in the dapagliflozin case studies, and it stayed below 0.03 in the other case studies. They grew when the target standard deviation was twice the source one, to 0.30 in the botox and 0.26 in the dapagliflozin case studies.

\begin{figure}[htbp]
    \centering
    \includegraphics[width=0.95\linewidth, height=0.95\textheight,keepaspectratio]{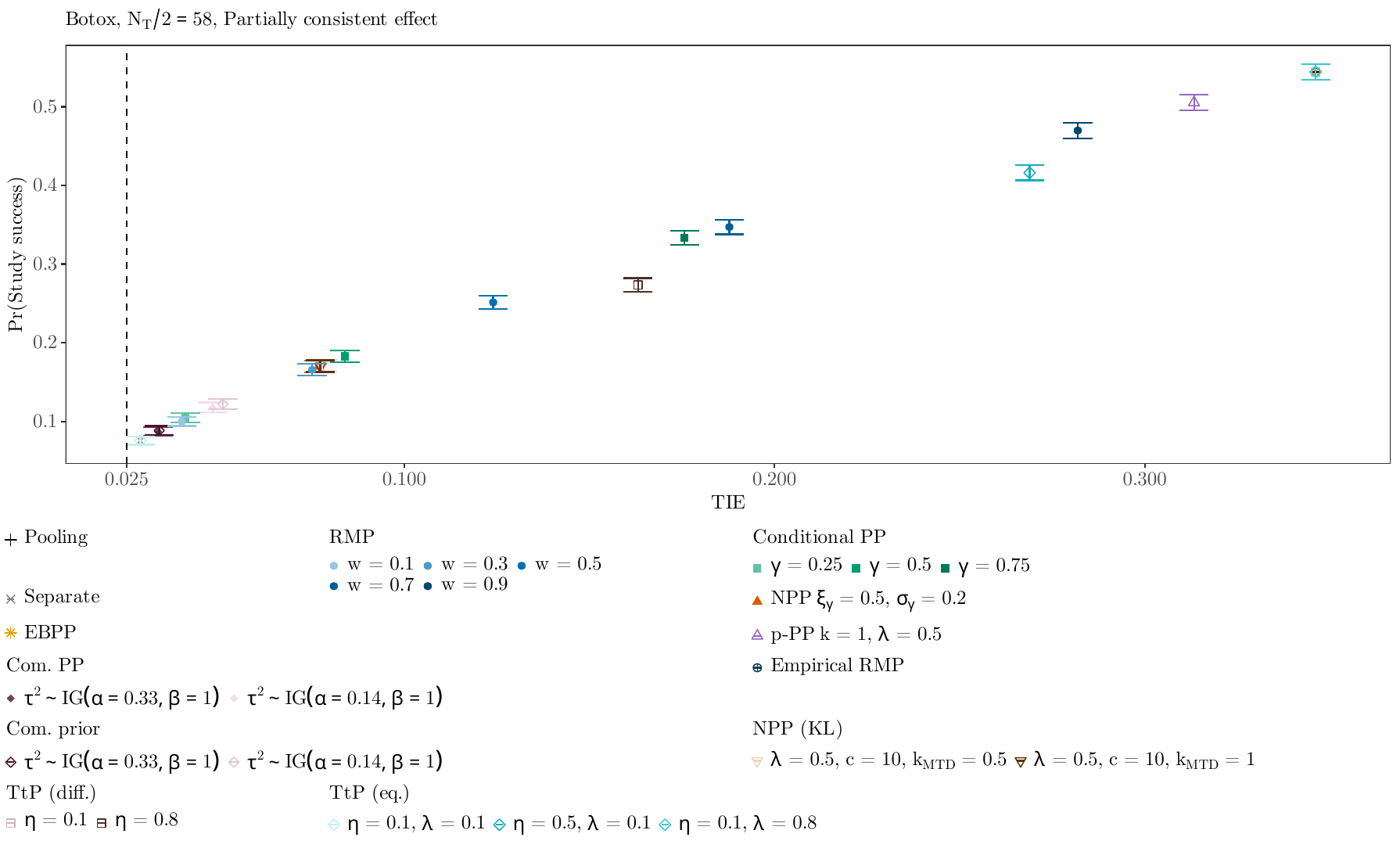}
    \caption{Probability of success as a function of the TIE. Each point represents one method-parameter combination in the botox case study with $58$ participants per arm. The treatment effect is partially consistent, the target to source standard deviation ratio is $1$. Error bars correspond to the 95\% confidence interval of the probability of success. Dashed vertical line represents the nominal TIE of $0.025$. Method abbreviations and parameters as in Table~\ref{tab:abbreviations}. For readability, only a subset of the parameter settings is shown, and the PDCCPP is omitted.}
    \label{fig:botox_success_proba_vs_tie_sample_size=58_treatment_effect=partially_consistent}
\end{figure}

Plotted against the TIE, the power of most configurations lay on a common curve, and in the botox case study with 58 participants per arm, TtP lay below it (Figure~\ref{fig:botox_success_proba_vs_tie_sample_size=58_treatment_effect=partially_consistent}): at a TIE of about 0.16, TtP with a difference test ($\eta = 0.8$) had a power of 0.27, against 0.33 to 0.35 for the CPP and the RMP at similar TIEs. Across all designs, under a partially consistent or a consistent treatment effect, 11\% of the configurations of TtP with a difference test and 6\% with an equivalence test fell more than 0.01 below the power--TIE frontier (Section~\ref{sec:methods_frontier}), by up to 0.12, mostly with small target samples (Table~\ref{tab:frontier_distance}). The PDCCPP fell below the frontier more often (31\% of its configurations), and every other method in at most 2\% of its configurations. For most configurations of every method, power was therefore determined by the TIE alone.

\subsection{Estimation and uncertainty under drift}
\label{sec:results_estimation}

\subsubsection{Bias, precision and MSE}
\label{sec:results_mse}

\begin{figure}[htbp]
     \centering
     \includegraphics[width=0.95\linewidth, height=0.95\textheight,keepaspectratio]{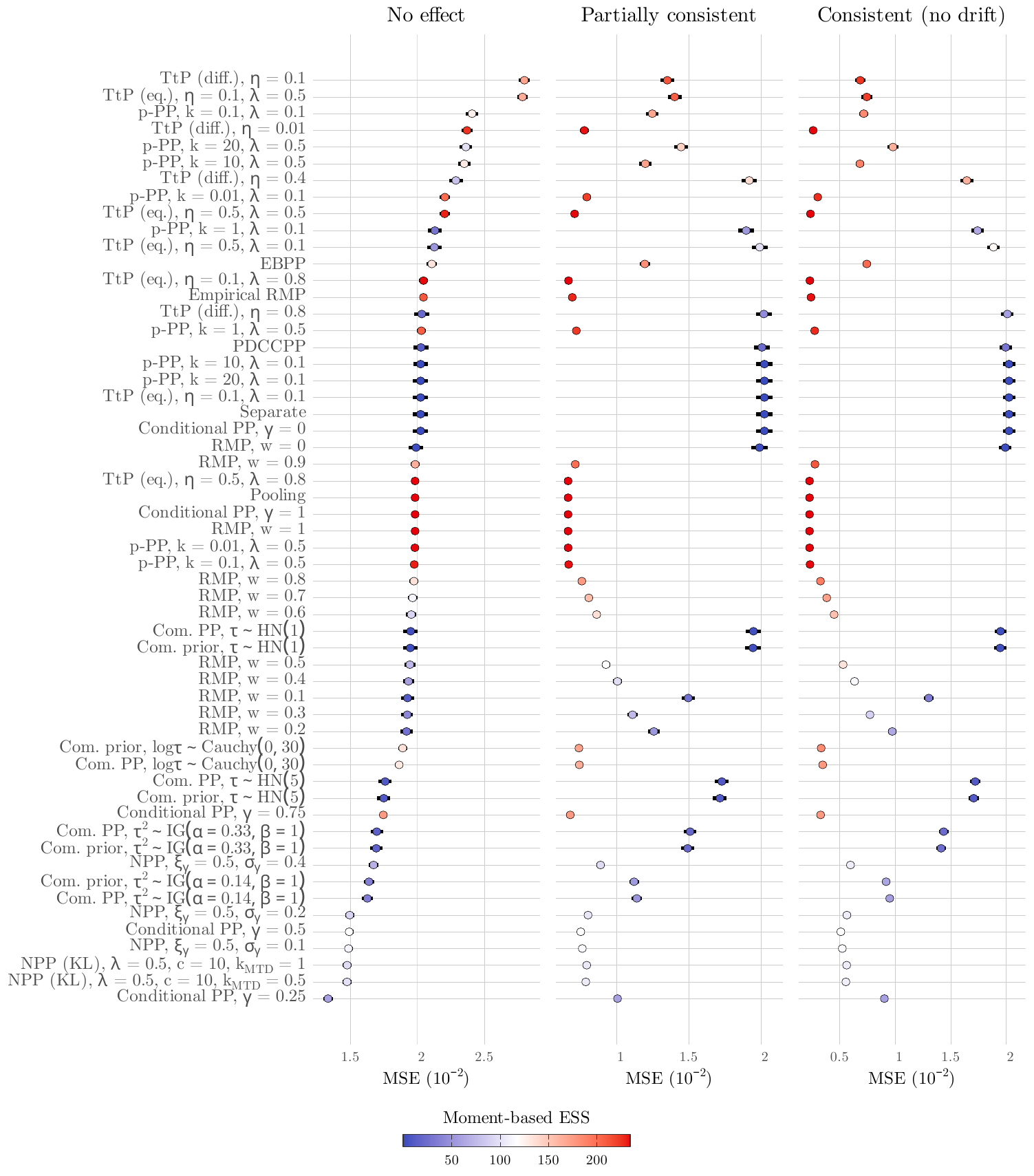}
     \caption{MSE for the three main treatment effects considered, in the botox case study ($n_T = 117$). Error bars are 95\% CI. Method abbreviations and parameters as in Table~\ref{tab:abbreviations}.}
     \label{fig:botox_mse_forest_plot_target_sample_size_per_arm_117}
 \end{figure}

Bias and precision are the two main components to consider when comparing methods on their estimation performance. Here, precision is measured as the half-width of the 95\% credible interval, which reflects the strength of the belief represented by the posterior distribution. The MSE combines bias and variance: a borrowing method with a lower MSE than the separate analysis compensates the bias due to drift by a reduction in variance. Figure~\ref{fig:botox_mse_forest_plot_target_sample_size_per_arm_117} compares the MSE of the different methods in the botox case study for the three main treatment effects considered (see also Figures \ref{fig:belimumab_mse_forest_plot_target_sample_size_per_arm_140}, \ref{fig:dapagliflozin_mse_forest_plot_target_sample_size_per_arm_66}, \ref{fig:mepolizumab_mse_forest_plot_target_sample_size_per_arm_68}, \ref{fig:teriflunomide_mse_forest_plot_target_sample_size_per_arm_185} and \ref{fig:aprepitant_mse_forest_plot_target_sample_size_per_arm_143} for the other case studies).

In the absence of drift, borrowing reduces the MSE, because it introduces no bias and improves precision (Figure~\ref{fig:belimumab_precision_forest_plot_target_sample_size_per_arm_140}): when the target treatment effect equalled the source estimate, the median configuration reduced the MSE to 33\% to 37\% of that of the separate analysis, depending on the case study, and the 95\% credible interval over-covered slightly. As drift increases, bias increases as well (Figure~\ref{fig:botox_bias_forest_plot_target_sample_size_per_arm_117}), although, for extreme drift, adaptive borrowing methods discard the source data and the bias decreases again.  Their precision also decreases, following their prior ESS. Under a partially consistent treatment effect, the median MSE ratio rose to between 0.52 and 1.07. In the absence of a treatment effect, where the source and target studies conflict, it was between 1.30 and 1.69 in the four case studies where the source estimate is far from the null, against 0.91 and 0.95 in the aprepitant and botox case studies.

The methods differed mainly in how they balance these two situations (Table~\ref{tab:method_comparison}). Methods that borrow fully achieved the largest reductions in MSE under consistency (median 8\% of the separate analysis) at the cost of the largest losses under conflict (median 2.7 times the MSE of the separate analysis). The data-adaptive methods (RMP, NPP, NPP (KL), EBPP, empirical RMP, p-PP and TtP) reduced the median MSE under consistency to between 9\% and 59\% of the separate analysis, but increased it by 22\% to 121\% under conflict. The commensurate priors behaved most robustly: they reduced the MSE under consistency (median ratio 0.69 to 0.70) without increasing it under conflict (median ratio 0.98). This robustness depended on the hyperprior. With an $\operatorname{HN}(5)$ or an $\operatorname{IG}(0.33, 1)$ hyperprior, their MSE never exceeded 1.04 and 1.08 times that of the separate analysis over the whole drift grid (median over designs), and their coverage never fell below 0.94. With a vaguer $\operatorname{IG}(0.14, 1)$ or $\operatorname{Cauchy}(0, 30)$ hyperprior, they borrowed more but reached about 1.35 and 2.5 times the MSE of the separate analysis. Among the configurations whose MSE never exceeded 1.10 times that of the separate analysis over the drift grid, the commensurate priors achieved the largest reduction under consistency (median best MSE ratio 0.81, against 0.89 for the p-PP, 0.94 for the PDCCPP, 0.96 to 0.99 for TtP and the RMP, and 1.00 for the CPP). The PDCCPP, by construction, departed little from the separate analysis (median ratios 0.85 to 1.05).

\paragraph{MSE at a matched TIE}

\begin{figure}[htbp]
    \centering
    \includegraphics[width=0.95\linewidth, height=0.95\textheight,keepaspectratio]{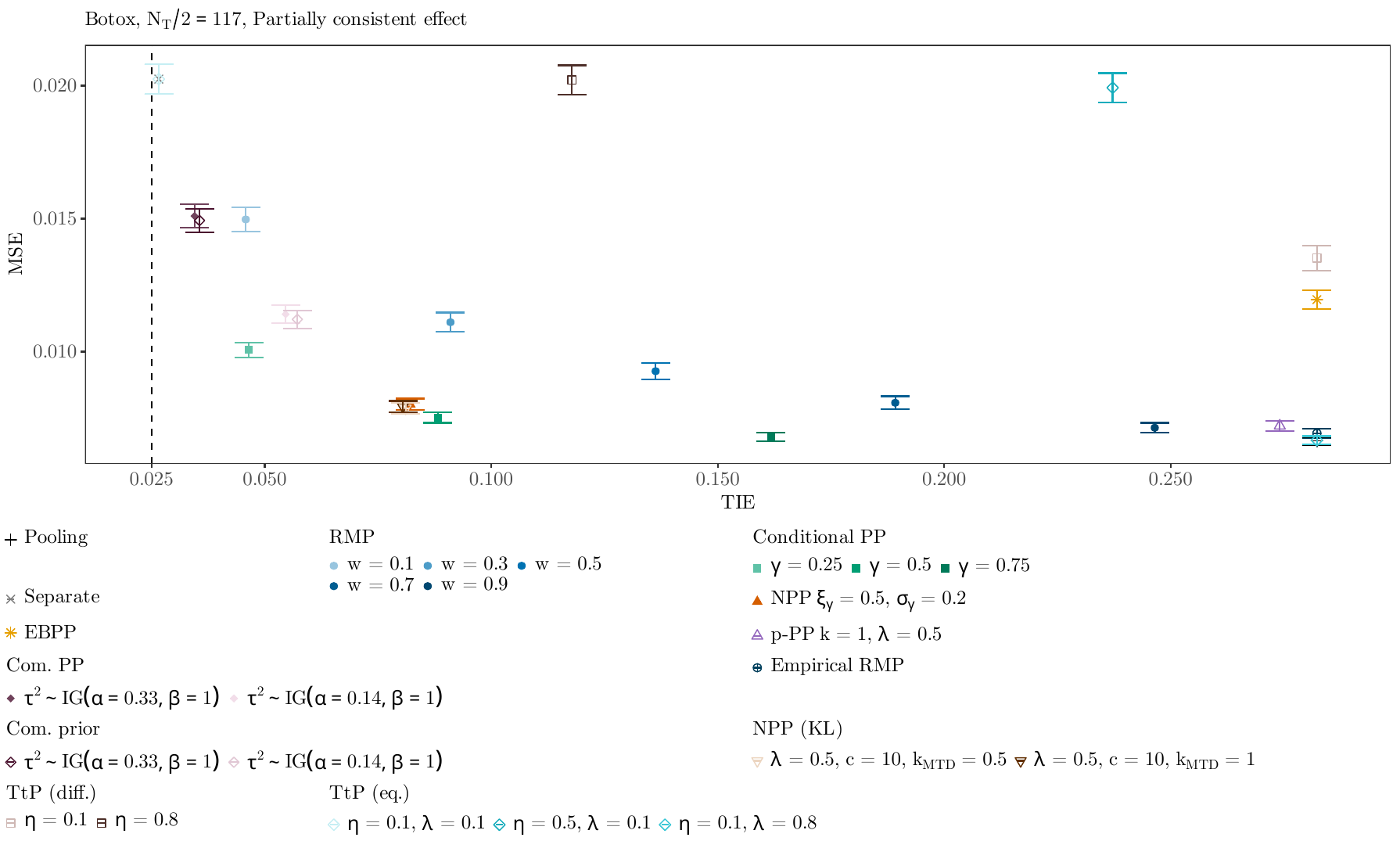}
    \caption{MSE as a function of the TIE. Each point represents one method-parameter combination in the botox case study with $117$ participants per arm. The treatment effect is  partially consistent, the target to source standard deviation ratio is $1$. Error bars correspond to the 95\% confidence interval of the MSE. Dashed vertical line represents the nominal TIE of $0.025$. Method abbreviations and parameters as in Table~\ref{tab:abbreviations}. For readability, only a subset of the parameter settings is shown, and the PDCCPP is omitted.}
    \label{fig:botox_mse_vs_tie_sample_size=117_treatment_effect=partially_consistent}
\end{figure}

Bias, precision and MSE all vary substantially with the tuning parameters of each method, which have no common scale across methods. We therefore indexed each method-parameter configuration by the TIE it produces and compared the MSE at a matched TIE (Section~\ref{sec:methods_frontier}). In the botox case study with 117 participants per arm and a partially consistent treatment effect (Figure~\ref{fig:botox_mse_vs_tie_sample_size=117_treatment_effect=partially_consistent}; see Figures~\ref{fig:dapagliflozin_mse_vs_tie_sample_size=33_treatment_effect=partially_consistent}, \ref{fig:belimumab_mse_vs_tie_sample_size=140_treatment_effect=partially_consistent}, \ref{fig:mepolizumab_mse_vs_tie_sample_size=68_treatment_effect=partially_consistent}, \ref{fig:teriflunomide_mse_vs_tie_sample_size=123_treatment_effect=partially_consistent} and \ref{fig:aprepitant_mse_vs_tie_sample_size=71_treatment_effect=partially_consistent} for the other case studies), TtP with a difference test ($\eta = 0.8$) had the MSE of the separate analysis at a TIE of 0.12, whereas the CPP with $\gamma = 0.75$ reduced the MSE by two thirds at a TIE of 0.16. Across all designs (Table~\ref{tab:frontier_distance}), the commensurate priors (median excess over the MSE--TIE frontier 0\%) and the CPP (1\%) were closest to the frontier, followed by the NPP (KL) (6\%) and the NPP (12\%). The RMP was not close to them (31\%), and TtP, the EBPP, the PDCCPP, the p-PP and the empirical RMP had much larger MSEs at a similar TIE. As a reference for Monte Carlo noise, the separate analysis lay 3\% above the frontier. The ordering was not the same in every case study: for example, the CPP was on the frontier in the aprepitant and botox case studies but 24\% to 31\% above it in the mepolizumab and teriflunomide case studies. Within a TIE budget of 0.05 or 0.10, the commensurate priors reached the smallest MSE, and within a budget of 0.20, the CPP did (Table~\ref{tab:tie_budget}). Results were similar in the sensitivity designs (Table~\ref{tab:frontier_distance_sensitivity}).

\paragraph{Break-even drift}

The break-even drift summarizes how much conflict each method tolerates: the smallest drift between the target and the source treatment effect beyond which its MSE exceeds that of the separate analysis, in units of the target standard error (Section~\ref{sec:methods_frontier}; Table~\ref{tab:break_even}). For a given method, it was strikingly stable across case studies, endpoints and sample sizes, and it was similar for most methods: towards the null, from 1.0 target standard errors in median for TtP with a difference test, which therefore lost its advantage earliest, to 1.6 for the NPP and the NPP (KL), with nearly the same values away from the null. The commensurate priors broke even later (2.2 target standard errors in median), and 15\% of their borrowing configurations did not break even within the explored drift. Among the settings with a median TIE of at most 0.05 that reduced the MSE by at least 10\% under consistency, the commensurate priors with an $\operatorname{HN}(5)$ hyperprior had the largest break-even drift (4.2 target standard errors). The coverage of the 95\% credible interval fell below 0.90 at a similar or smaller drift, except for the commensurate priors. In practice, for most methods, borrowing reduced the MSE only when the drift was smaller than about one and a half standard errors of the target estimate, a drift that the target trial alone can hardly detect.  

\subsubsection{Coverage of the 95\% credible interval and interval score}

The coverage probability of the 95\% credible interval measures the calibration of the uncertainty of a Bayesian method, and hence the trust that can be placed in the interpretation of its credible intervals. In the absence of drift, methods that borrow heavily from the source study show high precision and a coverage at or above 0.95 (Figure~\ref{fig:belimumab_coverage_forest_plot_target_sample_size_per_arm_140}; see also Figures \ref{fig:dapagliflozin_coverage_forest_plot_target_sample_size_per_arm_66}, \ref{fig:mepolizumab_coverage_forest_plot_target_sample_size_per_arm_68}, \ref{fig:teriflunomide_coverage_forest_plot_target_sample_size_per_arm_123} and \ref{fig:aprepitant_coverage_forest_plot_target_sample_size_per_arm_143} for the other case studies), but at the expense of a sharp drop in coverage when the target and source studies are inconsistent: in the absence of a treatment effect, the coverage of methods that borrow fell close to zero in the belimumab, mepolizumab and teriflunomide case studies. In the presence of drift, a larger target sample size improved the coverage probability, and for large drift, adaptive borrowing methods can discard the source data and thereby recover a good coverage probability.

In the belimumab case study with 140 participants per arm (Table~\ref{tab:belimumab_precision_ecp_table_target_sample_size_per_arm_140}), the coverage of pooling fell from 1 under a consistent treatment effect to 0 under no treatment effect, and that of the CPP with $\gamma = 0.5$ to 0.28. The coverage of the p-PP and of TtP depended strongly on their settings, from close to 0.95 for the settings that stop borrowing at the slightest conflict down to 0.04 for the p-PP and 0.07 for TtP with their most permissive settings. The commensurate priors with a half-normal or an $\operatorname{IG}(0.33, 1)$ hyperprior kept a coverage of 0.93 to 0.98 in all three scenarios, with credible intervals narrower than those of the separate analysis under consistency. The MSE and coverage of every case study are given in Supplementary Table~\ref{tab:supp_estimation}.

When the coverage was plotted against the TIE, the methods approximately lay on a common curve. Under no treatment effect, the coverage decreased steadily as the TIE increased, from 0.95 at the nominal TIE to about 0.70 at a TIE of 0.28 in the botox case study with 117 participants per arm (Figure~\ref{fig:botox_coverage_vs_tie_sample_size=117_treatment_effect=no_effect}; see Figures~\ref{fig:botox_coverage_vs_tie_sample_size=117_treatment_effect=partially_consistent} and \ref{fig:mepolizumab_coverage_vs_tie_sample_size=68_treatment_effect=partially_consistent_source_denominator_change_factor=1} for a partially consistent treatment effect).

Because a narrow interval is only informative if it covers the true effect, we also report the interval score \parencite{gneiting_strictly_2007}, which combines the width of the interval with a penalty for excluding the true treatment effect (Figures \ref{fig:botox_interval_score_forest_plot_target_sample_size_per_arm_117}, \ref{fig:botox_interval_score_vs_tie_sample_size=117_treatment_effect=no_effect}, \ref{fig:dapagliflozin_interval_score_forest_plot_target_sample_size_per_arm_66}, \ref{fig:dapagliflozin_interval_score_vs_tie_sample_size=66_treatment_effect=no_effect}, \ref{fig:belimumab_interval_score_forest_plot_target_sample_size_per_arm_140}, \ref{fig:belimumab_interval_score_vs_tie_sample_size=140_treatment_effect=no_effect}, \ref{fig:mepolizumab_interval_score_forest_plot_target_sample_size_per_arm_68}, \ref{fig:mepolizumab_interval_score_vs_tie_sample_size=68_treatment_effect=no_effect}, \ref{fig:teriflunomide_interval_score_forest_plot_target_sample_size_per_arm_123}, \ref{fig:teriflunomide_interval_score_vs_tie_sample_size=123_treatment_effect=no_effect}, \ref{fig:aprepitant_interval_score_forest_plot_target_sample_size_per_arm_143} and \ref{fig:aprepitant_interval_score_vs_tie_sample_size=143_treatment_effect=no_effect}). Compared with the configurations of similar TIE, the CPP, the NPP, the commensurate priors and the NPP (KL) had the lowest interval score under no treatment effect (median excess 0\% to 3\%), the p-PP and the RMP an intermediate one, and TtP, the PDCCPP, the EBPP and the empirical RMP the highest (Table~\ref{tab:frontier_distance}). Only the commensurate priors had, in median, an interval score below that of the separate analysis under no treatment effect (median ratio 0.99). Over all methods, configurations with an interval score below that of the separate analysis were mostly found in the aprepitant and botox case studies, where the source estimate is close to the null relative to the target standard error.

\subsection{Amount of borrowing}
\label{sec:results_ess}

\subsubsection{Impact of drift on the prior ESS}

Comparing the prior ESS with the sample size of the target study is a convenient way of comparing the amount of information borrowed from the source study with the information content of the target study. However, the impact of drift on the prior ESS is difficult to interpret when the summary measures are, e.g., risk ratios or odds ratios, as the standard deviation in the target study then depends on the drift. Similarly, an increase in the standard deviation of the target study relative to the source study would naturally increase the prior ESS. We therefore focused on the botox and dapagliflozin case studies (normally distributed endpoints), without a change in the standard deviation of the target study. In the range of drift values considered, the adaptiveness of the methods was limited (Figure~\ref{fig:botox_ess_moment_forest_plot_117}): in the botox case study with 117 participants per arm, the median moment-based ESS of the RMP fell from 134 under a consistent treatment effect to 74 under no effect, that of the EBPP from 202 to 132 and that of the p-PP from 187 to 127, whereas that of the NPP (106 to 91), the empirical RMP (233 to 209) and the commensurate priors (22--24 to 14--15) changed less. From a practical perspective, one may focus on methods and parameters for which the prior ESS is lower than $n_T$, as it may not be acceptable that the source trial provides more information for inference than the target trial. For very large drift, adaptive borrowing methods discard the source information, and their frequentist operating characteristics then approach those of the separate analysis (see e.g. Figure~\ref{fig:belimumab_RMP_mse_vs_drift_cat_parameters_sample_size=93_source_denominator_change_factor=1}).

\subsubsection{Comparison at a similar prior ESS}

The magnitude of the prior ESS relative to the target study sample size is a key consideration when selecting a prior. Therefore, one may wonder whether comparisons at similar prior ESS would agree with those at equivalent TIE. The two comparisons agreed for some methods but not for all (Figure~\ref{fig:botox_mse_vs_ess_moment_sample_size=117_treatment_effect=partially_consistent}): the commensurate priors, closest to the frontier at a matched TIE, were farther from the best MSE at a matched ESS, and the empirical RMP was closer. In the botox and dapagliflozin case studies, we compared each configuration with the configurations of similar moment-based ESS (Section~\ref{sec:methods_frontier}). The CPP, the NPP, the NPP (KL) and the empirical RMP had an MSE close to the smallest MSE at a similar ESS (median excess 0\% to 7\%), the commensurate priors (17\%) and the RMP (28\%) a larger one, and the p-PP, TtP, the EBPP and the PDCCPP a much larger one (36\% to 80\%; Table~\ref{tab:frontier_distance}).

\subsection{Sensitivity analyses}
\label{sec:results_sensitivity}

\subsubsection{Normal approximation for binary endpoints}
\label{sec:results_binary}


To assess how the analysis of a binary endpoint affects the results, we re-analyzed the aprepitant case study with the normal approximation used in the belimumab case study, that is, a normal likelihood for the estimated log odds ratio with its estimated standard error, in place of the binomial likelihoods of both arms on the risk-difference scale.

 Methods whose tuning parameters do not depend on the effect scale (pooling, the CPP, the RMP, the NPP and its KL-calibrated variant, the EBPP, the empirical RMP and the PDCCPP) had similar testing properties under both analyses: their TIEs differed by at most 0.055, and their coverage was close. The TIE, the power and the coverage do not change under a monotone change of the effect scale, so they can be compared directly between the two analyses. The MSE cannot: the MSE relative to the separate analysis compares the methods within one scale, and its value on the log odds ratio scale need not equal its value on the risk-difference scale. Because the two analyses also differ in the likelihood, the differences between them cannot be attributed to the normal approximation alone (Section~\ref{sec:discussion_binary}). The PDCCPP, whose calibration relies on normal approximations, exceeded its target TIE of 0.065 under the normal approximation (up to 0.075), but attained 0.062 to 0.064 with the binomial likelihood. By contrast, methods whose tuning parameters are expressed on the scale of the treatment effect changed markedly, because an equivalence margin or a hyperprior on the between-study heterogeneity does not have the same meaning on the log odds ratio and risk-difference scales. However, the main conclusion was unchanged. At a matched TIE, 93.1\% of the power differences were smaller than 0.01 under the normal approximation, and 97.1\% with the binomial likelihood. No configuration gained more than 0.02 under the normal approximation (one gained 0.038 with the binomial likelihood), and the largest losses were similar (TtP with a difference test, and the PDCCPP).
The ordering of the 56 configurations was only partly preserved: over sample sizes and scenarios, the Spearman correlation between the two analyses was 0.76 to 0.88 for the TIE and the power, 0.57 to 0.89 for the MSE, and 0.48 to 0.91 for the coverage.

The normal approximation itself was adequate for testing in the binary designs we considered: with control response rates of 0.24 to 0.55 and at least 47 participants per arm, the exact TIE of the one-sided Wald test at a nominal level of 0.025 was 0.024 to 0.027, on the risk-difference as on the log odds ratio scale.

\subsubsection{Time-to-event design (teriflunomide case study)}

In the teriflunomide case study, we examined how the time-to-event design affects the operating characteristics of the priors: loss to follow-up, Weibull event times and a placebo relapse rate in the target population two thirds of, or 1.5 times, that of the source trials change the number of observed events while keeping hazards proportional. 

Censoring and attrition affected the priors through the number of events they removed. Fewer events make the target trial less informative relative to the source trials, so the same source data count for more: under no treatment effect, the moment-based ESS of pooling rose from 514 to 676 participants under the lower placebo relapse rate, and under a conflicting treatment effect, the adaptive priors were pulled further towards the source trials, with a higher TIE (from 0.48 to 0.55 for the NPP with $\xi_\gamma = 0.5$ and $\sigma_\gamma = 0.2$) and a lower coverage. More events reversed these changes. The PDCCPP kept its TIE at 0.063 to 0.068 in every design, at a cost in power under a consistent treatment effect when fewer events were observed (from 0.76 to 0.67). 
The results of each design are given in Supplementary Table~\ref{tab:supp_teriflunomide}.

\subsubsection{Misspecification of the analysis model}
\label{sec:results_misspecification}

We asked whether the ordering of the borrowing methods survives a misspecification of the analysis model. In the teriflunomide case study, when the treatment effect started 12 or 24 weeks after randomization, the hazards were not proportional and the Cox model was misspecified. 
In the teriflunomide case study with 123 participants per arm, the ordering of the 56 configurations was preserved under the misspecified Cox model. Under a consistent or partially consistent treatment effect, the Spearman correlation between the primary design and the designs with a delayed effect was at least 0.998 for the power and the MSE. At the method level, the ranking of the excess MSE over the best MSE attainable at the same TIE was unchanged. Under no treatment effect, a delayed effect is no effect, so the TIE, the MSE and the coverage were those of the primary design. The conclusions at a matched TIE also held: 90\% to 91\% of the power differences with the reference were smaller than 0.01 (92.5\% in the primary design), no method lost more than 0.006, and the largest gain was 0.047 (0.039 in the primary design). The ordering was also preserved with Weibull event times. 
By contrast, in the aprepitant case study the ordering was only partly preserved between the normal and the binomial analyses, mainly because tuning parameters expressed on the scale of the treatment effect do not have the same meaning on the log odds ratio and risk-difference scales. However, the conclusions at a matched TIE were unchanged.

\section{Discussion}
\label{sec:discussion}

Despite interest in borrowing information to support trials with limited recruitment, applications that borrow directly on treatment effects in rare diseases or pediatric development remain limited \parencite{partington_design_2022}. 
One challenge is that analytical evaluation and calibration of the TIE are often difficult for dynamic borrowing procedures, in particular when using non-conjugate models (see however \textcite{nikolakopoulos_dynamic_2018, calderazzo_decisiontheoretic_2022a}).

The operating characteristics of a proposed borrowing procedure should therefore be evaluated in simulations tailored to the clinical setting and available source evidence. This is important both for comparing existing methods, selecting a prior and its hyperparameters (e.g., between-trials variance, power parameter, mixture weights), and assessing potential reductions in the required sample size. However, statistical recommendations are lacking for simulation studies specifically tailored to the problem of treatment effect borrowing.

In this paper, we report a simulation study motivated by published clinical examples and comparing several existing methods within a common framework, across target sample sizes, treatment effects, target variances, source control rates, endpoint types (continuous, binary, recurrent events and time to event) and model parameters, and, in the time-to-event case study, attrition, event-time distributions, delayed treatment effects and control-arm event rates. We did not study patient-level covariate adjustment, a different stage of the extrapolation problem: residual uncertainty about the transportability of the treatment effect may remain after adjustment, because effect modifiers may be unmeasured and adult and pediatric populations may have limited or no common support with respect to age. Extrapolation from adult to pediatric populations therefore requires structural assumptions beyond adjustment for observed covariates.

Restriction to aggregate data also limits the borrowing strategies that can be considered: it does not allow borrowing to be tailored to heterogeneous patterns of exchangeability with the target population, for example when the treatment effect depends on age and some age groups are exchangeable with adults whereas others are not. ICH E11A notes the advantages of using source population IPD where accessible.

Except in the aprepitant case study, which was analyzed with the binomial likelihoods of both arms, the methods were applied to a normal summary likelihood of the estimated treatment effect. These results therefore apply to settings in which the sampling distribution of the treatment effect estimator is adequately approximated by a normal distribution. In very small or sparse trials, where this approximation may fail, an analysis based on a Gaussian summary likelihood should be replaced by, or validated against, a full likelihood analysis. For binary endpoints, the comparison of the two analyses of the aprepitant case study shows that the overall conclusions were unchanged, but that individual methods could behave quite differently (Section~\ref{sec:discussion_binary}).

Some design choices should be kept in mind that may limit the scope of the results. First, the source evidence was fixed at its published estimate and standard error, as it is when a trial is analyzed. The operating characteristics are therefore conditional on the source data. Second, when several source studies were available (belimumab, teriflunomide), we followed the published case studies and combined them into one source, without a model of between-study heterogeneity. With heterogeneous source studies, this overstates the precision of the source, and every method would borrow more than it should. Third, the standard errors of the target estimates were estimated from the data and then treated as known. With small samples, this understates the uncertainty of every method in a similar way. Fourth, each method was evaluated on a finite grid of tuning parameters, and the comparisons at a matched TIE are comparisons between the configurations of this grid, not between the best tunings of each method (Section~\ref{sec:methods_frontier}).

\subsection{Probability of success of borrowing methods}

Borrowing treatment effects almost systematically inflated the TIE, to an extent that strongly depends on methods and method parameters, in agreement with previous literature \parencite{campbell_bayesian_2017}. The inflation was driven by the distance between the source estimate and the null hypothesis: where this distance was large, pooling rejected the null hypothesis almost always when it was true, and every method that borrowed substantially inherited part of this inflation. Methods that adapt the amount of borrowing to the observed conflict reduced the inflation but did not remove it: the conflict between the source and a null target effect is only detectable once the target data are informative enough, which is precisely when borrowing matters least.

Overall, static borrowing methods, in combination with calibration, provide a straightforward way to control the TIE to a pre-specified value. By contrast, it is more difficult to control the TIE of adaptive borrowing methods, although some methods, such as the PDCCPP \parencite{nikolakopoulos_dynamic_2018}, have been proposed that do so (see also \textcite{calderazzo_robust_2022}). The calibration of the PDCCPP relies on closed-form normal approximations, which were only approximately met with time-to-event, recurrent-event and small continuous trials, and its price was a loss of power with small target samples. 

Bayesian borrowing methods are sometimes motivated by potential power gains, with some authors suggesting, in the case of historical control borrowing, that this can be achieved at equivalent or lower TIE \parencite{viele_use_2014,yang_sam_2023}. However, \textcite{kopp-schneider_power_2020} (preceded by \textcite{psioda_bayesian_2019} in the Gaussian case) showed that, in terms of power gain, ``approaches adaptively discounting prior information do not offer any advantage over a fixed amount of borrowing, or no borrowing at all'', when a UMP test exists. Our results agree across six case studies, four endpoint types and fourteen methods: against a frequentist test of the target data alone calibrated to the TIE a borrowing method actually attains, the power gained by borrowing was almost entirely explained by the type I error the method spends, and losses were larger than gains.

The remaining gains do not contradict the result of \textcite{kopp-schneider_power_2020}, because our reference, the Wald test, is the standard analysis of the target data but not necessarily the most powerful test: with the nuisance parameters known, the most powerful test could be much more powerful, and no method exceeded it.
In the belimumab and teriflunomide case studies, most gains depended on the control-arm rate at which the TIE was evaluated. The TIE of the methods that gained power was low at the control rate of the design but higher at nearby control rates, and when the reference test was allowed this higher TIE, most of the gains disappeared. 
In the aprepitant case study, the largest gain was less sensitive to the control rate, but it also fell below 0.02 once the TIE was evaluated over control rates within 0.10 of that of the design.
In the mepolizumab case study, the gains depended less on the control rate: data-adaptive borrowing gained up to about 0.05 over the Wald test of the negative binomial rate ratio at the same largest TIE over the evaluated range of control rates.

Conversely, \textcite{kopp-schneider_simulating_2024} shows that, in some ``extreme borrowing'' cases, Bayesian borrowing leads to non-UMP tests, with a lower power than frequentist methods at equivalent TIE. We observed this for the methods that decide how much to borrow from a test or a threshold on the observed conflict (TtP, the p-PP and the PDCCPP), mostly with small target samples. TtP and the p-PP produced most of the few noticeable gains, together with the NPP (KL), the EBPP and the RMP in the recurrent-event and time-to-event case studies, but TtP, the p-PP and the PDCCPP also produced nearly all large losses.

This has practical implications. First, the power of a borrowing method should be compared at the TIE it actually incurs under the null hypothesis, not at the nominal level of the decision rule. Second, the reference must spend the same TIE as the method: with a $t$ reference distribution applied to approximately normal Wald statistics, we observed that the separate analysis itself would have a spurious power gain of up to 3.7 percentage points compared to the reference analysis. This was observed across the four case studies with non-continuous endpoint. Matching the actual, rather than the nominal, TIE of the reference removed this artifact.
Third, the TIE should be evaluated over the plausible range of the nuisance parameters, such as the control rate. Finally, when a confirmatory claim requires strict type I error control, these methods offer little beyond the separate analysis. Borrowing is therefore best understood as a choice of TIE, informed by the external data, rather than as a way of obtaining power for free. Whether that choice is acceptable is a regulatory question, and our results quantify its cost, as a function of the prior-data conflict, for each method and setting.


\subsection{Comparing the performance of partial extrapolation methods}

The parameters modulating information borrowing control the response of the operating characteristics to drift. For the RMP, the CPP, the commensurate power prior and TtP they span the range from no borrowing to pooling, whereas the NPP never fully pools the source and target data. Because these parameters have no common scale across methods, we compared the methods at a matched TIE (Section~\ref{sec:methods_frontier}), by analogy with comparisons of power at a common TIE \parencite{kopp-schneider_power_2020}.

The benefit of borrowing is clearer for estimation than for testing, but it is conditional on the consistency of the source and target studies. For most methods, borrowing was beneficial to estimation only while the drift was smaller than about one and a half target standard errors (about two for the commensurate priors), and under conflict most methods increased the MSE and strongly borrowing methods lost coverage almost entirely. Adaptive borrowing methods mitigate, but do not remove, this risk. The commensurate priors offered the most favorable balance, improving estimation under consistency without degrading it under conflict, although their gains were more modest than those of more aggressive methods, and this robustness was lost with a vague $\operatorname{Cauchy}(0, 30)$ hyperprior. The choice of a borrowing method is therefore mainly a choice of how much to rely on the consistency of the two studies, and should be justified by external evidence on that consistency rather than by the simulated operating characteristics alone.

Coverage and precision alone can be misleading for borrowing methods. Indeed, borrowing narrows the credible interval, which improves precision when the studies agree, but narrows interval on a biased estimate in case of prior-data conflict. The interval score \parencite{gneiting_strictly_2007} measures both properties: it adds a penalty for the width of the interval and a penalty when the interval excludes the true treatment effect. Under no treatment effect, only the commensurate priors had a median interval score below that of the separate analysis (median ratio 0.99). For all other methods, the narrower intervals did not compensate for the loss of coverage. Their median ratio was between 1.09 (CPP) and 5.4 (pooling). Most configurations with a lower interval score than the separate analysis were in the aprepitant and botox case studies. In these case studies, the source estimate is close to the null hypothesis, relative to the target standard error. At a matched TIE, the CPP, the NPP and the commensurate priors had the lowest interval score.

Overall, methods did not behave equally for a given increase in TIE. TtP had a larger MSE and interval score than the other methods at a similar TIE, the earliest break-even drift and, with small target samples, a lower power. 
At a similar moment-based ESS, the p-PP, TtP, the EBPP and the PDCCPP had a much larger MSE than the CPP and the NPP. These methods decide from the target data themselves how much to borrow, with full or almost full borrowing when the observed conflict is small and no or almost no borrowing otherwise. A method that borrows the same amount on average but uniformly across trials, such as the CPP, achieves a lower MSE.  These elements provide a strong argument against the use of TtP, and call for caution with the p-PP, the EBPP and the PDCCPP.

No method performed best on power, estimation accuracy and coverage at once. A surprising result is the good performance of the CPP, a fixed borrowing method: at a matched TIE, the CPP, the commensurate priors and, to a lesser extent, the NPP had the lowest MSE and interval score. 
However, under conflict, the intervals of the CPP were worse than those of the separate analysis.
 The leading method varied between case studies. One may expect adaptive methods to incur a lower MSE under drift, but a comparison at an equivalent TIE is performed after adaptation: it compares methods that, on average, borrow similar amounts under the null, which is not the same as comparing them at a similar prior ESS (the orderings at a matched ESS differ somewhat, Table~\ref{tab:frontier_distance}). The advantage of fixed borrowing does not extend to coverage under conflict, where the CPP with $\gamma = 0.5$ lost most of its coverage in the belimumab case study.

The NPP (KL) \parencite{shen_optimal_2024} and the empirical RMP \parencite{egidi_avoiding_2022}
are best described as hyperparameter-selection variants rather than as distinct borrowing families. The design-stage calibration did not change the behavior of the NPP materially, and the formal prior-predictive conflict rule of the empirical RMP did not protect better than a fixed mixture weight (Table~\ref{tab:method_comparison}).

Table~\ref{tab:recommendations} gathers these results into practical guidance for  statisticians, method by method. The choice of method and hyperparameters depends on the design of the study, and a simulation study tailored to the specific study is recommended.
Note also that, beyond their simulated performance, the methods rest on different assumptions: some, such as the CPP, assume that the source and target treatment effects are the same, whereas others include separate parameters for both and rely on their exchangeability.

\begin{table}[ht]
\centering
\scriptsize
\setlength{\tabcolsep}{3pt}
\renewcommand{\arraystretch}{1.15}

\begin{tabularx}{\textwidth}{@{}
    >{\raggedright\arraybackslash}p{1.9cm}
    >{\raggedright\arraybackslash}p{2.7cm}
    >{\raggedright\arraybackslash}X
    >{\raggedright\arraybackslash}p{2.0cm}
@{}}
\hline
\raggedright Method & \raggedright What it does & \raggedright When to use / avoid & \raggedright Recommended settings \tabularnewline
\hline
\raggedright Separate analysis & \raggedright Ignores the source study. & \raggedright Reference analysis; use when the relevance of the source is doubtful or the TIE must not exceed its nominal level. & \raggedright -- \tabularnewline
\raggedright Pooling; conditional PP & \raggedright Down-weights the source likelihood by a fixed power $\gamma$ ($\gamma = 1$: pooling). & \raggedright Use when the source is judged highly relevant and an increased TIE (of about 0.15) is acceptable. This is the most efficient method at matched TIE. Avoid when conflict is plausible, since it does not adapt. & \raggedright $\gamma \le 0.25$ \tabularnewline
\raggedright RMP & \raggedright Mixture of the informative prior (weight $w$) and a vague component. & \raggedright Use when a moderate, adaptive discount with a small TIE inflation (typically below 0.07) is wanted. Less efficient at matched TIE than the conditional PP or the commensurate priors. & \raggedright $w = 0.1$--$0.3$ \tabularnewline
\raggedright NPP; NPP (KL) & \raggedright Power parameter with its own prior (NPP), or set from a Kullback--Leibler criterion. & \raggedright Use only when a moderate TIE increase (at least 0.10) is acceptable. Close to the best MSE and interval score at matched TIE. Avoid when the TIE must stay near 0.05: no setting reaches it. & \raggedright -- \tabularnewline
\raggedright EBPP; empirical RMP (Egidi) & \raggedright Weight of the source estimated from the data. & \raggedright Avoid: large TIE with no setting to reduce it, and poor accuracy at matched TIE. & \raggedright -- \tabularnewline
\raggedright Commensurate PP; commensurate prior & \raggedright Hierarchical: the commensurability parameter, with its own hyperprior, controls borrowing. & \raggedright Use as the default when robustness to conflict matters more than the size of the gain. Avoid vague hyperpriors (Cauchy(0, 30), IG(0.14, 1)), which borrow more but lose this robustness. & \raggedright HN(5) or IG(0.33, 1) \tabularnewline
\raggedright $p$-value-based PP & \raggedright Power from the $p$-value of a test of conflict. & \raggedright Avoid, in particular with small target samples. & \raggedright ($k = 1$, $\lambda = 0.1$) \tabularnewline
\raggedright PDCCPP & \raggedright Calibrates borrowing so that the TIE reaches a target value. & \raggedright Use when a TIE fixed by design is required and the target sample is not small. Avoid with small target samples, due to power loss. & \raggedright target TIE as needed \tabularnewline
\hline
\end{tabularx}
\caption{Practical guidance for choosing a borrowing method.}
\label{tab:recommendations}
\end{table}

The amount of information in the target trial, and not only its sample size, determines the influence of the prior: in the teriflunomide case study, designs that observed fewer events made the source trials more influential and raised the TIE of the adaptive priors. 
 Holding the control-arm response fixed isolated the impact of treatment-effect drift, but the control-arm rate is a separate source of heterogeneity: even when it leaves the relative treatment effect unchanged, it alters the information observed in the target trial, and it can change the TIE of a borrowing method substantially, as in the belimumab and teriflunomide case studies.

\subsection{Effect of a normal approximation for binary endpoints}
\label{sec:discussion_binary}

For binary endpoints, the conclusion on the trade-off between power and TIE did not depend on how the data were analyzed: the aprepitant case study gave the same picture at a matched TIE whether it was analyzed with a normal approximation on the log odds ratio scale, like the belimumab case study, or with the binomial likelihoods of both arms on the risk-difference scale. The choice mattered for individual methods, in two ways. First, tuning parameters expressed on the scale of the treatment effect, such as equivalence margins or hyperpriors on the between-study heterogeneity, do not transfer from one effect measure to another, and must be chosen or calibrated for the measure used. The choice of measure is itself a substantive one, since relative measures such as the odds ratio are often more stable across populations with different baseline risks. 
Second, calibrated methods such as the PDCCPP attained their target TIE only with the exact likelihood. Because the two analyses differ both in the likelihood and in the effect scale, our comparison cannot separate these two sources of difference. We nevertheless recommend analyzing binary endpoints with the exact binomial likelihood whenever the method allows it, and we derived binomial versions of every method that previously relied on the normal approximation. 
Note also that, even when the analysis itself uses the binomial likelihood,  the tests that decide how much to borrow in TtP and the p-PP, and the rule that sets the power parameter of the PDCCPP, use the estimated treatment effects and their standard errors, as in the original methods.


\subsection*{Code availability}
 
The R code developed in this study is available as a GitHub repository at \url{https://github.com/TristanFauvel/BExTE}. It can be used to run and analyze simulation studies and for analyzing data using partial extrapolation.  
Whereas many of the methods investigated in this paper are also implemented in the \textit{hdbayes} package \parencite{alt_hdbayes_2025}, this implementation relies on MCMC, whereas for computational speed, we relied on analytical inference or numerical integration.
  
The code was reviewed by a team member who did not directly participate in the implementation. 
Statistical accuracy of the results was validated based on manual checks, involving a comparison of the figures produced with relevant published figures.  

\subsection*{Funding}

This project was funded through the reopening of competition no. 02 under framework contract following procurement procedure EMA/2020/46/TDA (Lot 3).

\subsection*{AI use statement}
LLM-based coding assistants were used for the optimization and review of the code. Additionally, commercially available LLMs were used for manuscript proofreading and for providing feedback on the manuscript.  

\subsection*{Acknowledgements}
We thank EMA experts Dr Juan José Abéllan Andrés and Dr Andrew Thompson for guidance and help in the planning, execution and reporting of this study.  

This work was granted access to the HPC/AI resources of the French National Computer Center for Higher Education  (CINES) under the allocation 2024-AD010315186 made by the Grand Equipement National de Calcul Intensif (GENCI).

\clearpage
\newpage

\printbibliography


\newpage

\appendix 
\section*{Supplementary material}
\label{appendix}
\renewcommand{\thefigure}{S\arabic{figure}}
\renewcommand{\thetable}{S\arabic{table}}

\setcounter{figure}{0}
\setcounter{table}{0}

\startcontents[appendices] 
\printcontents[appendices]{}{1}{\setcounter{tocdepth}{2}}

\section{Supplementary methods}
 \subsection{Definition of the drift ranges}
 \label{supp:definition_drift_ranges}

 With an adaptive borrowing method, the probability of meeting the decision criterion, $\operatorname{Pr}(\text{Study success}\mid\mathbf{D}_T = d_T, \mathbf{D}_S= d_S)$, is expected to reach a maximum at some drift value, beyond which the source data start being discarded. To study adaptive borrowing methods, the range of drift must therefore be wide enough for this discarding to be observed.

When the likelihood is Gaussian, we reasoned as follows: if the source and target estimates of the treatment effect barely overlap, the source study should be discarded. We quantified this overlap with the Hellinger distance between $\mathcal{N}\left(\hat{\theta}_S, \hat{\sigma}^2_{\theta_S}\right)$ and $\mathcal{N}\left(\hat{\theta}_S + \delta, (\hat{\sigma}^{\mathrm{pub}}_{\theta_T})^2\right)$, where $\hat{\sigma}_{\theta_S}$ is the standard error of the source estimate and $\hat{\sigma}^{\mathrm{pub}}_{\theta_T}$ is the standard error of the treatment effect in the published target trial of the case study (Section~\ref{sec:case_studies}). We determined the drift $\delta_{0.9} > 0$ at which this distance reaches 0.9. Beyond such a drift, borrowing from the source data can be considered futile. Since the Hellinger distance is symmetric in $\delta$, this defines the symmetric range $\mathscr{R} = [-\delta_{0.9}, \delta_{0.9}]$.

We are interested in a range where the target treatment effect lies between the source estimate and the null value, corresponding to drifts between $0$ and $\theta_0 - \hat{\theta}_S$. This interval is negative when the null hypothesis is $\theta_T \leq \theta_0$ (botox, dapagliflozin, belimumab, aprepitant) and positive when it is $\theta_T \geq \theta_0$ (teriflunomide, mepolizumab). To guarantee that it is covered, we used the smallest interval containing both $\mathscr{R}$ and this range:
\[
\left[\min\left(-\delta_{0.9},\ \theta_0 - \hat{\theta}_S\right),\ \max\left(\delta_{0.9},\ \theta_0 - \hat{\theta}_S\right)\right].
\]
In all case studies considered, the range of interest was already contained in $\mathscr{R}$.

Because $\hat{\sigma}^{\mathrm{pub}}_{\theta_T}$ is taken from the published target trial rather than from each simulated design, a single drift range was used for all scenarios of a given case study. 

When the treatment effect is a difference of response rates, $\theta_T = p_T^{(t)} - p_T^{(c)}$, where $p_T^{(a)}$ is the response rate in arm $a$ of the target trial, $\theta_T$ lies in $[-1,1]$, so the drift lies in $[-1-\hat{\theta}_S,\ 1-\hat{\theta}_S]$. Both response rates must also lie in $[0,1]$. Since we assume $p_T^{(c)} = \hat{p}_S^{(c)}$, we have
\[
p_T^{(t)} = \theta_T + p_T^{(c)} = \delta + \hat{\theta}_S + \hat{p}_S^{(c)} = \delta + \hat{p}_S^{(t)},
\]
which implies $-\hat{p}_S^{(t)} \leq \delta \leq 1 - \hat{p}_S^{(t)}$. Combining the two constraints, the drift range is
\[
\left[\max\left(-1-\hat{\theta}_S,\ -\hat{p}_S^{(t)}\right),\ \min\left(1-\hat{\theta}_S,\ 1 - \hat{p}_S^{(t)}\right)\right],
\]
and the Hellinger criterion is not used.

In each case, we considered 30 evenly spaced drift values over the range, to which we added the drift values of the three principal scenarios (no treatment effect, partially consistent and consistent treatment effect), for 33 values in total. The resulting ranges are listed in Supplementary Table~\ref{tab:drift_ranges_and_sample_sizes}.

\subsection{KL-calibrated normalized power prior}
\label{supp:npp_kl}

This section gives the calibration criterion of the KL-calibrated normalized power prior \parencite{shen_optimal_2024}, summarized in Section~\ref{subsec:NPP} of the main text.

In this variant, the shape parameters $a_\gamma$ and $b_\gamma$ of the beta prior $\gamma\sim\operatorname{Beta}(a_\gamma,b_\gamma)$ are selected by minimizing: 

$$
K(a_\gamma,b_\gamma)
=
\lambda_{\mathrm{KL}}\,
\operatorname{KL}\!\left[
p_{0}(\gamma\mid a_\gamma,b_\gamma),
\operatorname{Beta}(c,1)
\right]
+
(1-\lambda_{\mathrm{KL}})\,
\operatorname{KL}\!\left[
p_{\mathrm{MTD}}(\gamma\mid a_\gamma,b_\gamma),
\operatorname{Beta}(1,c)
\right],
$$

where $p_0(\gamma\mid a_\gamma,b_\gamma)$ and $p_{\mathrm{MTD}}(\gamma\mid a_\gamma,b_\gamma)$ denote the hypothetical marginal posterior distributions of $\gamma$ under absence of observed drift, and a prespecified maximum tolerable discrepancy (MTD), respectively. For a normally distributed source treatment effect estimate, these distributions are given, for $d\in\{0,d_{\mathrm{MTD}}\}$, by: 

$$
p_d(\gamma\mid a_\gamma,b_\gamma)
\propto
\mathcal{N}\!\left(
\hat{\theta}_S-b_{\mathrm{benefit}}d
\,\middle|\,
\hat{\theta}_S,\,
\sigma_{\theta_T}^2+\frac{\hat{\sigma}_{\theta_S}^2}{\gamma}
\right)
\operatorname{Beta}(\gamma\mid a_\gamma,b_\gamma),
$$

where $\hat{\sigma}_{\theta_S}$ is the standard error of the source estimate and $\sigma_{\theta_T}$ the design-expected standard error of the target estimate. $b_{\mathrm{benefit}}=1$ when larger treatment effects are beneficial and $b_{\mathrm{benefit}}=-1$ when smaller effects are beneficial. 

Thus, $p_0$ is obtained by setting $\hat{\theta}_T=\hat{\theta}_S$, whereas $p_{\mathrm{MTD}}$ is obtained by setting $\hat{\theta}_T=\hat{\theta}_S-b_{\mathrm{benefit}}d_{\mathrm{MTD}}$. The target distributions $\operatorname{Beta}(c,1)$ and $\operatorname{Beta}(1,c)$ represent the desired concentration of the posterior discounting parameter near one under absence of drift and near zero at the maximum tolerable discrepancy, respectively.

We used the default values $\lambda_{\mathrm{KL}}=0.5$ and $c=10$ proposed by \textcite{shen_optimal_2024}. We defined the maximum tolerable discrepancy as the distance between the source estimate and the null value,

$$
d_{\mathrm{MTD}}= k_\text{MTD}
\left|\hat{\theta}_S-\theta_0\right|,
$$
with $k_\text{MTD} \in \{0.5,1\}$.
With $k_\text{MTD} = 1$, the hypothetical target estimate under the maximum tolerable discrepancy corresponds to the null value. Therefore, calibration implies that there will be almost no borrowing when there is no treatment effect in the target population. $k_\text{MTD}$ = 0.5 yields more conservative borrowing.

The objective function $K$ was evaluated by numerical integration and minimized over $a_\gamma,b_\gamma>0$. For each design scenario, the beta parameters were calibrated once using the design-expected target standard error and were subsequently held fixed across simulation replicates.

\subsection{Empirical Bayes power prior in closed form}
\label{supp:ebpp_closed_form}

\textcite{gravestock_adaptive_2017} derive the empirical Bayes estimate in closed form for a normal likelihood with a flat initial prior on $\theta_T$:

\begin{equation}
\hat{\gamma}=\frac{\hat{\sigma}_{\theta_S}^2}{\max \left\{\left(\hat{\theta}_T-\hat{\theta}_S\right)^2, \hat{\sigma}_{\theta_T}^2+ \hat{\sigma}_{\theta_S}^2\right\}-\hat{\sigma}_{\theta_T}^2},
\end{equation}

where the max is required to restrict $\hat{\gamma} \leq 1$.  Under the same prior and the same likelihood, the empirical Bayes posterior distribution is given by:

\begin{equation}
p\left(\theta_T \mid \hat{\theta}_T,\hat{\theta}_S, \gamma=\hat{\gamma}\right) \propto \begin{cases}\mathcal{N}\left(\theta_T \mid \hat{\theta}_T, \hat{\sigma}_{\theta_T}^2\right) \times \mathcal{N}\left(\theta_T \mid\hat{\theta}_S,\left(\hat{\theta}_T-\hat{\theta}_S\right)^2-\hat{\sigma}_{\theta_T}^2\right) & \text { if }\left(\hat{\theta}_S-\hat{\theta}_T\right)^2>\hat{\sigma}_{\theta_T}^2+\hat{\sigma}_{\theta_S}^2 \\ \mathcal{N}\left(\theta_T \mid \hat{\theta}_T, \hat{\sigma}_{\theta_T}^2\right) \times \mathcal{N}\left(\theta_T \mid\hat{\theta}_S, \hat{\sigma}_{\theta_S}^2\right) & \text { otherwise. }\end{cases}
\end{equation}

\subsection{Empirical robust mixture prior}
\label{supp:egidi}

With the prior $\pi_{\psi}(\theta_T) = (1-\psi)\pi_{\mathrm{S}}(\theta_T) + \psi\pi_{\mathrm{W}}(\theta_T)$ of the main text, the weight $\psi$ of the weakly informative component is selected as follows; $1 - \psi$ plays the role of the weight $w$ of the informative component in the RMP. Let $T(\mathbf{D}_T)=t_{\mathrm{obs}}$ denote a sufficient statistic for the target data. The prior-predictive distributions of this statistic under the informative and weakly informative components are, respectively, $m_{\mathrm{S}}(t) =
\int p(t\mid\theta_T)\pi_{\mathrm{S}}(\theta_T)\,d\theta_T
$, and $m_{\mathrm{W}}(t)
= \int p(t\mid\theta_T)\pi_{\mathrm{W}}(\theta_T)\,d\theta_T$.
The prior-predictive distribution under the mixture prior is therefore $m_{\psi}(t)
= (1-\psi)m_{\mathrm{S}}(t) + \psi m_{\mathrm{W}}(t)$.

For a given value of $\psi$, prior-data conflict was quantified using the prior-predictive $p$-value

$$
P_{\psi}(t_{\mathrm{obs}})
=
\Pr_{T\sim m_{\psi}}
\left(m_{\psi}(T)
\leq
m_{\psi}(t_{\mathrm{obs}})
\right).
$$

The weak-component weight was selected as the smallest value for which the observed target statistic was no longer considered to conflict with the mixture prior:

$$
\widehat{\psi}
=
\inf\left\{
\psi\in[0,1]:
P_{\psi}(t_{\mathrm{obs}})
\geq\alpha_{\mathrm{PC}}
\right\}.
$$

We used $\alpha_{\mathrm{PC}}=0.05$, as in \textcite{egidi_avoiding_2022}. Consequently, the method retained the source-based prior in the absence of prior-data conflict and introduced only the minimum amount of the weakly informative component required to satisfy the conflict criterion. When the criterion could not be satisfied for any $\psi<1$, we set $\widehat{\psi}=1$.

If $\pi_{\mathrm{S}}(\theta_T)
= \mathcal{N}(\mu_{\mathrm{S}},v_{\mathrm{S}})$  and $\pi_{\mathrm{W}}(\theta_T)
= \mathcal{N}(\mu_{\mathrm{W}},v_{\mathrm{W}})$, the corresponding component prior-predictive distributions are:

$$
m_{\mathrm{S}}(t)
=
\mathcal{N}\!\left(
t\mid\mu_{\mathrm{S}},
\hat{\sigma}_{\theta_T}^2+v_{\mathrm{S}}
\right) \quad \text{and} \quad m_{\mathrm{W}}(t)
=
\mathcal{N}\!\left(
t\mid\mu_{\mathrm{W}},
\hat{\sigma}_{\theta_T}^2+v_{\mathrm{W}}
\right).
$$

The prior-predictive $p$-value and $\widehat{\psi}$ were computed numerically.  $\widehat{\psi}$ was estimated separately in every simulation replicate because it is explicitly a function of the observed target data.
After selecting $\widehat{\psi}$, the resulting mixture prior was updated using the target likelihood. In particular, the posterior weight assigned to the informative component was:

$$
\tilde{w}
=
\frac{
(1-\widehat{\psi})m_{\mathrm{S}}(t_{\mathrm{obs}})
}{
(1-\widehat{\psi})m_{\mathrm{S}}(t_{\mathrm{obs}})
+
\widehat{\psi}m_{\mathrm{W}}(t_{\mathrm{obs}})
}.
$$

Thus, $1-\widehat{\psi}$ represents the empirically selected prior weight of the informative component, whereas $\tilde{w}$ is its posterior weight after updating with the target data.

\subsection{Computation of the posterior distributions}
\label{supp:posterior_computation}

With a normal prior, the posterior is normal and available in closed form. This covers the separate and pooled analyses, the conditional power prior, the test-then-pool variants, the p-value-based power prior, the empirical Bayes power prior and the PDCCPP, whose data-dependent power parameters only set the variance of a normal prior. The robust mixture prior and its variant with an empirically selected weight \parencite{egidi_avoiding_2022} have normal mixture priors, whose posterior is again a normal mixture with closed-form component weights. For the normalized power prior, the prior on the power parameter was integrated by Gauss-Jacobi quadrature with 60 nodes, which turns the prior into a finite normal mixture (Supplementary Section~\ref{supp:normalized_power_prior}). For the commensurate power prior and the commensurate prior, the commensurability parameter $\tau$ was integrated by quadrature with 48 nodes (144 for the $\operatorname{Cauchy}(0, 30)$ prior on $\log\tau$, whose scale is wider) and, for the commensurate power prior, the power parameter given $\tau$ with 12 nodes, again yielding a finite normal mixture. Quantiles of normal mixtures, and hence credible intervals, were obtained by a safeguarded Newton iteration on the mixture distribution function, to an absolute tolerance of $10^{-12}$.

\subsection{Binomial model (aprepitant)}

Figure \ref{fig:graphical_model_binomial_likelihood} represents the structure of the model used in the aprepitant case study.

\begin{figure}[htbp]
    \centering
    \tikzset{every picture/.style={line width=0.75pt}} 

\begin{tikzpicture}[x=0.75pt,y=0.75pt,yscale=-1,xscale=1]

\draw    (105, 194.5) circle [x radius= 90.51, y radius= 21.92]   ;
\draw (42,181.4) node [anchor=north west][inner sep=0.75pt]    {$y_{T}^{(c)} \sim \operatorname{Bin}\left( N_{T}^{(c)}, p_{T}^{(c)}\right)$};
\draw    (105.5, 65.5) circle [x radius= 71.42, y radius= 17.68]   ;
\draw (56,55.4) node [anchor=north west][inner sep=0.75pt]    {$p_{T}^{(c)} \sim \mathcal{U}( 0,1)$};
\draw    (299.5, 64) circle [x radius= 64.35, y radius= 15.56]   ;
\draw (255,55.4) node [anchor=north west][inner sep=0.75pt]    {$\theta _{T} \sim \pi( \theta_T \mid \mathbf{D}_{S})$};
\draw    (300, 195.5) circle [x radius= 80.61, y radius= 17.68]   ;
\draw (244,185.4) node [anchor=north west][inner sep=0.75pt]    {$p_{T}^{(t)} = p_{T}^{(c)} + \theta _{T}$};
\draw    (300, 273.5) circle [x radius= 90.51, y radius= 21.92]   ;
\draw (237,260.4) node [anchor=north west][inner sep=0.75pt]    {$y_{T}^{(t)} \sim \operatorname{Bin}\left( N_{T}^{(t)}, p_{T}^{(t)}\right)$};
\draw    (105.93,84.38) -- (105.6,169.58) ;
\draw [shift={(105.59,172.58)}, rotate = 270.22] [fill={rgb, 255:red, 0; green, 0; blue, 0 }  ][line width=0.08]  [draw opacity=0] (8.93,-4.29) -- (0,0) -- (8.93,4.29) -- cycle    ;
\draw    (300.12,214.38) -- (300.34,248.58) ;
\draw [shift={(300.36,251.58)}, rotate = 269.63] [fill={rgb, 255:red, 0; green, 0; blue, 0 }  ][line width=0.08]  [draw opacity=0] (8.93,-4.29) -- (0,0) -- (8.93,4.29) -- cycle    ;
\draw    (131.65,83.19) -- (271.53,176.92) ;
\draw [shift={(274.02,178.59)}, rotate = 213.83] [fill={rgb, 255:red, 0; green, 0; blue, 0 }  ][line width=0.08]  [draw opacity=0] (8.93,-4.29) -- (0,0) -- (8.93,4.29) -- cycle    ;
\draw    (300,79.56) -- (300,174.62) ;
\draw [shift={(300,177.62)}, rotate = 270] [fill={rgb, 255:red, 0; green, 0; blue, 0 }  ][line width=0.08]  [draw opacity=0] (8.93,-4.29) -- (0,0) -- (8.93,4.29) -- cycle    ;

\end{tikzpicture}
\caption{Structure of the binomial model of the aprepitant case study. In the separate analysis, $\pi(\theta_T \mid \mathbf{D}_S)$ is the initial prior; the other methods use the joint priors described in Section~\ref{sec:binomial_adaptation}.}
    \label{fig:graphical_model_binomial_likelihood}
\end{figure}

\subsection{Normalized power prior}
\label{supp:normalized_power_prior}

\subsubsection{Gaussian normalized power prior}

\paragraph{Prior.} The Gaussian normalized power prior is (\textcite{pawel_normalized_2023},  appendix A of \textcite{gravestock_adaptive_2017}):

$$
\pi\left(\theta_T, \gamma \mid \mathbf{D}_S\right)=\frac{\mathscr{L}\left(\theta_T \mid \mathbf{D}_S\right)^\gamma \pi(\gamma)}{\int_{-\infty}^{+\infty} \mathscr{L}\left(\theta_T^{\prime} \mid \mathbf{D}_S\right)^\gamma d \theta_T^{\prime}}=\mathcal{N}\left(\theta_T \mid \hat{\theta}_S, \hat{\sigma}_{\theta_S}^2 / \gamma\right) \operatorname{Beta}(\gamma \mid a_\gamma, b_\gamma)
$$
The marginal prior on $\theta_T$ is:

\begin{equation}
\begin{aligned}
\pi\left(\theta_T\mid \mathbf{D}_S\right) &= \int_0^1 \mathcal{N}\left(\theta_T \mid \hat{\theta}_S, \hat{\sigma}_{\theta_S}^2 / \gamma\right) \operatorname{Beta}(\gamma \mid a_\gamma, b_\gamma)d\gamma \\
&= \frac{\mathrm{B}\left(a_\gamma+\frac{1}{2}, b_\gamma\right)}{\mathrm{B}(a_\gamma, b_\gamma)\sqrt{2 \pi \hat{\sigma}_{\theta_S}^2}}\,
\mathrm{M}\left(a_\gamma+\frac{1}{2},\, a_\gamma+b_\gamma+\frac{1}{2},\, -\frac{\left(\hat{\theta}_S-\theta_T\right)^2}{2 \hat{\sigma}_{\theta_S}^2}\right),
\end{aligned}
\end{equation}

where $\mathrm{M}(\alpha, \beta, z)=\frac{\Gamma(\beta)}{\Gamma(\alpha)\,\Gamma(\beta-\alpha)} \int_0^1 e^{z t}\, t^{\alpha-1}(1-t)^{\beta-\alpha-1}\, d t$ is Kummer's confluent hypergeometric function, which is implemented in standard numerical libraries.

\paragraph{Posterior.}
Combining the joint prior $\pi\left(\theta_T, \gamma \mid \mathbf{D}_S\right)$ with the likelihood of the target study data produces a joint posterior for $\theta_T$ and $\gamma$, that is,
\begin{equation}
\begin{aligned}
\pi\left(\theta_T, \gamma \mid \mathbf{D}_T, \mathbf{D}_S\right) &= \frac{\mathscr{L}(\theta_T \mid \mathbf{D}_T) \pi\left(\theta_T, \gamma \mid \mathbf{D}_S\right)}{\int_0^1 \int_{-\infty}^{\infty} \mathscr{L}\left(\theta_T^{\prime} \mid \mathbf{D}_T\right) \pi\left(\theta_T^{\prime}, \gamma^{\prime} \mid \mathbf{D}_S\right) d \theta_T^{\prime} d \gamma^{\prime}} \\
&=\frac{\mathcal{N}\left(\hat{\theta}_T \mid \theta_T, \hat{\sigma}_{\theta_T}^2\right) \mathcal{N}\left(\theta_T \mid \hat{\theta}_S, \hat{\sigma}_{\theta_S}^2 / \gamma\right) \operatorname{Beta}(\gamma \mid a_\gamma, b_\gamma)}{\int_0^1 \mathcal{N}\left(\hat{\theta}_T \mid \hat{\theta}_S, \hat{\sigma}_{\theta_T}^2+\hat{\sigma}_{\theta_S}^2 / \gamma^{\prime}\right) \operatorname{Beta}\left(\gamma^{\prime} \mid a_\gamma, b_\gamma\right) d \gamma^{\prime}},
\end{aligned}
\end{equation}

from which a marginal posterior for $\gamma$ can be obtained by integrating out $\theta_T$, that is,
\begin{equation}
    \begin{aligned}
        \pi\left(\gamma \mid \mathbf{D}_T, \mathbf{D}_S\right) &=\int_{-\infty}^{+\infty} \pi\left(\theta_T, \gamma \mid \mathbf{D}_T, \mathbf{D}_S\right) d \theta_T \\
&= \frac{\mathcal{N}\left(\hat{\theta}_T \mid \hat{\theta}_S, \hat{\sigma}_{\theta_T}^2+\hat{\sigma}_{\theta_S}^2 / \gamma\right) \operatorname{Beta}(\gamma \mid a_\gamma, b_\gamma)}{\int_0^1 \mathcal{N}\left(\hat{\theta}_T \mid \hat{\theta}_S, \hat{\sigma}_{\theta_T}^2+\hat{\sigma}_{\theta_S}^2 / \gamma^{\prime}\right) \operatorname{Beta}\left(\gamma^{\prime} \mid a_\gamma, b_\gamma\right) d \gamma^{\prime}} \\
&\propto \mathcal{N}\left(\hat{\theta}_T \mid \hat{\theta}_S, \hat{\sigma}_{\theta_T}^2+\hat{\sigma}_{\theta_S}^2 / \gamma\right) \operatorname{Beta}(\gamma \mid a_\gamma, b_\gamma).
    \end{aligned}
\end{equation}

The posterior distribution of the power parameter can therefore be approximated using numerical integration.

Moreover, integrating the joint posterior over $\gamma$ gives the marginal posterior of the treatment effect \parencite{gravestock_adaptive_2017}:
\begin{equation}
\begin{aligned}
\pi\left(\theta_T\mid \mathbf{D}_T, \mathbf{D}_S\right)
&= \frac{1}{m(\mathbf{D}_T)}\,\mathcal{N}\left(\hat{\theta}_T \mid \theta_T, \hat{\sigma}_{\theta_T}^2\right)\int_0^1 \mathcal{N}\left(\theta_T \mid \hat{\theta}_S, \hat{\sigma}_{\theta_S}^2 / \gamma\right) \operatorname{Beta}(\gamma \mid a_\gamma, b_\gamma)\, d\gamma\\
&\propto \exp\left(- \frac{( \hat{\theta}_T-  \theta_T)^2}{2\hat{\sigma}_{\theta_T}^2} \right) \mathrm{M}\left(\frac{1}{2}+ a_\gamma, \frac{1}{2}+ a_\gamma+b_\gamma,-\frac{\left(\hat{\theta}_S-\theta_T\right)^2}{2 \hat{\sigma}_{\theta_S}^2}\right),
\end{aligned}
\end{equation}
where
\begin{equation}
m(\mathbf{D}_T) = \int_0^1 \mathcal{N}\left(\hat{\theta}_T \mid \hat{\theta}_S, \hat{\sigma}_{\theta_T}^2+\hat{\sigma}_{\theta_S}^2 / \gamma\right) \operatorname{Beta}(\gamma \mid a_\gamma, b_\gamma)\, d\gamma
\end{equation}
is the marginal likelihood of the target data under the normalized power prior.

\paragraph{Computation.}

When computing the prior distribution $\pi\left(\theta_T \mid \mathbf{D}_S\right)$, marginalizing over $\gamma$ yields a continuous normal mixture:

\begin{equation}
    \begin{aligned}
        \pi\left(\theta_T \mid \mathbf{D}_S\right) &= \int_0^1 \mathcal{N}\left(\theta_T \mid \hat{\theta}_S, \hat{\sigma}_{\theta_S}^2 / \gamma\right) \operatorname{Beta}(\gamma \mid a_\gamma, b_\gamma) d\gamma,
    \end{aligned}
\end{equation}
 The Beta density is $$  f_\gamma(\gamma) = \frac{\gamma^{a_\gamma-1}(1-\gamma)^{b_\gamma-1}}{\mathrm{B}(a_\gamma,b_\gamma)},$$
We set $\gamma=u^2$, the change of variable leads to:
\begin{equation}
    \begin{aligned}
       f_\gamma(u^2)2udu &= \frac{2}{\mathrm{B}(a_\gamma,b_\gamma)}u^{2a_\gamma-1}(1-u^2)^{b_\gamma-1}du \\
       &= \frac{2}{\mathrm{B}(a_\gamma,b_\gamma)}u^{2a_\gamma-1}(1-u)^{b_\gamma-1}(1+u)^{b_\gamma-1}du
    \end{aligned}
\end{equation}
We then map $u\in(0,1)$ to $x=2u-1\in(-1,1)$:
\begin{equation}
    \begin{aligned}
        f_\gamma(\gamma)d\gamma &= \frac{2^{2-2a_\gamma-b_\gamma}}{\mathrm{B}(a_\gamma,b_\gamma)}(1-x)^{b_\gamma-1}(1+x)^{2a_\gamma-1}\left( 1 + \frac{1+x}{2} \right)^{b_\gamma-1}dx \\
    \end{aligned}
\end{equation}

So that we can approximate the mixture by a finite mixture using Gauss-Jacobi quadrature with weight proportional to $(1-x)^{b_\gamma-1}(1+x)^{2a_\gamma-1}$. We used $K= 60$ nodes.

For quadrature nodes $x_k$ and associated Gauss-Jacobi weights $\nu_k$, we define:
\[
u_k=\frac{1+x_k}{2},\qquad
\gamma_k=u_k^2,\qquad
\omega_k=
\frac{\nu_k(1+u_k)^{b_\gamma-1}}
{\sum_{j=1}^{K}\nu_j(1+u_j)^{b_\gamma-1}}.
\]
The resulting approximation is
\[
\pi_{\mathrm{NPP}}(\theta_T\mid \mathbf{D}_S)
\approx
\sum_{k=1}^{K}\omega_k\,
\mathcal{N}\!\left(\theta_T \mid \hat\theta_S,\frac{\hat{\sigma}_{\theta_S}^2}{\gamma_k}
\right).
\]
This mixture was computed once per source-prior configuration and reused across target-trial replicates. Under the Gaussian target summary likelihood $\hat\theta_{T,r}\mid\theta_T\sim
\mathcal{N}(\theta_T,\hat{\sigma}_{\theta_T,r}^{2})$, each component updates conjugately. In replicate $r$, its posterior weight is:
\[
W_{rk}=
\frac{\omega_k\,
\mathcal{N}\!\left(\hat\theta_{T,r} \mid 
\hat\theta_S,\hat{\sigma}_{\theta_S}^2/\gamma_k+
\hat{\sigma}_{\theta_T,r}^{2}\right)}
{\sum_{j=1}^{K}\omega_j\,
\mathcal{N}\!\left(\hat\theta_{T,r} \mid 
\hat\theta_S,\hat{\sigma}_{\theta_S}^2/\gamma_j+
\hat{\sigma}_{\theta_T,r}^{2}\right)}.
\]

\subsubsection{Normalized power prior with binomial likelihoods}
\label{supp:normalized_power_prior_binomial}

In the aprepitant case study, the source study has $y_S^{(c)}$ responders out of $N_S^{(c)}$ patients in the control arm and $y_S^{(t)}$ out of $N_S^{(t)}$ in the treatment arm. $y_T^{(c)}$, $N_T^{(c)}$, $y_T^{(t)}$ and $N_T^{(t)}$ are the same counts in the target study. The model has three parameters: the control response rates $p_S^{(c)}$ and $p_T^{(c)}$ of the source and target studies, and the risk difference $\theta_T$, shared by the two studies, so that the treatment response rates are $p_S^{(c)} + \theta_T$ and $p_T^{(c)} + \theta_T$. Both treatment rates must lie in $[0, 1]$, so the parameter space is
\begin{equation}
\Omega = \left\{ (p_S^{(c)}, p_T^{(c)}, \theta_T) \in [0,1]^2 \times \mathbb{R} : \underline{\theta}(p_S^{(c)}, p_T^{(c)}) \leq \theta_T \leq \overline{\theta}(p_S^{(c)}, p_T^{(c)}) \right\},
\end{equation}
with $\underline{\theta}(p_S^{(c)}, p_T^{(c)}) = -\min(p_S^{(c)}, p_T^{(c)})$ and $\overline{\theta}(p_S^{(c)}, p_T^{(c)}) = 1 - \max(p_S^{(c)}, p_T^{(c)})$. The initial prior is uniform on each control rate and, given the control rates, uniform on the admissible risk differences:
\begin{equation}
\pi_0(p_S^{(c)}, p_T^{(c)}, \theta_T) = \frac{1}{\overline{\theta}(p_S^{(c)}, p_T^{(c)}) - \underline{\theta}(p_S^{(c)}, p_T^{(c)})} = \frac{1}{1 - |p_S^{(c)} - p_T^{(c)}|}, \qquad (p_S^{(c)}, p_T^{(c)}, \theta_T) \in \Omega.
\end{equation}
The log-likelihoods of the source and target studies are
\begin{align}
\log \mathscr{L}_S(p_S^{(c)}, \theta_T)
&= y_S^{(c)}\log p_S^{(c)}
+ (N_S^{(c)}-y_S^{(c)})\log(1-p_S^{(c)}) \\
&\quad + y_S^{(t)}\log(p_S^{(c)}+\theta_T)
+ (N_S^{(t)}-y_S^{(t)})\log(1-p_S^{(c)}-\theta_T), \\
\log \mathscr{L}_T(p_T^{(c)}, \theta_T)
&= y_T^{(c)}\log p_T^{(c)}
+ (N_T^{(c)}-y_T^{(c)})\log(1-p_T^{(c)}) \\
&\quad + y_T^{(t)}\log(p_T^{(c)}+\theta_T)
+ (N_T^{(t)}-y_T^{(t)})\log(1-p_T^{(c)}-\theta_T).
\end{align}
where the binomial coefficients are omitted: they are constant in the parameters, and cancel between the numerator and the denominator of every ratio below.

\paragraph{Prior.} The normalized power prior is
\begin{equation}
\pi(p_S^{(c)}, p_T^{(c)}, \theta_T, \gamma \mid \mathbf{D}_S) = \frac{\mathscr{L}_S(p_S^{(c)}, \theta_T)^{\gamma} \, \pi_0(p_S^{(c)}, p_T^{(c)}, \theta_T)}{Z_S(\gamma)} \, \pi(\gamma)
\end{equation}
with:
\begin{equation}
Z_S(\gamma) = \int_\Omega \mathscr{L}_S(p_S^{(c)}, \theta_T)^{\gamma} \, \pi_0(p_S^{(c)}, p_T^{(c)}, \theta_T) \, \mathrm{d}p_S^{(c)} \, \mathrm{d}p_T^{(c)} \, \mathrm{d}\theta_T,
\end{equation}
with $\gamma \sim \operatorname{Beta}(a_\gamma, b_\gamma)$ as in the normal case. Unlike the normal case with a flat initial prior, $\pi_0$ is proper here, so $Z_S(\gamma)$ is finite for every $\gamma \in [0, 1]$, including $Z_S(0) = 1$.

The target control rate $p_T^{(c)}$ does not appear in $\mathscr{L}_S$, but it cannot be dropped: it sets the admissible range of $\theta_T$, and therefore enters $\pi_0$. It can, however, be integrated out in closed form. Given $p_S^{(c)}$ and $\theta_T$, the admissible values of $p_T^{(c)}$ are those that keep $p_T^{(c)}$ and $p_T^{(c)} + \theta_T$ in $[0,1]$, that is, $p_T^{(c)} \in [a(\theta_T), b(\theta_T)]$ with $a(\theta_T) = \max(0, -\theta_T)$ and $b(\theta_T) = \min(1, 1 - \theta_T)$; the same interval contains $p_S^{(c)}$. Splitting the integral at $p_T^{(c)} = p_S^{(c)}$ gives
\begin{equation}
\chi(p_S^{(c)}, \theta_T) = \int_{a(\theta_T)}^{b(\theta_T)} \frac{\mathrm{d}p_T^{(c)}}{1 - |p_S^{(c)} - p_T^{(c)}|} = -\log\bigl(1 - p_S^{(c)} + a(\theta_T)\bigr) - \log\bigl(1 + p_S^{(c)} - b(\theta_T)\bigr),
\end{equation}
so that
\begin{equation}
Z_S(\gamma) = \iint \mathscr{L}_S(p_S^{(c)}, \theta_T)^{\gamma} \, \chi(p_S^{(c)}, \theta_T) \, \mathrm{d}p_S^{(c)} \, \mathrm{d}\theta_T,
\end{equation}
over $p_S^{(c)} \in [0,1]$ and $p_S^{(c)} + \theta_T \in [0,1]$. In terms of the source treatment rate $p_S^{(c)} + \theta_T$, $\mathscr{L}_S^{\gamma}$ is a product of two beta kernels; without the weight $w$, $Z_S(\gamma)$ would be the product of two beta functions, as for a single proportion. The weight $w$, which comes from the dependence of the support of $\theta_T$ on the control rates, is what makes $Z_S(\gamma)$ a two-dimensional integral without a closed form.

\paragraph{Posterior.} Given the target data, the joint posterior is
\begin{equation}
\pi(p_S^{(c)}, p_T^{(c)}, \theta_T, \gamma \mid \mathbf{D}_S, \mathbf{D}_T) \propto \mathscr{L}_T(p_T^{(c)}, \theta_T) \, \frac{\mathscr{L}_S(p_S^{(c)}, \theta_T)^{\gamma} \, \pi_0(p_S^{(c)}, p_T^{(c)}, \theta_T)}{Z_S(\gamma)} \, \pi(\gamma).
\end{equation}
Define the source kernel, which depends on the source data only,
\begin{equation}
g_\gamma(p_T^{(c)}, \theta_T) = \int \mathscr{L}_S(p_S^{(c)}, \theta_T)^{\gamma} \, \pi_0(p_S^{(c)}, p_T^{(c)}, \theta_T) \, \mathrm{d}p_S^{(c)},
\end{equation}
where the integral runs over the $p_S^{(c)}$ for which $(p_S^{(c)}, p_T^{(c)}, \theta_T) \in \Omega$, and the marginal likelihood of the target data given $\gamma$,
\begin{equation}
Z_T(\gamma) = \iint \mathscr{L}_T(p_T^{(c)}, \theta_T) \, g_\gamma(p_T^{(c)}, \theta_T) \, \mathrm{d}p_T^{(c)} \, \mathrm{d}\theta_T.
\end{equation}
The posterior of the power parameter and the posterior of the risk difference given $\gamma$ are then
\begin{align}
\pi(\gamma \mid \mathbf{D}_S, \mathbf{D}_T) &\propto \pi(\gamma) \, \frac{Z_T(\gamma)}{Z_S(\gamma)}, \\
\pi(\theta_T \mid \gamma, \mathbf{D}_S, \mathbf{D}_T) &= \frac{1}{Z_T(\gamma)} \int \mathscr{L}_T(p_T^{(c)}, \theta_T) \, g_\gamma(p_T^{(c)}, \theta_T) \, \mathrm{d}p_T^{(c)},
\end{align}
and the posterior of the treatment effect is the mixture
\begin{equation}
\pi(\theta_T \mid \mathbf{D}_S, \mathbf{D}_T) = \int_0^1 \pi(\theta_T \mid \gamma, \mathbf{D}_S, \mathbf{D}_T) \, \pi(\gamma \mid \mathbf{D}_S, \mathbf{D}_T) \, \mathrm{d}\gamma.
\end{equation}
The ratio $Z_T(\gamma)/Z_S(\gamma)$ is the prior-predictive probability of the target data under the power prior with power parameter $\gamma$: the posterior of $\gamma$ concentrates on large values when the target data agree with the source data, and on small values when they conflict.

\paragraph{Computation.} The rates and the risk difference are discretized on the lattice of Section~\ref{supp:binomial_lattice}. Integrated over the power parameter, the normalized power prior is a fixed prior on the target control rate and the risk difference: on the lattice, the prior weight of a source control rate $r_i$ and a risk difference $q/M$ is the initial prior mass times
\begin{equation}
h\bigl(\ell(i, q)\bigr) = \int_0^1 \operatorname{Beta}(\gamma \mid a_\gamma, b_\gamma)\, \frac{e^{\gamma \ell(i, q)}}{Z_S(\gamma)}\, \mathrm{d}\gamma,
\end{equation}
where $\ell(i, q)$ is the source log likelihood. Because $h$ is a function of a single scalar, it was tabulated once on 20,001 values of $\ell$ and interpolated. The integral over $\gamma$ was computed on the logit scale, $\gamma = 1/(1 + e^{-t})$, by the trapezoidal rule on $t \in [-30, 30]$ with step $0.05$, the prior mass beyond these limits being assigned to $\gamma = 0$ and $\gamma = 1$; on this scale the integrand is smooth and decays exponentially in both tails for all beta priors, including U-shaped ones, so the rule converges geometrically. $Z_S(\gamma)$ was computed exactly on the lattice at each node. The posterior moments of $\gamma$ follow from the same construction with $h$ weighted by $\gamma$ and $\gamma^2$. The prior is computed once per source and prior setting, and each target dataset then costs a single weighted sum over the lattice. Halving the step in $t$ changed posterior summaries by less than $4 \times 10^{-6}$, and without target data the posterior mean and standard deviation of $\gamma$ equal those of its prior.

\subsubsection{KL-calibrated normalized power prior with binomial likelihoods}
\label{supp:npp_kl_binomial}

In the aprepitant case study, the beta prior of the power parameter was calibrated with the same criterion as in the normal case, which compares the posterior of the power parameter under two hypothetical target results, one compatible with the source and one at the maximum tolerable discrepancy, with two reference beta distributions. With binomial likelihoods, the hypothetical results are the expected responder counts of the design: the control arm at the design's control rate, and the treatment arm at that rate plus the source risk difference (compatible result) or plus the source risk difference shifted by the maximum tolerable discrepancy. For each hypothetical result, the posterior of the power parameter is its beta prior times the marginal likelihood $m(\gamma) = Z_T(\gamma)/Z_S(\gamma)$ of the binomial power prior (Supplementary Section~\ref{supp:empirical_bayes_power_prior_binomial}), which does not depend on the beta prior; it was therefore tabulated once per design and interpolated during the optimization over the shape parameters. The target data were then analyzed with the binomial normalized power prior of Section~\ref{supp:normalized_power_prior_binomial} under the calibrated prior. Because the calibration uses the design's expected counts, the prior does not depend on the true treatment effect of the simulated scenario.

\subsection{Empirical Bayes power prior with binomial likelihoods}
\label{supp:empirical_bayes_power_prior_binomial}

\paragraph{Prior.} The empirical Bayes power prior \parencite{gravestock_adaptive_2017} sets the power parameter to the value that maximizes the marginal likelihood of the target data under the normalized power prior with a fixed power parameter,
\begin{equation}
\hat{\gamma} = \operatorname*{arg\,max}_{\gamma \in [0, 1]} m(\gamma), \qquad m(\gamma) = \int_\Omega \mathscr{L}_T(p_T^{(c)}, \theta_T) \, \frac{\mathscr{L}_S(p_S^{(c)}, \theta_T)^{\gamma} \, \pi_0(p_S^{(c)}, p_T^{(c)}, \theta_T)}{Z_S(\gamma)} \, \mathrm{d}p_S^{(c)} \, \mathrm{d}p_T^{(c)} \, \mathrm{d}\theta_T = \frac{Z_T(\gamma)}{Z_S(\gamma)},
\end{equation}
with the model, the initial prior $\pi_0$ and the normalizing constants $Z_S(\gamma)$ and $Z_T(\gamma)$ of Section~\ref{supp:normalized_power_prior_binomial}.

\paragraph{Posterior.} The posterior is then that of the conditional power prior with $\gamma = \hat{\gamma}$. The marginal likelihood $m(\gamma)$ is the prior-predictive probability of the target data: it is largest at $\gamma = 1$ when the target data agree with the source data, and decreases with $\gamma$ when they conflict. Because the initial prior is proper, $m(0)$ is finite, so $\hat{\gamma} = 0$ is attained when the conflict is strong enough, whereas with a normal likelihood and a flat initial prior $m(\gamma) \to 0$ as $\gamma \to 0$ and $\hat{\gamma} > 0$.

\paragraph{Computation.} The rates and the risk difference are discretized on the lattice of Section~\ref{supp:binomial_lattice}. Writing $\ell(i, q) = \log \mathscr{L}_S(r_i, q/M)$ for the source log likelihood and
\begin{equation}
\Lambda_T(i, q) = \sum_{j} \frac{\mathscr{L}_T(r_j, q/M)}{M - |i - j|},
\end{equation}
where the sum runs over the target control rates $r_j$ for which $r_j + q/M \in [0, 1]$, and which depends on the target data only, both the marginal likelihood and the posterior of the risk difference reduce to sums over the source control rate and the risk difference:
\begin{equation}
m(\gamma) \propto \frac{\sum_{i, q} \Lambda_T(i, q)\, e^{\gamma \ell(i, q)}}{Z_S(\gamma)}, \qquad \pi(\theta_T = q/M \mid \mathbf{D}_S, \mathbf{D}_T, \hat{\gamma}) \propto \sum_{i} \Lambda_T(i, q)\, e^{\hat{\gamma} \ell(i, q)}.
\end{equation}
$\Lambda_T$ is computed once per target dataset. Because $\ell(i, q)$ is the sum of the two source arms' log likelihoods, the numerator is a bilinear form in their discounted likelihoods, so each evaluation of $m(\gamma)$ costs two matrix--vector products. We maximized $\log m(\gamma)$ by evaluating it on a grid of 21 equally spaced values of $\gamma$ and refining around the best grid point by Brent's method, keeping the end points $\gamma = 0$ and $\gamma = 1$ as candidates. The marginal likelihood agreed with brute-force numerical integration to $10^{-6}$.

\subsection{Calibrated power prior with binomial likelihoods}
\label{supp:pdccpp_binomial}

With a normal likelihood, the power parameter of the PDCCPP \parencite{nikolakopoulos_dynamic_2018} (Equation~\ref{eq:PDCCPP}) has a closed form. Writing $\hat{\delta} = \hat{\theta}_T - \hat{\theta}_S$ for the estimated drift and $z_{\mathrm{PPP}}$ for the critical value of the prior-predictive test, related to the threshold $c_{\mathrm{PPP}}$ by $c_{\mathrm{PPP}} = 2\{1 - \Phi(z_{\mathrm{PPP}})\}$,
\begin{equation}
\hat{\gamma} =
\begin{cases}
1 & \text{if } |\hat{\delta}| \leq z_{\mathrm{PPP}} \sqrt{\hat{\sigma}_{\theta_S}^2 + \hat{\sigma}_{\theta_T}^2}, \\[2pt]
\dfrac{\hat{\sigma}_{\theta_S}^2}{(\hat{\delta}/z_{\mathrm{PPP}})^2 - \hat{\sigma}_{\theta_T}^2} & \text{otherwise,}
\end{cases}
\end{equation}
since $\hat{\gamma}$ is the largest power parameter for which $|\hat{\delta}| \leq z_{\mathrm{PPP}} \sqrt{\hat{\sigma}_{\theta_T}^2 + \hat{\sigma}_{\theta_S}^2/\gamma}$. Larger values of $z_{\mathrm{PPP}}$ discount the source less, and $z_{\mathrm{PPP}}$ is calibrated so that the design attains a prespecified type I error rate $\alpha^\star$, here 0.065. In the normal case, the type I error rate is available in closed form for any $z_{\mathrm{PPP}}$, and $z_{\mathrm{PPP}}$ is found by bisection.

In the aprepitant case study, the power parameter was given by the same rule, applied to the estimated risk differences and their standard errors, and the posterior was that of the binomial conditional power prior with this power parameter (Supplementary Section~\ref{supp:normalized_power_prior_binomial}, with $\gamma$ fixed). The calibration parameter was chosen on the exact type I error rate of this binomial analysis. Under the null hypothesis, with the target control rate of the design $p_T^{(c)}$ and the treatment rate $p_T^{(c)} + \theta_0$, the type I error rate is the sum, over all pairs of responder counts, of their binomial probability times the indicator that the analysis rejects the null hypothesis:
\begin{equation}
\begin{aligned}
    \alpha(z_{\mathrm{PPP}})
    &= \sum_{y_T^{(c)} = 0}^{N_T^{(c)}} \sum_{y_T^{(t)} = 0}^{N_T^{(t)}}
    \Bigl\{
    \operatorname{Bin}\bigl(y_T^{(c)} \mid N_T^{(c)}, p_T^{(c)}\bigr)\,
    \operatorname{Bin}\bigl(y_T^{(t)} \mid N_T^{(t)}, p_T^{(c)} + \theta_0\bigr) \\
    &\quad \times \mathbb{1}\Bigl\{
    \operatorname{Pr}\bigl(\theta_T > \theta_0 \mid y_T^{(c)}, y_T^{(t)},
    \mathbf{D}_S, \hat{\gamma}(z_{\mathrm{PPP}}, y_T^{(c)}, y_T^{(t)})\bigr)
    > 0.975
    \Bigr\}
    \Bigr\},
\end{aligned}
\end{equation}
where pairs of counts of total probability below $10^{-10}$ were omitted. For each pair of counts, the set of power parameters for which the analysis rejects the null hypothesis was computed once: the posterior probability was evaluated on 21 equally spaced power parameters, and each change of decision was located by root finding. $\alpha(z_{\mathrm{PPP}})$ could then be evaluated for any $z_{\mathrm{PPP}}$ at negligible cost. Because the decisions are discrete, $\alpha(z_{\mathrm{PPP}})$ is a step function of $z_{\mathrm{PPP}}$. $z_{\mathrm{PPP}}$ was located on a logarithmic grid over $[10^{-3}, 10^{3}]$ and refined by bisection to the last step on which $\alpha(z_{\mathrm{PPP}}) \leq \alpha^\star$. We took $z_{\mathrm{PPP}}$ in the geometric middle of this step rather than at its upper edge: at the edge, the decision for some pair of counts is borderline, and a numerical error smaller than the root-finding tolerance could make the analysis spend the type I error of the next step. The resulting exact type I error rates were 0.063 and 0.064 for designs with 71 and 143 patients per arm. These decision sets agreed with the decisions of the analysis computed directly for every pair checked.

\subsection{Commensurate priors with binomial likelihoods}
\label{supp:commensurate_binomial}

With binomial likelihoods, the source and target studies have their own risk differences, $\theta_S$ and $\theta_T$, and their own control rates, $p_S^{(c)}$ and $p_T^{(c)}$. As for the other binomial models, the control rates have uniform priors and, given $p_S^{(c)}$, the initial prior of $\theta_S$ is uniform on $(-p_S^{(c)}, 1 - p_S^{(c)})$, so that both source response rates are uniform. The source likelihood discounted by the power parameter $\gamma$ defines the normalized power posterior
\begin{equation}
\pi(\theta_S \mid \mathbf{D}_S, \gamma) = \frac{1}{Z_S(\gamma)} \int_0^1 \mathscr{L}_S(p_S^{(c)}, \theta_S)^{\gamma}\, \mathbb{1}\{0 \leq p_S^{(c)} + \theta_S \leq 1\}\, \mathrm{d}p_S^{(c)},
\end{equation}
where, because the two source arms are independent under the initial prior, $Z_S(\gamma)$ is the product of the two arms' integrals $\int_0^1 r^{\gamma s}(1-r)^{\gamma(n-s)}\,\mathrm{d}r$. The target risk difference is linked to the source one by the commensurability parameter $\tau$,
\begin{equation}
\theta_T \mid \theta_S, \tau, p_T^{(c)} \sim \mathcal{N}\bigl(\theta_S, 1/\tau\bigr) \text{ truncated to } (-p_T^{(c)}, 1 - p_T^{(c)}),
\end{equation}
renormalized for each $(p_T^{(c)}, \theta_S, \tau)$, as for the robust mixture prior. The prior of $\tau$ is one of the five used in the normal case. For the commensurate power prior, $\gamma \mid \tau \sim \operatorname{Beta}(g(\tau), 1)$ with the same function $g$ as in the normal case; for the commensurate prior, $\gamma = 1$. Integrating over $\theta_S$, $\gamma$ and $\tau$ gives a prior on $(p_T^{(c)}, \theta_T)$, which was tabulated on the lattice of Section~\ref{supp:binomial_lattice}: the integral over $\gamma$ given $\tau$ was computed as for the normalized power prior, and the integral over $\tau$ on the quadrature nodes of its prior used in the normal case. With binomial likelihoods, the source information is not summarized by a normal distribution, so the prior of $\theta_T$ given $\tau$ is not the normal $\mathcal{N}(\hat{\theta}_S, 1/\tau + \hat{\sigma}_{\theta_S}^2/\gamma)$ of the normal case, but the source power posterior convolved with the truncated commensurability kernel. At fixed $\tau$, the lattice agreed with brute-force numerical integration to five to six significant digits, including under strong prior-data conflict.

\subsection{Robust mixture priors with binomial likelihoods}
\label{supp:rmp_binomial}

With binomial likelihoods, both components of the robust mixture prior are priors on the target control rate $p_T^{(c)}$ and the risk difference $\theta_T$. The informative component is the source posterior of the risk difference carried to the target study:
\begin{equation}
\pi_{\mathrm{S}}(p_T^{(c)}, \theta_T \mid \mathbf{D}_S) \propto \int \mathscr{L}_S(p_S^{(c)}, \theta_T)\, \pi_0(p_S^{(c)}, p_T^{(c)}, \theta_T)\, \mathrm{d}p^{(c)}_S
\end{equation}
The weak component $\pi_{\mathrm{W}}$ gives both target response rates independent uniform priors, which is the prior of the separate analysis. The robust mixture prior is
\begin{equation}
\pi(p_T^{(c)}, \theta_T \mid \mathbf{D}_S) = w\, \pi_{\mathrm{S}}(p_T^{(c)}, \theta_T \mid \mathbf{D}_S) + (1 - w)\, \pi_{\mathrm{W}}(p_T^{(c)}, \theta_T),
\end{equation}
and the posterior weight of the informative component is $w\, m_{\mathrm{S}} / \{w\, m_{\mathrm{S}} + (1 - w)\, m_{\mathrm{W}}\}$, where $m_{\mathrm{S}}$ and $m_{\mathrm{W}}$ are the marginal likelihoods of the target data under the two components. For the empirical mixture prior of \textcite{egidi_avoiding_2022}, the weight of the weak component was chosen for each dataset from the prior-predictive conflict $p$-value of the observed responder counts, computed exactly from the prior-predictive probabilities of all pairs of counts under the two components; under $\pi_{\mathrm{W}}$ every pair of counts is equally likely a priori, with probability $1/\{(N_T^{(c)} + 1)(N_T^{(t)} + 1)\}$. Both components were tabulated on the lattice of Section~\ref{supp:binomial_lattice}; the posterior, the posterior weight and the prior-predictive probabilities then follow from weighted sums over the lattice. With $w = 0$ the posterior agreed with the exact separate analysis to $2 \times 10^{-6}$, and with $w = 1$ it was the conditional power prior with $\gamma = 1$.

\subsection{Most powerful tests with known nuisance parameters}
\label{supp:most_powerful_test}

To bound the power that any test of the target data can reach at a given type I error rate $\alpha$ in a design, we computed the most powerful test of $H_0: \theta_T = \theta_0$ against a single alternative $\theta_T = \theta_1$, with the nuisance parameters fixed at their true values in the simulation. By the Neyman--Pearson lemma, this test rejects for large values of the likelihood ratio of the two hypotheses, and it is randomized at its boundary when the statistic is discrete, so that its size is exactly $\alpha$. Such a test cannot be used in practice, because it needs the true nuisance parameters, but no test of the target data can be more powerful in that design. We computed it at the type I error rate of each borrowing method, for the designs and settings checked in Section~\ref{sec:methods_reference}.

\paragraph{Binary endpoints.} With the control rate $p_T^{(c)}$ known, the control count $y_T^{(c)}$ has the same distribution under $H_0$ and $H_1$, so it does not enter the likelihood ratio. The likelihood ratio is then a function of the treatment count alone: $y_T^{(t)} \sim \operatorname{Bin}(N_T^{(t)}, p^{(t)})$, with $p^{(t)} = p_T^{(c)} + \theta$ for the risk difference (aprepitant) and $\operatorname{logit} p^{(t)} = \operatorname{logit} p_T^{(c)} + \theta$ for the log odds ratio (belimumab). The binomial family has a monotone likelihood ratio in $y_T^{(t)}$, so the test rejects for large treatment counts and is the same for every alternative on the same side of $\theta_0$: it is uniformly most powerful with $p_T^{(c)}$ known. Its size and power were computed exactly from the binomial probabilities, with randomization at the boundary count.

\paragraph{Recurrent events.} With the control rate and the dispersion $\phi$ known, the counts of the treatment arm are independent negative binomial with mean $\mu^{(t)}$ and size $\phi$, and their total over the $n_T$ patients of the arm is negative binomial with mean $n_T\mu^{(t)}$ and size $n_T\phi$. This total is sufficient for $\mu^{(t)}$ given $\phi$, and the family has a monotone likelihood ratio in it. Under the null, $\mu^{(t)}$ equals the control rate, and the alternative of the mepolizumab case study has a lower rate. The size and power of the test were computed exactly from the negative binomial distribution.

\paragraph{Time to event.} With the control hazard and the censoring distribution known, and exponential event times (primary design), the likelihood of the treatment arm is $\lambda^{E} \exp(-\lambda V)$, where $E$ is the number of events and $V$ the total follow-up time in that arm. The censoring distribution does not depend on $\lambda$ and cancels from the likelihood ratio. The log likelihood ratio of $\lambda_1 = \lambda_0 e^{\theta_1}$ against $\lambda_0$, the hazard under the null, is $E \log(\lambda_1/\lambda_0) - (\lambda_1 - \lambda_0) V$. The test rejects when this statistic is large. Because $(E, V)$ is two-dimensional, the test depends on $\lambda_1$, so each alternative has its own most powerful test. Its critical value was the $(1-\alpha)$ empirical quantile of the statistic over 200{,}000 trials simulated under the null with the study's own sampler of the treatment arm (accrual, administrative censoring, no loss to follow-up), and its power was estimated on 200{,}000 trials simulated under the alternative. With 200{,}000 trials, the Monte Carlo standard error of this power is at most 0.0011.

\subsection{Numerical integration for the binomial models (aprepitant case study)}
\label{supp:binomial_lattice}

 \paragraph{Binomial likelihood (aprepitant case study).} Under the arm-level binomial likelihood (Figure~\ref{fig:graphical_model_binomial_likelihood}), every method was analyzed with the binomial likelihoods of both arms. For the separate and pooled analyses, and therefore the test-then-pool variants, the uniform priors are conjugate: the two arm response rates have independent beta posteriors, and the distribution of their difference was computed by Gauss--Legendre quadrature over the control-arm rate. All other methods were computed on the lattice defined below. Moments, quantiles and posterior probabilities were obtained from the resulting density of the risk difference by the trapezoidal rule.

\paragraph{Lattice definition.} The response rates were discretized on the midpoints $r_i = (i - 1/2)/M$, $i = 1, \dots, M$, of a uniform lattice on $[0, 1]$, with $M = 1000$.
The risk difference is discretized on a grid with the same spacing as the rates, by discretizing it using the multiples $q/M$ of the $1/M$ spacing ($q = -(M-1), \dots, M-1$). This ensures that every treatment rate $r_i + q/M$ inside $[0, 1]$ is a lattice point. On this lattice, the uniform prior of the risk difference given two control rates $r_i$ and $r_j$ becomes the discrete uniform distribution on the $M - |i - j|$ differences that keep both treatment rates in $[0, 1]$. 
Every model considered is a prior on the target control rate and the risk difference, possibly after integrating out source parameters and hyperparameters. Once this prior is computed at every point on the lattice, the posterior of a target dataset is its product with the target likelihood, summed over the target control rate. 
Prior quantities that depend only on the source data and on the hyperparameters were therefore computed once, and each target dataset cost a single weighted sum. Moments of hyperparameters (power parameter, commensurability parameter) were obtained in the same way from the prior weighted by the hyperparameter or its square.


\newpage

\begin{landscape}
\section{Supplementary tables}

\begingroup
\small
\setlength{\tabcolsep}{4pt}
\renewcommand{\arraystretch}{1.1}

\begin{center}
  \begin{tabularx}{\linewidth}{|X|C|C|C|C|C|C|}
    \hline 
    \textbf{Disease} & Lower limb spasticity & Type-2 diabetes & Postoperative nausea and vomiting & Systemic Lupus Erythematosus (SLE) & Multiple Sclerosis & Severe Eosinophilic Asthma \\
    \hline
    \textbf{Drug} & \textbf{Botox vs placebo} & \textbf{Dapagliflozin vs placebo (+ Metformin)} & \textbf{Aprepitant vs ondansetron} & \textbf{Belimumab vs placebo} & \textbf{Teriflunomide vs placebo} & \textbf{Mepolizumab vs placebo} \\
    \hline
    \textbf{Endpoint} & Disease severity score & Glycated hemoglobin HbA1c & Absence of vomiting and rescue therapy 0-24h after surgery & SLE Responder Index & Time to first relapse & Number of clinically significant exacerbations. \\
    \hline
    \textbf{Endpoint type} & \textbf{Continuous} & \textbf{Continuous} & \textbf{Binary} & \textbf{Binary} & \textbf{Time to event} & \textbf{Recurrent event} \\
    \hline
    \textbf{Summary measure} & Difference in mean scores between the two arms & Difference between the two arms in change in HbA(1c) scores from baseline to week 24/26 & Difference in response rates between the two arms & Log odds ratio for active treatment compared to placebo & Log hazard ratio for active treatment compared to placebo & Log exacerbation rate ratio for active treatment compared to placebo \\
    \hline
    \textbf{Treatment effect distribution} & Normal & Normal & Integral of the product of binomials  & Normal (approximation for the log OR) & Normal (approximation for the log HR) & Normal (approximation for the log rate ratio) \\
    \hline
    \textbf{$N_T$: ctrl/trt/tot} & 130/126/256 & 76/81/157 & 52/57/109 & 39/53/92 & 57/109/166 & 9/16/25 \\
    \hline
    \textbf{$N_S$: ctrl/trt/tot} & 235/233/468 & 134/133/267 & 280/293/573 & 562/563/1125 & 752/731/1483 & 182/369/551 \\
    \hline
    \textbf{Target estimate / data} & 0.10 (0.10) & 1.03 (95\% CI, 0.49-1.57) (at week 26) & Treatment : 48/57, control: 42/52 & Treatment : 28 /53 Placebo: 17/39 & HR : 0.66 (95\% CI, 0.39-1.11) & Rate ratio : 0.67 (0.17, 2.68) \\
    \hline
    \textbf{Source estimate / data} & 0.20 (0.10) & 0.36 (0.102) (at week 24) & Treatment : 184/293 Control : 154/280 & Treatment: 285/563 Placebo: 218/562 & HR : 0.68 (95\% CI, 0.58-0.79) & Rate ratio : 0.50 (0.39, 0.64) \\
    \hline
    \textbf{Reference} & \textcite{wang_bayesian_2022} & \textcite{shehadeh_dapagliflozin_2023}, \textcite{bailey_effect_2010} & \textcite{jin_bayesian_2021}, \textcite{salman_pharmacokinetics_2019}, \textcite{diemunsch_singledose_2007} & \textcite{best_classical_2025}, \textcite{psioda_bayesian_2020}, \textcite{brunner_safety_2020}, \textcite{brunner_efficacy_2021} & \textcite{bovis_reinterpreting_2022} & \textcite{best_assessing_2021}, MENSA trial \parencite{ortega_mepolizumab_2014}, \textcite{keene_use_2020} \\
    \hline
  \end{tabularx}
\captionof{table}{Case studies used to design the simulation study. Mepolizumab: the treatment group combines the 75 mg intravenous and 100 mg subcutaneous arms of the MENSA trial; the source study is the adult subgroup (the 25 adolescents form the target study), and the source estimate is that of the adults \parencite{best_assessing_2021}.}
\label{tab:RCTs_selection}
\end{center}
\endgroup
\end{landscape}

\section{Parameters of the case studies}

\begin{table}[htbp]
\centering 
\setlength{\tabcolsep}{3pt}
\renewcommand{\arraystretch}{1.3}
\begin{tabular}{|l|c|c|c|c|c|c|c|}
\hline
\textbf{Case study} &
\textbf{Drift range} &
\shortstack{\textbf{Drift when}\\$\theta_T=\theta_0$} &
\shortstack{\textbf{Treatment}\\\textbf{effect range}} &
$\mathbf{N_S}$ & $\mathbf{N_S/2}$ & $\mathbf{N_S/4}$ & $\mathbf{N_S/6}$ \\
\hline
Belimumab     & $[-1.02,1.02]$   & $-0.48$  & $[-0.541,1.5]$   & 1125 & 562 & 281 & 187 \\
Botox         & $[-0.365,0.365]$ & $-0.2$   & $[-0.165,0.565]$ & 468 & 234 & 117 & 78 \\
Dapagliflozin & $[-0.707,0.707]$ & $-0.36$  & $[-0.347,1.07]$  & 267 & 133 & 66  & 44 \\
Mepolizumab   & $[-1.53,1.53]$   & $0.693$  & $[-2.23,0.839]$  & 551 & 275 & 137 & 91 \\
Aprepitant    & $[-0.628,0.372]$ & $-0.078$ & $[-0.55,0.45]$ & 573 & 286 & 143 & 95 \\
Teriflunomide & $[-0.652, 0.652]$ & $0.393$  & $[-1.05, 0.259]$ & 1483 & 741 & 370 & 247 \\
\hline
\end{tabular}
\caption{Drift and treatment effect ranges, and target study sample sizes considered for each case study. $N_S$ denotes the total source-study sample size, and the $N_S/f$ columns give
the corresponding nominal total target-study sample size, $\lfloor N_S/f \rfloor$, for $f \in \{1, 2, 4, 6\}$.
In the simulation, both arms of the target study had
$\lfloor N_S/(2f) \rfloor$ participants, so the total target sample size actually
simulated, $2\lfloor N_S/(2f) \rfloor$, can be one or two participants lower than
the value shown (e.g.\ 116 rather than 117 for Botox with $f = 4$). Sample sizes
per arm quoted in the figure captions ($n_T = n_T$) refer to the simulated values.}
\label{tab:drift_ranges_and_sample_sizes}
\end{table}

\begin{landscape}
\begin{table}[p]
\centering
\footnotesize
\setlength{\tabcolsep}{5pt}
\renewcommand{\arraystretch}{1.2}

\begin{tabularx}{\linewidth}{|
  >{\centering\arraybackslash\hsize=.85\hsize}X|
  >{\centering\arraybackslash\hsize=1.15\hsize}X|
  >{\centering\arraybackslash}X|
  >{\centering\arraybackslash}X|}
\hline
\textbf{Method}
& \textbf{Fixed parameters/Priors}
& \textbf{Parameters to vary}
& \textbf{Range of variation} \\
\hline

Test-then-pool, equivalence test (TtP (eq))
& None
& Significance level of the equivalence test $\eta$. Equivalence margin $\lambda$.
& $\eta \in \{0.1, 0.5\}$, $\lambda \in \{0.1, 0.5, 0.8\}$ \\
\hline

Test-then-pool, difference test (TtP (diff))
& None
& Significance level of the difference test $\eta$
& $\eta \in \{0.01, 0.1, 0.4, 0.8\}$ \\
\hline

Conditional power prior (CPP)
& Initial prior on $\theta$
& Power parameter $\gamma$
& $\gamma \in \{0, 0.25, 0.5, 0.75, 1\}$ \\
\hline

Normalized PP (NPP)
& Initial prior on $\theta$;
  $\gamma \sim \operatorname{Beta}(a_\gamma, b_\gamma)$ with mean $\xi_\gamma = 0.5$
& Standard deviation $\sigma_\gamma$ of the beta prior
& $\sigma_\gamma \in \{0.1, 0.2, 0.4\}$ \\
\hline

KL-calibrated NPP (NPP (KL))
& Initial prior on $\theta$; source-to-null calibration with
  $\lambda_{\mathrm{KL}}=0.5$ and $c=10$
& Discrepancy multiplier $k_{\mathrm{MTD}}$
& $k_{\mathrm{MTD}} \in \{0.5, 1\}$ \\
\hline

Empirical Bayes PP (EBPP)
& Initial prior on $\theta$
& None
& None \\
\hline

Prior-data conflict calibrated PP (PDCCPP)
& Initial prior on $\theta$; target type I error rate 0.065, at the decision threshold $\rho = 0.975$ of the analysis
& None
& None \\
\hline

\textit{p}-value-based PP (p-PP)
& Initial prior on $\theta$
& Shape parameter $k$. Equivalence margin $\lambda$.
& $k \in \{0.01, 0.1, 1, 10, 20\}$, $\lambda \in \{0.1, 0.5\}$ \\
\hline

Commensurate PP (Com.\ PP)
& Initial prior on $\theta_S$
& Prior on the commensurability parameter $\tau$
& $\tau^2 \sim \operatorname{IG}(\alpha, 1)$,
  $\alpha \in \{1/3, 1/7\}$;
  $\tau \sim \operatorname{HN}(\sigma)$,
  $\sigma \in \{1, 5\}$;
  $\log\tau \sim \operatorname{Cauchy}(0, 30)$ \\
\hline

Commensurate prior (Com.\ prior)
& Initial prior on $\theta_S$; power parameter fixed at 1
& Prior on $\tau$
& Same five priors as commensurate PP \\
\hline

Robust mixture prior (RMP)
& Variance of the vague component
& Mixture weight $w$
& $w$ in a grid of values ranging from 0 to 1 in steps of 0.1 \\
\hline

Empirical RMP
& Same components as robust mixture prior. Conflict level
  $\alpha_{\mathrm{PC}}=0.05$
& None
& None \\
\hline
\end{tabularx}

\caption[Methods and parameters considered in the simulation study]{
Methods and parameters considered in the simulation study.
When the method is based on a consistency assumption
($\theta_T = \theta_S$), we denote the treatment effect as $\theta$.
}
\label{tab:study_protocol_methods_range}
\end{table}
\end{landscape}

\afterpage{%
\begingroup
\setlength{\tabcolsep}{8pt}
\renewcommand{\arraystretch}{1.5}

\begin{landscape}
\begin{table}[htbp]
\centering
\begin{tabularx}{1.4\textwidth}{|>{\hsize=1.45\hsize}C|>{\hsize=1\hsize}C|>{\hsize=1.1\hsize}C|>{\hsize=1.05\hsize}C|>{\hsize=1\hsize}C|>{\hsize=0.7\hsize}C|>{\hsize=1.15\hsize}C|>{\hsize=1\hsize}C|>{\hsize=0.9\hsize}C|>{\hsize=0.65\hsize}C|}
\hline
\textbf{Case study}
& \textbf{Endpoint}
& \textbf{Effect measure}
& \textbf{Likelihood}
& \textbf{Operating characteristics}
& \textbf{\# drift values}
& \textbf{$n_T$ (per arm)}
& \textbf{$N_S/N_T$}
& \textbf{Denom.\ change factor}
& \textbf{$\sigma_T/\sigma_S$} \\
\hline
Botox & Continuous & Mean difference & Normal & 10,000 & 33 & 39, 58, 117, 234 & 1, 2, 4, 6 & NA & 1, 2 \\
\hline
Dapagliflozin & Continuous & Mean difference & Normal & 10,000 & 33 & 22, 33, 66, 133 & 1, 2, 4, 6 & NA & 1, 2 \\
\hline
Belimumab & Binary & Log odds ratio & Normal & Exact & 33 & 93, 140, 281, 562 & 1, 2, 4, 6 & 1/2, 1, 3/2 & NA \\
\hline
Aprepitant & Binary & Response rates (binomial) & Binomial & Exact & 33 & 47, 71, 143, 286 & 1, 2, 4, 6 & NA & NA \\
\hline
Mepolizumab & Recurrent events & Log rate ratio & Normal & 10,000 & 33 & 45, 68, 137, 275 & 1, 2, 4, 6 & 1/2, 1, 3/2 & NA \\
\hline
Teriflunomide & Time to event & Log hazard ratio & Normal & 10,000 & 33 & 123, 185, 370, 741 & 1, 2, 4, 6 & 1/2, 1, 3/2 & NA \\
\hline
\end{tabularx}
\caption{Simulation configuration. All 14 methods were simulated with the same configuration in a case study. Operating characteristics are exact (enumeration of all possible trial outcomes) or estimated from 10,000 simulated trials. \# drift values: number of target treatment effects for each design. $N_S/N_T$: mean source arm size over the target arm size. Denom.\ change factor: factor applied to the source control odds (binary endpoint) or control event rate (other endpoints), with the same source treatment effect. $\sigma_T/\sigma_S$: target to source standard deviation ratio. NA: not applicable. In the teriflunomide case study, 7 more designs changed one factor at a time from the design with $n_T = 123$ per arm ($N_S/N_T = 6$, denominator change factor 1): loss to follow-up of 5\% or 10\%; Weibull event times; treatment effect delayed by 12 or 24 weeks; placebo event rate multiplied by 0.67 or 1.50 ($\kappa = -0.405$ or $+0.405$).}
\label{tab:simulation_configuration}
\end{table}
\end{landscape}
\endgroup
}

\clearpage

\section{Supplementary results tables}
\label{sec:supp_results_details}

\begin{table}[ht]
\centering
\scriptsize
\setlength{\tabcolsep}{3pt}
\begin{tabular}{lrrrrrrrrrr}
\hline
 & \multicolumn{2}{c}{MSE, partial} & \multicolumn{2}{c}{Power} & \multicolumn{2}{c}{Coverage, null} & \multicolumn{2}{c}{Interval score, null} & \multicolumn{2}{c}{MSE at matched ESS} \\
\cline{2-3}\cline{4-5}\cline{6-7}\cline{8-9}\cline{10-11}
Method & Excess \% & Near & Short \% & Max & Shortfall & Below & Excess \% & Near & Excess \% & Near \\
\hline
Separate & 2 (1--4) & 75 & 0 & 0.004 & 0.003 & 0 & 3 (1--7) & 69 & 182 (76--359) & 0 \\
Pooling & 44 (20--77) & 16 & 0 & 0.000 & 0.136 & 0 & 48 (26--115) & 25 & 0 (0--0) & 100 \\
CPP & 6 (0--35) & 49 & 0 & 0.005 & 0.042 & 0 & 2 (0--29) & 58 & 0 (0--1) & 80 \\
RMP & 37 (21--62) & 10 & 2 & 0.044 & 0.019 & 0 & 24 (10--49) & 20 & 61 (30--112) & 10 \\
NPP & 17 (4--25) & 27 & 0 & 0.004 & 0.016 & 0 & 0 (0--16) & 59 & 1 (1--19) & 54 \\
NPP (KL) & 12 (2--21) & 34 & 0 & 0.008 & 0.030 & 0 & 5 (0--11) & 53 & 1 (0--3) & 94 \\
EBPP & 49 (42--75) & 0 & 8 & 0.052 & 0.029 & 0 & 34 (10--66) & 19 & 224 (128--347) & 0 \\
Empirical RMP & 52 (35--80) & 0 & 0 & 0.008 & 0.070 & 0 & 111 (30--127) & 3 & 16 (7--39) & 12 \\
Com.\ PP & 0 (0--19) & 66 & 8 & 0.047 & 0.000 & 0 & 0 (0--12) & 66 & 51 (23--79) & 12 \\
Com.\ prior & 0 (0--20) & 65 & 8 & 0.042 & 0.002 & 0 & 1 (0--12) & 66 & 49 (24--73) & 8 \\
p-PP & 60 (21--90) & 17 & 3 & 0.094 & 0.030 & 0 & 28 (4--77) & 26 & 150 (51--348) & 8 \\
TtP (diff) & 80 (34--126) & 12 & 9 & 0.114 & 0.029 & 0 & 53 (14--128) & 16 & 235 (97--550) & 0 \\
TtP (eq) & 81 (14--121) & 16 & 12 & 0.115 & 0.038 & 1 & 51 (12--117) & 21 & 325 (99--495) & 4 \\
PDCCPP & 43 (34--74) & 0 & 23 & 0.067 & 0.018 & 0 & 21 (7--41) & 19 & 241 (142--375) & 0 \\
\hline
\end{tabular}
\caption{As Table~\ref{tab:frontier_distance}, for the sensitivity designs (target-to-source standard deviation ratio of 2 for botox and dapagliflozin; source denominator factor of 0.5 or 1.5 for the other case studies). The columns are defined as in Table~\ref{tab:frontier_distance}.}
\label{tab:frontier_distance_sensitivity}
\end{table}

\begin{table}[ht]
\centering
\scriptsize
\setlength{\tabcolsep}{3pt} 
\begin{tabular}{lrrrrrrrrrr}
\hline
 & \multicolumn{2}{c}{MSE, partial} & \multicolumn{2}{c}{Power} & \multicolumn{2}{c}{Coverage, null} & \multicolumn{2}{c}{Interval score, null} & \multicolumn{2}{c}{MSE at matched ESS} \\
\cline{2-3}\cline{4-5}\cline{6-7}\cline{8-9}\cline{10-11}
Method & Excess \% & Near & Short \% & Max & Shortfall & Below & Excess \% & Near & Excess \% & Near \\
\hline
Separate & 3 (1--6) & 62 & 0 & 0.002 & 0.003 & 0 & 4 (2--6) & 67 & 41 (21--59) & 0 \\
Pooling & 31 (2--58) & 33 & 0 & 0.000 & 0.091 & 0 & 20 (2--51) & 46 & 0 (0--0) & 100 \\
CPP & 1 (0--24) & 59 & 0 & 0.002 & 0.016 & 0 & 0 (0--7) & 72 & 0 (0--0) & 80 \\
RMP & 31 (17--53) & 14 & 1 & 0.024 & 0.018 & 0 & 20 (7--41) & 19 & 28 (12--37) & 14 \\
NPP & 12 (2--23) & 35 & 0 & 0.004 & 0.007 & 0 & 2 (0--15) & 58 & 3 (1--13) & 54 \\
NPP (KL) & 6 (0--18) & 48 & 0 & 0.007 & 0.011 & 0 & 3 (0--6) & 71 & 4 (0--5) & 75 \\
EBPP & 53 (43--80) & 0 & 0 & 0.006 & 0.032 & 0 & 29 (16--42) & 12 & 59 (46--93) & 0 \\
Empirical RMP & 37 (21--61) & 12 & 0 & 0.008 & 0.035 & 0 & 48 (9--124) & 12 & 7 (4--12) & 38 \\
Com.\ PP & 0 (0--18) & 62 & 2 & 0.027 & 0.002 & 0 & 2 (0--18) & 61 & 17 (6--30) & 25 \\
Com.\ prior & 0 (0--19) & 61 & 2 & 0.026 & 0.003 & 0 & 2 (0--19) & 60 & 17 (4--29) & 28 \\
p-PP & 41 (4--71) & 27 & 2 & 0.059 & 0.030 & 0 & 14 (2--44) & 35 & 36 (9--77) & 21 \\
TtP (diff) & 77 (29--119) & 11 & 11 & 0.124 & 0.043 & 0 & 34 (16--68) & 14 & 80 (35--162) & 0 \\
TtP (eq) & 52 (4--94) & 27 & 6 & 0.101 & 0.038 & 1 & 24 (3--60) & 33 & 46 (6--106) & 25 \\
PDCCPP & 44 (36--94) & 0 & 31 & 0.083 & 0.020 & 0 & 28 (9--38) & 17 & 67 (43--112) & 0 \\
\hline
\end{tabular}
\caption{Operating characteristics at matched TIE (and matched prior ESS), primary designs (six case studies, 24 case study--sample size designs, standard deviation ratio and source denominator factor of 1). MSE, partial: relative excess (\%) of the MSE under a partially consistent effect (half the source estimate) over the Pareto frontier, i.e.\ the smallest MSE reached by any of the 56 method-setting configurations of the design with a TIE no larger than the configuration's own, interpolated linearly in TIE between Pareto-optimal configurations; median (interquartile range) over the method's settings and the designs, and share (\%) within 5\% of the frontier (Near). Power: share (\%) of configurations whose probability of success under a partially consistent or consistent effect falls more than 0.01 below the Pareto frontier of power against TIE, and largest shortfall. Coverage, null: median shortfall from the highest coverage among configurations whose TIE lies within a factor 1.25 of the configuration's own (see interval score), and share (\%) of configurations more than 0.02 below a common decreasing (isotonic) curve of coverage against TIE fitted to the 56 configurations of each design. Interval score, null: relative excess (\%) over the smallest interval score among configurations whose TIE lies within a factor 1.25 of the configuration's own (a two-sided window: under no treatment effect, the interval score worsens with the TIE, so a one-sided frontier would only reproduce the separate analysis); Near: within 5\%. MSE at matched ESS (botox and dapagliflozin only): relative excess of the MSE under a partially consistent effect over the smallest MSE among the configurations of the same design whose moment-based prior ESS differs from the configuration's own ESS by at most 10\% of the ESS of pooling (a window centered on each configuration, whose half-width is set by the full source information).}
\label{tab:frontier_distance}
\end{table}

\begin{table}[ht]
\centering
\small 
\begin{tabular}{lrrr}
\hline
Method & TIE $\le$ 0.05 & TIE $\le$ 0.10 & TIE $\le$ 0.20 \\
\hline
Separate & 31 (24) & 46 (24) & 66 (24) \\
Pooling & -- & -- & 2 (1) \\
CPP & 11 (24) & 9 (24) & 0 (24) \\
RMP & 15 (24) & 20 (24) & 22 (24) \\
NPP & 0 (1) & 2 (12) & 13 (23) \\
NPP (KL) & -- & 5 (8) & 17 (14) \\
EBPP & 38 (2) & 40 (6) & 50 (12) \\
Empirical RMP & -- & 122 (1) & 101 (4) \\
Com.\ PP & 1 (24) & 1 (24) & 13 (24) \\
Com.\ prior & 0 (24) & 1 (24) & 13 (24) \\
p-PP & 31 (24) & 46 (24) & 49 (24) \\
TtP (diff) & 15 (14) & 33 (17) & 52 (23) \\
TtP (eq) & 27 (22) & 46 (24) & 57 (24) \\
PDCCPP & 38 (2) & 45 (24) & 63 (24) \\
\hline
\end{tabular}
\caption{Smallest MSE under a partially consistent effect that each method reaches with a TIE no larger than 0.05, 0.10 or 0.20, as a relative excess (\%) over the smallest MSE reached by any method within the same TIE budget. Median over the primary designs (standard deviation ratio and source denominator factor of 1) in which the method has at least one setting within the budget, with number of such designs (out of 24) in parentheses. Methods with many settings are favored, since a finer grid is more likely to have a setting just under the budget.}
\label{tab:tie_budget}
\end{table}

\begin{table}[ht]
\centering
\scriptsize
\setlength{\tabcolsep}{3pt} 
\begin{tabular}{lrrrrrrp{2.1cm}p{2.1cm}r}
\hline
 & & \multicolumn{3}{c}{Break-even, towards null} & Break-even & Coverage & \multicolumn{2}{c}{Best setting (break-even)} & Worst \\
\cline{3-5}\cline{9-10}
Method & Borrowing & Median & Range & \% $>$max & away & $<$ 0.90 & TIE $\le$ 0.05 & TIE $\le$ 0.10 & MSE ratio \\
\hline
Pooling & 100 & 1.3 & 1.1--1.9 & 0 & 1.3 & 0.9 & -- & -- & 14.06 \\
CPP & 82 & 1.5 & 1.1--$>$max & 1 & 1.6 & 1.2 & -- & -- & 8.57 \\
RMP & 92 & 1.4 & 1.1--$>$max & 1 & 1.4 & 1.2 & 1.6 ($w=$0.1) & 1.6 ($w=$0.1) & 1.85 \\
NPP & 100 & 1.6 & 1.3--2.5 & 0 & 1.6 & 1.3 & -- & -- & 2.26 \\
NPP (KL) & 100 & 1.6 & 1.3--2.4 & 0 & 1.7 & 1.3 & -- & -- & 2.71 \\
EBPP & 100 & 1.3 & 1.2--1.5 & 0 & 1.3 & 0.9 & -- & -- & 1.40 \\
Empirical RMP & 100 & 1.3 & 1.1--1.8 & 0 & 1.3 & 0.9 & -- & -- & 3.41 \\
Com.\ PP & 86 & 2.2 & 1.2--$>$max & 15 & 2.3 & $>$max & 4.2 (HN(5)) & 4.2 (HN(5)) & 1.19 \\
Com.\ prior & 86 & 2.2 & 1.2--$>$max & 15 & 2.2 & $>$max & 4.2 (HN(5)) & 4.2 (HN(5)) & 1.22 \\
p-PP & 82 & 1.2 & 0.9--5.3 & 0 & 1.2 & 1.0 & 1.0 ($k=$1, $\lambda=$0.1) & 1.0 ($k=$1, $\lambda=$0.1) & 2.14 \\
TtP (diff) & 75 & 1.0 & 0.8--1.7 & 0 & 1.0 & 0.8 & -- & -- & 1.84 \\
TtP (eq) & 74 & 1.2 & 0.8--1.9 & 0 & 1.2 & 1.0 & -- & -- & 3.25 \\
PDCCPP & 75 & 1.4 & 1.2--6.4 & 0 & 1.4 & 1.2 & -- & 1.3 & 1.14 \\
\hline
\end{tabular}
\caption{Break-even drift: smallest distance between the target and the source treatment effect beyond which the MSE of a method exceeds that of the separate analysis in the same design, by linear interpolation on the drift grid, in units of the target standard error. The designs include the primary designs (with standard deviation ratio and source denominator factor of 1). Configurations (setting $\times$ design) that reduce the MSE by at least 5\% under consistency (Borrowing: their share, \%). Towards null: drift in the direction of no treatment effect; away: drift in the other direction. Median and range over these configurations; \% $>$max: no break-even within the explored drift (between 1.5 and 15.4 target standard errors depending on the design, median 5.4). Coverage $<$ 0.90: median drift (towards null) at which the coverage of the 95\% interval falls below 0.90. Best setting: largest median break-even (towards null) among the settings with median TIE $\le$ 0.05 or $\le$ 0.10 across designs and a median MSE ratio to the separate analysis of at most 0.90 under consistency (settings that do borrow); --: no such setting. Worst MSE ratio: median over settings and designs of the largest MSE ratio to the separate analysis over the drift grid (towards null).}
\label{tab:break_even}
\end{table}
\begin{table}[ht]
\centering
\footnotesize
\setlength{\tabcolsep}{3pt}
\renewcommand{\arraystretch}{1.15}

\resizebox{\textwidth}{!}{%
\begin{tabular}{lcccccc}
\hline
 & \multicolumn{5}{c}{TIE} & Power \\
Case study & Separate & Pooling & PDCCPP & Median
 & $\leq 0.03$ & Separate / pooling \\
\hline
Botox
 & 0.025--0.030 & 0.217--0.383 & 0.064--0.070
 & 0.080--0.127 & 10--11 & 0.13--0.51 / 0.70--0.89 \\
Dapagliflozin
 & 0.026--0.032 & 0.768--0.992 & 0.064--0.077
 & 0.116--0.215 & 1--11 & 0.31--0.94 / 1.00 \\
Belimumab
 & 0.023--0.025 & 0.895--1.000 & 0.063--0.066
 & 0.083--0.250 & 8--12 & 0.36--0.98 / 1.00 \\
Aprepitant
 & 0.024--0.026 & 0.189--0.302 & 0.062--0.064
 & 0.060--0.090 & 9--17 & 0.11--0.47 / 0.60--0.84 \\
Mepolizumab
 & 0.025--0.030 & 0.985--1.000 & 0.046--0.072
 & 0.042--0.146 & 1--22 & 0.62--1.00 / 1.00 \\
Teriflunomide
 & 0.022--0.026 & 0.930--1.000 & 0.047--0.065
 & 0.047--0.211 & 9--16 & 0.60--1.00 / 1.00 \\
\hline
\end{tabular}%
}

\caption{Type I error rate (TIE) and power by case study, primary designs (target-to-source standard deviation ratio and source denominator factor of 1); ranges over the four target sample sizes per arm ($n_T$). TIE of the separate analysis, of pooling and of the PDCCPP (target 0.065); median TIE over the 56 method-setting configurations, and number of configurations with a TIE of at most 0.03. Power at the consistent treatment effect: separate analysis / pooling. TIEs at or below that of the separate analysis occurred only for settings that reduce to the separate analysis under conflict, and, with a target standard deviation twice the source one, in the botox case study.}
\label{tab:supp_tie_power}
\end{table}

\begin{table}[ht]
\centering
\small
\begin{tabular}{lcccc}
\hline
 & \multicolumn{3}{c}{Primary designs} & Sensitivity designs \\
Case study & $|\text{difference}| < 0.01$ (\%) & Largest gain & Largest loss & Largest gain \\
\hline
Botox & 87.3 & $+0.007$ (RMP, 58) & $-0.241$ (TtP (diff), 39) & $+0.007$ \\
Dapagliflozin & 97.1 & $+0.013$ (Com.\ prior, 66) & $-0.085$ (PDCCPP, 22) & $+0.011$ \\
Belimumab & 99.1 & $+0.027$ (TtP (eq), 93) & $-0.025$ (TtP (diff), 93) & $+0.109$ \\
Aprepitant & 97.1 & $+0.038$ (p-PP, 47) & $-0.280$ (TtP (diff), 47) & -- \\
Mepolizumab & 95.2 & $+0.046$ (TtP (diff), 45) & $-0.005$ (TtP (eq), 45) & $+0.062$ \\
Teriflunomide & 95.9 & $+0.040$ (TtP (diff), 185) & $-0.005$ (TtP (eq), 123) & $+0.045$ \\
\hline
\end{tabular}
\caption{Comparison with a frequentist test of the target data alone at a matched TIE, by case study. Primary designs: share (\%) of power differences smaller than 0.01 in absolute value, over all method-setting configurations, target sample sizes and treatment effects in the alternative; largest gain and largest loss (method, participants per arm). Sensitivity designs (standard deviation ratio 2, or source denominator factor 0.5 or 1.5): largest gain.}
\label{tab:supp_matched_tie}
\end{table}

\begin{table}[ht]
\centering
\small
\begin{tabular}{llccccc}
\hline
 & & \multicolumn{3}{c}{MSE relative to separate} & \multicolumn{2}{c}{Coverage} \\
Case study & Treatment effect & Smallest & Median & Largest & Median & Smallest \\
\hline
Botox & Consistent & 0.02 & 0.35 & 1.00 & 0.98 & 0.94 \\
 & Partially consistent & 0.14 & 0.53 & 1.01 & 0.96 & 0.93 \\
 & No effect & 0.40 & 0.95 & 1.62 & 0.86 & 0.59 \\
Dapagliflozin & Consistent & 0.02 & 0.36 & 1.00 & 0.98 & 0.93 \\
 & Partially consistent & 0.35 & 0.82 & 1.45 & 0.90 & 0.77 \\
 & No effect & 0.76 & 1.30 & 3.54 & 0.81 & 0.01 \\
Belimumab & Consistent & 0.02 & 0.34 & 1.00 & 0.99 & 0.95 \\
 & Partially consistent & 0.39 & 0.95 & 1.62 & 0.85 & 0.59 \\
 & No effect & 0.78 & 1.45 & 4.26 & 0.83 & 0.00 \\
Aprepitant & Consistent & 0.02 & 0.36 & 1.00 & 0.99 & 0.95 \\
 & Partially consistent & 0.13 & 0.52 & 1.01 & 0.97 & 0.94 \\
 & No effect & 0.38 & 0.91 & 1.58 & 0.91 & 0.67 \\
Mepolizumab & Consistent & 0.02 & 0.37 & 1.00 & 0.98 & 0.94 \\
 & Partially consistent & 0.61 & 1.07 & 2.41 & 0.82 & 0.09 \\
 & No effect & 0.91 & 1.43 & 7.43 & 0.91 & 0.00 \\
Teriflunomide & Consistent & 0.03 & 0.33 & 1.00 & 0.99 & 0.95 \\
 & Partially consistent & 0.58 & 1.07 & 2.08 & 0.81 & 0.27 \\
 & No effect & 0.89 & 1.69 & 6.02 & 0.86 & 0.00 \\
\hline
\end{tabular}
\caption{Estimation by case study, primary designs: MSE relative to the separate analysis (smallest, median and largest over the 55 borrowing configurations and the four target sample sizes) and coverage of the 95\% credible interval (median and smallest), under a treatment effect consistent with the source estimate, partially consistent (half of it) and null.}
\label{tab:supp_estimation}
\end{table}

\begin{table}[ht]
\centering
\scriptsize
\begin{tabular}{lcccccccccc}
\hline
 & Events & ESS & \multicolumn{4}{c}{TIE} & \multicolumn{3}{c}{Probability of success (consistent)} & Mean \\
Design & C / T & pooling & NPP & Empirical RMP & RMP & PDCCPP & Separate & PDCCPP & Median & $|\Delta|$ \\
\hline
Primary & 73 / 57 & 514 & 0.48 & 0.56 & 0.15 & 0.064 & 0.60 & 0.76 & 0.93 & 0.000 \\
Dropout 5\% & 72 / 55 & 524 & 0.48 & 0.57 & 0.14 & 0.065 & 0.59 & 0.76 & 0.93 & 0.002 \\
Dropout 10\% & 70 / 54 & 536 & 0.48 & 0.57 & 0.15 & 0.066 & 0.58 & 0.75 & 0.93 & 0.004 \\
Weibull & 66 / 50 & 572 & 0.51 & 0.60 & 0.16 & 0.068 & 0.56 & 0.73 & 0.92 & 0.009 \\
Placebo $\kappa=-0.405$ & 56 / 42 & 676 & 0.55 & 0.65 & 0.17 & 0.063 & 0.48 & 0.67 & 0.90 & 0.025 \\
Placebo $\kappa=+0.405$ & 91 / 74 & 413 & 0.43 & 0.49 & 0.13 & 0.063 & 0.71 & 0.84 & 0.96 & 0.018 \\
Delay 12 weeks & 73 / 55 & 514 & 0.48 & 0.56 & 0.15 & 0.064 & 0.60 & 0.76 & 0.93 & 0.001 \\
Delay 24 weeks & 73 / 54 & 514 & 0.48 & 0.56 & 0.15 & 0.064 & 0.59 & 0.76 & 0.93 & 0.002 \\
\hline
\end{tabular}
\caption{Sensitivity analyses in the teriflunomide case study, 123 participants per arm: primary design; loss to follow-up of 5\% or 10\%; Weibull event times; placebo relapse rate two thirds of, or one and a half times, that of the source trials ($\kappa = \mp 0.405$); treatment effect delayed by 12 or 24 weeks. Expected events per arm (control / treatment) under a consistent treatment effect; moment-based ESS of pooling under no treatment effect; TIE of the NPP ($\xi_\gamma = 0.5$, $\sigma_\gamma = 0.2$), the empirical RMP, the RMP ($w = 0.5$) and the PDCCPP; probability of success under a consistent treatment effect of the separate analysis, the PDCCPP and the median borrowing configuration; mean absolute change in the probability of success from the primary design, over all configurations and treatment effects. With a delayed effect, the treatment effect of a scenario is the limit of the Cox estimate.}
\label{tab:supp_teriflunomide}
\end{table}

\begin{table}[ht]
\centering
\footnotesize
\begin{tabular}{lcccccccc}
\hline
 & \multicolumn{4}{c}{Belimumab} & \multicolumn{4}{c}{Aprepitant} \\
 & \multicolumn{2}{c}{$\pm 0.05$} & \multicolumn{2}{c}{$\pm 0.10$} & \multicolumn{2}{c}{$\pm 0.05$} & \multicolumn{2}{c}{$\pm 0.10$} \\
Method & Median & Max & Median & Max & Median & Max & Median & Max \\
\hline
Separate & 0.000 & 0.001 & 0.001 & 0.002 & 0.001 & 0.001 & 0.001 & 0.001 \\
Pooling & 0.000 & 0.036 & 0.001 & 0.089 & 0.001 & 0.003 & 0.001 & 0.019 \\
CPP & 0.002 & 0.043 & 0.007 & 0.115 & 0.001 & 0.016 & 0.001 & 0.016 \\
RMP & 0.003 & 0.036 & 0.009 & 0.090 & 0.001 & 0.016 & 0.002 & 0.026 \\
NPP & 0.008 & 0.041 & 0.019 & 0.093 & 0.001 & 0.006 & 0.006 & 0.009 \\
NPP (KL) & 0.011 & 0.033 & 0.027 & 0.071 & 0.002 & 0.005 & 0.004 & 0.009 \\
EBPP & 0.006 & 0.031 & 0.021 & 0.076 & 0.001 & 0.016 & 0.001 & 0.016 \\
Empirical RMP & 0.009 & 0.041 & 0.034 & 0.110 & 0.001 & 0.016 & 0.001 & 0.016 \\
Com.\ PP & 0.001 & 0.029 & 0.002 & 0.048 & 0.001 & 0.002 & 0.001 & 0.005 \\
Com.\ prior & 0.001 & 0.020 & 0.002 & 0.048 & 0.001 & 0.007 & 0.001 & 0.007 \\
p-PP & 0.001 & 0.035 & 0.004 & 0.097 & 0.001 & 0.020 & 0.001 & 0.026 \\
TtP (diff) & 0.005 & 0.045 & 0.015 & 0.115 & 0.001 & 0.005 & 0.001 & 0.019 \\
TtP (eq) & 0.001 & 0.036 & 0.002 & 0.089 & 0.001 & 0.009 & 0.001 & 0.019 \\
PDCCPP & 0.000 & 0.003 & 0.001 & 0.005 & 0.005 & 0.008 & 0.005 & 0.008 \\
\hline
\end{tabular}
\caption{Sensitivity of the TIE to the true target control rate, binary case studies (computed exactly). Increase of the largest TIE over control rates within 0.05 and 0.10 of the control rate of the design (belimumab 0.24 to 0.49 depending on the source denominator factor; aprepitant 0.55) above the TIE at the control rate of the design: median and maximum over the method's settings, the target sample sizes and, for belimumab, the source denominator factors.}
\label{tab:supp_tie_control_rate}
\end{table}

\clearpage
\newpage

\section{Supplementary figures}


\begin{figure}[ht]
    \centering

\tikzset{every picture/.style={line width=0.75pt}} 
\begin{adjustbox}{max width=\linewidth,max totalheight=0.75\textheight}
\begin{tikzpicture}[x=0.75pt,y=0.75pt,yscale=-1,xscale=1]

\draw   (24,446.08) .. controls (24,417.32) and (47.32,394) .. (76.08,394) -- (453.92,394) .. controls (482.68,394) and (506,417.32) .. (506,446.08) -- (506,602.32) .. controls (506,631.08) and (482.68,654.4) .. (453.92,654.4) -- (76.08,654.4) .. controls (47.32,654.4) and (24,631.08) .. (24,602.32) -- cycle ;

\draw  [fill={rgb, 255:red, 208; green, 2; blue, 27 }  ,fill opacity=1 ]  (394.66,107) -- (516.66,107) -- (516.66,154) -- (394.66,154) -- cycle  ;
\draw (397.66,111) node [anchor=north west][inner sep=0.75pt]   [align=left] {\begin{minipage}[lt]{82.74pt}\setlength\topsep{0pt}
\begin{center}
Source study\\($\hat{\theta}_S, N_S, \sigma_{\theta_S}$)
\end{center}

\end{minipage}};
\draw    (385.58,302) -- (532.58,302) -- (532.58,348) -- (385.58,348) -- cycle  ;
\draw (388.58,306) node [anchor=north west][inner sep=0.75pt]   [align=left] {\begin{minipage}[lt]{100.28pt}\setlength\topsep{0pt}
\begin{center}
Prior on the target TE\\ $p(\theta_T|\mathbf{D}_S)$
\end{center}

\end{minipage}};
\draw    (250,510) -- (343,510) -- (343,556) -- (250,556) -- cycle  ;
\draw (253,514) node [anchor=north west][inner sep=0.75pt]   [align=left] {\begin{minipage}[lt]{63.25pt}\setlength\topsep{0pt}
\begin{center}
Inference\\$p(\theta_T |\mathbf{D}_S, \mathbf{D}_T)$
\end{center}

\end{minipage}};
\draw    (65,510) -- (158,510) -- (158,555) -- (65,555) -- cycle  ;
\draw (68,514) node [anchor=north west][inner sep=0.75pt]   [align=left] {\begin{minipage}[lt]{61.15pt}\setlength\topsep{0pt}
\begin{center}
Target\\ data sample
\end{center}

\end{minipage}};
\draw    (400,551) -- (486,551) -- (486,576) -- (400,576) -- cycle  ;
\draw (403,555) node [anchor=north west][inner sep=0.75pt]   [align=left] {\begin{minipage}[lt]{56.6pt}\setlength\topsep{0pt}
\begin{center}
Save summaries
\end{center}

\end{minipage}};
\draw    (376,689) -- (512,689) -- (512,714) -- (376,714) -- cycle  ;
\draw (379,693) node [anchor=north west][inner sep=0.75pt]   [align=left] {\begin{minipage}[lt]{90.07pt}\setlength\topsep{0pt}
\begin{center}
Replicates analysis
\end{center}

\end{minipage}};
\draw    (163,574) -- (301,574) -- (301,599) -- (163,599) -- cycle  ;
\draw (166,578) node [anchor=north west][inner sep=0.75pt]   [align=left] {\begin{minipage}[lt]{92.92pt}\setlength\topsep{0pt}
\begin{center}
Check accuracy
\end{center}

\end{minipage}};
\draw    (195,619) -- (267,619) -- (267,644) -- (195,644) -- cycle  ;
\draw (198,623) node [anchor=north west][inner sep=0.75pt]   [align=left] {\begin{minipage}[lt]{47.53pt}\setlength\topsep{0pt}
\begin{center}
Save logs
\end{center}

\end{minipage}};
\draw    (259,725) -- (413,725) -- (413,771) -- (259,771) -- cycle  ;
\draw (262,729) node [anchor=north west][inner sep=0.75pt]   [align=left] {\begin{minipage}[lt]{102.01pt}\setlength\topsep{0pt}
\begin{center}
Frequentist operating \\characteristics
\end{center}

\end{minipage}};
\draw    (386,797) -- (490,797) -- (490,843) -- (386,843) -- cycle  ;
\draw (389,801) node [anchor=north west][inner sep=0.75pt]   [align=left] {\begin{minipage}[lt]{68.51pt}\setlength\topsep{0pt}
Save analysis 
\begin{center}
results
\end{center}

\end{minipage}};
\draw    (333,873.67) -- (543,873.67) -- (543,898.67) -- (333,898.67) -- cycle  ;
\draw (336,877.67) node [anchor=north west][inner sep=0.75pt]   [align=left] {\begin{minipage}[lt]{139.98pt}\setlength\topsep{0pt}
\begin{center}
Computation of Bayesian OCs
\end{center}

\end{minipage}};
\draw (42,622) node [anchor=north west][inner sep=0.75pt]   [align=left] {$N_{\text{sims}}$ replicates};
\draw    (525.52,179) -- (580.52,179) -- (580.52,225) -- (525.52,225) -- cycle  ;
\draw (528.52,183) node [anchor=north west][inner sep=0.75pt]   [align=left] {\begin{minipage}[lt]{35.05pt}\setlength\topsep{0pt}
\begin{center}
Source\\ data \ 
\end{center}

\end{minipage}};
\draw  [fill={rgb, 255:red, 84; green, 120; blue, 222 }  ,fill opacity=1 ]  (138,107) -- (261,107) -- (261,153) -- (138,153) -- cycle  ;
\draw (141,111) node [anchor=north west][inner sep=0.75pt]   [align=left] {\begin{minipage}[lt]{83.23pt}\setlength\topsep{0pt}
\begin{center}
Target study\\($\theta_T, N_T, \sigma^2_T$)
\end{center}

\end{minipage}};
\draw  [fill={rgb, 255:red, 68; green, 116; blue, 14 }  ,fill opacity=0.53 ]  (572.04,347) -- (630.04,347) -- (630.04,372) -- (572.04,372) -- cycle  ;
\draw (575.04,351) node [anchor=north west][inner sep=0.75pt]   [align=left] {\begin{minipage}[lt]{36.77pt}\setlength\topsep{0pt}
\begin{center}
Method
\end{center}

\end{minipage}};
\draw    (307.72,181) -- (429.72,181) -- (429.72,227) -- (307.72,227) -- cycle  ;
\draw (310.72,185) node [anchor=north west][inner sep=0.75pt]   [align=left] {\begin{minipage}[lt]{80.44pt}\setlength\topsep{0pt}
\begin{center}
Source likelihood\\function
\end{center}

\end{minipage}};
\draw    (286,617) -- (418,617) -- (418,642) -- (286,642) -- cycle  ;
\draw (289,621) node [anchor=north west][inner sep=0.75pt]   [align=left] {\begin{minipage}[lt]{89.44pt}\setlength\topsep{0pt}
\begin{center}
Compute prior ESS
\end{center}

\end{minipage}};
\draw    (437,737) -- (626,737) -- (626,762) -- (437,762) -- cycle  ;
\draw (440,741) node [anchor=north west][inner sep=0.75pt]   [align=left] {\begin{minipage}[lt]{125.81pt}\setlength\topsep{0pt}
\begin{center}
Metrics related to inference
\end{center}

\end{minipage}};
\draw    (561.46,274) -- (638.46,274) -- (638.46,320) -- (561.46,320) -- cycle  ;
\draw (564.46,278) node [anchor=north west][inner sep=0.75pt]   [align=left] {\begin{minipage}[lt]{49.8pt}\setlength\topsep{0pt}
\begin{center}
Initial prior\\$\pi_0(\theta_T)$
\end{center}

\end{minipage}};
\draw    (35,426) -- (193,426) -- (193,451) -- (35,451) -- cycle  ;
\draw (38,430) node [anchor=north west][inner sep=0.75pt]   [align=left] {\begin{minipage}[lt]{104.87pt}\setlength\topsep{0pt}
\begin{center}
Target data generation
\end{center}

\end{minipage}};
\draw    (233,313) -- (360,313) -- (360,359) -- (233,359) -- cycle  ;
\draw (236,317) node [anchor=north west][inner sep=0.75pt]   [align=left] {\begin{minipage}[lt]{83.86pt}\setlength\topsep{0pt}
\begin{center}
Target\\likelihood function
\end{center}

\end{minipage}};
\draw    (30,238) -- (202,238) -- (202,283) -- (30,283) -- cycle  ;
\draw (33,242) node [anchor=north west][inner sep=0.75pt]   [align=left] {\begin{minipage}[lt]{115.07pt}\setlength\topsep{0pt}
\begin{center}
True target\\ data generating process
\end{center}

\end{minipage}};
\draw    (8,304) -- (221,304) -- (221,349) -- (8,349) -- cycle  ;
\draw (11,308) node [anchor=north west][inner sep=0.75pt]   [align=left] {\begin{minipage}[lt]{142.3pt}\setlength\topsep{0pt}
\begin{center}
Data generation approximation\\(TRUE/FALSE) 
\end{center}

\end{minipage}};
\draw    (159,533) -- (247,533) ;
\draw [shift={(250,533)}, rotate = 180] [fill={rgb, 255:red, 0; green, 0; blue, 0 }  ][line width=0.08]  [draw opacity=0] (8.93,-4.29) -- (0,0) -- (8.93,4.29) -- cycle    ;
\draw    (347,543.2) -- (397.06,553.73) ;
\draw [shift={(400,554.35)}, rotate = 191.88] [fill={rgb, 255:red, 0; green, 0; blue, 0 }  ][line width=0.08]  [draw opacity=0] (8.93,-4.29) -- (0,0) -- (8.93,4.29) -- cycle    ;
\draw    (443.55,576) -- (443.94,686) ;
\draw [shift={(443.95,689)}, rotate = 269.79] [fill={rgb, 255:red, 0; green, 0; blue, 0 }  ][line width=0.08]  [draw opacity=0] (8.93,-4.29) -- (0,0) -- (8.93,4.29) -- cycle    ;
\draw    (400,568.25) -- (305.98,578.53) ;
\draw [shift={(303,578.85)}, rotate = 353.76] [fill={rgb, 255:red, 0; green, 0; blue, 0 }  ][line width=0.08]  [draw opacity=0] (8.93,-4.29) -- (0,0) -- (8.93,4.29) -- cycle    ;
\draw    (232.72,599) -- (232.34,616) ;
\draw [shift={(232.28,619)}, rotate = 271.27] [fill={rgb, 255:red, 0; green, 0; blue, 0 }  ][line width=0.08]  [draw opacity=0] (8.93,-4.29) -- (0,0) -- (8.93,4.29) -- cycle    ;
\draw    (414.97,714) -- (392.17,723.81) ;
\draw [shift={(389.42,725)}, rotate = 336.71] [fill={rgb, 255:red, 0; green, 0; blue, 0 }  ][line width=0.08]  [draw opacity=0] (8.93,-4.29) -- (0,0) -- (8.93,4.29) -- cycle    ;
\draw    (368.58,771) -- (402.97,795.27) ;
\draw [shift={(405.42,797)}, rotate = 215.22] [fill={rgb, 255:red, 0; green, 0; blue, 0 }  ][line width=0.08]  [draw opacity=0] (8.93,-4.29) -- (0,0) -- (8.93,4.29) -- cycle    ;
\draw    (438,843) -- (438,870.67) ;
\draw [shift={(438,873.67)}, rotate = 270] [fill={rgb, 255:red, 0; green, 0; blue, 0 }  ][line width=0.08]  [draw opacity=0] (8.93,-4.29) -- (0,0) -- (8.93,4.29) -- cycle    ;
\draw    (488.67,154) -- (523.12,179.7) ;
\draw [shift={(525.52,181.49)}, rotate = 216.72] [fill={rgb, 255:red, 0; green, 0; blue, 0 }  ][line width=0.08]  [draw opacity=0] (8.93,-4.29) -- (0,0) -- (8.93,4.29) -- cycle    ;
\draw    (428.88,154) -- (398.7,179.08) ;
\draw [shift={(396.39,181)}, rotate = 320.27] [fill={rgb, 255:red, 0; green, 0; blue, 0 }  ][line width=0.08]  [draw opacity=0] (8.93,-4.29) -- (0,0) -- (8.93,4.29) -- cycle    ;
\draw    (386.27,227) -- (441.7,299.62) ;
\draw [shift={(443.52,302)}, rotate = 232.64] [fill={rgb, 255:red, 0; green, 0; blue, 0 }  ][line width=0.08]  [draw opacity=0] (8.93,-4.29) -- (0,0) -- (8.93,4.29) -- cycle    ;
\draw    (535.83,225) -- (480.07,299.6) ;
\draw [shift={(478.27,302)}, rotate = 306.78] [fill={rgb, 255:red, 0; green, 0; blue, 0 }  ][line width=0.08]  [draw opacity=0] (8.93,-4.29) -- (0,0) -- (8.93,4.29) -- cycle    ;
\draw    (572.04,352.35) -- (539.49,344.33) ;
\draw [shift={(536.58,343.61)}, rotate = 13.85] [fill={rgb, 255:red, 0; green, 0; blue, 0 }  ][line width=0.08]  [draw opacity=0] (8.93,-4.29) -- (0,0) -- (8.93,4.29) -- cycle    ;
\draw    (443.1,348) -- (318.33,507.64) ;
\draw [shift={(316.48,510)}, rotate = 308.01] [fill={rgb, 255:red, 0; green, 0; blue, 0 }  ][line width=0.08]  [draw opacity=0] (8.93,-4.29) -- (0,0) -- (8.93,4.29) -- cycle    ;
\draw    (426.45,576) -- (372.96,615.23) ;
\draw [shift={(370.55,617)}, rotate = 323.75] [fill={rgb, 255:red, 0; green, 0; blue, 0 }  ][line width=0.08]  [draw opacity=0] (8.93,-4.29) -- (0,0) -- (8.93,4.29) -- cycle    ;
\draw    (286,630.61) -- (272,630.84) ;
\draw [shift={(269,630.89)}, rotate = 359.06] [fill={rgb, 255:red, 0; green, 0; blue, 0 }  ][line width=0.08]  [draw opacity=0] (8.93,-4.29) -- (0,0) -- (8.93,4.29) -- cycle    ;
\draw    (514.92,762) -- (470.9,795.19) ;
\draw [shift={(468.5,797)}, rotate = 322.98] [fill={rgb, 255:red, 0; green, 0; blue, 0 }  ][line width=0.08]  [draw opacity=0] (8.93,-4.29) -- (0,0) -- (8.93,4.29) -- cycle    ;
\draw    (466.79,714) -- (506.08,735.56) ;
\draw [shift={(508.71,737)}, rotate = 208.75] [fill={rgb, 255:red, 0; green, 0; blue, 0 }  ][line width=0.08]  [draw opacity=0] (8.93,-4.29) -- (0,0) -- (8.93,4.29) -- cycle    ;
\draw    (561.46,304.76) -- (539.52,309.18) ;
\draw [shift={(536.58,309.78)}, rotate = 348.6] [fill={rgb, 255:red, 0; green, 0; blue, 0 }  ][line width=0.08]  [draw opacity=0] (8.93,-4.29) -- (0,0) -- (8.93,4.29) -- cycle    ;
\draw    (113.74,451) -- (112.53,508) ;
\draw [shift={(112.49,510)}, rotate = 271.21] [color={rgb, 255:red, 0; green, 0; blue, 0 }  ][line width=0.75]    (10.93,-3.29) .. controls (6.95,-1.4) and (3.31,-0.3) .. (0,0) .. controls (3.31,0.3) and (6.95,1.4) .. (10.93,3.29)   ;
\draw    (296.73,359) -- (298.25,508) ;
\draw [shift={(298.27,510)}, rotate = 269.42] [color={rgb, 255:red, 0; green, 0; blue, 0 }  ][line width=0.75]    (10.93,-3.29) .. controls (6.95,-1.4) and (3.31,-0.3) .. (0,0) .. controls (3.31,0.3) and (6.95,1.4) .. (10.93,3.29)   ;
\draw    (114.4,350) -- (114.07,424) ;
\draw [shift={(114.06,426)}, rotate = 270.26] [color={rgb, 255:red, 0; green, 0; blue, 0 }  ][line width=0.75]    (10.93,-3.29) .. controls (6.95,-1.4) and (3.31,-0.3) .. (0,0) .. controls (3.31,0.3) and (6.95,1.4) .. (10.93,3.29)   ;
\draw    (115.8,284) -- (115.26,302) ;
\draw [shift={(115.2,304)}, rotate = 271.74] [color={rgb, 255:red, 0; green, 0; blue, 0 }  ][line width=0.75]    (10.93,-3.29) .. controls (6.95,-1.4) and (3.31,-0.3) .. (0,0) .. controls (3.31,0.3) and (6.95,1.4) .. (10.93,3.29)   ;
\draw    (186.16,153) -- (132.42,236.32) ;
\draw [shift={(131.34,238)}, rotate = 302.82] [color={rgb, 255:red, 0; green, 0; blue, 0 }  ][line width=0.75]    (10.93,-3.29) .. controls (6.95,-1.4) and (3.31,-0.3) .. (0,0) .. controls (3.31,0.3) and (6.95,1.4) .. (10.93,3.29)   ;
\draw    (211.66,153) -- (285,311.19) ;
\draw [shift={(285.84,313)}, rotate = 245.13] [color={rgb, 255:red, 0; green, 0; blue, 0 }  ][line width=0.75]    (10.93,-3.29) .. controls (6.95,-1.4) and (3.31,-0.3) .. (0,0) .. controls (3.31,0.3) and (6.95,1.4) .. (10.93,3.29)   ;
\draw    (394.66,130.38) -- (266,130.13) ;
\draw [shift={(264,130.12)}, rotate = 0.11] [color={rgb, 255:red, 0; green, 0; blue, 0 }  ][line width=0.75]    (10.93,-3.29) .. controls (6.95,-1.4) and (3.31,-0.3) .. (0,0) .. controls (3.31,0.3) and (6.95,1.4) .. (10.93,3.29)   ;

\end{tikzpicture}

\end{adjustbox}

\caption[Simulation pipeline]{Summary of the simulation study pipeline. Colored boxes correspond to components of the configuration that will be varied.} 
\label{fig:simulation_pipeline}
\end{figure}

\clearpage
\subsection{Probability of study success as a function of the drift in treatment effect}

\begin{figure}[htbp]
    \centering
    \includegraphics[width=0.95\linewidth, height=0.95\textheight,keepaspectratio]{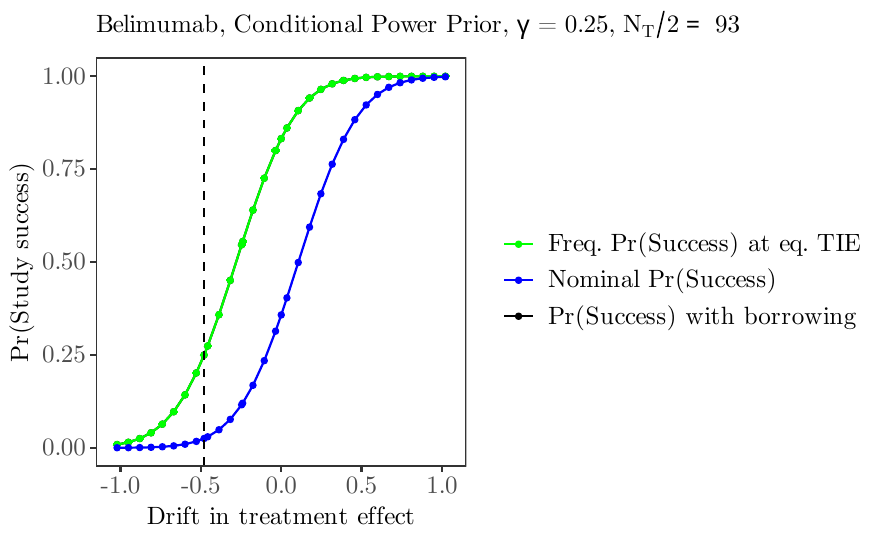}
    \caption{Probability of success of the CPP with $\gamma = 0.25$ as a function of the drift in treatment effect (black) in the belimumab case study at a sample size per arm of 93, without change introduced in the denominator of the source study summary measure. The probability of success of the frequentist test of the target data alone (the randomized Wald test) at a nominal type I error rate of $0.025$ as a function of drift is displayed in blue. The probability of success of the same test at a type I error rate equal to that of the CPP is displayed in green.
    Borrowing of external data that favors the alternative hypothesis also implies that the probability of success of the borrowing method is always larger, in the alternative hypothesis space, than the probability of success of the frequentist method at the nominal type I error rate of 0.025. At an equivalent type I error rate, the power curves of the CPP and of the frequentist test are almost the same.
$\theta_T = \theta_0$ is indicated by a dashed line. Error bars correspond to the 95\% CI of the metric.}
    \label{fig:power_equivalent_tie}
\end{figure}

\begin{figure}[htbp]
    \centering
    \includegraphics[width=0.95\linewidth, height=0.95\textheight,keepaspectratio]{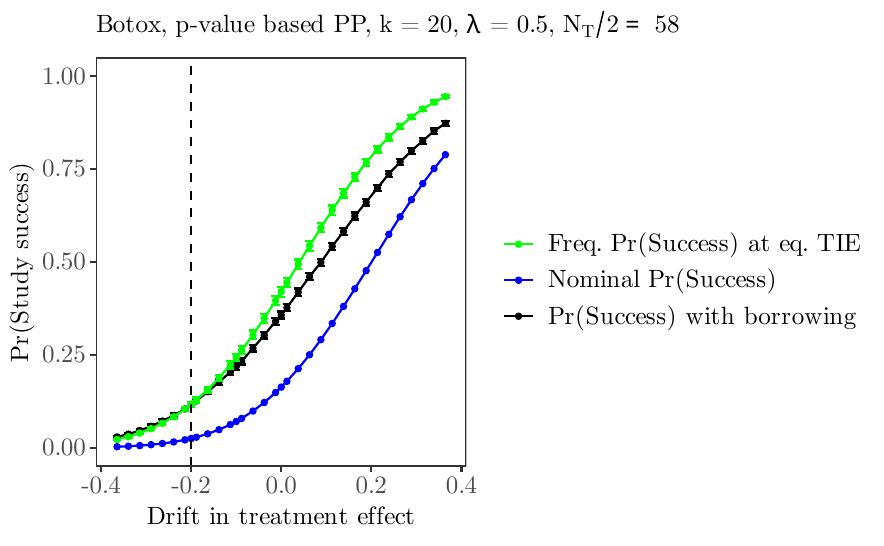}
    \caption{Probability of success of the p-PP with parameters $k= 20$ and $\lambda = 0.5$ as a function of the drift in treatment effect (black) in the botox case study at a sample size per arm of 58, with a sampling standard deviation equal between the source and target study. The probability of success of the t-test at a nominal type I error rate of $0.025$ as a function of drift is displayed in blue. The probability of success of the t-test at a type I error rate equal to the p-PP type I error rate is displayed in green. $\theta_T = \theta_0$ is indicated by a dashed line. In this example, the power of the p-value based power prior is lower than the power of the frequentist t-test at equivalent type I error rate in the whole alternative hypothesis space. Error bars correspond to the 95\% CI of the probability of success.}
    \label{fig:power_loss}
\end{figure}

\clearpage

\subsection{Supplementary figures for the botox case study (continuous endpoint).}

\begin{figure}[htbp]
    \centering
    \includegraphics[width=0.95\linewidth, height=0.95\textheight,keepaspectratio]{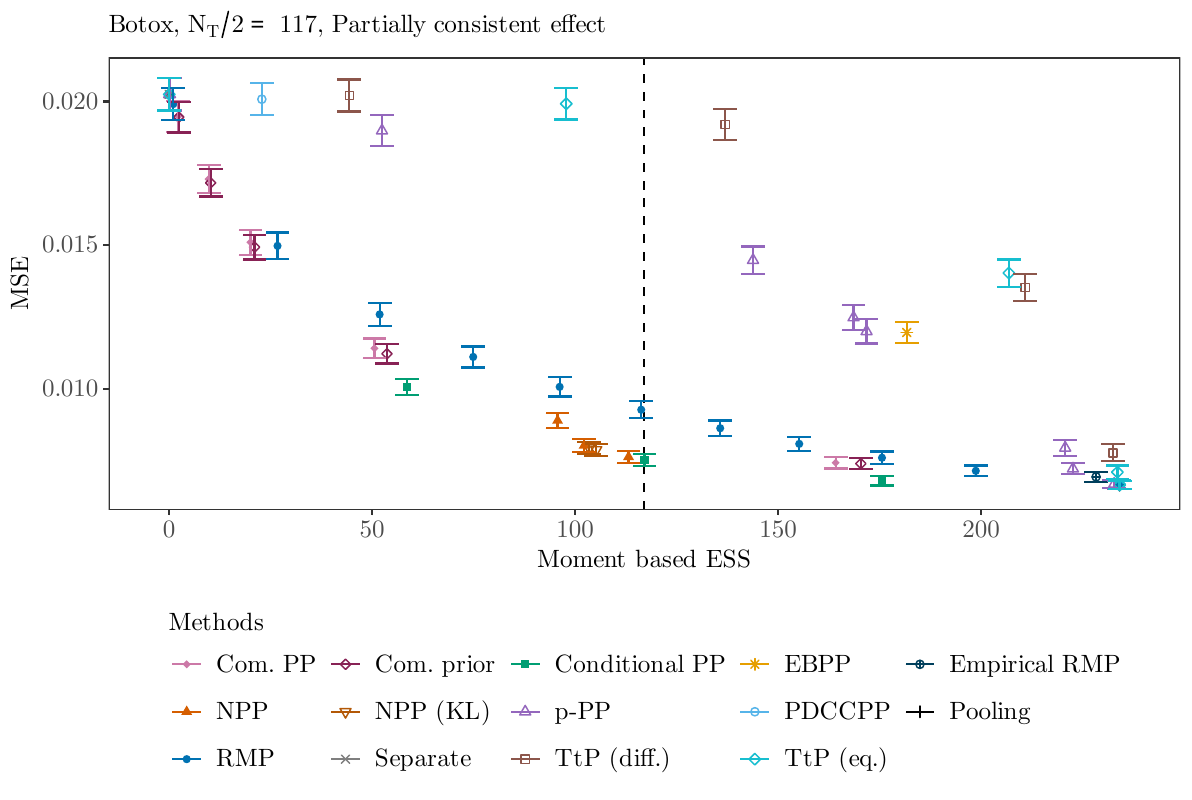}
    \caption{MSE as a function of the mean moment-based ESS in the botox case study with a sample size per arm of 117. Error bars correspond to the 95\% CI of the MSE. Method abbreviations and parameters as in Table~\ref{tab:abbreviations}.}
    \label{fig:botox_mse_vs_ess_moment_sample_size=117_treatment_effect=partially_consistent}
\end{figure}

\begin{figure}[h!]
\centering
\includegraphics[width=0.95\linewidth, height=0.95\textheight,keepaspectratio]{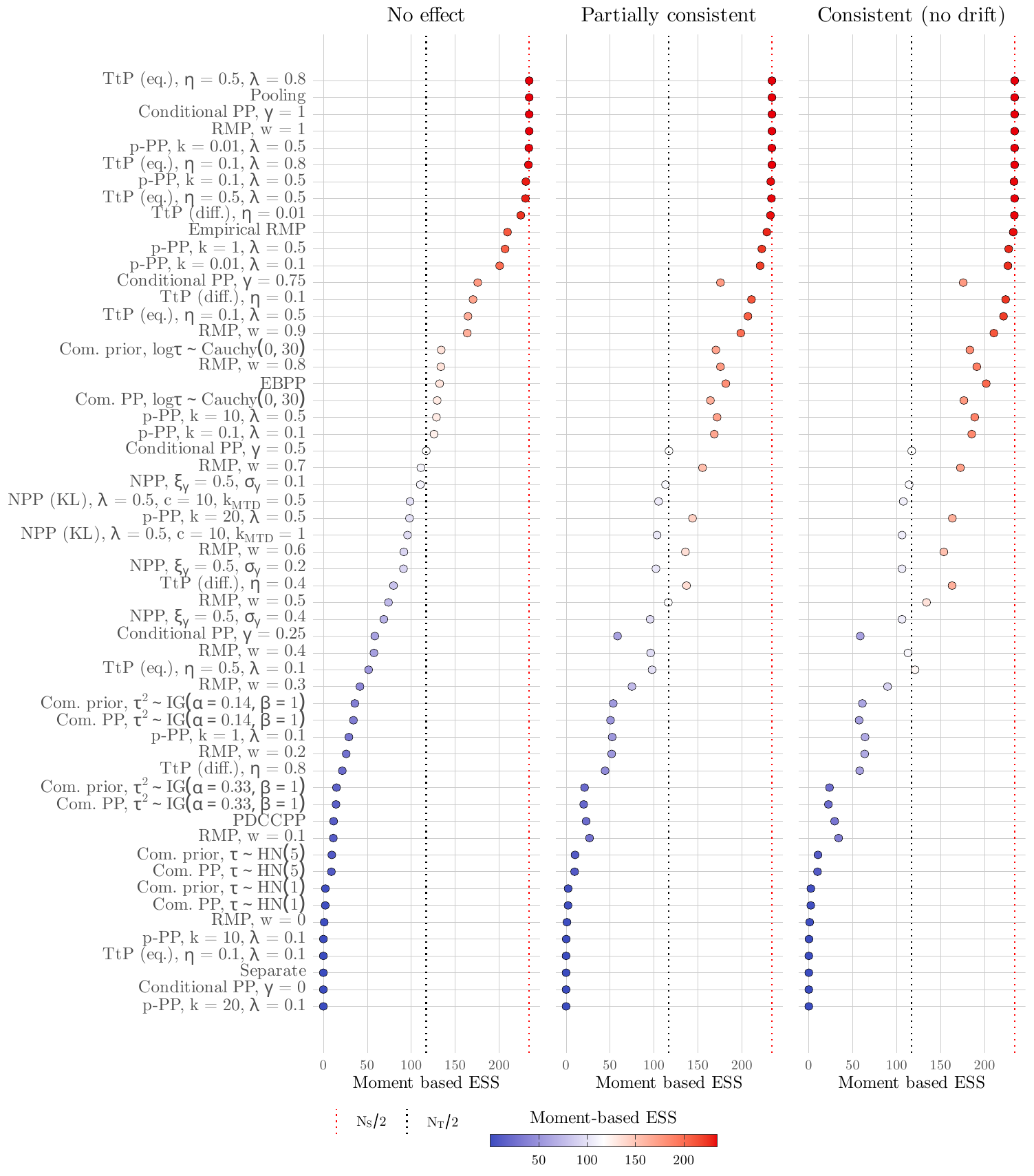}
\caption{Moment-based ESS across methods and for the three main treatment effects considered in the botox case study with $n_T = 117$. The red dashed line corresponds to an ESS equal to the source study sample size per arm, and the black dashed line corresponds to the target study sample size per arm. Method abbreviations and parameters as in Table~\ref{tab:abbreviations}.}
\label{fig:botox_ess_moment_forest_plot_117}
\end{figure}

\begin{figure}[htbp]
    \centering
    \includegraphics[width=0.95\linewidth, height=0.95\textheight,keepaspectratio]{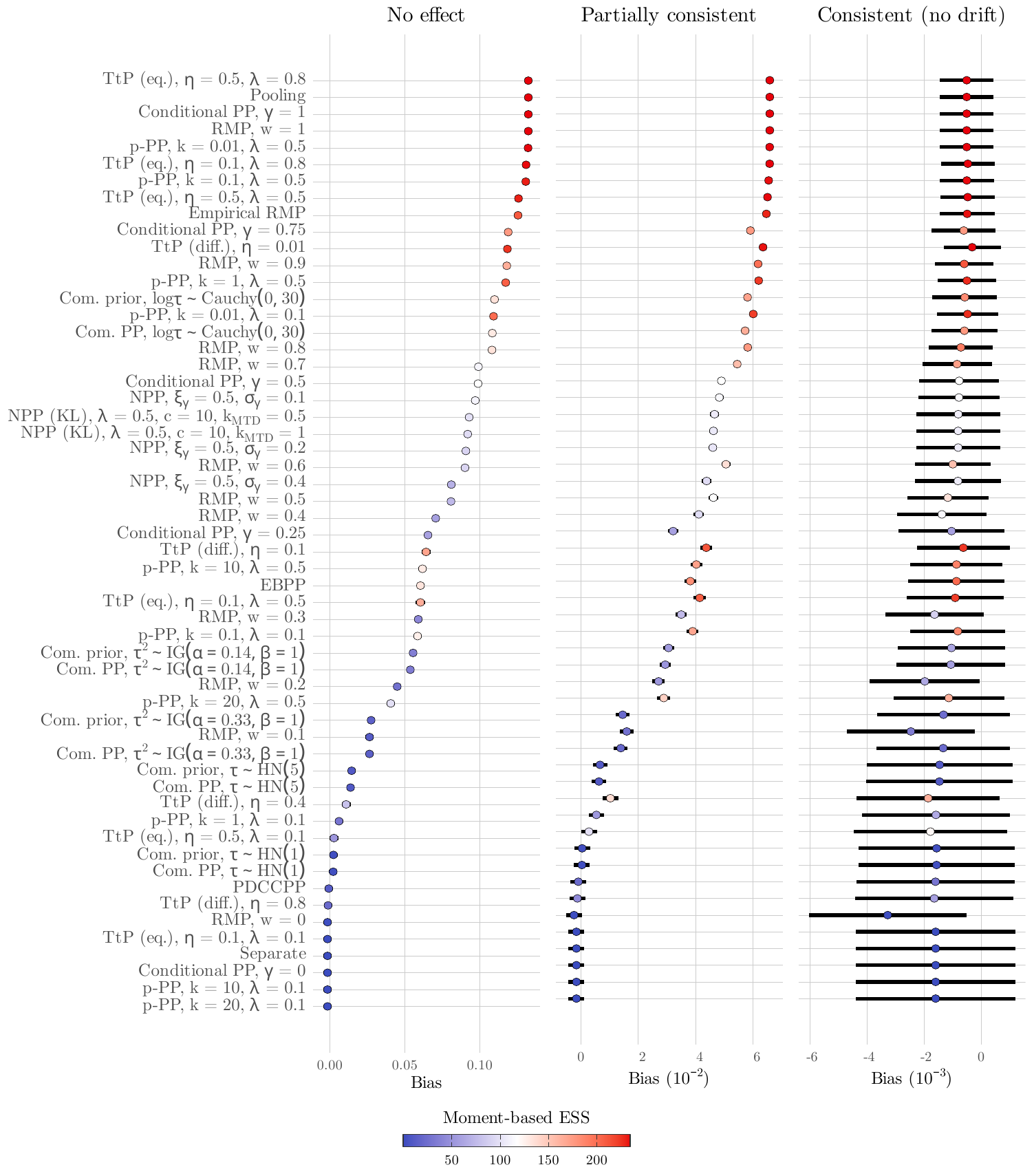}
    \caption{Bias and associated 95\% CI, for the three main treatment effects considered, in the botox case study ($n_T = 117$). Method abbreviations and parameters as in Table~\ref{tab:abbreviations}.}
    \label{fig:botox_bias_forest_plot_target_sample_size_per_arm_117}
\end{figure}

\begin{figure}[htbp]
    \centering
    \includegraphics[width=0.95\linewidth, height=0.95\textheight,keepaspectratio]{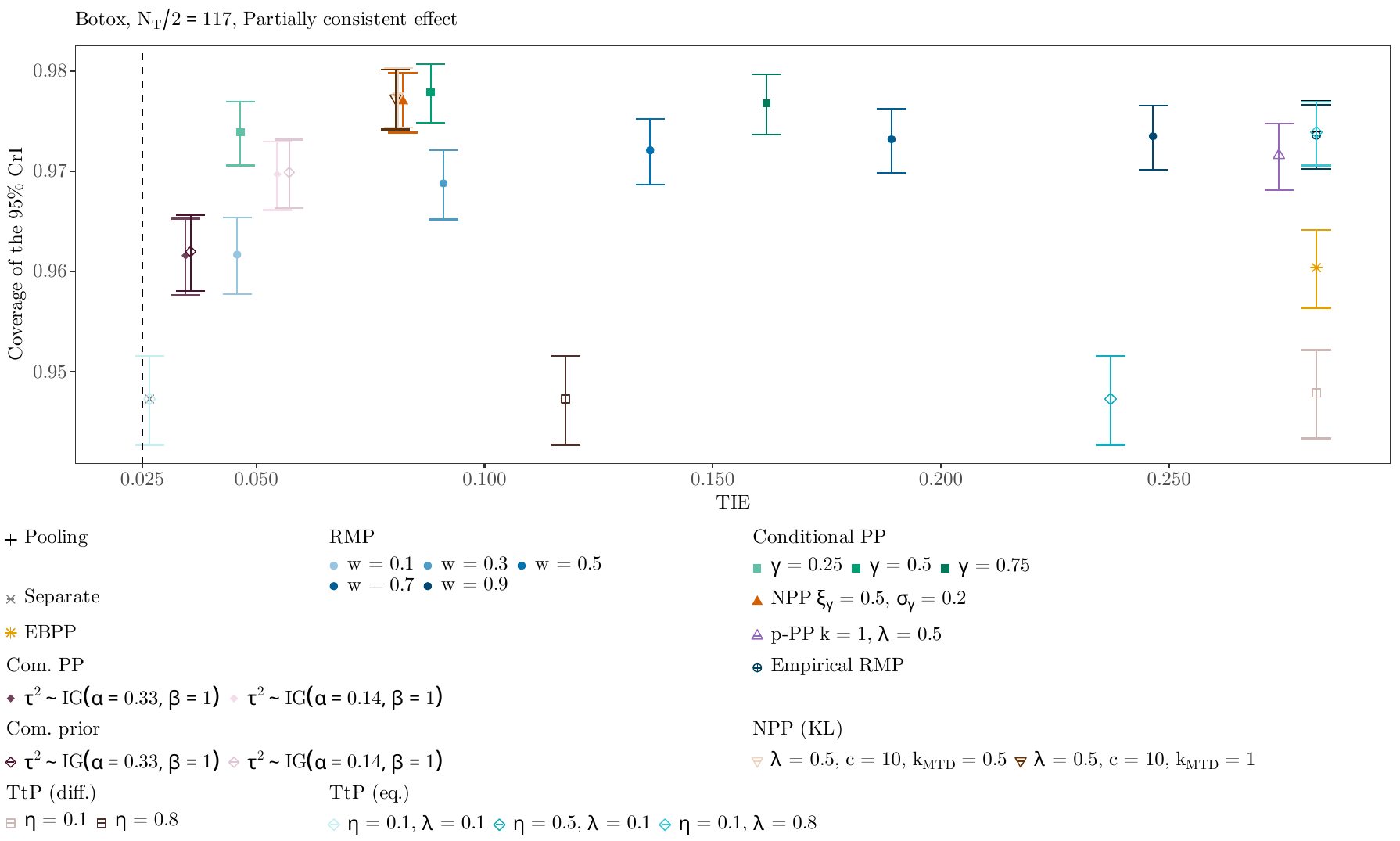}
    \caption{Coverage of the 95\% credible interval as a function of type I error rate. Each point represents one method-parameter combination in the botox case study with a sample size per arm of $117$, across all the methods and parameters, with partially consistent treatment effect. The target to source standard deviation ratio is $1$. Error bars correspond to the 95\% CI of the Coverage and type I error rate. Dashed vertical line represents the nominal type I error rate of $0.025$. Method abbreviations and parameters as in Table~\ref{tab:abbreviations}.}\label{fig:botox_coverage_vs_tie_sample_size=117_treatment_effect=partially_consistent}
\end{figure}

\begin{figure}[htbp]
    \centering
    \includegraphics[width=0.95\linewidth, height=0.95\textheight,keepaspectratio]{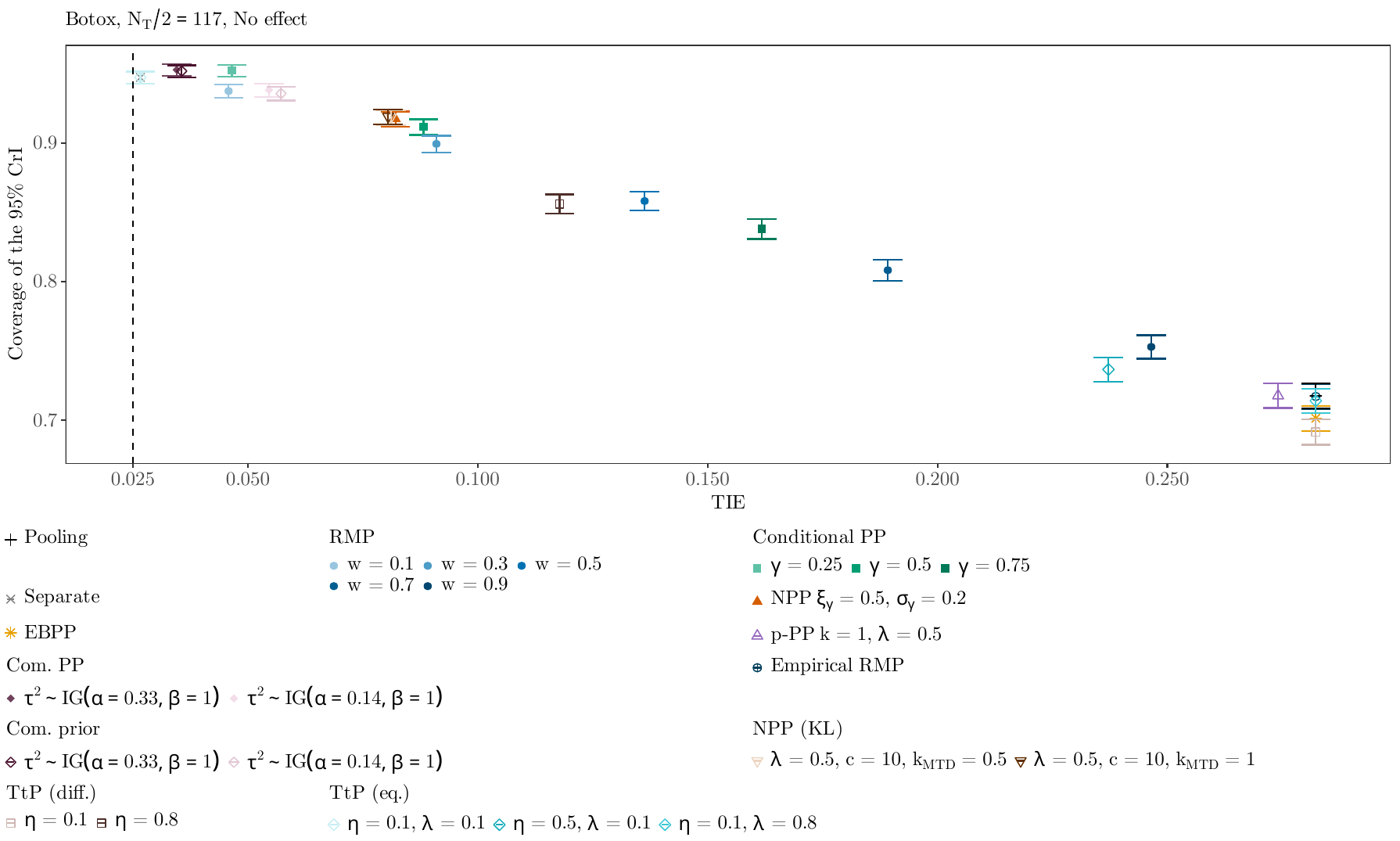}
    \caption{Coverage of the 95\% credible interval as a function of the TIE, in the botox case study with 117 participants per arm and no treatment effect (target to source standard deviation ratio of 1). Each point is one method-parameter configuration. Error bars are 95\% Monte Carlo confidence intervals. The dashed vertical line is the nominal TIE of 0.025. Method abbreviations as in Table~\ref{tab:abbreviations}.}
    \label{fig:botox_coverage_vs_tie_sample_size=117_treatment_effect=no_effect}
\end{figure}

\begin{figure}
    \centering
    \includegraphics[width=0.95\linewidth, height=0.95\textheight,keepaspectratio]{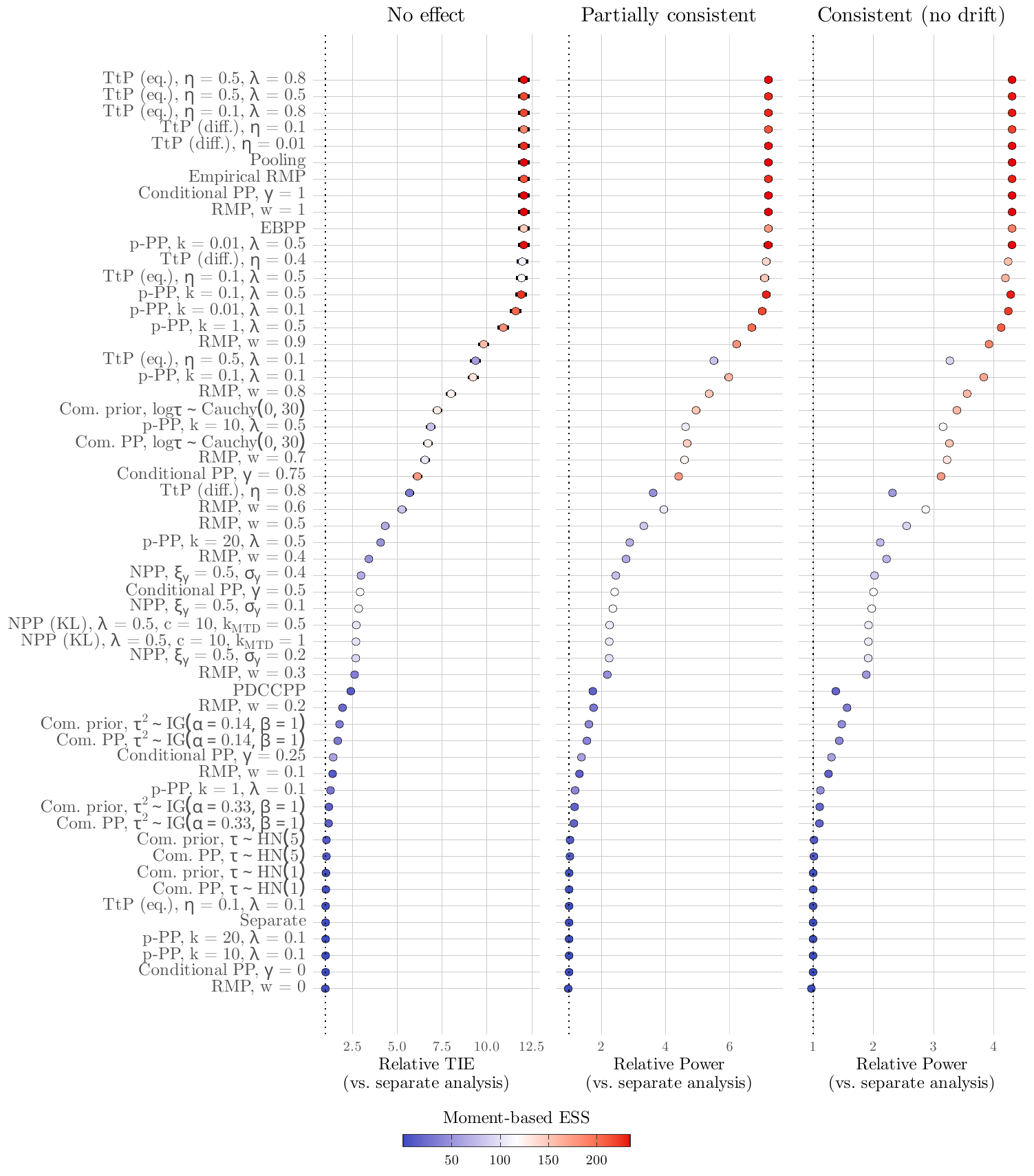}
    \caption{TIE and power relative to a separate analysis for the three main treatment effects considered, in the botox case study ($n_T = 58$). Method abbreviations and parameters as in Table~\ref{tab:abbreviations}.}
    \label{fig:botox_relative_success_proba_forest_plot_target_sample_size_per_arm_58}
\end{figure}

\begin{figure}[p]
  \centering
  \includegraphics[width=\textwidth]{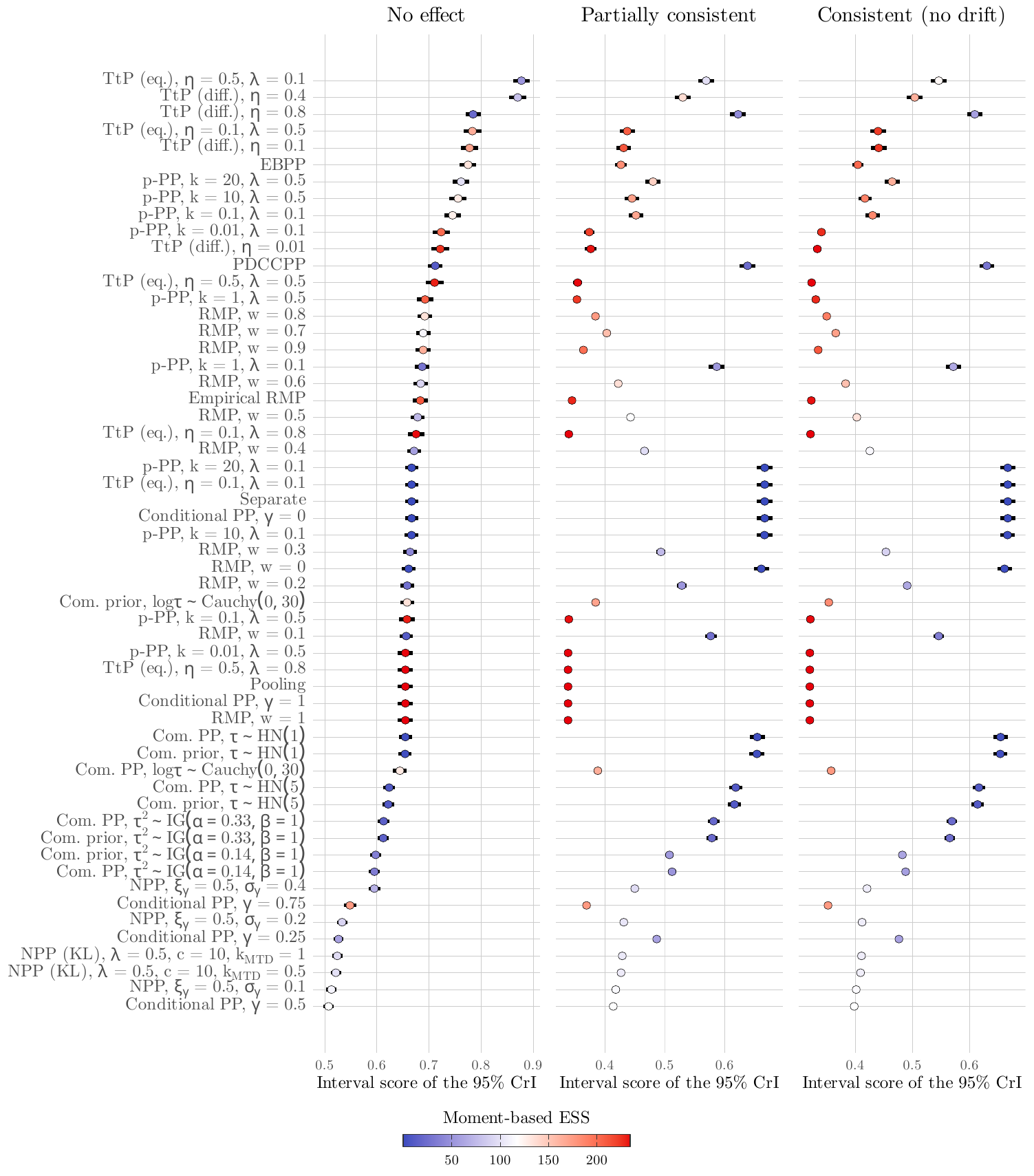}
  \caption{Interval score of the 95\% credible interval for the three main treatment effect values considered in the botox case study ($n_T = 117$), with a target to source standard deviation ratio of 1. For an interval $[l, u]$ and true treatment effect $\theta_T$, the interval score is $(u - l) + \frac{2}{\alpha_{\mathrm{IS}}}(l - \theta_T)\mathbb{1}\{\theta_T < l\} + \frac{2}{\alpha_{\mathrm{IS}}}(\theta_T - u)\mathbb{1}\{\theta_T > u\}$, with $\alpha_{\mathrm{IS}} = 0.05$ \parencite{gneiting_strictly_2007}: it adds to the interval width a penalty proportional to how far the interval misses the true value. Smaller is better. Error bars correspond to the 95\% CI.}
  \label{fig:botox_interval_score_forest_plot_target_sample_size_per_arm_117}
\end{figure}

\begin{figure}[p]
  \centering
  \includegraphics[width=\textwidth]{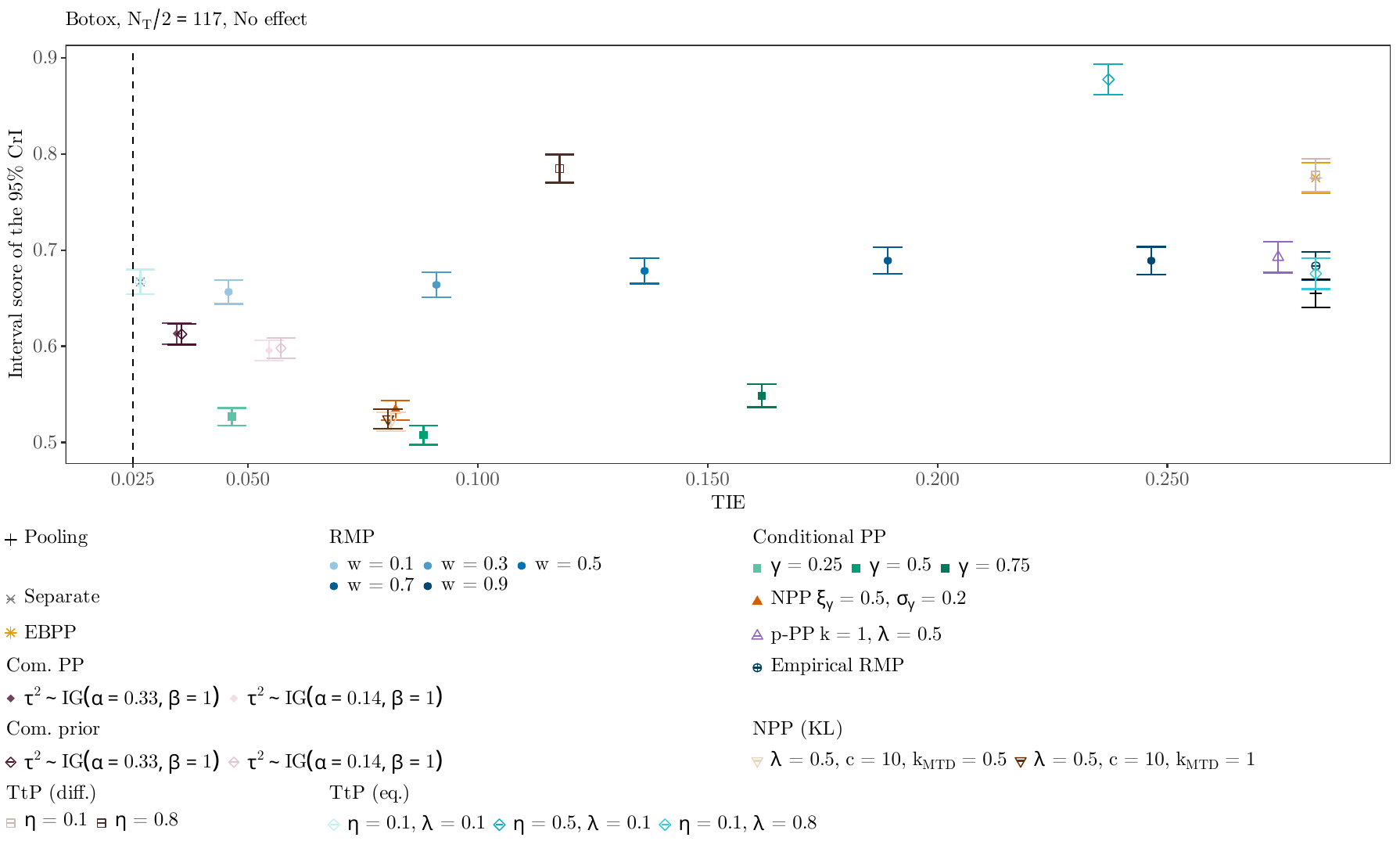}
  \caption{Interval score of the 95\% credible interval as a function of type I error rate. Each point represents one method-parameter combination in the botox case study with a sample size per arm of 117, without treatment effect. The target to source standard deviation ratio is 1. Smaller is better. Error bars correspond to the 95\% CI of the interval score and type I error rate. Dashed vertical line represents the nominal type I error rate of 0.025. } \label{fig:botox_interval_score_vs_tie_sample_size=117_treatment_effect=no_effect}
\end{figure}

\begin{figure}[htbp]
\centering
\includegraphics[width=0.95\linewidth, height=0.95\textheight,keepaspectratio]{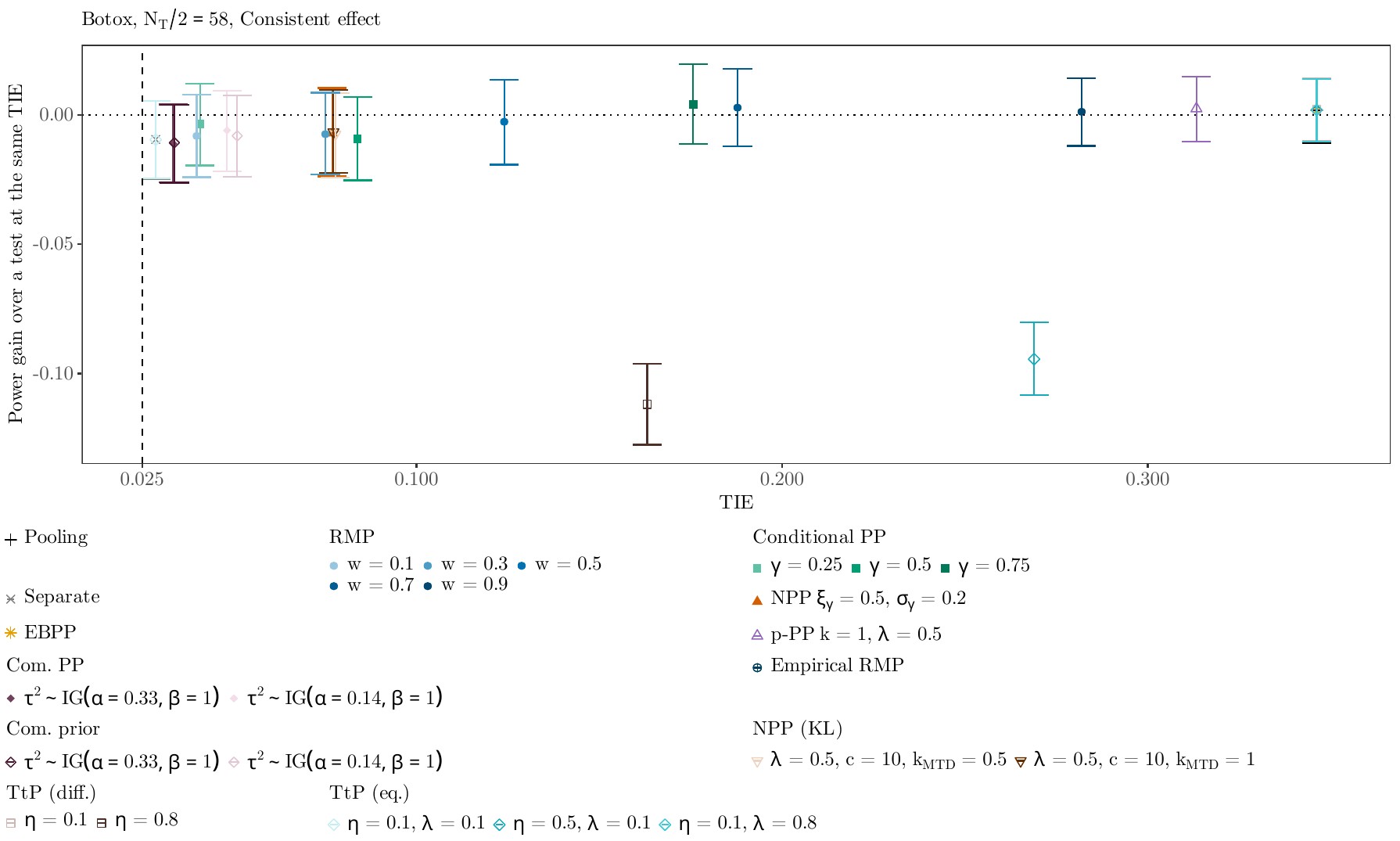}
\caption{Power gain over a frequentist test of the target data alone performed at the same TIE, versus TIE, in the botox case study, with 58 participants per arm and a consistent treatment effect. Error bars are 95\% intervals for the difference. For readability, only a subset of the parameter settings is shown, and the PDCCPP is omitted.}
\label{fig:botox_power_gain_vs_tie_sample_size=58_treatment_effect=consistent_target_to_source_std_ratio=1}
\end{figure}

\clearpage
\subsection{Supplementary figures for the dapagliflozin case study (continuous endpoint).}

\begin{figure}[htbp]
    \centering
    \includegraphics[width=0.95\linewidth, height=0.95\textheight,keepaspectratio]{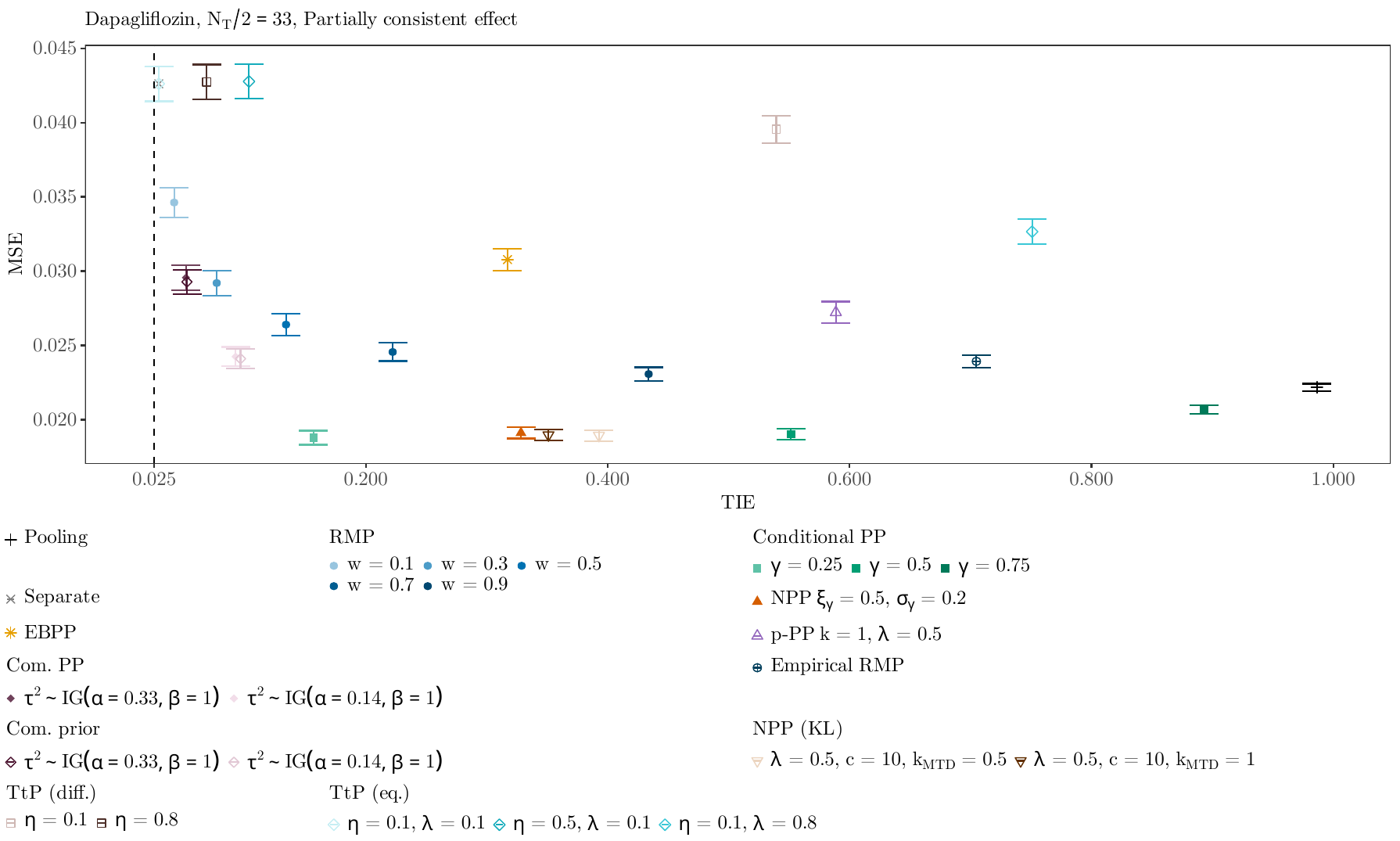}
    \caption{MSE as a function of type I error rate. Each point represents one method-parameter combination in the dapagliflozin case study with a sample size per arm of $33$, across all the methods and parameters, with a partially consistent treatment effect. Error bars correspond to the 95\% CI. Dashed vertical line represents the nominal type I error rate of $0.025$. Method abbreviations and parameters as in Table~\ref{tab:abbreviations}.}
\label{fig:dapagliflozin_mse_vs_tie_sample_size=33_treatment_effect=partially_consistent}
\end{figure}
 
 \begin{figure}[htbp]
\centering
\includegraphics[width=0.95\linewidth, height=0.95\textheight,keepaspectratio]{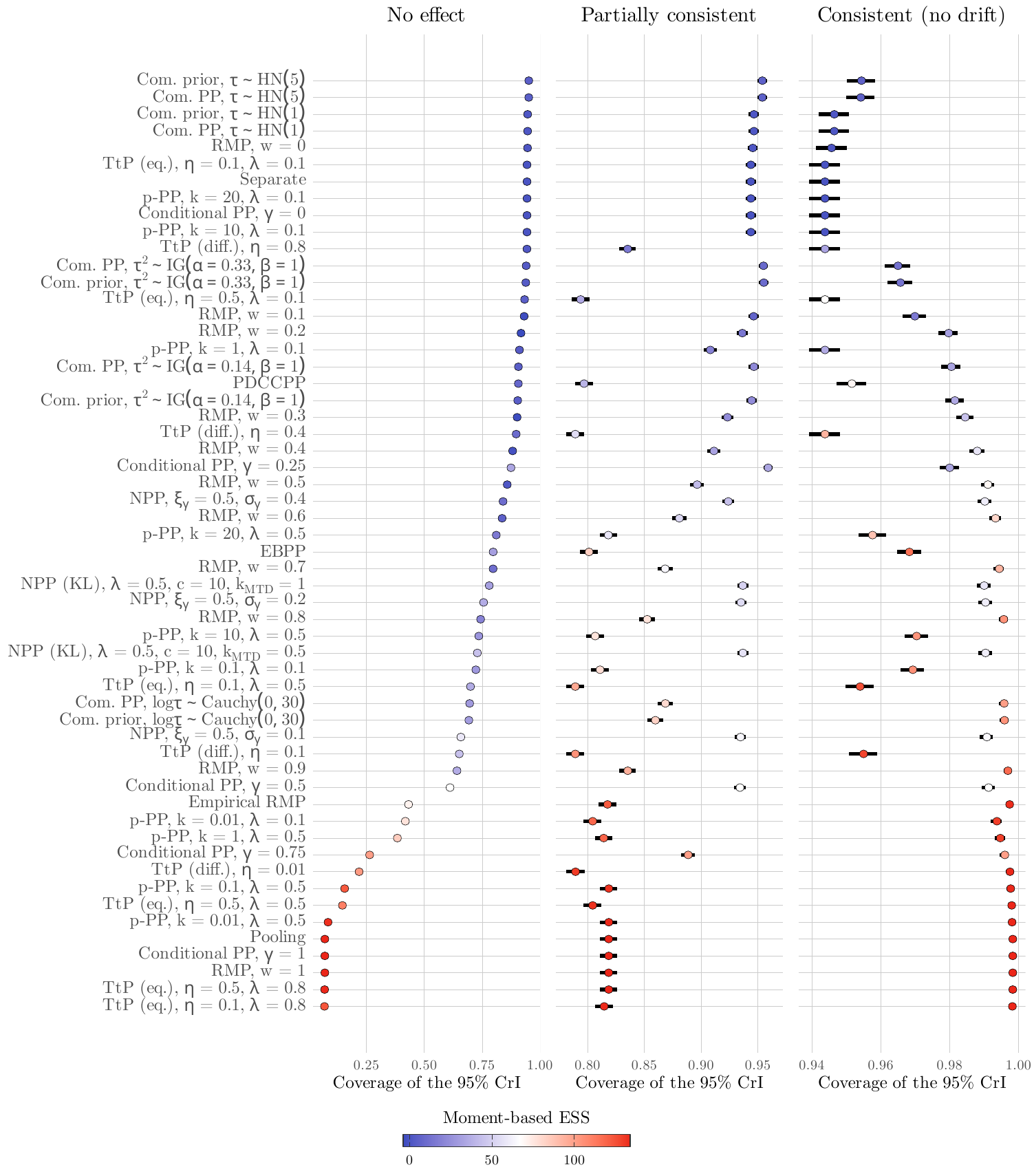}
\caption{Coverage probability of the 95\% credible interval for the three main treatment effect values considered in the dapagliflozin case study ($n_T = 66$). Error bars correspond to the 95\% CI on the coverage probability. Method abbreviations and parameters as in Table~\ref{tab:abbreviations}.}
\label{fig:dapagliflozin_coverage_forest_plot_target_sample_size_per_arm_66}
\end{figure}

\begin{figure}[htbp]
\centering
\includegraphics[width=0.95\linewidth, height=0.95\textheight,keepaspectratio]{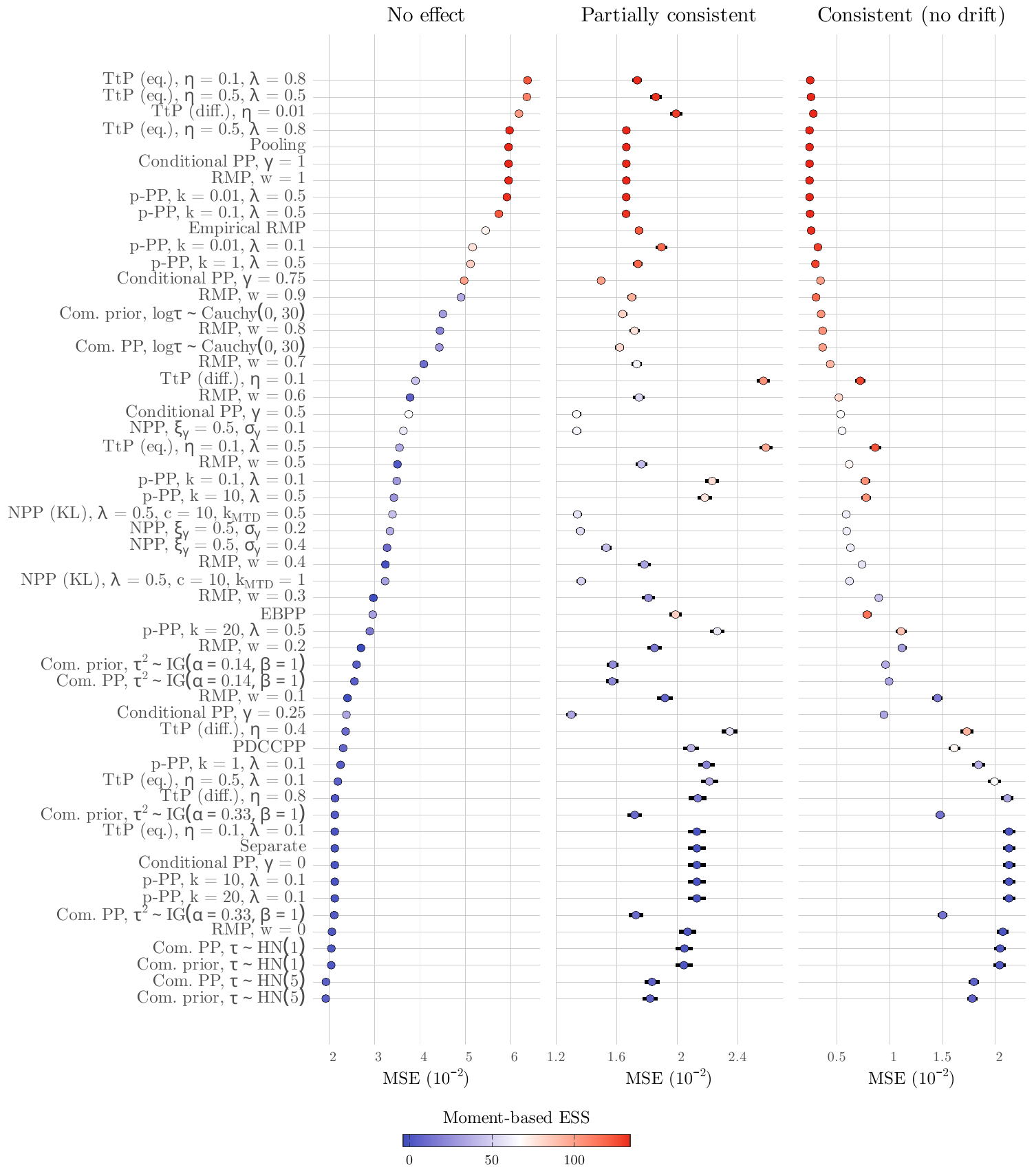}
\caption{MSE for the three main treatment effect values considered in the dapagliflozin case study ($n_T = 66$). Method abbreviations and parameters as in Table~\ref{tab:abbreviations}.}
\label{fig:dapagliflozin_mse_forest_plot_target_sample_size_per_arm_66}
\end{figure} 

\begin{figure}[htbp]
\centering
\includegraphics[width=0.95\linewidth, height=0.95\textheight,keepaspectratio]{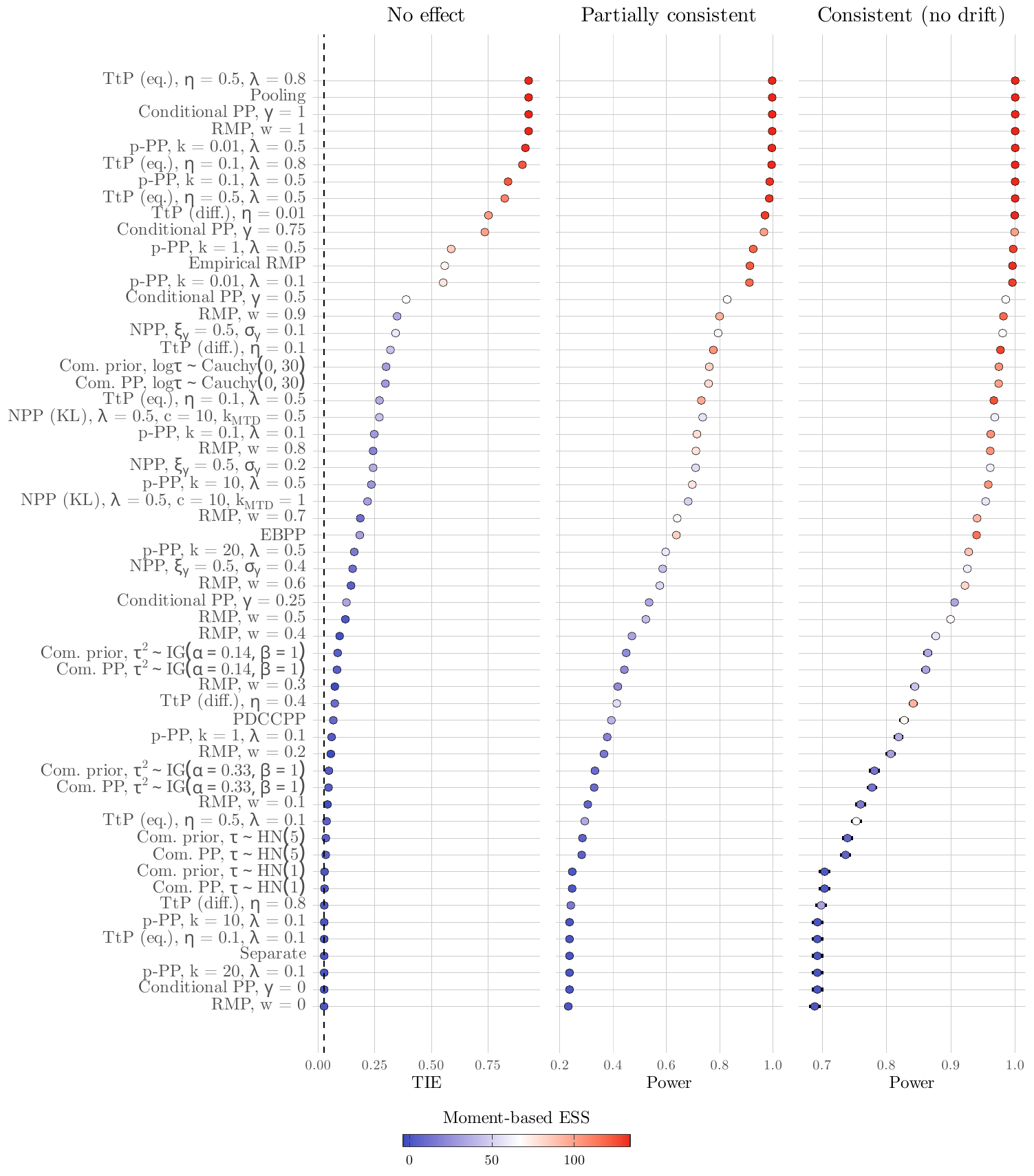}
\caption{Probability of study success for the three main treatment effect values considered in the dapagliflozin case study ($n_T = 66$). Method abbreviations and parameters as in Table~\ref{tab:abbreviations}.}
\label{fig:dapagliflozin_success_proba_forest_plot_target_sample_size_per_arm_66}
\end{figure} 

\begin{figure}[htbp]
\centering
\includegraphics[width=0.95\linewidth, height=0.95\textheight,keepaspectratio]{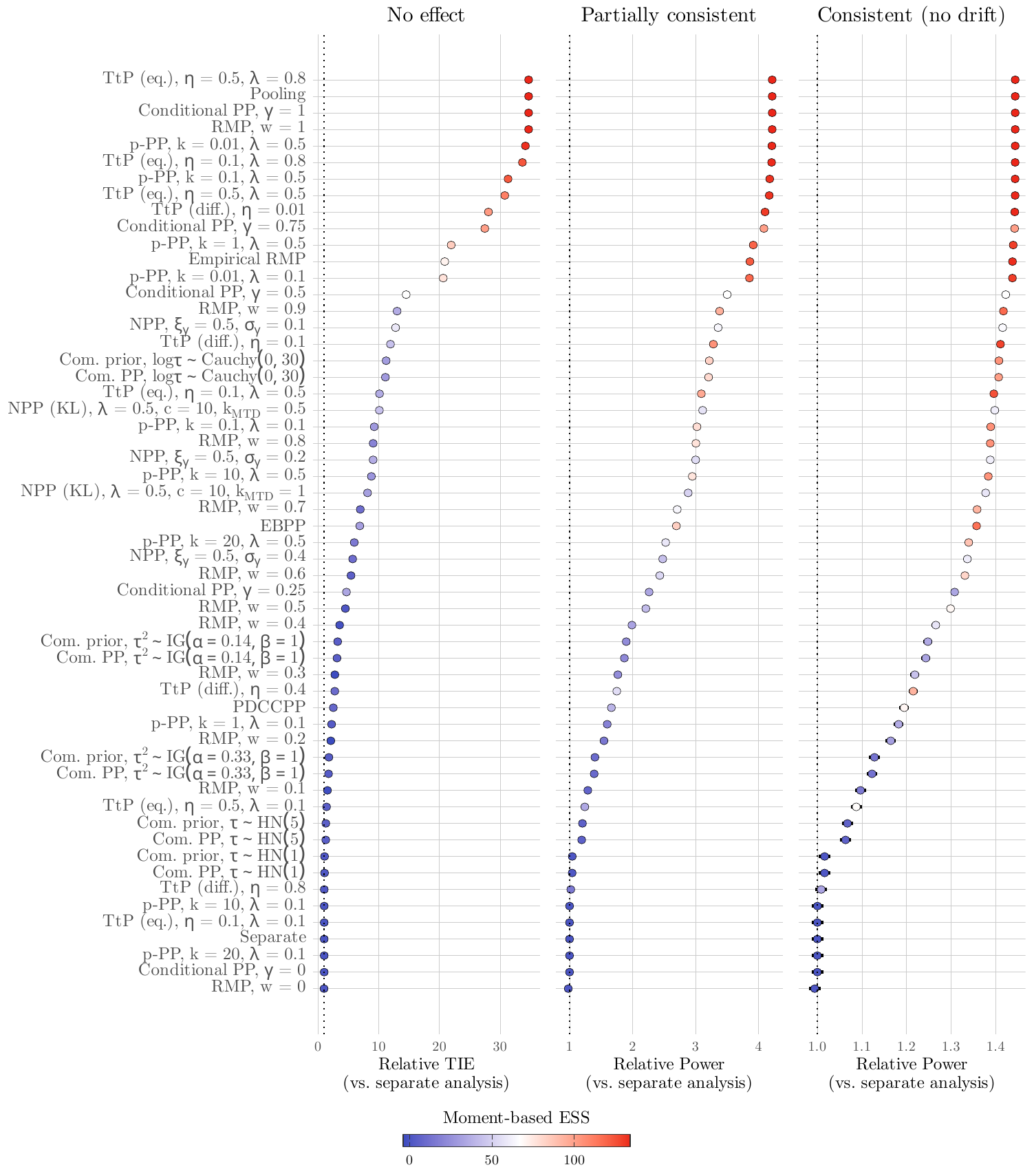}
\caption{Probability of study success for the three main treatment effect values considered in the dapagliflozin case study ($n_T = 66$), relative to the probability of success for a separate analysis of the target study data. Method abbreviations and parameters as in Table~\ref{tab:abbreviations}.}
\label{fig:dapagliflozin_relative_success_proba_forest_plot_target_sample_size_per_arm_66}
\end{figure} 

\begin{figure}[htbp]
  \centering
  \includegraphics[width=\textwidth]{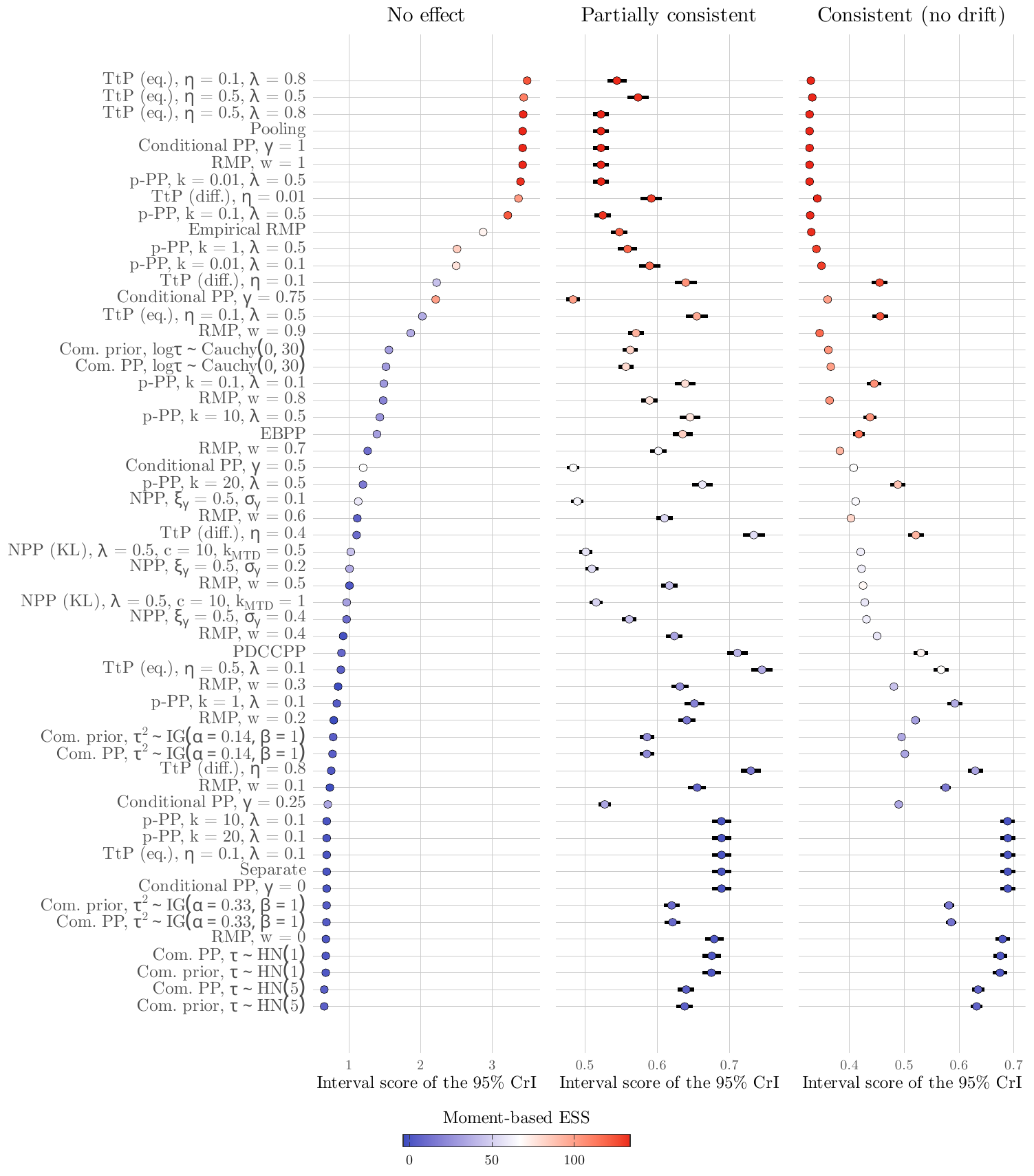}
  \caption{Interval score of the 95\% credible interval for the three main treatment effect values considered in the dapagliflozin case study ($n_T = 66$). Smaller is better. Error bars correspond to the 95\% CI.}
  \label{fig:dapagliflozin_interval_score_forest_plot_target_sample_size_per_arm_66}
\end{figure}

\begin{figure}[htbp]
  \centering
  \includegraphics[width=\textwidth]{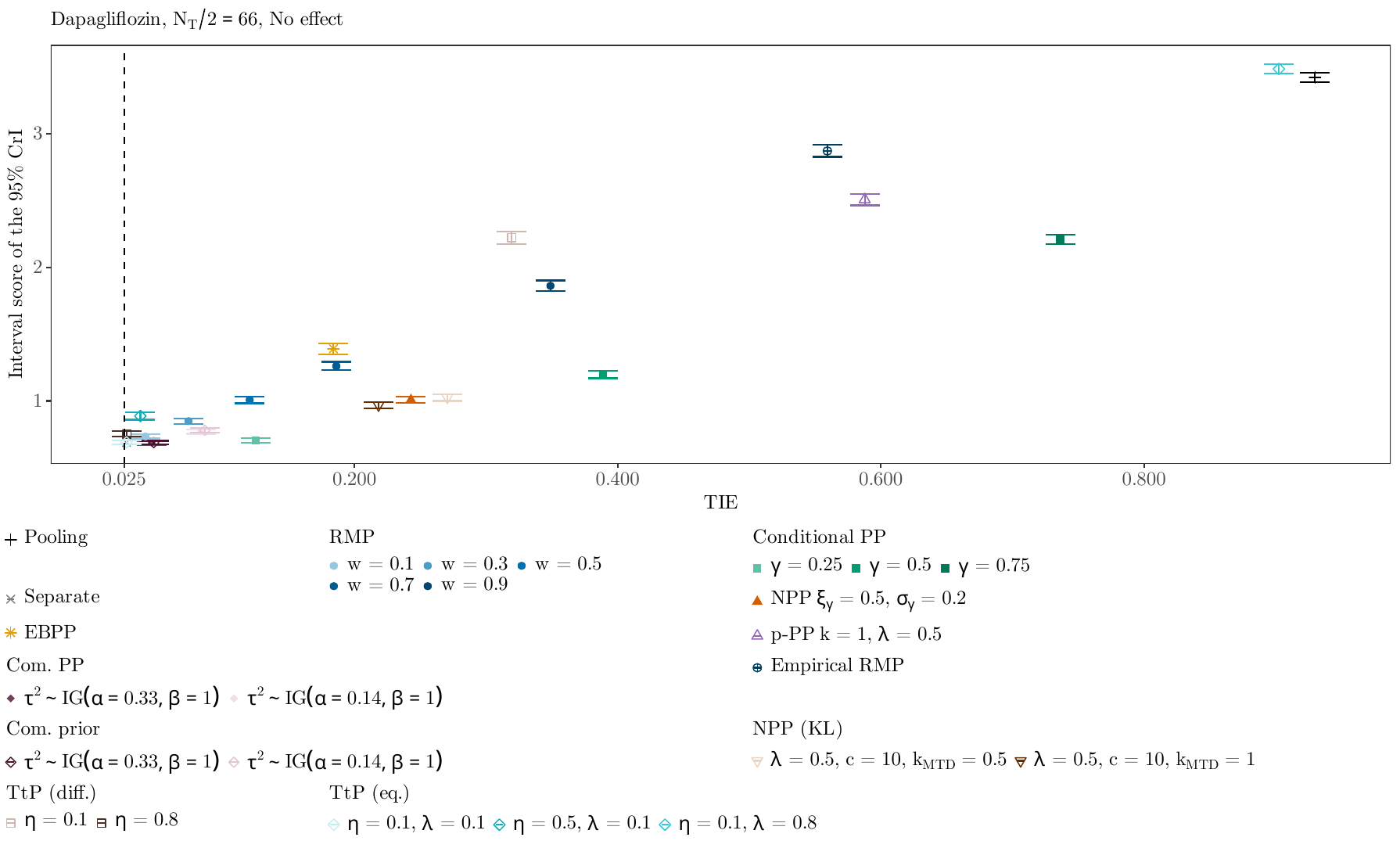}
  \caption{Interval score of the 95\% credible interval as a function of type I error rate in the dapagliflozin case study. Each point represents one method-parameter combination with a sample size per arm of 66, without treatment effect. The target to source standard deviation ratio is 1. Smaller is better. Error bars correspond to the 95\% CI of the interval score and type I error rate. Dashed vertical line represents the nominal type I error rate of 0.025. } \label{fig:dapagliflozin_interval_score_vs_tie_sample_size=66_treatment_effect=no_effect}
\end{figure}

\clearpage
\subsection{Supplementary figures for the belimumab case study (binary endpoint with normal approximation).}

\begin{figure}[htbp]
    \centering
    \includegraphics[width=0.95\linewidth, height=0.95\textheight,keepaspectratio]{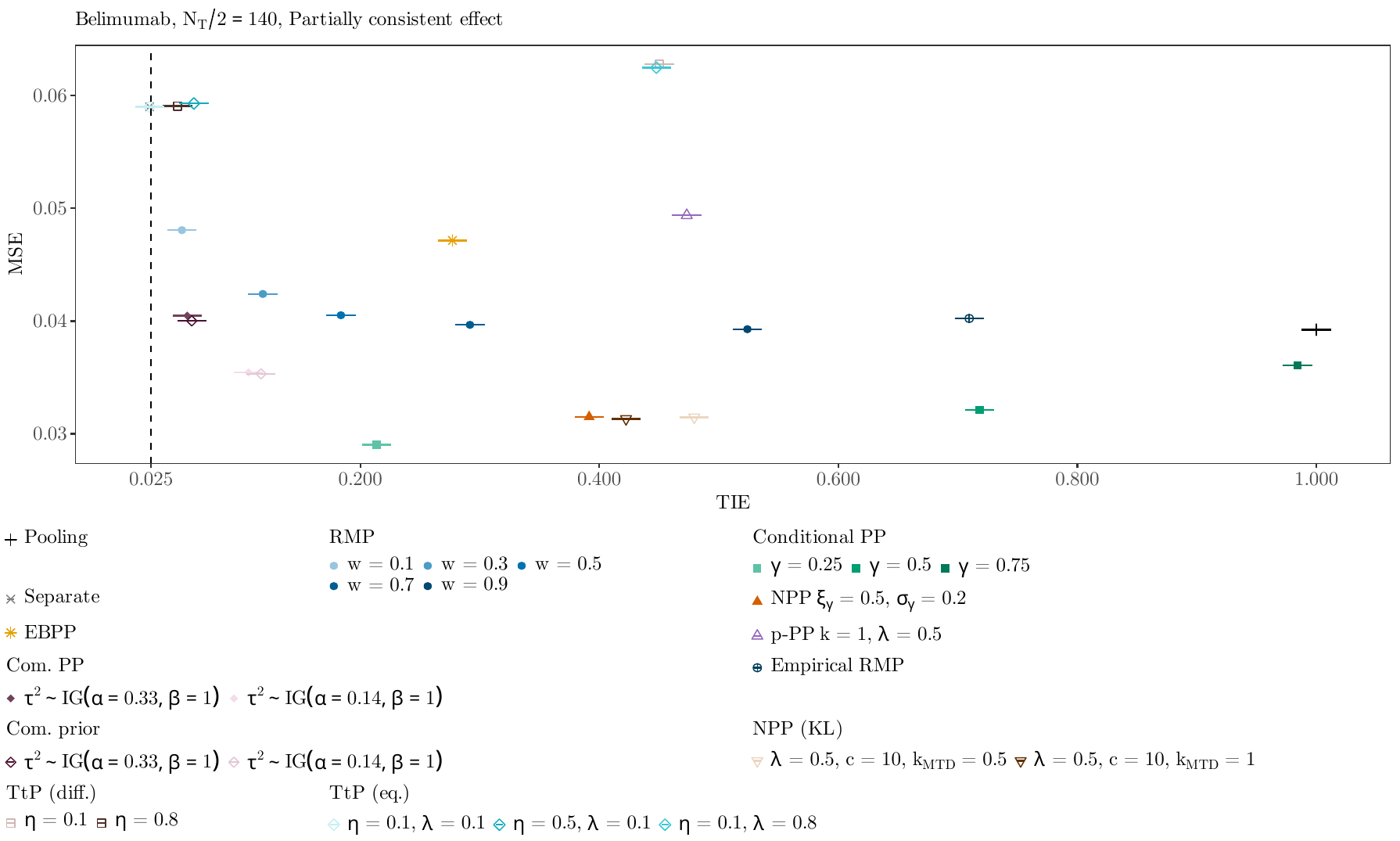}
    \caption{MSE as a function of type I error rate in the belimumab case study. Each point represents one method-parameter combination with a sample size per arm of $140$, across all the methods and parameters, with a partially consistent treatment effect. Error bars correspond to the 95\% CI. Dashed vertical line represents the nominal type I error rate of $0.025$. Method abbreviations and parameters as in Table~\ref{tab:abbreviations}.}
    \label{fig:belimumab_mse_vs_tie_sample_size=140_treatment_effect=partially_consistent}
\end{figure}

\begin{figure}[htbp]
    \centering
    \includegraphics[width=0.95\linewidth, height=0.95\textheight,keepaspectratio]{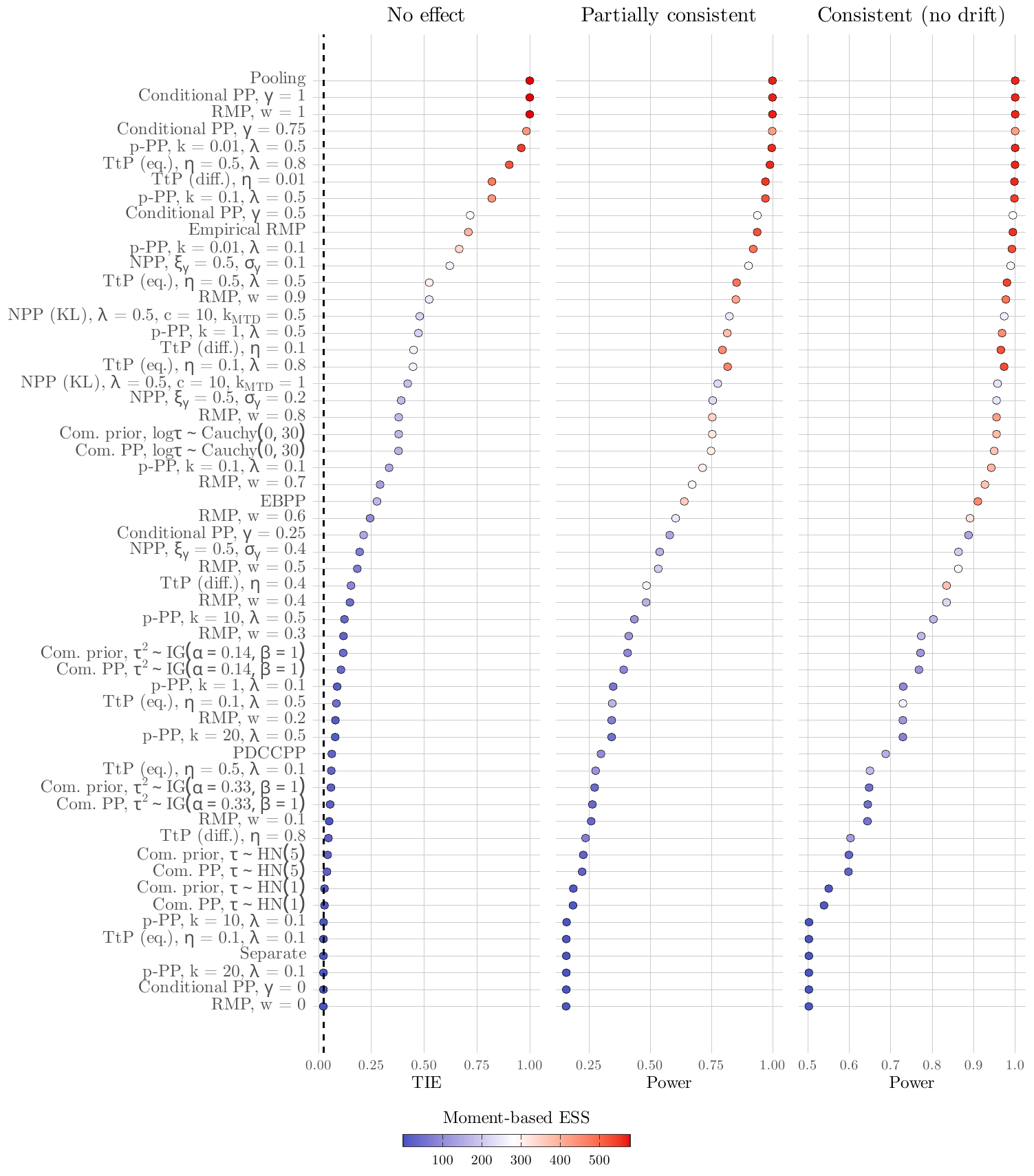}
    \caption{Probability of study success for the three main treatment effect values considered in the belimumab case study ($n_T = 140$). Method abbreviations and parameters as in Table~\ref{tab:abbreviations}.}
    \label{fig:belimumab_success_proba_forest_plot_target_sample_size_per_arm_140}
 \end{figure} 

\begin{figure}[htbp]
    \centering
    \includegraphics[width=0.95\linewidth, height=0.95\textheight,keepaspectratio]{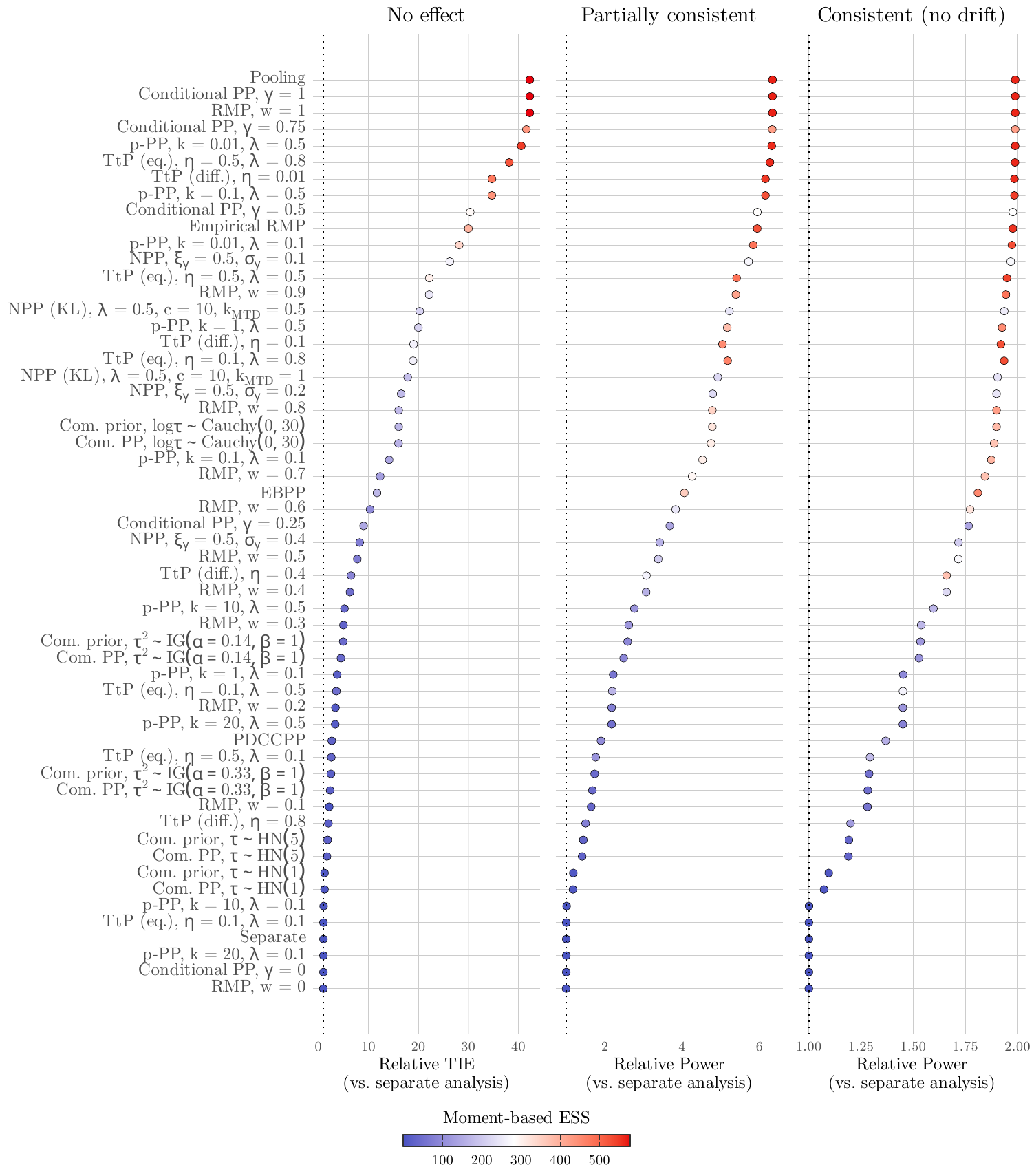}
    \caption{Probability of study success relative to the separate analysis of target study data for the three main treatment effect values considered in the belimumab case study ($n_T = 140$). Method abbreviations and parameters as in Table~\ref{tab:abbreviations}.}
    \label{fig:belimumab_relative_success_proba_forest_plot_target_sample_size_per_arm_140}
 \end{figure} 
 
\begin{figure}[htbp]
    \centering
    \includegraphics[width=0.95\linewidth, height=0.95\textheight,keepaspectratio]{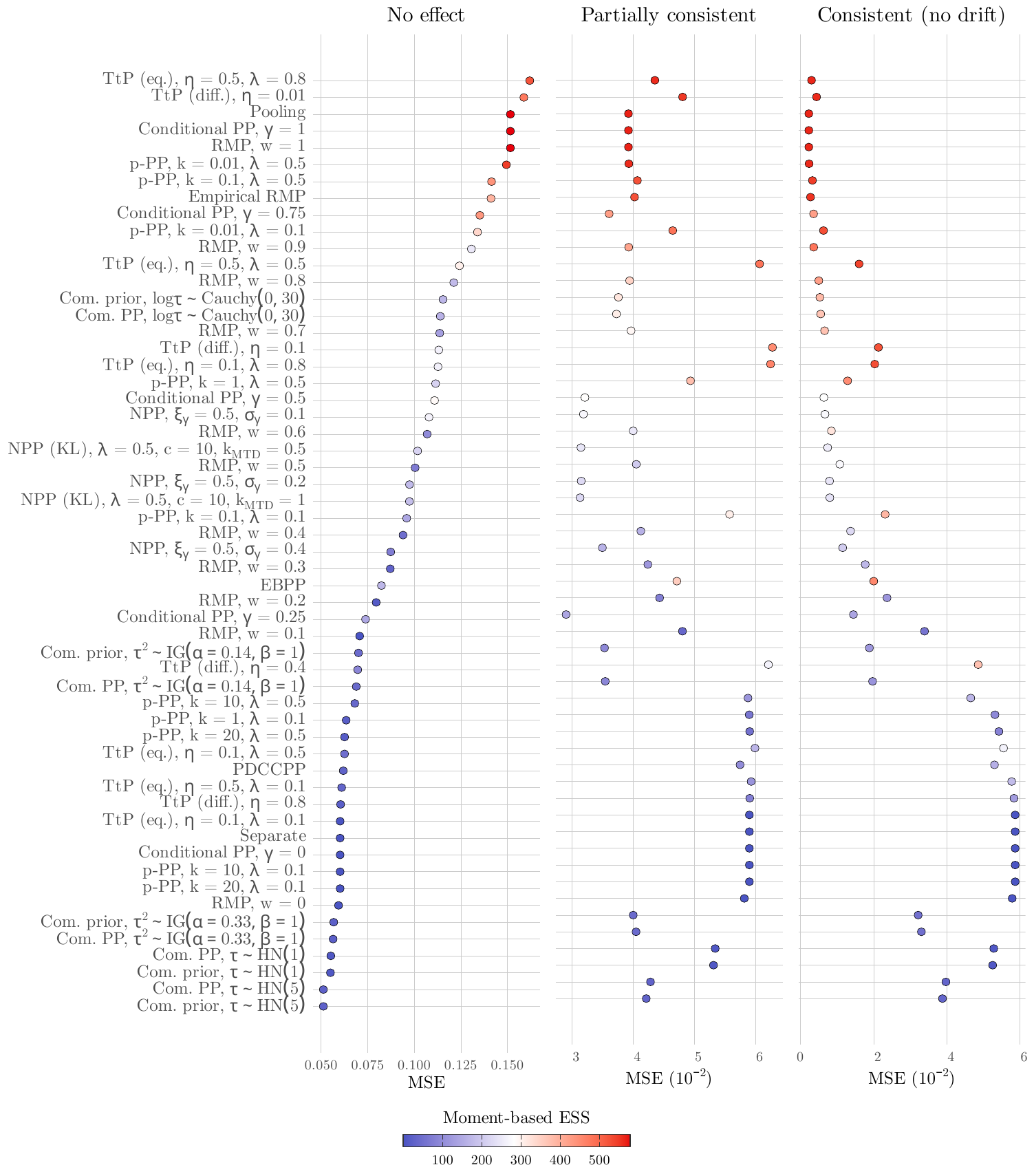}
    \caption{MSE for the three main treatment effect values considered in the belimumab case study ($n_T = 140$). Method abbreviations and parameters as in Table~\ref{tab:abbreviations}.}
    \label{fig:belimumab_mse_forest_plot_target_sample_size_per_arm_140}
 \end{figure} 
 
\begin{figure}[htbp]
    \centering
    \includegraphics[width=0.95\linewidth, height=0.95\textheight,keepaspectratio]{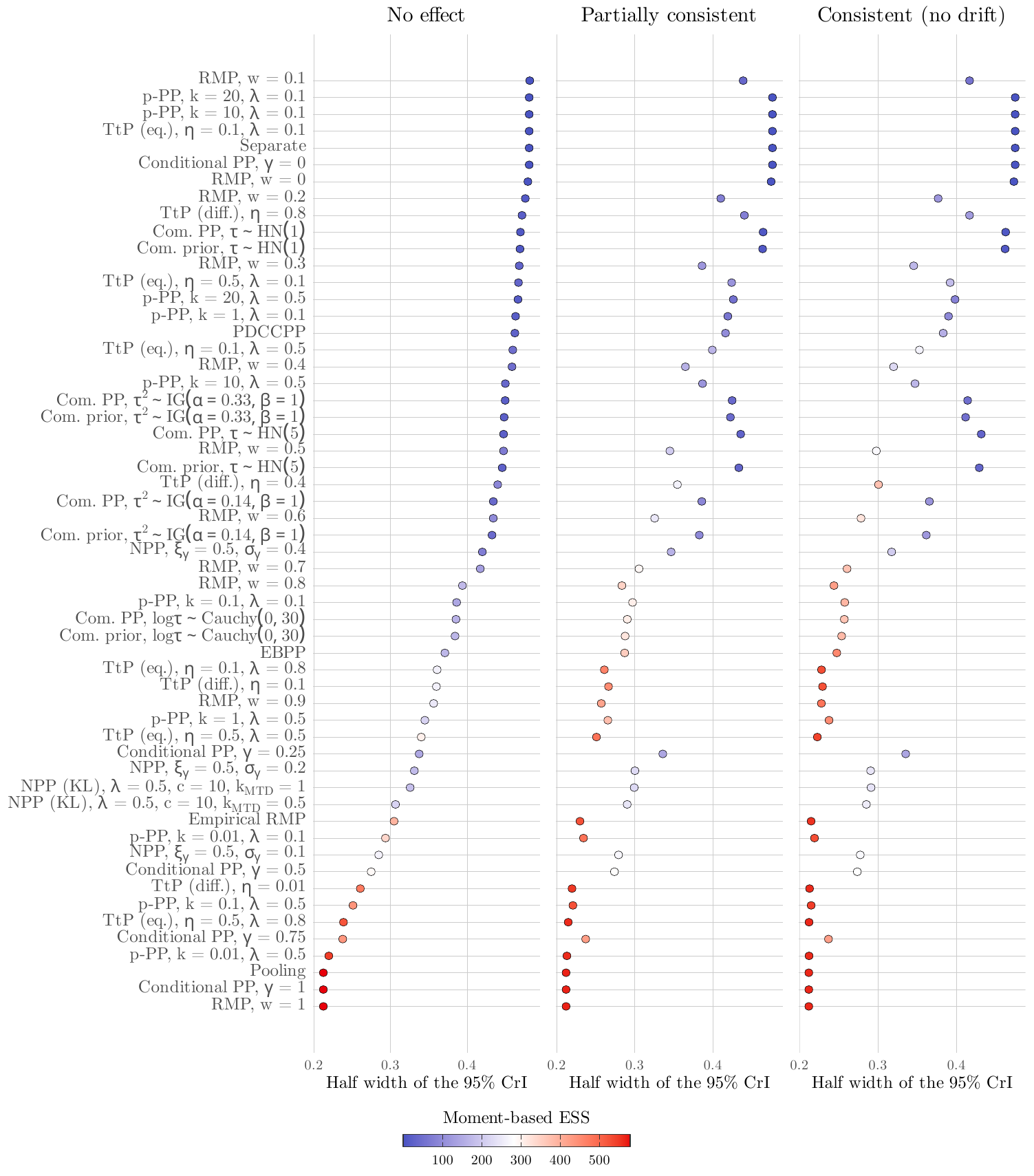}
    \caption{Precision, measured as the mean half-width of the 95\% credible interval,  for the three main treatment effect values considered in the belimumab case study with a sample size per arm of 140 subjects. Error bars correspond to the 95\% CI. Method abbreviations and parameters as in Table~\ref{tab:abbreviations}.}
    \label{fig:belimumab_precision_forest_plot_target_sample_size_per_arm_140}
 \end{figure}

\begin{figure}[htbp]
        \centering
        \includegraphics[width=0.95\linewidth, height=0.95\textheight,keepaspectratio]{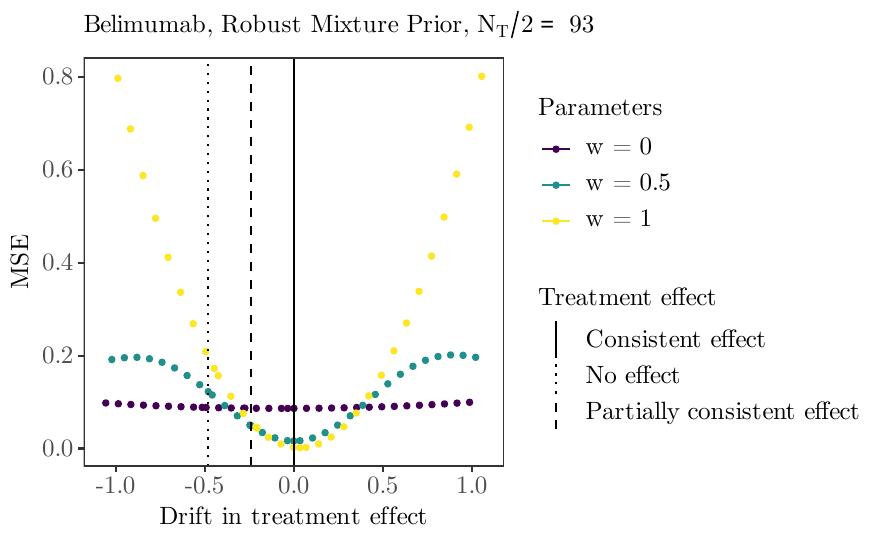}
        \caption{MSE as a function of drift for the RMP in the belimumab case study, with a sample size per arm in the target study of 93 patients, for different values of the weight of the informative component $w$. Error bars correspond to the 95\% CI of the MSE.}
        \label{fig:belimumab_RMP_mse_vs_drift_cat_parameters_sample_size=93_source_denominator_change_factor=1}
\end{figure}

 \begin{figure}[htbp]
\centering
\includegraphics[width=0.95\linewidth, height=0.95\textheight,keepaspectratio]{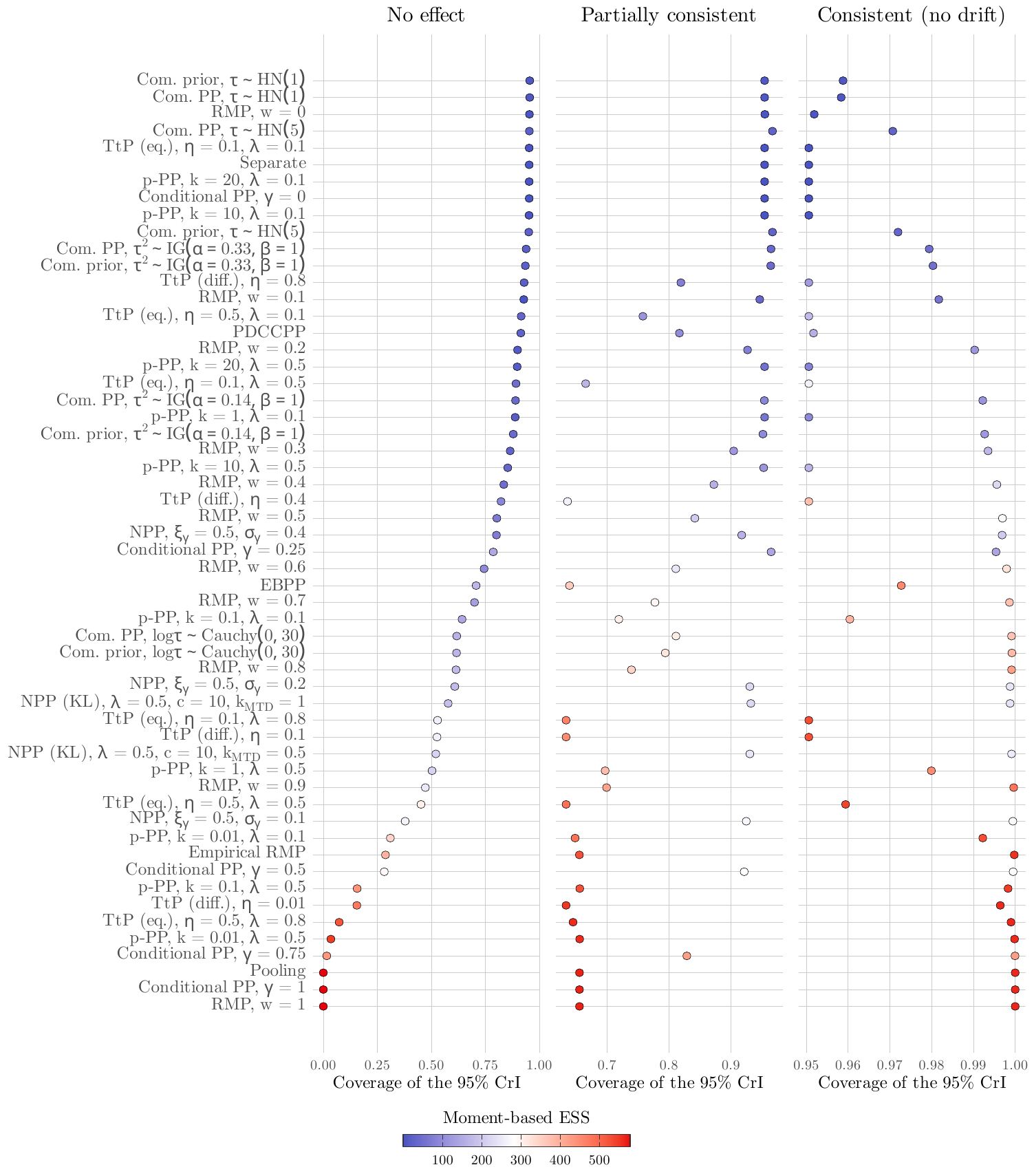}
\caption{Coverage probability of the 95\% credible interval for the three main treatment effect values considered in the belimumab case study with a sample size per arm of 140 patients. Error bars correspond to the 95\% CI on the coverage probability. Method abbreviations and parameters as in Table~\ref{tab:abbreviations}.}
\label{fig:belimumab_coverage_forest_plot_target_sample_size_per_arm_140}
\end{figure}

\begin{figure}[p]
  \centering
  \includegraphics[width=\textwidth]{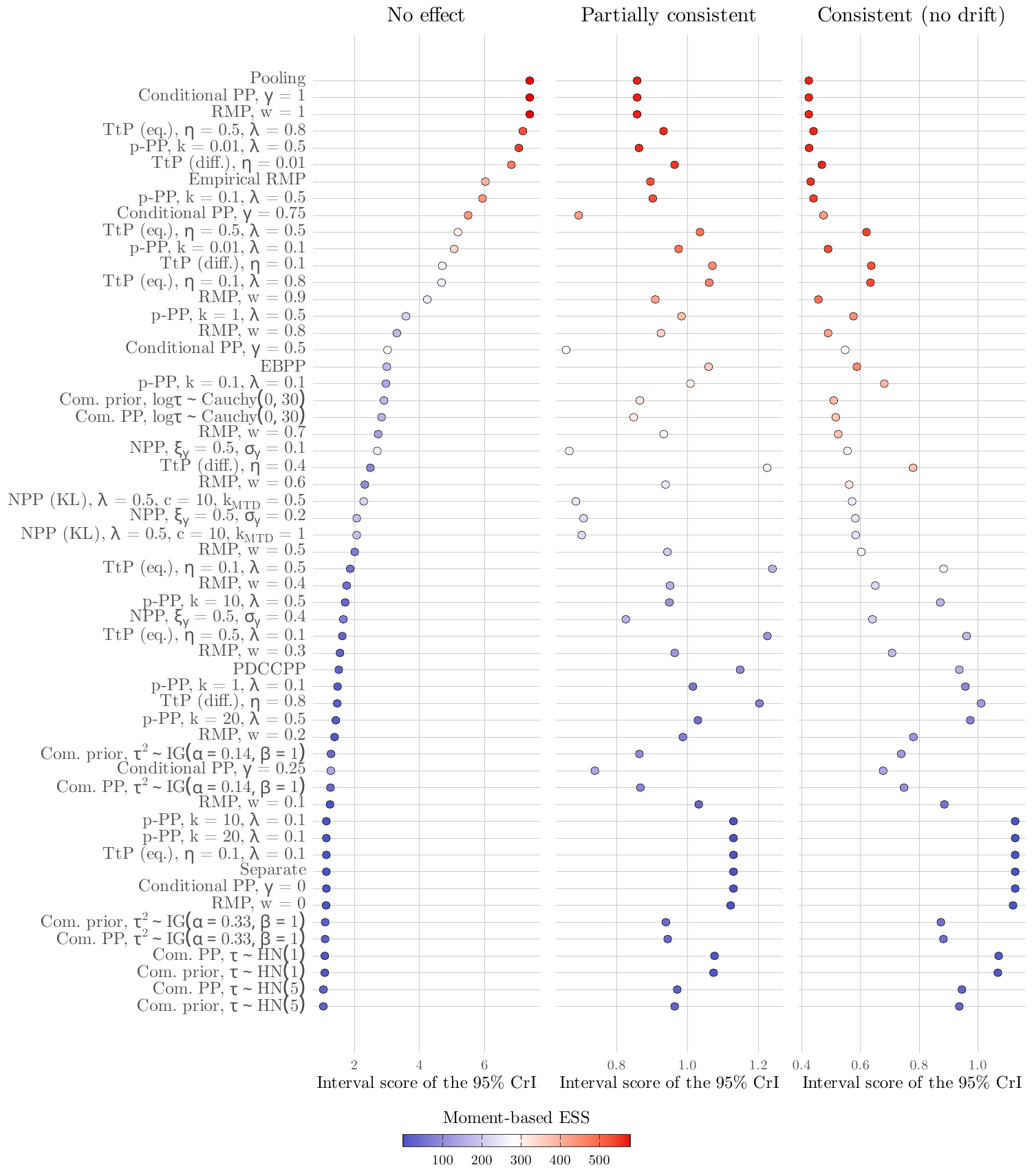}
  \caption{Interval score of the 95\% credible interval for the three main treatment effect values considered in the belimumab case study with a sample size per arm of 140 subjects. Smaller is better. Error bars correspond to the 95\% CI.}
  \label{fig:belimumab_interval_score_forest_plot_target_sample_size_per_arm_140}
\end{figure}

\begin{figure}[p]
\centering
\includegraphics[width=\textwidth]{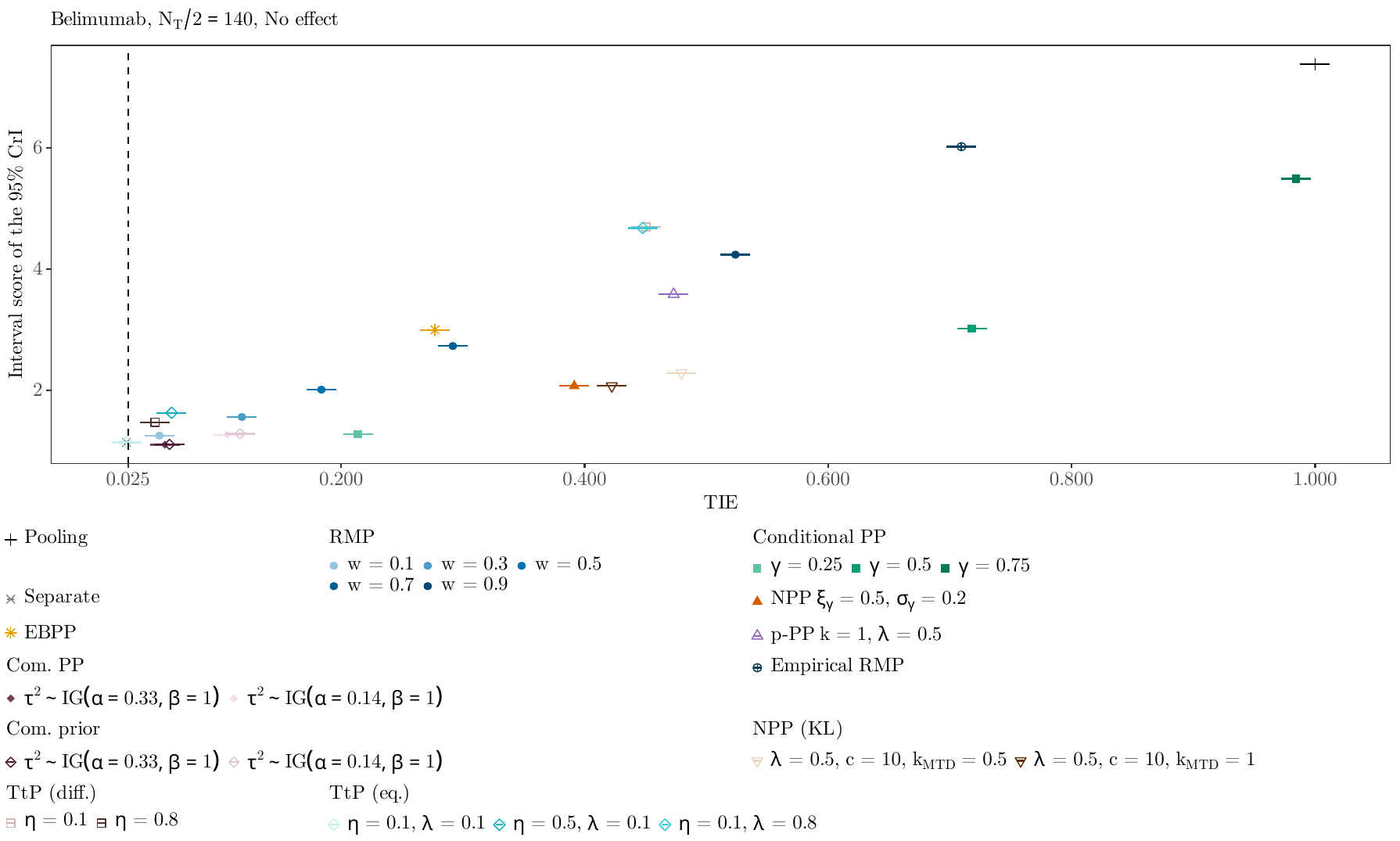}
\caption{Interval score of the 95\% credible interval as a function of type I error rate in the belimumab case study. Each point represents one method-parameter combination with a sample size per arm of 140, without treatment effect. Smaller is better. Error bars correspond to the 95\% CI of the interval score and type I error rate. Dashed vertical line represents the nominal type I error rate of 0.025. } \label{fig:belimumab_interval_score_vs_tie_sample_size=140_treatment_effect=no_effect}
\end{figure}
 

  \centering
 
\begin{longtable}[t]{ccccc}
\caption{Belimumab: mean half-width of the 95\% credible interval (precision), empirical coverage probability (ECP) and interval score, $n_T = 140$}\label{tab:belimumab_precision_ecp_table_target_sample_size_per_arm_140}\\
\toprule
\textbf{Method} & \textbf{Treatment effect} & \textbf{Precision} & \textbf{ECP} & \textbf{Interval score}\\
\midrule
\cellcolor{gray!10}{Com. PP, $\log \tau \sim Cauchy(0, 30)$} & \cellcolor{gray!10}{No treatment effect} & \cellcolor{gray!10}{0.385} & \cellcolor{gray!10}{0.617} & \cellcolor{gray!10}{2.835}\\
Com. PP, $\log \tau \sim Cauchy(0, 30)$ & Partially consistent & 0.290 & 0.811 & 0.848\\
\cellcolor{gray!10}{Com. PP, $\log \tau \sim Cauchy(0, 30)$} & \cellcolor{gray!10}{Consistent} & \cellcolor{gray!10}{0.257} & \cellcolor{gray!10}{0.999} & \cellcolor{gray!10}{0.516}\\
Com. PP, $\tau \sim HN(1)$ & No treatment effect & 0.469 & 0.955 & 1.094\\
\cellcolor{gray!10}{Com. PP, $\tau \sim HN(1)$} & \cellcolor{gray!10}{Partially consistent} & \cellcolor{gray!10}{0.464} & \cellcolor{gray!10}{0.954} & \cellcolor{gray!10}{1.076}\\
\addlinespace
Com. PP, $\tau \sim HN(1)$ & Consistent & 0.463 & 0.958 & 1.071\\
\cellcolor{gray!10}{Com. PP, $\tau \sim HN(5)$} & \cellcolor{gray!10}{No treatment effect} & \cellcolor{gray!10}{0.447} & \cellcolor{gray!10}{0.953} & \cellcolor{gray!10}{1.050}\\
Com. PP, $\tau \sim HN(5)$ & Partially consistent & 0.435 & 0.967 & 0.971\\
\cellcolor{gray!10}{Com. PP, $\tau \sim HN(5)$} & \cellcolor{gray!10}{Consistent} & \cellcolor{gray!10}{0.432} & \cellcolor{gray!10}{0.971} & \cellcolor{gray!10}{0.946}\\
Com. PP, $\tau^2 \sim IG(\alpha = 0.14, \beta = 1)$ & No treatment effect & 0.434 & 0.889 & 1.266\\
\addlinespace
\cellcolor{gray!10}{Com. PP, $\tau^2 \sim IG(\alpha = 0.14, \beta = 1)$} & \cellcolor{gray!10}{Partially consistent} & \cellcolor{gray!10}{0.386} & \cellcolor{gray!10}{0.954} & \cellcolor{gray!10}{0.867}\\
Com. PP, $\tau^2 \sim IG(\alpha = 0.14, \beta = 1)$ & Consistent & 0.365 & 0.992 & 0.748\\
\cellcolor{gray!10}{Com. PP, $\tau^2 \sim IG(\alpha = 0.33, \beta = 1)$} & \cellcolor{gray!10}{No treatment effect} & \cellcolor{gray!10}{0.450} & \cellcolor{gray!10}{0.939} & \cellcolor{gray!10}{1.102}\\
Com. PP, $\tau^2 \sim IG(\alpha = 0.33, \beta = 1)$ & Partially consistent & 0.425 & 0.965 & 0.944\\
\cellcolor{gray!10}{Com. PP, $\tau^2 \sim IG(\alpha = 0.33, \beta = 1)$} & \cellcolor{gray!10}{Consistent} & \cellcolor{gray!10}{0.414} & \cellcolor{gray!10}{0.979} & \cellcolor{gray!10}{0.883}\\
\addlinespace
Com. prior, $\log \tau \sim Cauchy(0, 30)$ & No treatment effect & 0.384 & 0.616 & 2.910\\
\cellcolor{gray!10}{Com. prior, $\log \tau \sim Cauchy(0, 30)$} & \cellcolor{gray!10}{Partially consistent} & \cellcolor{gray!10}{0.288} & \cellcolor{gray!10}{0.794} & \cellcolor{gray!10}{0.866}\\
Com. prior, $\log \tau \sim Cauchy(0, 30)$ & Consistent & 0.254 & 0.999 & 0.509\\
\cellcolor{gray!10}{Com. prior, $\tau \sim HN(1)$} & \cellcolor{gray!10}{No treatment effect} & \cellcolor{gray!10}{0.469} & \cellcolor{gray!10}{0.955} & \cellcolor{gray!10}{1.092}\\
Com. prior, $\tau \sim HN(1)$ & Partially consistent & 0.464 & 0.954 & 1.073\\
\addlinespace
\cellcolor{gray!10}{Com. prior, $\tau \sim HN(1)$} & \cellcolor{gray!10}{Consistent} & \cellcolor{gray!10}{0.462} & \cellcolor{gray!10}{0.959} & \cellcolor{gray!10}{1.068}\\
Com. prior, $\tau \sim HN(5)$ & No treatment effect & 0.446 & 0.951 & 1.049\\
\cellcolor{gray!10}{Com. prior, $\tau \sim HN(5)$} & \cellcolor{gray!10}{Partially consistent} & \cellcolor{gray!10}{0.433} & \cellcolor{gray!10}{0.967} & \cellcolor{gray!10}{0.964}\\
Com. prior, $\tau \sim HN(5)$ & Consistent & 0.429 & 0.972 & 0.937\\
\cellcolor{gray!10}{Com. prior, $\tau^2 \sim IG(\alpha = 0.14, \beta = 1)$} & \cellcolor{gray!10}{No treatment effect} & \cellcolor{gray!10}{0.432} & \cellcolor{gray!10}{0.879} & \cellcolor{gray!10}{1.284}\\
\addlinespace
Com. prior, $\tau^2 \sim IG(\alpha = 0.14, \beta = 1)$ & Partially consistent & 0.383 & 0.952 & 0.864\\
\cellcolor{gray!10}{Com. prior, $\tau^2 \sim IG(\alpha = 0.14, \beta = 1)$} & \cellcolor{gray!10}{Consistent} & \cellcolor{gray!10}{0.362} & \cellcolor{gray!10}{0.993} & \cellcolor{gray!10}{0.739}\\
Com. prior, $\tau^2 \sim IG(\alpha = 0.33, \beta = 1)$ & No treatment effect & 0.448 & 0.935 & 1.105\\
\cellcolor{gray!10}{Com. prior, $\tau^2 \sim IG(\alpha = 0.33, \beta = 1)$} & \cellcolor{gray!10}{Partially consistent} & \cellcolor{gray!10}{0.422} & \cellcolor{gray!10}{0.964} & \cellcolor{gray!10}{0.939}\\
Com. prior, $\tau^2 \sim IG(\alpha = 0.33, \beta = 1)$ & Consistent & 0.412 & 0.980 & 0.874\\
\addlinespace
\cellcolor{gray!10}{CPP, $\gamma$ = 0} & \cellcolor{gray!10}{No treatment effect} & \cellcolor{gray!10}{0.481} & \cellcolor{gray!10}{0.953} & \cellcolor{gray!10}{1.141}\\
CPP, $\gamma$ = 0 & Partially consistent & 0.476 & 0.954 & 1.130\\
\cellcolor{gray!10}{CPP, $\gamma$ = 0} & \cellcolor{gray!10}{Consistent} & \cellcolor{gray!10}{0.475} & \cellcolor{gray!10}{0.951} & \cellcolor{gray!10}{1.127}\\
CPP, $\gamma$ = 0.25 & No treatment effect & 0.337 & 0.786 & 1.277\\
\cellcolor{gray!10}{CPP, $\gamma$ = 0.25} & \cellcolor{gray!10}{Partially consistent} & \cellcolor{gray!10}{0.336} & \cellcolor{gray!10}{0.965} & \cellcolor{gray!10}{0.739}\\
\addlinespace
CPP, $\gamma$ = 0.25 & Consistent & 0.335 & 0.995 & 0.677\\
\cellcolor{gray!10}{CPP, $\gamma$ = 0.5} & \cellcolor{gray!10}{No treatment effect} & \cellcolor{gray!10}{0.275} & \cellcolor{gray!10}{0.282} & \cellcolor{gray!10}{3.017}\\
CPP, $\gamma$ = 0.5 & Partially consistent & 0.274 & 0.921 & 0.657\\
\cellcolor{gray!10}{CPP, $\gamma$ = 0.5} & \cellcolor{gray!10}{Consistent} & \cellcolor{gray!10}{0.274} & \cellcolor{gray!10}{0.999} & \cellcolor{gray!10}{0.548}\\
CPP, $\gamma$ = 0.75 & No treatment effect & 0.238 & 0.016 & 5.492\\
\addlinespace
\cellcolor{gray!10}{CPP, $\gamma$ = 0.75} & \cellcolor{gray!10}{Partially consistent} & \cellcolor{gray!10}{0.237} & \cellcolor{gray!10}{0.829} & \cellcolor{gray!10}{0.693}\\
CPP, $\gamma$ = 0.75 & Consistent & 0.237 & 1.000 & 0.474\\
\cellcolor{gray!10}{CPP, $\gamma$ = 1} & \cellcolor{gray!10}{No treatment effect} & \cellcolor{gray!10}{0.212} & \cellcolor{gray!10}{0.000} & \cellcolor{gray!10}{7.382}\\
CPP, $\gamma$ = 1 & Partially consistent & 0.212 & 0.655 & 0.858\\
\cellcolor{gray!10}{CPP, $\gamma$ = 1} & \cellcolor{gray!10}{Consistent} & \cellcolor{gray!10}{0.212} & \cellcolor{gray!10}{1.000} & \cellcolor{gray!10}{0.424}\\
\addlinespace
EBPP & No treatment effect & 0.371 & 0.707 & 2.997\\
\cellcolor{gray!10}{EBPP} & \cellcolor{gray!10}{Partially consistent} & \cellcolor{gray!10}{0.287} & \cellcolor{gray!10}{0.639} & \cellcolor{gray!10}{1.060}\\
EBPP & Consistent & 0.248 & 0.973 & 0.588\\
\cellcolor{gray!10}{Empirical RMP} & \cellcolor{gray!10}{No treatment effect} & \cellcolor{gray!10}{0.305} & \cellcolor{gray!10}{0.288} & \cellcolor{gray!10}{6.021}\\
Empirical RMP & Partially consistent & 0.230 & 0.655 & 0.895\\
\addlinespace
\cellcolor{gray!10}{Empirical RMP} & \cellcolor{gray!10}{Consistent} & \cellcolor{gray!10}{0.215} & \cellcolor{gray!10}{1.000} & \cellcolor{gray!10}{0.430}\\
NPP (KL), $\lambda_{\mathrm{KL}}$ = 0.5, $c$ = 10, $k_{\mathrm{MTD}}$ = 0.5 & No treatment effect & 0.307 & 0.521 & 2.286\\
\cellcolor{gray!10}{NPP (KL), $\lambda_{\mathrm{KL}}$ = 0.5, $c$ = 10, $k_{\mathrm{MTD}}$ = 0.5} & \cellcolor{gray!10}{Partially consistent} & \cellcolor{gray!10}{0.290} & \cellcolor{gray!10}{0.930} & \cellcolor{gray!10}{0.685}\\
NPP (KL), $\lambda_{\mathrm{KL}}$ = 0.5, $c$ = 10, $k_{\mathrm{MTD}}$ = 0.5 & Consistent & 0.285 & 0.999 & 0.571\\
\cellcolor{gray!10}{NPP (KL), $\lambda_{\mathrm{KL}}$ = 0.5, $c$ = 10, $k_{\mathrm{MTD}}$ = 1} & \cellcolor{gray!10}{No treatment effect} & \cellcolor{gray!10}{0.326} & \cellcolor{gray!10}{0.578} & \cellcolor{gray!10}{2.072}\\
\addlinespace
NPP (KL), $\lambda_{\mathrm{KL}}$ = 0.5, $c$ = 10, $k_{\mathrm{MTD}}$ = 1 & Partially consistent & 0.299 & 0.932 & 0.702\\
\cellcolor{gray!10}{NPP (KL), $\lambda_{\mathrm{KL}}$ = 0.5, $c$ = 10, $k_{\mathrm{MTD}}$ = 1} & \cellcolor{gray!10}{Consistent} & \cellcolor{gray!10}{0.291} & \cellcolor{gray!10}{0.999} & \cellcolor{gray!10}{0.584}\\
NPP, $\xi_\gamma$ = 0.5, $\sigma_\gamma$ = 0.1 & No treatment effect & 0.285 & 0.379 & 2.705\\
\cellcolor{gray!10}{NPP, $\xi_\gamma$ = 0.5, $\sigma_\gamma$ = 0.1} & \cellcolor{gray!10}{Partially consistent} & \cellcolor{gray!10}{0.279} & \cellcolor{gray!10}{0.925} & \cellcolor{gray!10}{0.667}\\
NPP, $\xi_\gamma$ = 0.5, $\sigma_\gamma$ = 0.1 & Consistent & 0.277 & 0.999 & 0.555\\
\addlinespace
\cellcolor{gray!10}{NPP, $\xi_\gamma$ = 0.5, $\sigma_\gamma$ = 0.2} & \cellcolor{gray!10}{No treatment effect} & \cellcolor{gray!10}{0.331} & \cellcolor{gray!10}{0.608} & \cellcolor{gray!10}{2.073}\\
NPP, $\xi_\gamma$ = 0.5, $\sigma_\gamma$ = 0.2 & Partially consistent & 0.300 & 0.930 & 0.707\\
\cellcolor{gray!10}{NPP, $\xi_\gamma$ = 0.5, $\sigma_\gamma$ = 0.2} & \cellcolor{gray!10}{Consistent} & \cellcolor{gray!10}{0.291} & \cellcolor{gray!10}{0.999} & \cellcolor{gray!10}{0.583}\\
NPP, $\xi_\gamma$ = 0.5, $\sigma_\gamma$ = 0.4 & No treatment effect & 0.420 & 0.801 & 1.662\\
\cellcolor{gray!10}{NPP, $\xi_\gamma$ = 0.5, $\sigma_\gamma$ = 0.4} & \cellcolor{gray!10}{Partially consistent} & \cellcolor{gray!10}{0.346} & \cellcolor{gray!10}{0.917} & \cellcolor{gray!10}{0.826}\\
\addlinespace
NPP, $\xi_\gamma$ = 0.5, $\sigma_\gamma$ = 0.4 & Consistent & 0.317 & 0.997 & 0.641\\
\cellcolor{gray!10}{p-PP, $k$ = 0.01, $\lambda$ = 0.1} & \cellcolor{gray!10}{No treatment effect} & \cellcolor{gray!10}{0.293} & \cellcolor{gray!10}{0.310} & \cellcolor{gray!10}{5.064}\\
p-PP, $k$ = 0.01, $\lambda$ = 0.1 & Partially consistent & 0.234 & 0.648 & 0.975\\
\cellcolor{gray!10}{p-PP, $k$ = 0.01, $\lambda$ = 0.1} & \cellcolor{gray!10}{Consistent} & \cellcolor{gray!10}{0.219} & \cellcolor{gray!10}{0.992} & \cellcolor{gray!10}{0.490}\\
p-PP, $k$ = 0.01, $\lambda$ = 0.5 & No treatment effect & 0.220 & 0.035 & 7.049\\
\addlinespace
\cellcolor{gray!10}{p-PP, $k$ = 0.01, $\lambda$ = 0.5} & \cellcolor{gray!10}{Partially consistent} & \cellcolor{gray!10}{0.213} & \cellcolor{gray!10}{0.656} & \cellcolor{gray!10}{0.863}\\
p-PP, $k$ = 0.01, $\lambda$ = 0.5 & Consistent & 0.212 & 1.000 & 0.425\\
\cellcolor{gray!10}{p-PP, $k$ = 0.1, $\lambda$ = 0.1} & \cellcolor{gray!10}{No treatment effect} & \cellcolor{gray!10}{0.386} & \cellcolor{gray!10}{0.642} & \cellcolor{gray!10}{2.971}\\
p-PP, $k$ = 0.1, $\lambda$ = 0.1 & Partially consistent & 0.297 & 0.719 & 1.008\\
\cellcolor{gray!10}{p-PP, $k$ = 0.1, $\lambda$ = 0.1} & \cellcolor{gray!10}{Consistent} & \cellcolor{gray!10}{0.258} & \cellcolor{gray!10}{0.960} & \cellcolor{gray!10}{0.681}\\
\addlinespace
p-PP, $k$ = 0.1, $\lambda$ = 0.5 & No treatment effect & 0.251 & 0.156 & 5.935\\
\cellcolor{gray!10}{p-PP, $k$ = 0.1, $\lambda$ = 0.5} & \cellcolor{gray!10}{Partially consistent} & \cellcolor{gray!10}{0.221} & \cellcolor{gray!10}{0.656} & \cellcolor{gray!10}{0.902}\\
p-PP, $k$ = 0.1, $\lambda$ = 0.5 & Consistent & 0.215 & 0.998 & 0.440\\
\cellcolor{gray!10}{p-PP, $k$ = 1, $\lambda$ = 0.1} & \cellcolor{gray!10}{No treatment effect} & \cellcolor{gray!10}{0.463} & \cellcolor{gray!10}{0.888} & \cellcolor{gray!10}{1.485}\\
p-PP, $k$ = 1, $\lambda$ = 0.1 & Partially consistent & 0.419 & 0.954 & 1.015\\
\addlinespace
\cellcolor{gray!10}{p-PP, $k$ = 1, $\lambda$ = 0.1} & \cellcolor{gray!10}{Consistent} & \cellcolor{gray!10}{0.390} & \cellcolor{gray!10}{0.951} & \cellcolor{gray!10}{0.958}\\
p-PP, $k$ = 1, $\lambda$ = 0.5 & No treatment effect & 0.345 & 0.503 & 3.585\\
\cellcolor{gray!10}{p-PP, $k$ = 1, $\lambda$ = 0.5} & \cellcolor{gray!10}{Partially consistent} & \cellcolor{gray!10}{0.265} & \cellcolor{gray!10}{0.697} & \cellcolor{gray!10}{0.983}\\
p-PP, $k$ = 1, $\lambda$ = 0.5 & Consistent & 0.238 & 0.980 & 0.576\\
\cellcolor{gray!10}{p-PP, $k$ = 10, $\lambda$ = 0.1} & \cellcolor{gray!10}{No treatment effect} & \cellcolor{gray!10}{0.481} & \cellcolor{gray!10}{0.952} & \cellcolor{gray!10}{1.141}\\
\addlinespace
p-PP, $k$ = 10, $\lambda$ = 0.1 & Partially consistent & 0.476 & 0.954 & 1.130\\
\cellcolor{gray!10}{p-PP, $k$ = 10, $\lambda$ = 0.1} & \cellcolor{gray!10}{Consistent} & \cellcolor{gray!10}{0.475} & \cellcolor{gray!10}{0.951} & \cellcolor{gray!10}{1.127}\\
p-PP, $k$ = 10, $\lambda$ = 0.5 & No treatment effect & 0.450 & 0.853 & 1.719\\
\cellcolor{gray!10}{p-PP, $k$ = 10, $\lambda$ = 0.5} & \cellcolor{gray!10}{Partially consistent} & \cellcolor{gray!10}{0.387} & \cellcolor{gray!10}{0.953} & \cellcolor{gray!10}{0.949}\\
p-PP, $k$ = 10, $\lambda$ = 0.5 & Consistent & 0.347 & 0.951 & 0.872\\
\addlinespace
\cellcolor{gray!10}{p-PP, $k$ = 20, $\lambda$ = 0.1} & \cellcolor{gray!10}{No treatment effect} & \cellcolor{gray!10}{0.481} & \cellcolor{gray!10}{0.953} & \cellcolor{gray!10}{1.141}\\
p-PP, $k$ = 20, $\lambda$ = 0.1 & Partially consistent & 0.476 & 0.954 & 1.130\\
\cellcolor{gray!10}{p-PP, $k$ = 20, $\lambda$ = 0.1} & \cellcolor{gray!10}{Consistent} & \cellcolor{gray!10}{0.475} & \cellcolor{gray!10}{0.951} & \cellcolor{gray!10}{1.127}\\
p-PP, $k$ = 20, $\lambda$ = 0.5 & No treatment effect & 0.466 & 0.897 & 1.431\\
\cellcolor{gray!10}{p-PP, $k$ = 20, $\lambda$ = 0.5} & \cellcolor{gray!10}{Partially consistent} & \cellcolor{gray!10}{0.426} & \cellcolor{gray!10}{0.954} & \cellcolor{gray!10}{1.029}\\
\addlinespace
p-PP, $k$ = 20, $\lambda$ = 0.5 & Consistent & 0.398 & 0.951 & 0.974\\
\cellcolor{gray!10}{PDCCPP} & \cellcolor{gray!10}{No treatment effect} & \cellcolor{gray!10}{0.462} & \cellcolor{gray!10}{0.914} & \cellcolor{gray!10}{1.522}\\
PDCCPP & Partially consistent & 0.416 & 0.817 & 1.149\\
\cellcolor{gray!10}{PDCCPP} & \cellcolor{gray!10}{Consistent} & \cellcolor{gray!10}{0.383} & \cellcolor{gray!10}{0.952} & \cellcolor{gray!10}{0.937}\\
Pooling & No treatment effect & 0.212 & 0.000 & 7.382\\
\addlinespace
\cellcolor{gray!10}{Pooling} & \cellcolor{gray!10}{Partially consistent} & \cellcolor{gray!10}{0.212} & \cellcolor{gray!10}{0.655} & \cellcolor{gray!10}{0.858}\\
Pooling & Consistent & 0.212 & 1.000 & 0.424\\
\cellcolor{gray!10}{RMP, $w$ = 0} & \cellcolor{gray!10}{No treatment effect} & \cellcolor{gray!10}{0.479} & \cellcolor{gray!10}{0.954} & \cellcolor{gray!10}{1.133}\\
RMP, $w$ = 0 & Partially consistent & 0.474 & 0.955 & 1.122\\
\cellcolor{gray!10}{RMP, $w$ = 0} & \cellcolor{gray!10}{Consistent} & \cellcolor{gray!10}{0.473} & \cellcolor{gray!10}{0.952} & \cellcolor{gray!10}{1.120}\\
\addlinespace
RMP, $w$ = 0.1 & No treatment effect & 0.482 & 0.928 & 1.253\\
\cellcolor{gray!10}{RMP, $w$ = 0.1} & \cellcolor{gray!10}{Partially consistent} & \cellcolor{gray!10}{0.439} & \cellcolor{gray!10}{0.946} & \cellcolor{gray!10}{1.032}\\
RMP, $w$ = 0.1 & Consistent & 0.417 & 0.982 & 0.886\\
\cellcolor{gray!10}{RMP, $w$ = 0.2} & \cellcolor{gray!10}{No treatment effect} & \cellcolor{gray!10}{0.476} & \cellcolor{gray!10}{0.899} & \cellcolor{gray!10}{1.392}\\
RMP, $w$ = 0.2 & Partially consistent & 0.410 & 0.927 & 0.987\\
\addlinespace
\cellcolor{gray!10}{RMP, $w$ = 0.2} & \cellcolor{gray!10}{Consistent} & \cellcolor{gray!10}{0.377} & \cellcolor{gray!10}{0.990} & \cellcolor{gray!10}{0.781}\\
RMP, $w$ = 0.3 & No treatment effect & 0.468 & 0.865 & 1.560\\
\cellcolor{gray!10}{RMP, $w$ = 0.3} & \cellcolor{gray!10}{Partially consistent} & \cellcolor{gray!10}{0.386} & \cellcolor{gray!10}{0.904} & \cellcolor{gray!10}{0.964}\\
RMP, $w$ = 0.3 & Consistent & 0.346 & 0.993 & 0.708\\
\cellcolor{gray!10}{RMP, $w$ = 0.4} & \cellcolor{gray!10}{No treatment effect} & \cellcolor{gray!10}{0.459} & \cellcolor{gray!10}{0.835} & \cellcolor{gray!10}{1.765}\\
\addlinespace
RMP, $w$ = 0.4 & Partially consistent & 0.365 & 0.872 & 0.951\\
\cellcolor{gray!10}{RMP, $w$ = 0.4} & \cellcolor{gray!10}{Consistent} & \cellcolor{gray!10}{0.320} & \cellcolor{gray!10}{0.996} & \cellcolor{gray!10}{0.650}\\
RMP, $w$ = 0.5 & No treatment effect & 0.447 & 0.803 & 2.011\\
\cellcolor{gray!10}{RMP, $w$ = 0.5} & \cellcolor{gray!10}{Partially consistent} & \cellcolor{gray!10}{0.345} & \cellcolor{gray!10}{0.842} & \cellcolor{gray!10}{0.943}\\
RMP, $w$ = 0.5 & Consistent & 0.298 & 0.997 & 0.603\\
\addlinespace
\cellcolor{gray!10}{RMP, $w$ = 0.6} & \cellcolor{gray!10}{No treatment effect} & \cellcolor{gray!10}{0.434} & \cellcolor{gray!10}{0.744} & \cellcolor{gray!10}{2.322}\\
RMP, $w$ = 0.6 & Partially consistent & 0.325 & 0.811 & 0.938\\
\cellcolor{gray!10}{RMP, $w$ = 0.6} & \cellcolor{gray!10}{Consistent} & \cellcolor{gray!10}{0.278} & \cellcolor{gray!10}{0.998} & \cellcolor{gray!10}{0.561}\\
RMP, $w$ = 0.7 & No treatment effect & 0.417 & 0.700 & 2.731\\
\cellcolor{gray!10}{RMP, $w$ = 0.7} & \cellcolor{gray!10}{Partially consistent} & \cellcolor{gray!10}{0.305} & \cellcolor{gray!10}{0.777} & \cellcolor{gray!10}{0.933}\\
\addlinespace
RMP, $w$ = 0.7 & Consistent & 0.261 & 0.999 & 0.524\\
\cellcolor{gray!10}{RMP, $w$ = 0.8} & \cellcolor{gray!10}{No treatment effect} & \cellcolor{gray!10}{0.394} & \cellcolor{gray!10}{0.614} & \cellcolor{gray!10}{3.305}\\
RMP, $w$ = 0.8 & Partially consistent & 0.283 & 0.739 & 0.925\\
\cellcolor{gray!10}{RMP, $w$ = 0.8} & \cellcolor{gray!10}{Consistent} & \cellcolor{gray!10}{0.244} & \cellcolor{gray!10}{0.999} & \cellcolor{gray!10}{0.490}\\
RMP, $w$ = 0.9 & No treatment effect & 0.356 & 0.472 & 4.239\\
\addlinespace
\cellcolor{gray!10}{RMP, $w$ = 0.9} & \cellcolor{gray!10}{Partially consistent} & \cellcolor{gray!10}{0.257} & \cellcolor{gray!10}{0.699} & \cellcolor{gray!10}{0.909}\\
RMP, $w$ = 0.9 & Consistent & 0.228 & 1.000 & 0.457\\
\cellcolor{gray!10}{RMP, $w$ = 1} & \cellcolor{gray!10}{No treatment effect} & \cellcolor{gray!10}{0.212} & \cellcolor{gray!10}{0.000} & \cellcolor{gray!10}{7.382}\\
RMP, $w$ = 1 & Partially consistent & 0.212 & 0.655 & 0.858\\
\cellcolor{gray!10}{RMP, $w$ = 1} & \cellcolor{gray!10}{Consistent} & \cellcolor{gray!10}{0.212} & \cellcolor{gray!10}{1.000} & \cellcolor{gray!10}{0.424}\\
\addlinespace
Separate & No treatment effect & 0.481 & 0.953 & 1.141\\
\cellcolor{gray!10}{Separate} & \cellcolor{gray!10}{Partially consistent} & \cellcolor{gray!10}{0.476} & \cellcolor{gray!10}{0.954} & \cellcolor{gray!10}{1.130}\\
Separate & Consistent & 0.475 & 0.951 & 1.127\\
\cellcolor{gray!10}{TtP (diff), $\eta$ = 0.01} & \cellcolor{gray!10}{No treatment effect} & \cellcolor{gray!10}{0.261} & \cellcolor{gray!10}{0.155} & \cellcolor{gray!10}{6.820}\\
TtP (diff), $\eta$ = 0.01 & Partially consistent & 0.220 & 0.634 & 0.964\\
\addlinespace
\cellcolor{gray!10}{TtP (diff), $\eta$ = 0.01} & \cellcolor{gray!10}{Consistent} & \cellcolor{gray!10}{0.213} & \cellcolor{gray!10}{0.996} & \cellcolor{gray!10}{0.468}\\
TtP (diff), $\eta$ = 0.1 & No treatment effect & 0.360 & 0.526 & 4.697\\
\cellcolor{gray!10}{TtP (diff), $\eta$ = 0.1} & \cellcolor{gray!10}{Partially consistent} & \cellcolor{gray!10}{0.266} & \cellcolor{gray!10}{0.634} & \cellcolor{gray!10}{1.070}\\
TtP (diff), $\eta$ = 0.1 & Consistent & 0.229 & 0.951 & 0.637\\
\cellcolor{gray!10}{TtP (diff), $\eta$ = 0.4} & \cellcolor{gray!10}{No treatment effect} & \cellcolor{gray!10}{0.440} & \cellcolor{gray!10}{0.822} & \cellcolor{gray!10}{2.491}\\
\addlinespace
TtP (diff), $\eta$ = 0.4 & Partially consistent & 0.354 & 0.636 & 1.225\\
\cellcolor{gray!10}{TtP (diff), $\eta$ = 0.4} & \cellcolor{gray!10}{Consistent} & \cellcolor{gray!10}{0.301} & \cellcolor{gray!10}{0.951} & \cellcolor{gray!10}{0.779}\\
TtP (diff), $\eta$ = 0.8 & No treatment effect & 0.472 & 0.929 & 1.474\\
\cellcolor{gray!10}{TtP (diff), $\eta$ = 0.8} & \cellcolor{gray!10}{Partially consistent} & \cellcolor{gray!10}{0.440} & \cellcolor{gray!10}{0.819} & \cellcolor{gray!10}{1.203}\\
TtP (diff), $\eta$ = 0.8 & Consistent & 0.417 & 0.951 & 1.011\\
\addlinespace
\cellcolor{gray!10}{TtP (eq), $\eta$ = 0.1, $\lambda$ = 0.1} & \cellcolor{gray!10}{No treatment effect} & \cellcolor{gray!10}{0.481} & \cellcolor{gray!10}{0.953} & \cellcolor{gray!10}{1.141}\\
TtP (eq), $\eta$ = 0.1, $\lambda$ = 0.1 & Partially consistent & 0.476 & 0.954 & 1.130\\
\cellcolor{gray!10}{TtP (eq), $\eta$ = 0.1, $\lambda$ = 0.1} & \cellcolor{gray!10}{Consistent} & \cellcolor{gray!10}{0.475} & \cellcolor{gray!10}{0.951} & \cellcolor{gray!10}{1.127}\\
TtP (eq), $\eta$ = 0.1, $\lambda$ = 0.5 & No treatment effect & 0.459 & 0.892 & 1.873\\
\cellcolor{gray!10}{TtP (eq), $\eta$ = 0.1, $\lambda$ = 0.5} & \cellcolor{gray!10}{Partially consistent} & \cellcolor{gray!10}{0.399} & \cellcolor{gray!10}{0.665} & \cellcolor{gray!10}{1.240}\\
\addlinespace
TtP (eq), $\eta$ = 0.1, $\lambda$ = 0.5 & Consistent & 0.353 & 0.951 & 0.883\\
\cellcolor{gray!10}{TtP (eq), $\eta$ = 0.1, $\lambda$ = 0.8} & \cellcolor{gray!10}{No treatment effect} & \cellcolor{gray!10}{0.361} & \cellcolor{gray!10}{0.529} & \cellcolor{gray!10}{4.679}\\
TtP (eq), $\eta$ = 0.1, $\lambda$ = 0.8 & Partially consistent & 0.261 & 0.634 & 1.061\\
\cellcolor{gray!10}{TtP (eq), $\eta$ = 0.1, $\lambda$ = 0.8} & \cellcolor{gray!10}{Consistent} & \cellcolor{gray!10}{0.228} & \cellcolor{gray!10}{0.951} & \cellcolor{gray!10}{0.634}\\
TtP (eq), $\eta$ = 0.5, $\lambda$ = 0.1 & No treatment effect & 0.467 & 0.916 & 1.629\\
\addlinespace
\cellcolor{gray!10}{TtP (eq), $\eta$ = 0.5, $\lambda$ = 0.1} & \cellcolor{gray!10}{Partially consistent} & \cellcolor{gray!10}{0.424} & \cellcolor{gray!10}{0.758} & \cellcolor{gray!10}{1.226}\\
TtP (eq), $\eta$ = 0.5, $\lambda$ = 0.1 & Consistent & 0.392 & 0.951 & 0.962\\
\cellcolor{gray!10}{TtP (eq), $\eta$ = 0.5, $\lambda$ = 0.5} & \cellcolor{gray!10}{No treatment effect} & \cellcolor{gray!10}{0.340} & \cellcolor{gray!10}{0.452} & \cellcolor{gray!10}{5.178}\\
TtP (eq), $\eta$ = 0.5, $\lambda$ = 0.5 & Partially consistent & 0.251 & 0.634 & 1.035\\
\cellcolor{gray!10}{TtP (eq), $\eta$ = 0.5, $\lambda$ = 0.5} & \cellcolor{gray!10}{Consistent} & \cellcolor{gray!10}{0.223} & \cellcolor{gray!10}{0.959} & \cellcolor{gray!10}{0.620}\\
\addlinespace
TtP (eq), $\eta$ = 0.5, $\lambda$ = 0.8 & No treatment effect & 0.239 & 0.073 & 7.173\\
\cellcolor{gray!10}{TtP (eq), $\eta$ = 0.5, $\lambda$ = 0.8} & \cellcolor{gray!10}{Partially consistent} & \cellcolor{gray!10}{0.215} & \cellcolor{gray!10}{0.645} & \cellcolor{gray!10}{0.932}\\
TtP (eq), $\eta$ = 0.5, $\lambda$ = 0.8 & Consistent & 0.212 & 0.999 & 0.440\\
\bottomrule
\end{longtable}  
 


\begin{figure}[htbp]
\centering
\includegraphics[width=0.85\linewidth, height=0.85\textheight,keepaspectratio]{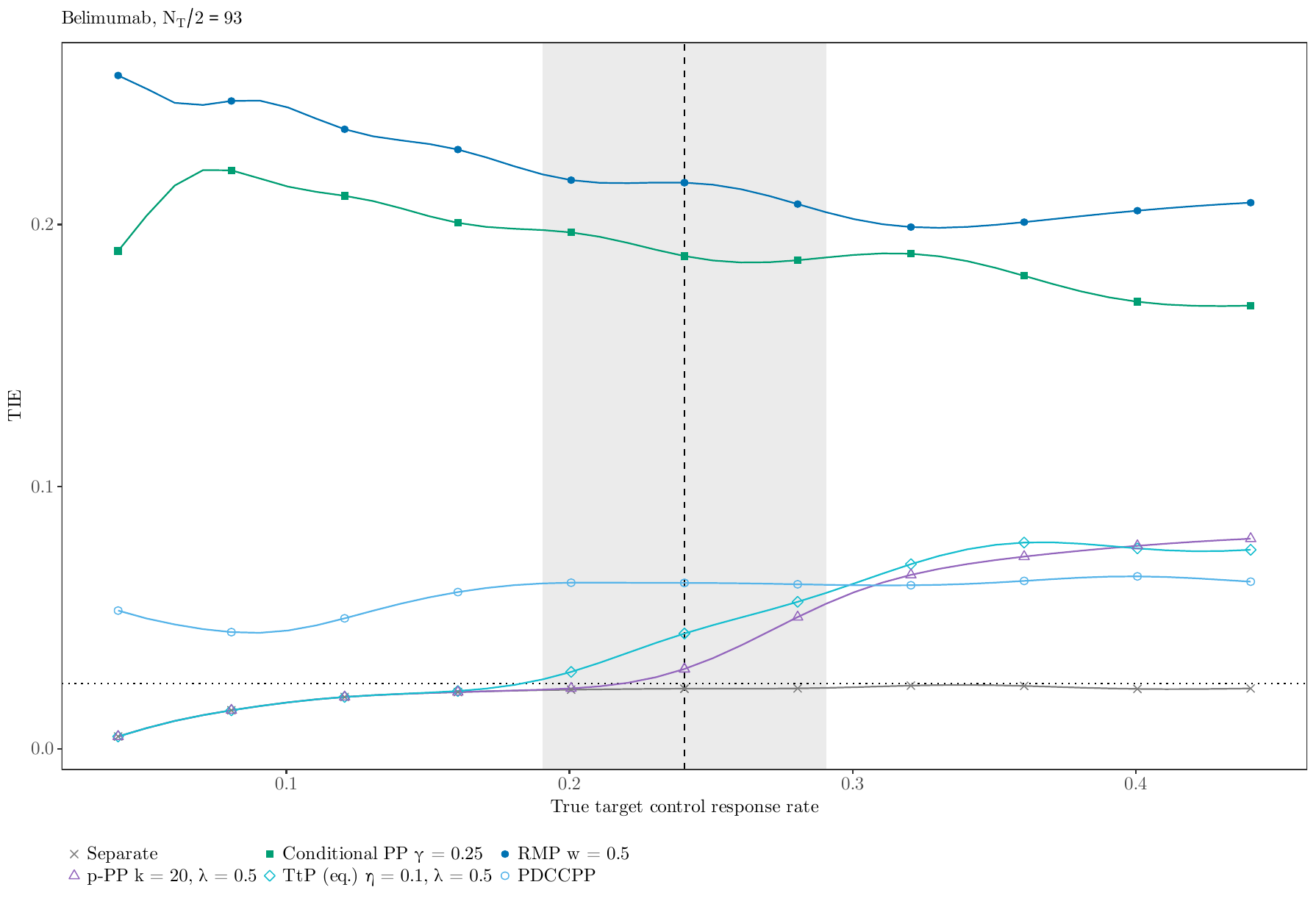}
\caption{TIE as a function of the true target control response rate, in the belimumab case study with 93 participants per arm and a lower source control rate (source denominator factor of 0.5), computed exactly for each method as specified for the design. The dashed vertical line is the control rate of the design, at which the simulation study evaluates the TIE, and the shaded band covers control rates within 0.05 of it. The dotted horizontal line is the nominal level of 0.025. p-PP: $p$-value-based power prior ($k = 20$, $\lambda = 0.5$), the configuration with the largest power gain at a matched TIE in the study. TtP (eq): test-then-pool with an equivalence test. CPP: conditional power prior. RMP: robust mixture prior. PDCCPP: prior-data conflict calibrated power prior.}
\label{fig:belimumab_tie_vs_control_rate_sample_size=93_source_denominator_change_factor=0.5}
\end{figure}

\clearpage
\subsection{Supplementary figures for the mepolizumab case study (recurrent event endpoint).}

\begin{figure}[htbp]
    \centering
    \includegraphics[width=0.95\linewidth, height=0.95\textheight,keepaspectratio]{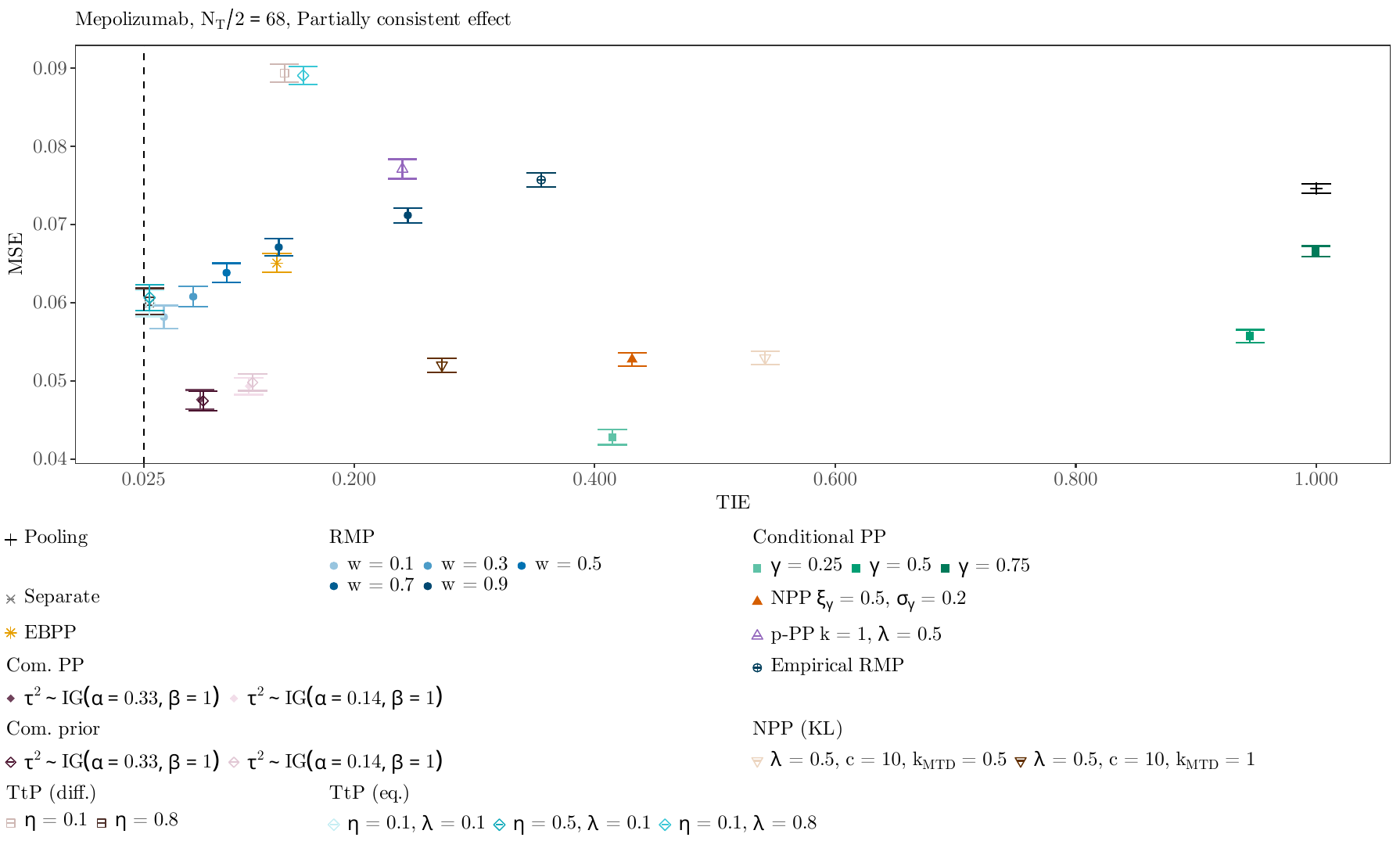}
    \caption{MSE as a function of type I error rate. Each point represents one method-parameter combination in the mepolizumab case study ($n_T = 68$), with partially consistent treatment effect. Error bars correspond to the 95\% CI. Dashed vertical line represents the nominal type I error rate of $0.025$. Method abbreviations and parameters as in Table~\ref{tab:abbreviations}.}
    \label{fig:mepolizumab_mse_vs_tie_sample_size=68_treatment_effect=partially_consistent}
\end{figure}

\begin{figure}[htbp]
    \centering
    \includegraphics[width=0.95\linewidth, height=0.95\textheight,keepaspectratio]{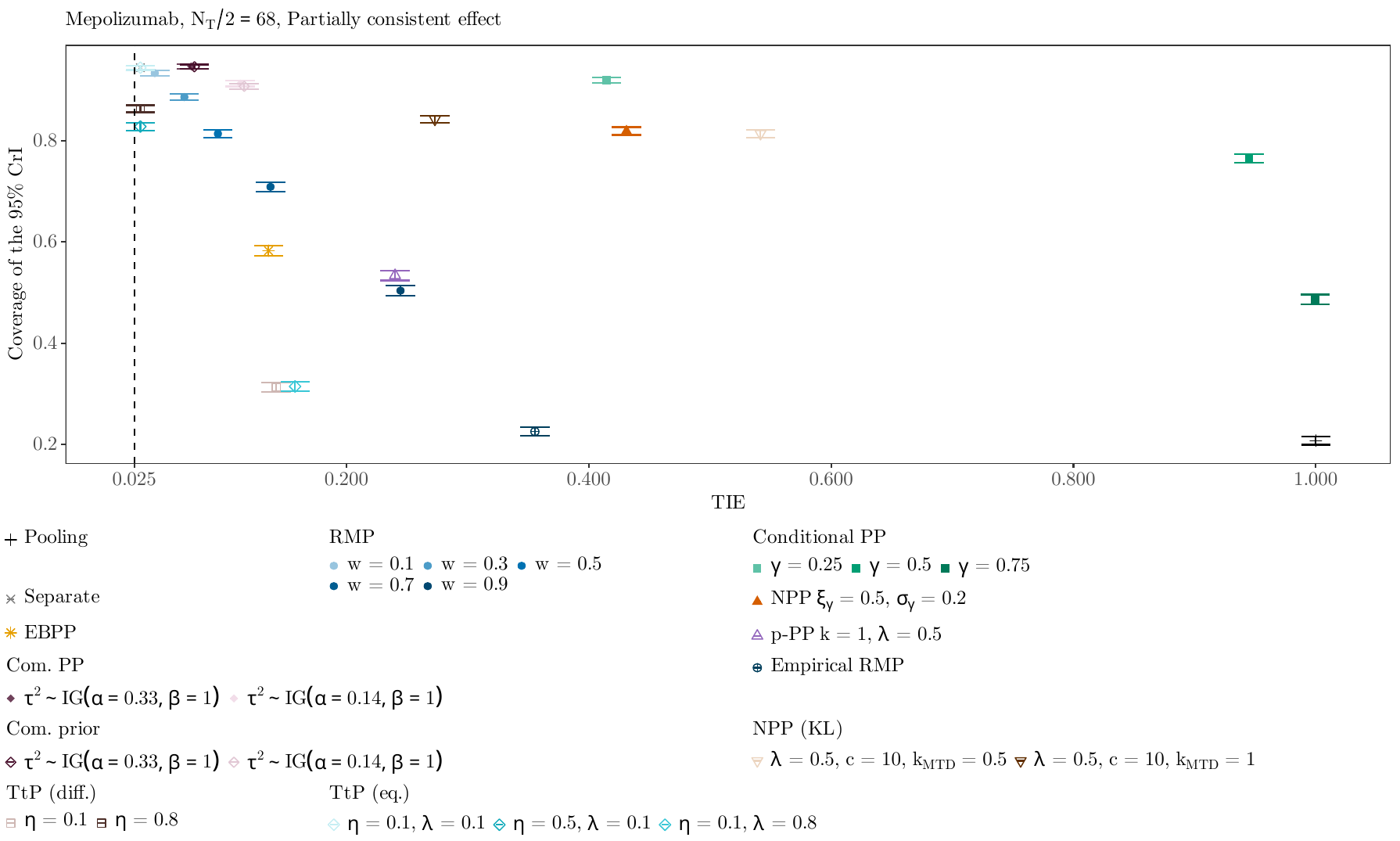}
    \caption{Coverage of the 95\% credible interval as a function of type I error rate. Each point represents one method-parameter combination in the mepolizumab case study with a sample size per arm of $68$, across all the methods and parameters, with a partially consistent treatment effect. Error bars correspond to the 95\% CI of the Coverage and type I error rate. Dashed vertical line represents the nominal type I error rate of $0.025$. Method abbreviations and parameters as in Table~\ref{tab:abbreviations}.}    \label{fig:mepolizumab_coverage_vs_tie_sample_size=68_treatment_effect=partially_consistent_source_denominator_change_factor=1}
\end{figure}

\begin{figure}
    \centering
    \includegraphics[width=0.95\linewidth, height=0.95\textheight,keepaspectratio]{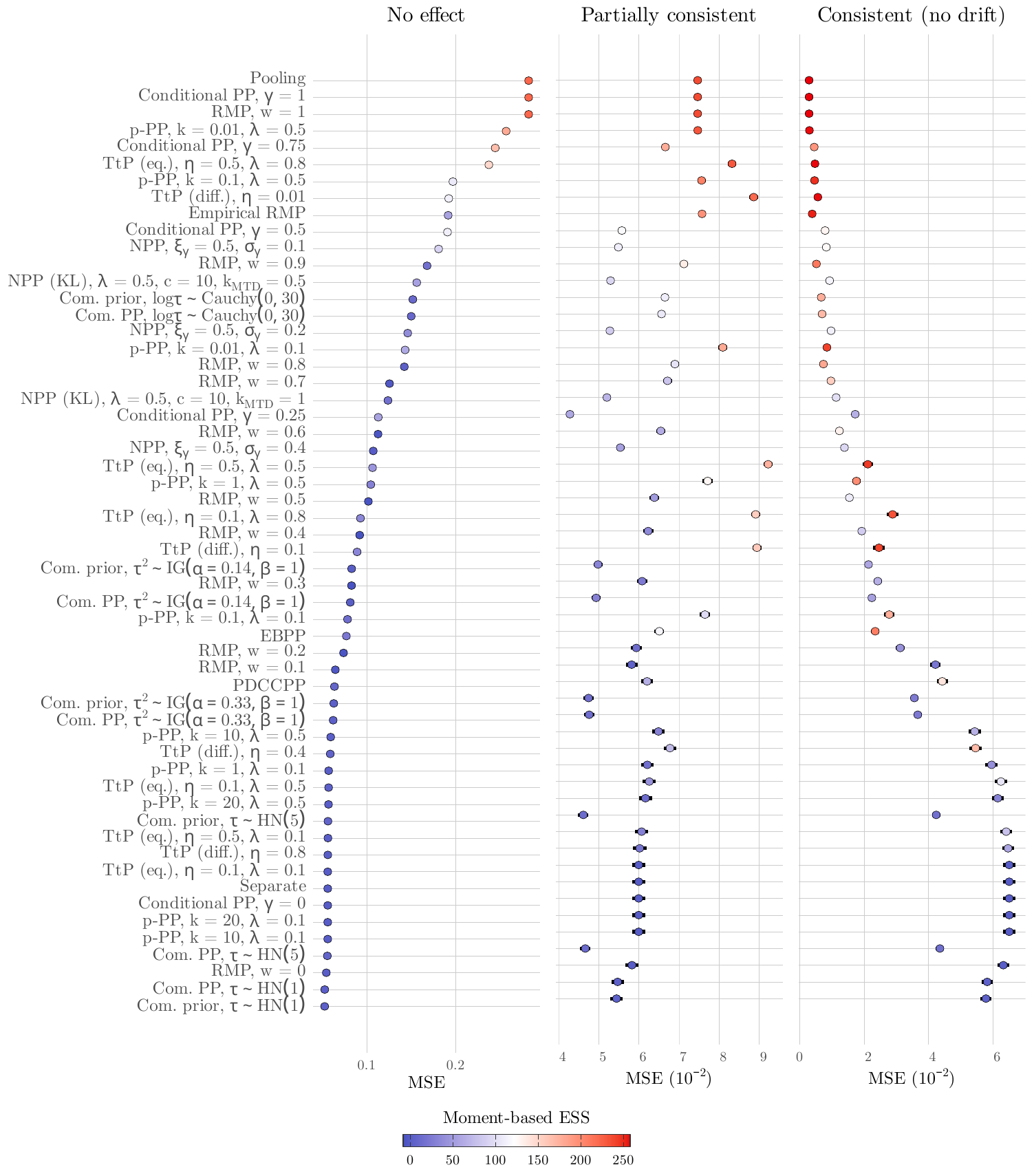}
    \caption{MSE for the three main treatment effect values considered in the mepolizumab case study ($n_T = 68$). Method abbreviations and parameters as in Table~\ref{tab:abbreviations}.}
    \label{fig:mepolizumab_mse_forest_plot_target_sample_size_per_arm_68}
\end{figure}

\begin{figure}
    \centering
    \includegraphics[width=0.95\linewidth, height=0.95\textheight,keepaspectratio]{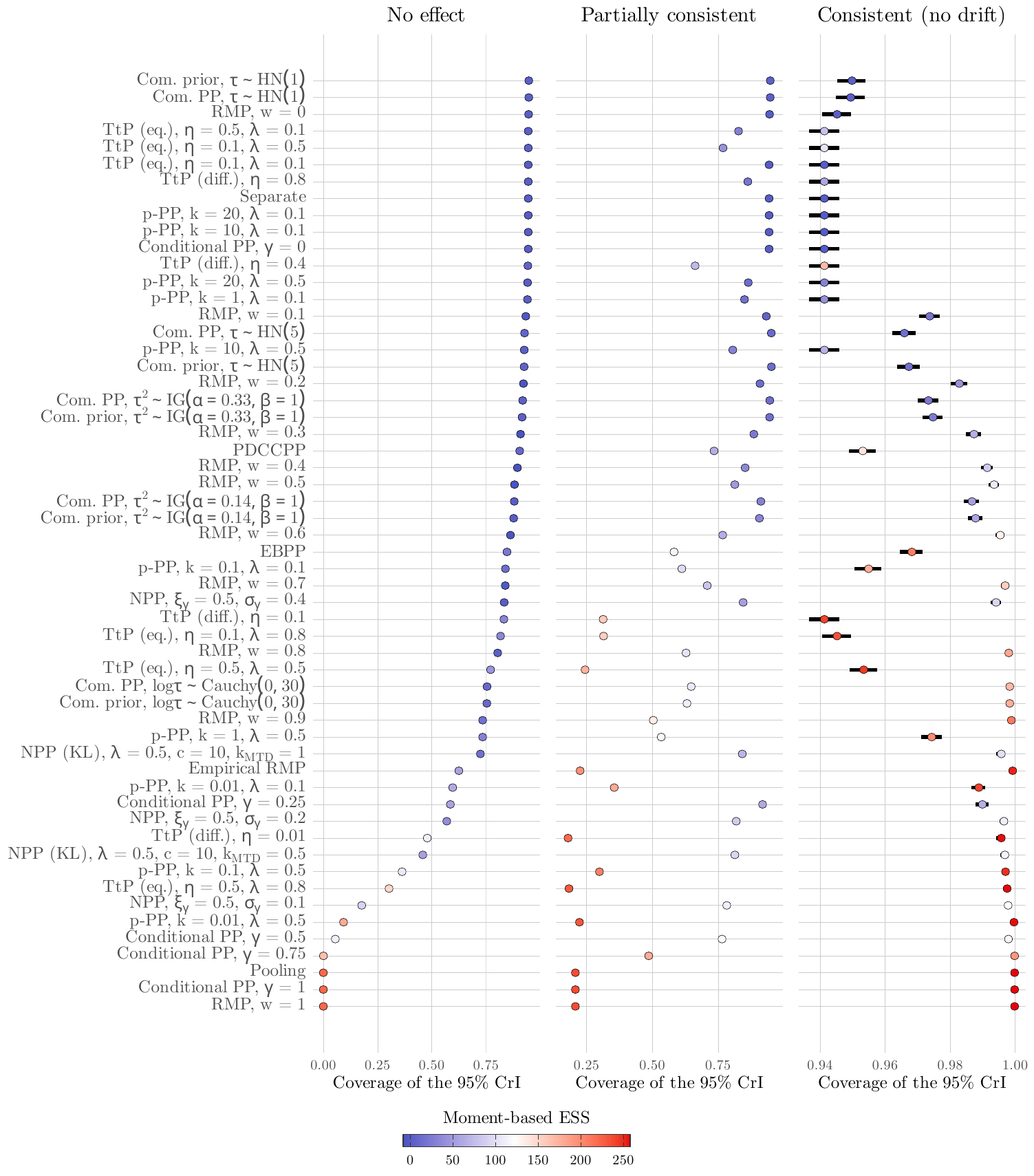}
    \caption{Empirical coverage probability  for the three main treatment effect values considered in the mepolizumab case study ($n_T = 68$). Error bars correspond to the 95\% CI. Method abbreviations and parameters as in Table~\ref{tab:abbreviations}.}
    \label{fig:mepolizumab_coverage_forest_plot_target_sample_size_per_arm_68}
\end{figure}

\begin{figure}
    \centering
    \includegraphics[width=0.95\linewidth, height=0.95\textheight,keepaspectratio]{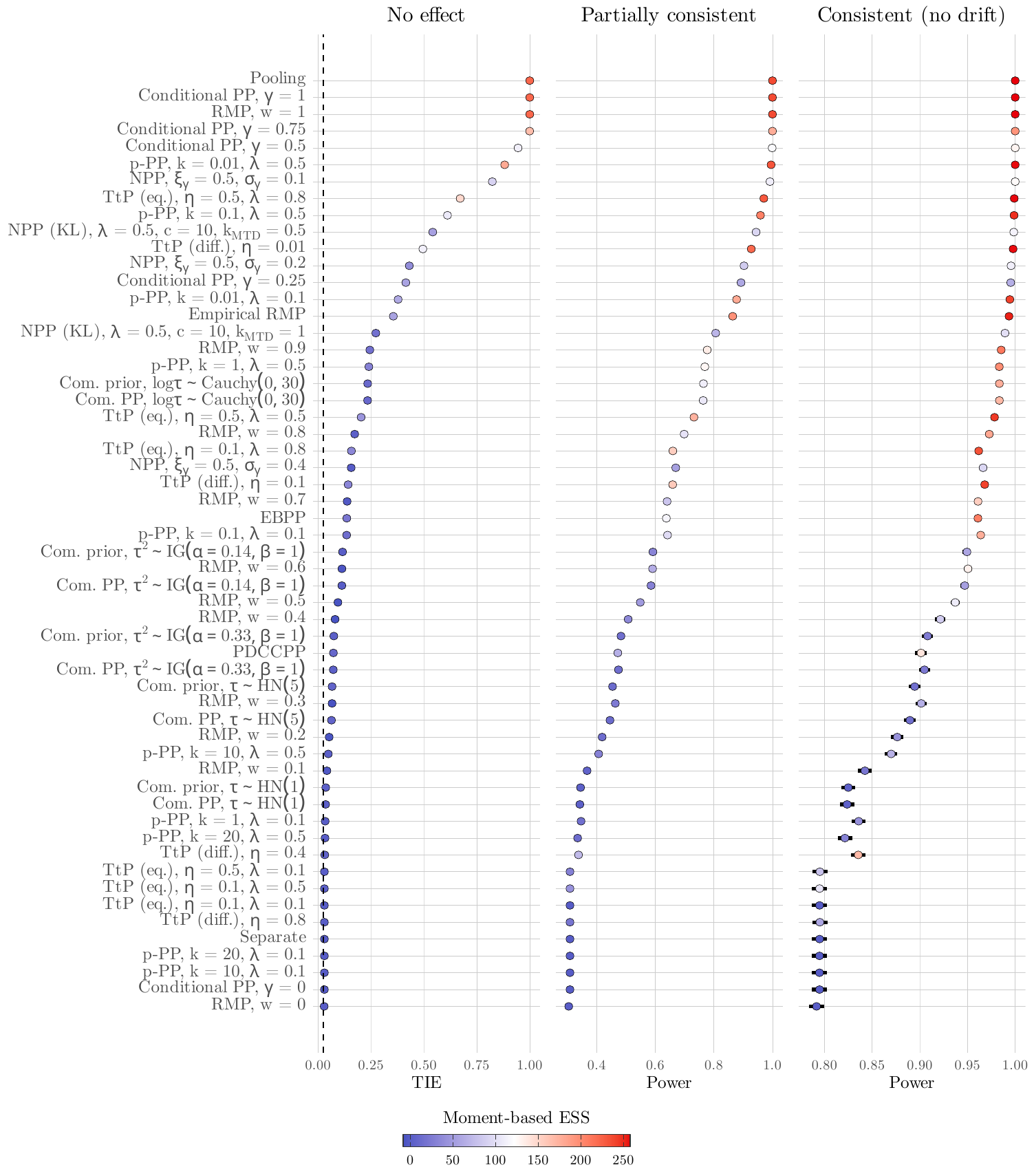}
    \caption{Probability of study success for the three main treatment effect values considered in the mepolizumab case study ($n_T = 68$). Method abbreviations and parameters as in Table~\ref{tab:abbreviations}.}    \label{fig:mepolizumab_success_proba_forest_plot_target_sample_size_per_arm_68}
\end{figure}

 \begin{figure}
     \centering
     \includegraphics[width=0.95\linewidth, height=0.95\textheight,keepaspectratio]{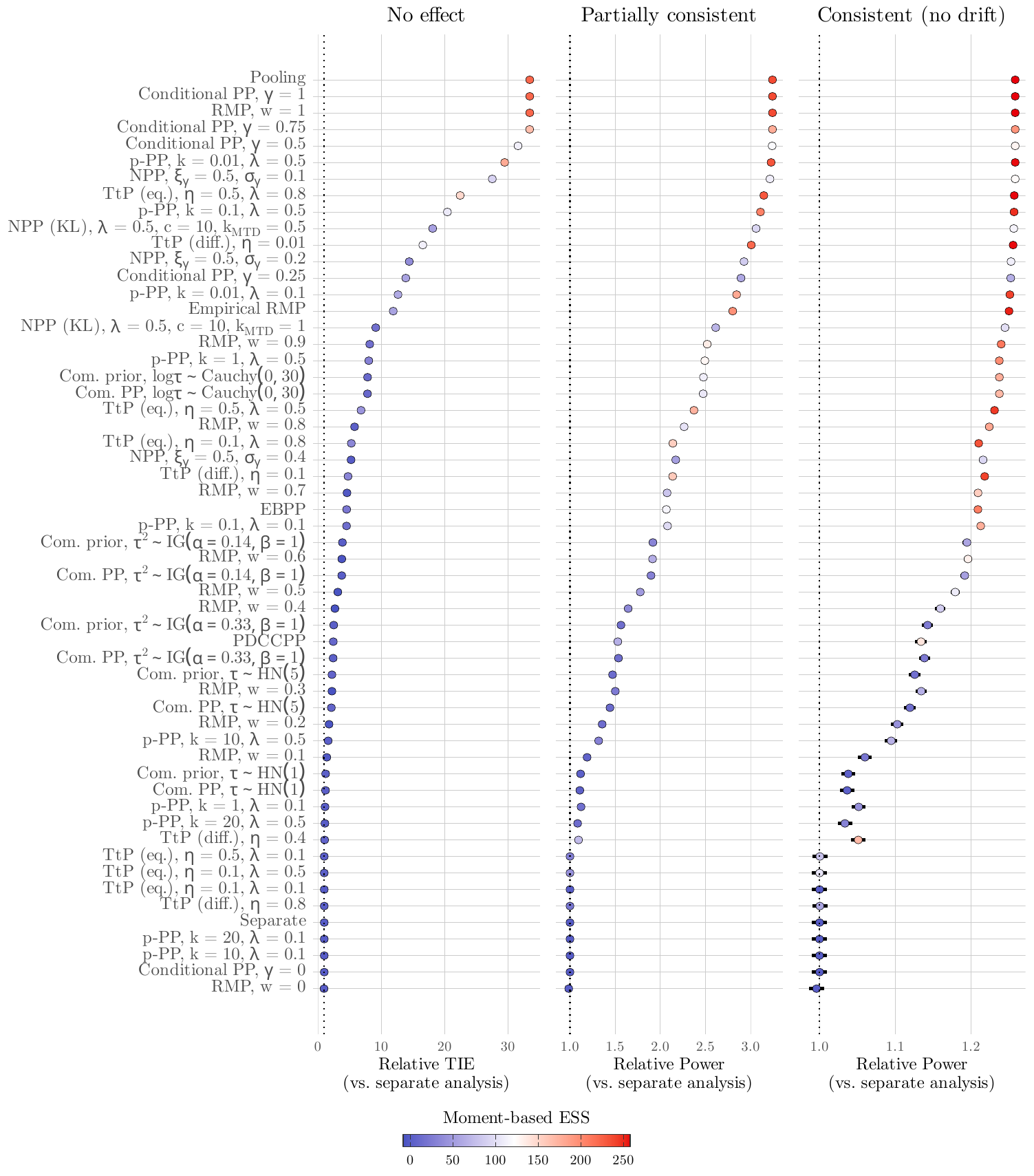}
     \caption{Probability of study success for the different methods for the three main treatment effect values considered in the mepolizumab case study ($n_T = 68$), relative to the probability of success in the case of a separate analysis. Method abbreviations and parameters as in Table~\ref{tab:abbreviations}.}     \label{fig:mepolizumab_relative_success_proba_forest_plot_target_sample_size_per_arm_68}
 \end{figure}

\begin{figure}[p]
  \centering
  \includegraphics[width=\textwidth]{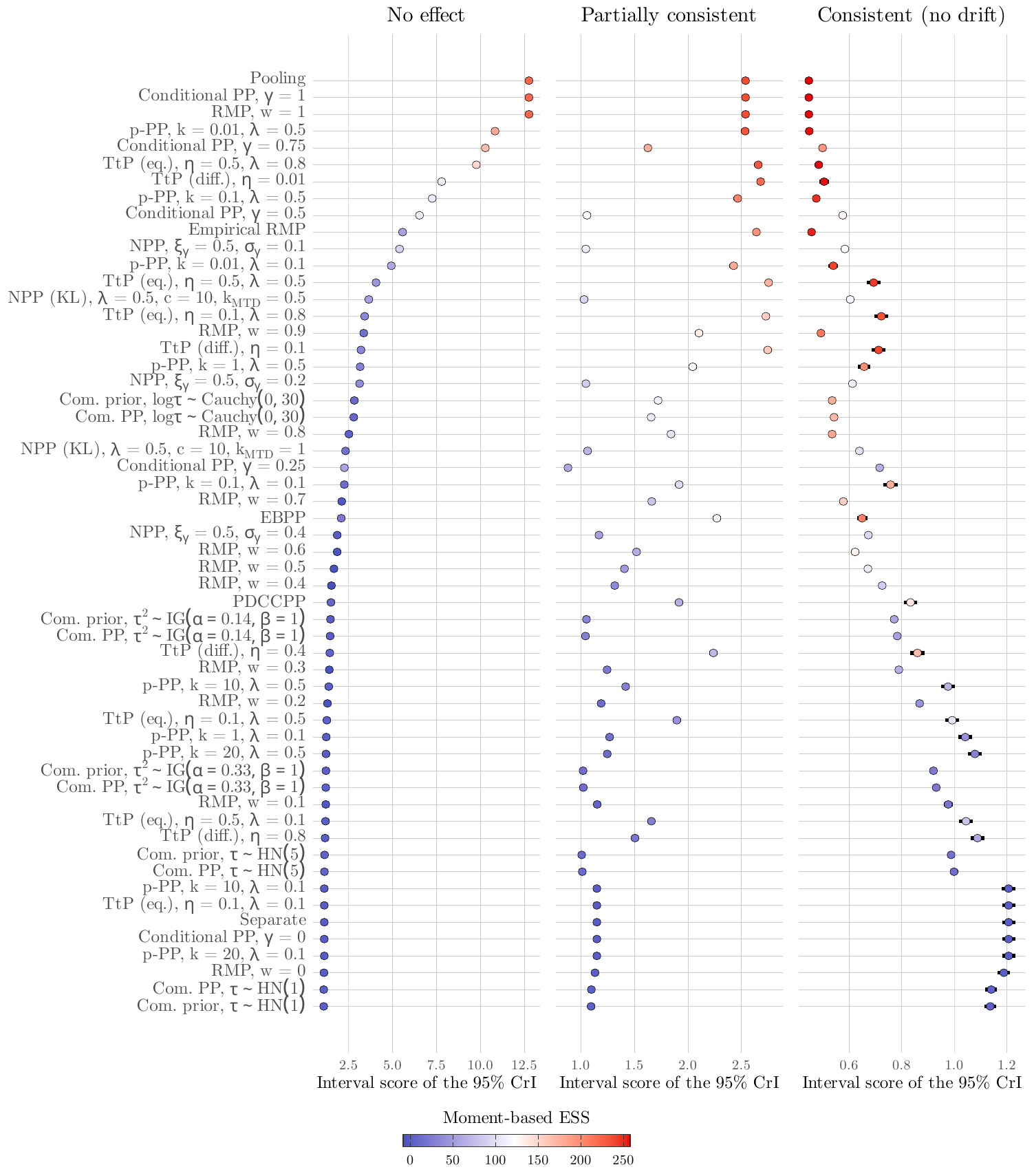}
  \caption{Interval score of the 95\% credible interval for the three main treatment effect values considered in the mepolizumab case study ($n_T = 68$). Smaller is better. Error bars correspond to the 95\% CI.}
  \label{fig:mepolizumab_interval_score_forest_plot_target_sample_size_per_arm_68}
\end{figure}

\begin{figure}[p]
  \centering
  \includegraphics[width=\textwidth]{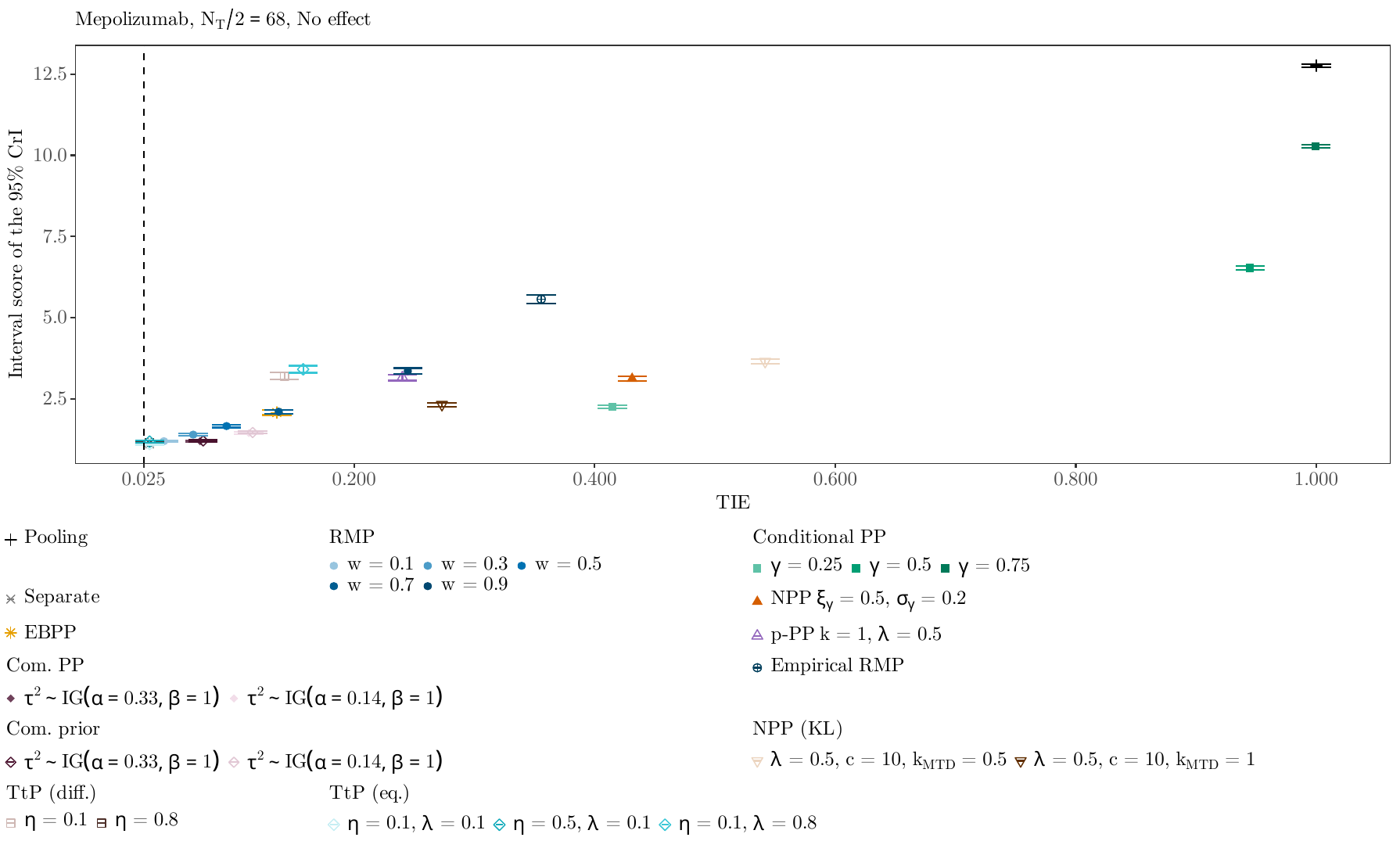}
  \caption{Interval score of the 95\% credible interval as a function of type I error rate. Each point represents one method-parameter combination in the mepolizumab case study with a sample size per arm of 68, without treatment effect. Smaller is better. Error bars correspond to the 95\% CI of the interval score and type I error rate. Dashed vertical line represents the nominal type I error rate of 0.025.}  \label{fig:mepolizumab_interval_score_vs_tie_sample_size=68_treatment_effect=no_effect}
\end{figure}

\begin{figure}[htbp]
\centering
\includegraphics[width=0.95\linewidth, height=0.95\textheight,keepaspectratio]{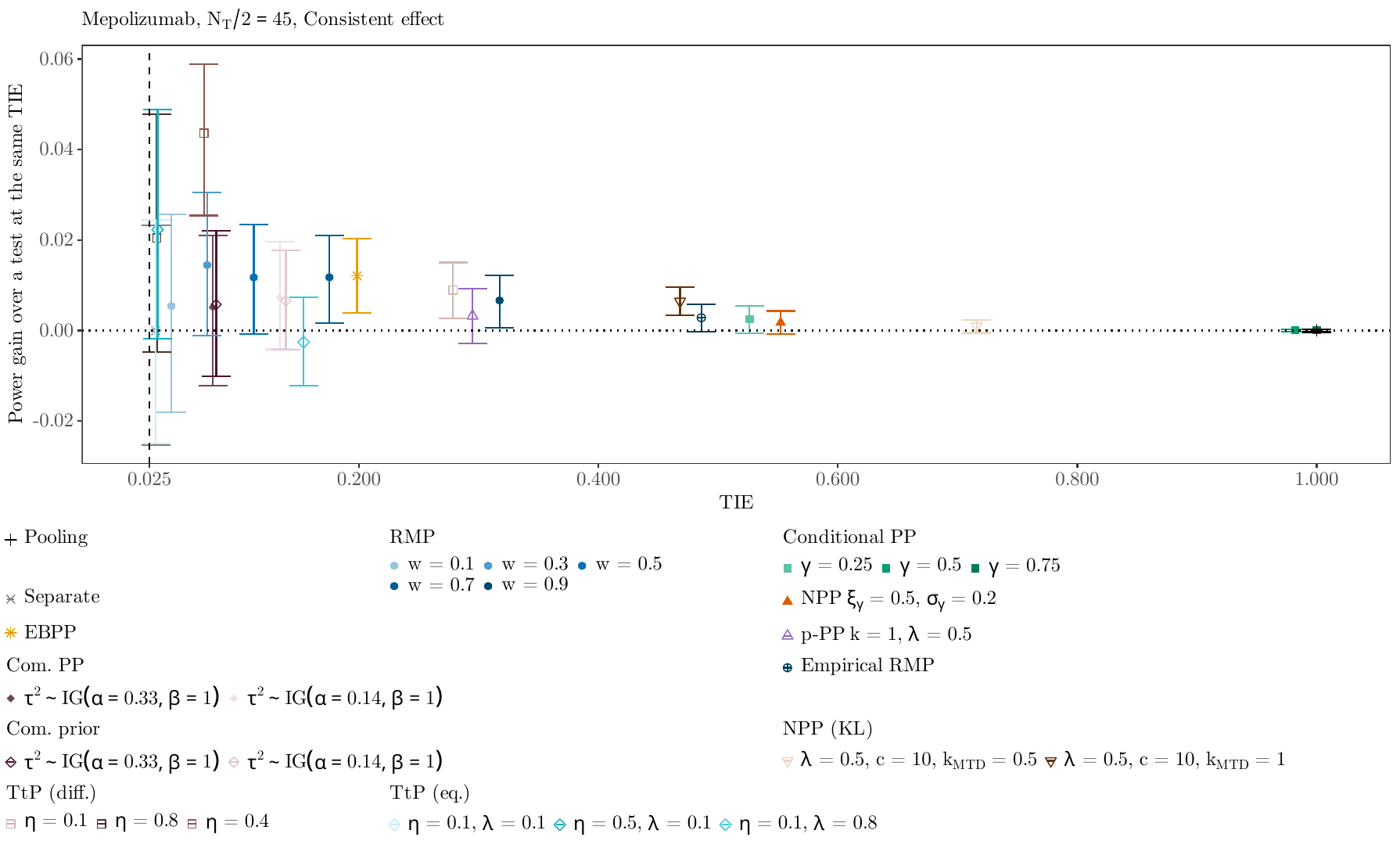}
\caption{Power gain over a frequentist test of the target data alone performed at the same TIE, versus TIE, in the mepolizumab case study, with 45 participants per arm and a consistent treatment effect. Error bars are 95\% intervals for the difference. For readability, only a subset of the parameter settings is shown, to which test-then-pool with a difference test at $\eta = 0.4$, the setting with the largest gain, was added; the PDCCPP is omitted.}
\label{fig:mepolizumab_power_gain_vs_tie_sample_size=45_treatment_effect=consistent_source_denominator_change_factor=1}
\end{figure}

\clearpage
 \subsection{Supplementary figures for the teriflunomide case study (time-to-event endpoint).}

\begin{figure}[htbp]
    \centering
    \includegraphics[width=0.95\linewidth, height=0.95\textheight,keepaspectratio]{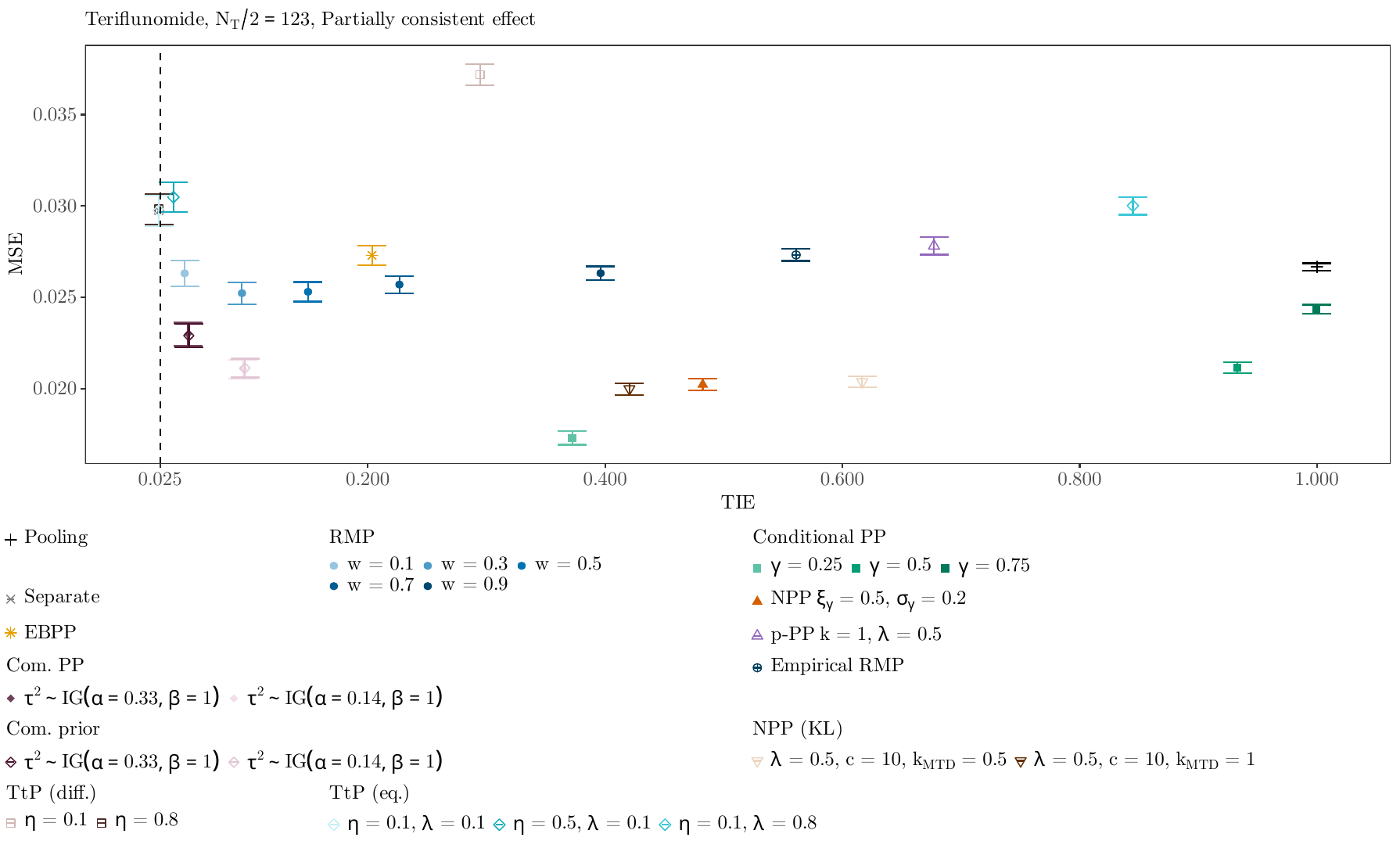}
    \caption{MSE as a function of type I error rate. Each point represents one method-parameter combination in the teriflunomide case study ($n_T = 123$), with partially consistent treatment effect. Error bars correspond to the 95\% CI. Dashed vertical line represents the nominal type I error rate of $0.025$. Method abbreviations and parameters as in Table~\ref{tab:abbreviations}.}
    \label{fig:teriflunomide_mse_vs_tie_sample_size=123_treatment_effect=partially_consistent}
\end{figure}

\begin{figure}[htbp]
    \centering
    \includegraphics[width=0.90\linewidth, height=0.95\textheight,keepaspectratio]{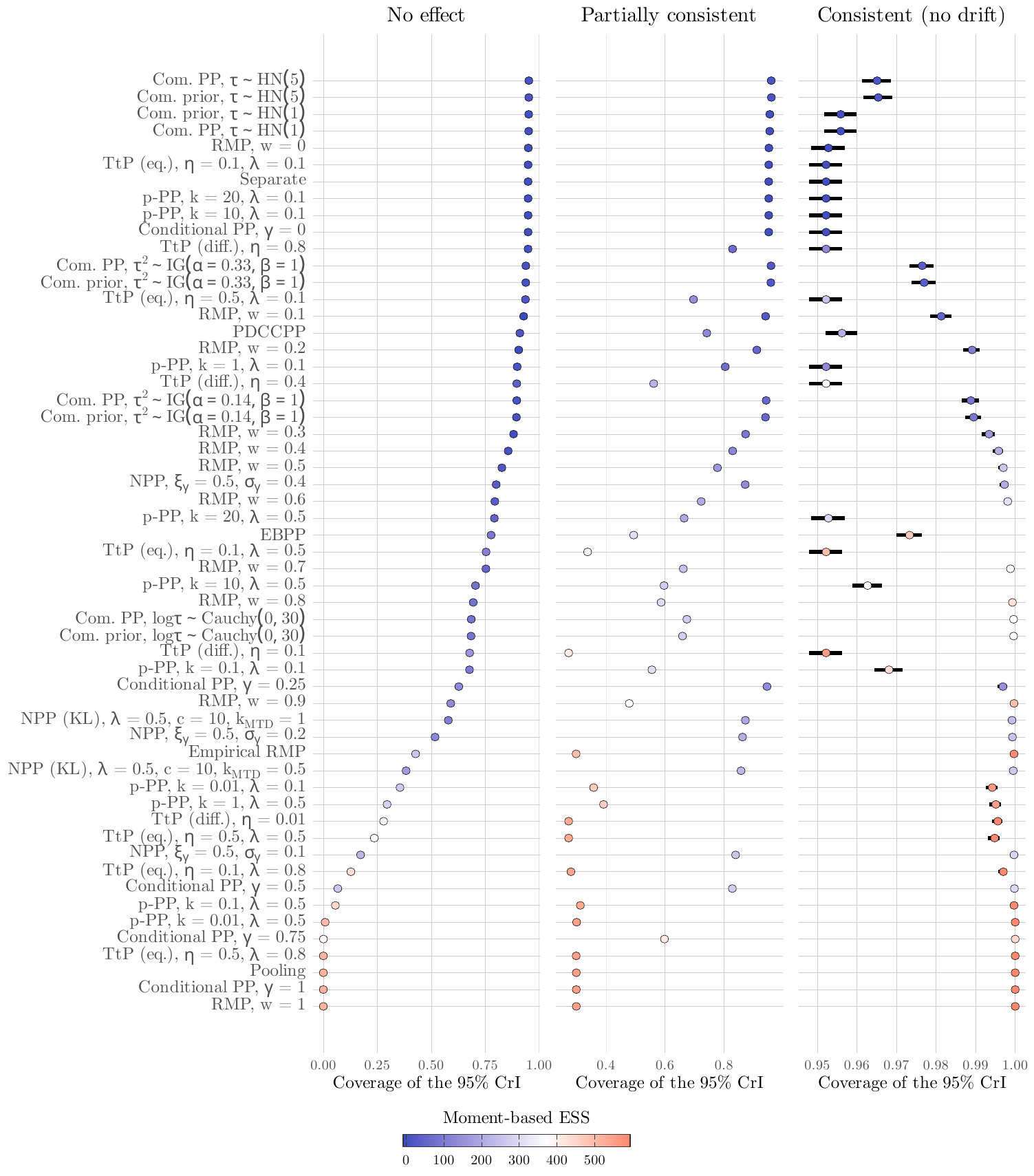}
    \caption{Coverage probability of the 95\% credible interval for the three main treatment effect values considered in the teriflunomide case study ($n_T = 123$), under the primary time-to-event design and without change in the denominator of the source study summary measure. Error bars correspond to the 95\% CI on the coverage probability. Method abbreviations and parameters as in Table~\ref{tab:abbreviations}.}
    \label{fig:teriflunomide_coverage_forest_plot_target_sample_size_per_arm_123}
\end{figure} 

\begin{figure}[htbp]
    \centering
    \includegraphics[width=0.95\linewidth, height=0.95\textheight,keepaspectratio]{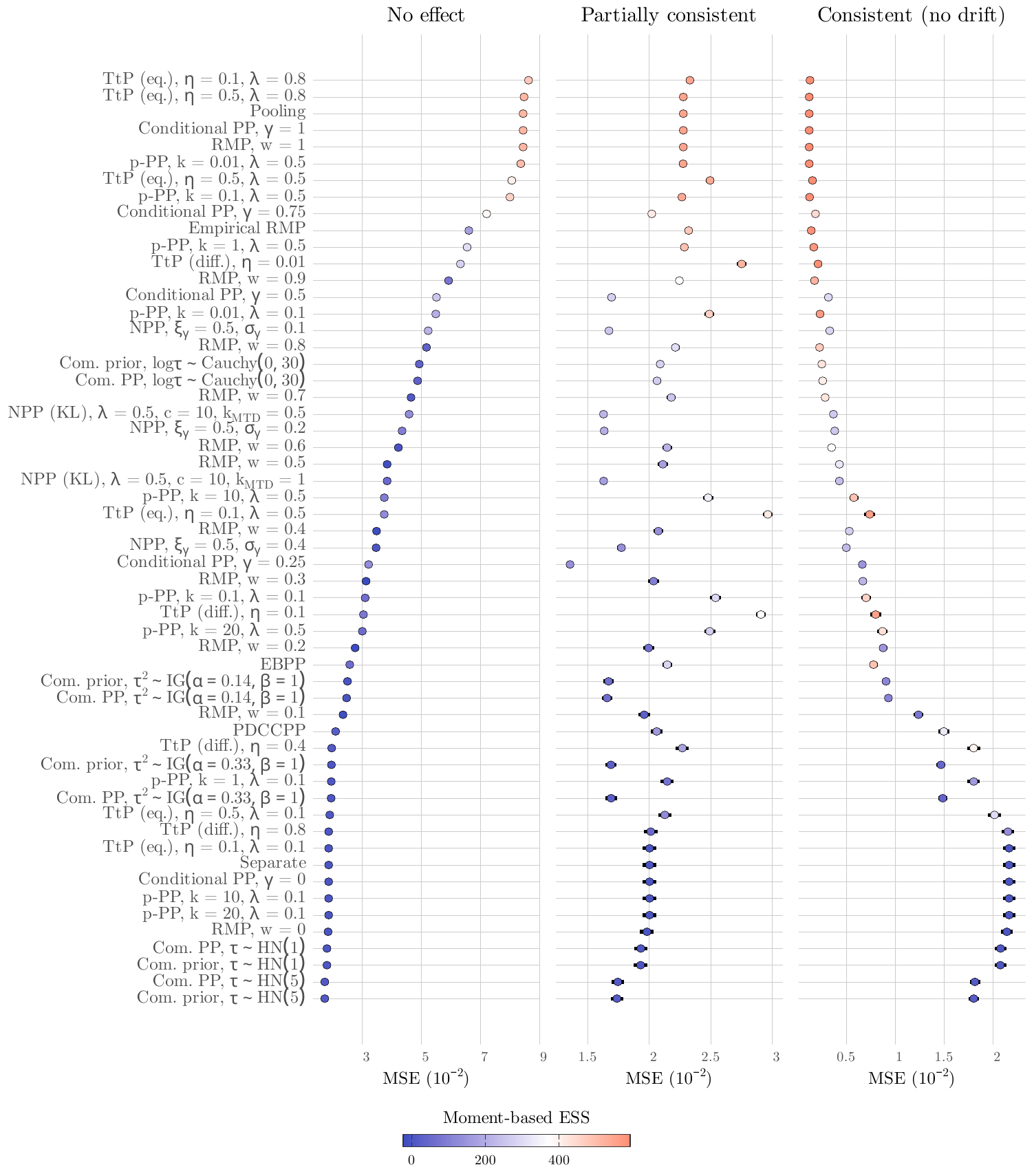}
    \caption{MSE for the three main treatment effect values considered in the teriflunomide case study ($n_T = 185$). Method abbreviations and parameters as in Table~\ref{tab:abbreviations}.}    \label{fig:teriflunomide_mse_forest_plot_target_sample_size_per_arm_185}
\end{figure}
\clearpage

 \begin{figure}[htbp]
     \centering
     \includegraphics[width=0.95\linewidth, height=0.95\textheight,keepaspectratio]{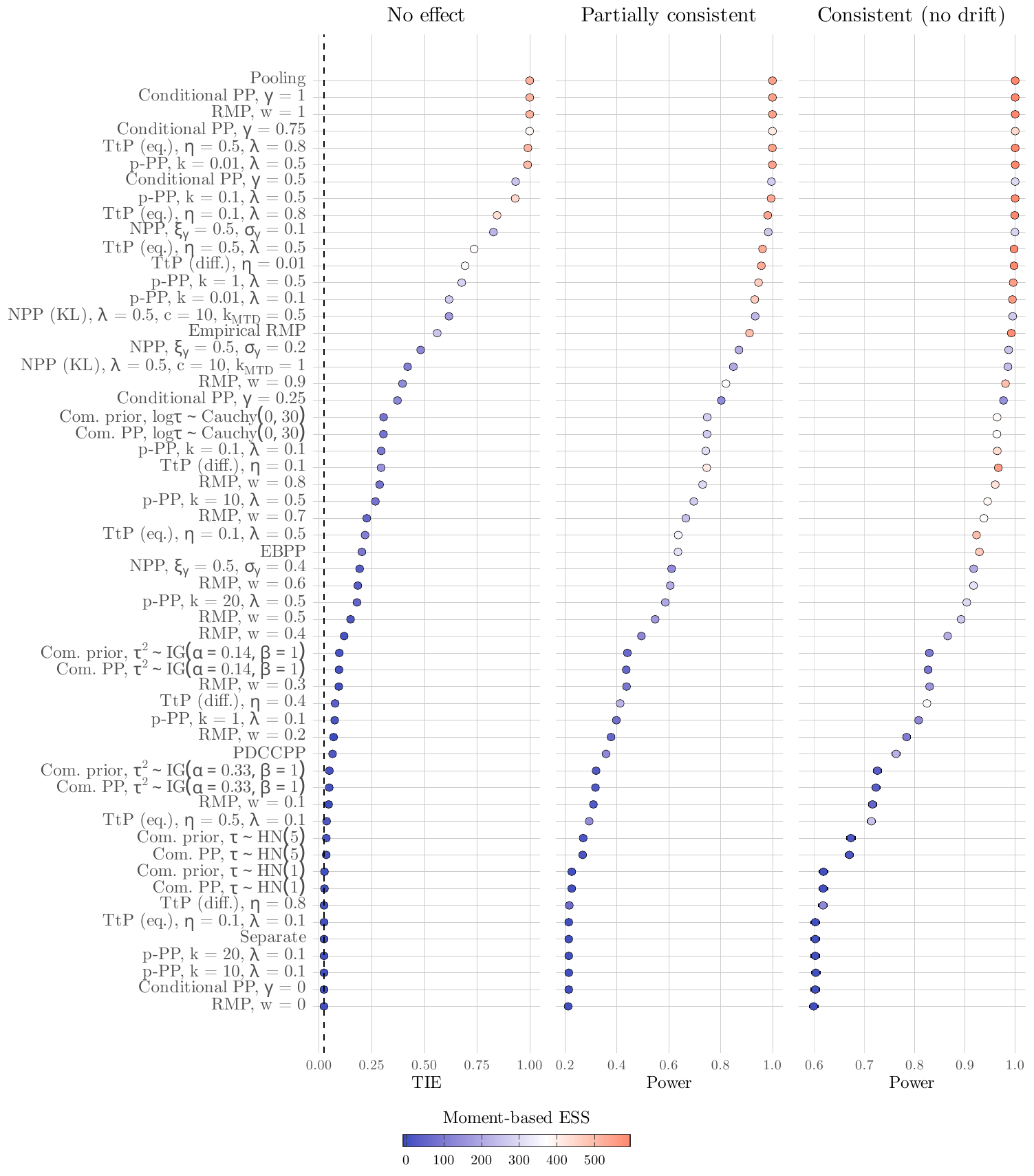}
     \caption{Probability of study success  for the three main treatment effect values considered in the teriflunomide case study ($n_T = 123$). Method abbreviations and parameters as in Table~\ref{tab:abbreviations}.}     \label{fig:teriflunomide_success_proba_forest_plot_target_sample_size_per_arm_123}
 \end{figure}

 \begin{figure}[htbp]
     \centering
     \includegraphics[width=0.95\linewidth, height=0.95\textheight,keepaspectratio]{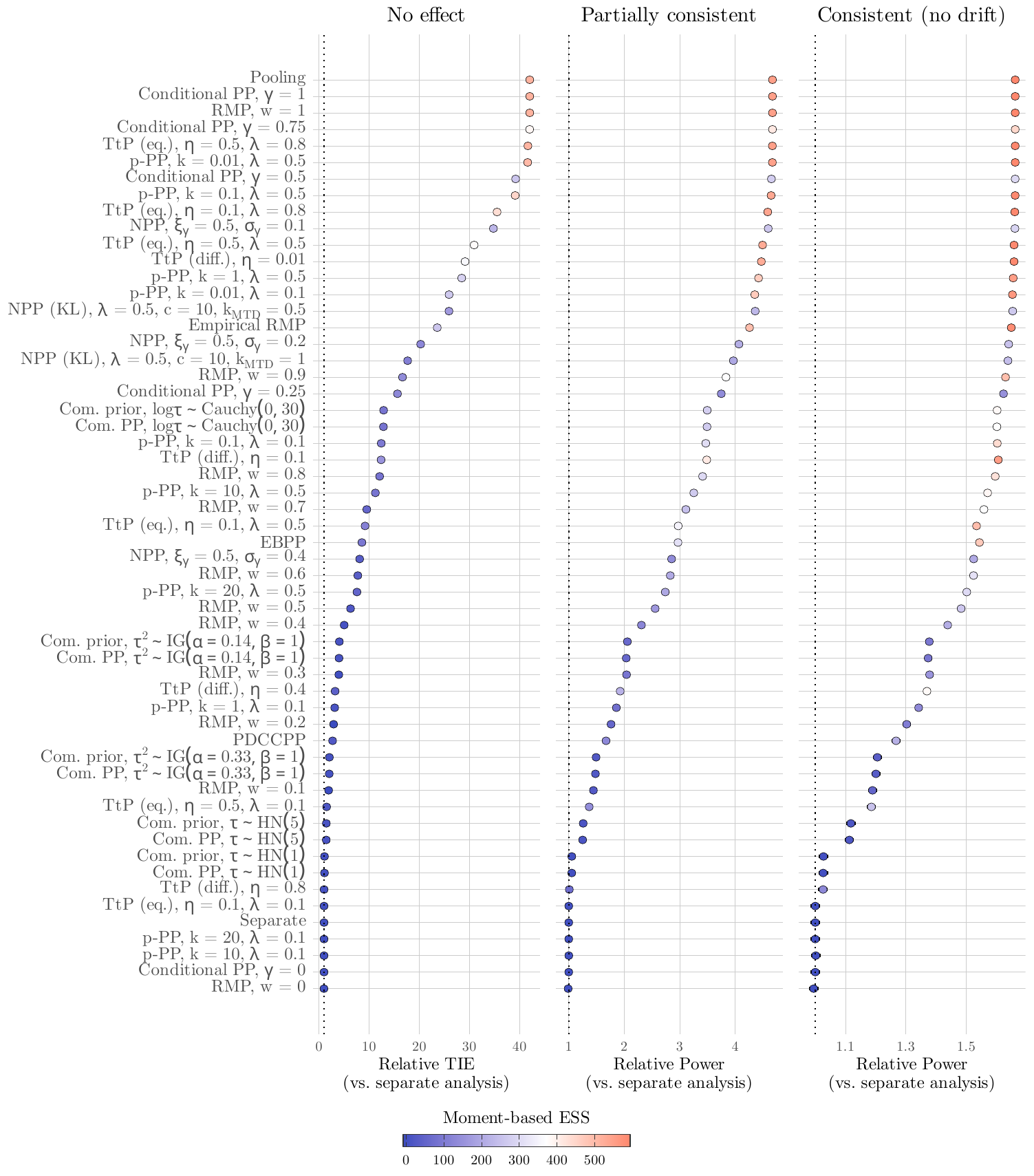}
     \caption{Probability of study success  for the three main treatment effect values considered in the teriflunomide case study ($n_T = 123$), relative to the probability of success of a separate analysis. Method abbreviations and parameters as in Table~\ref{tab:abbreviations}.}     \label{fig:teriflunomide_relative_success_proba_forest_plot_target_sample_size_per_arm_123}
 \end{figure}

\begin{figure}[htbp]
\centering
\includegraphics[width=\textwidth]{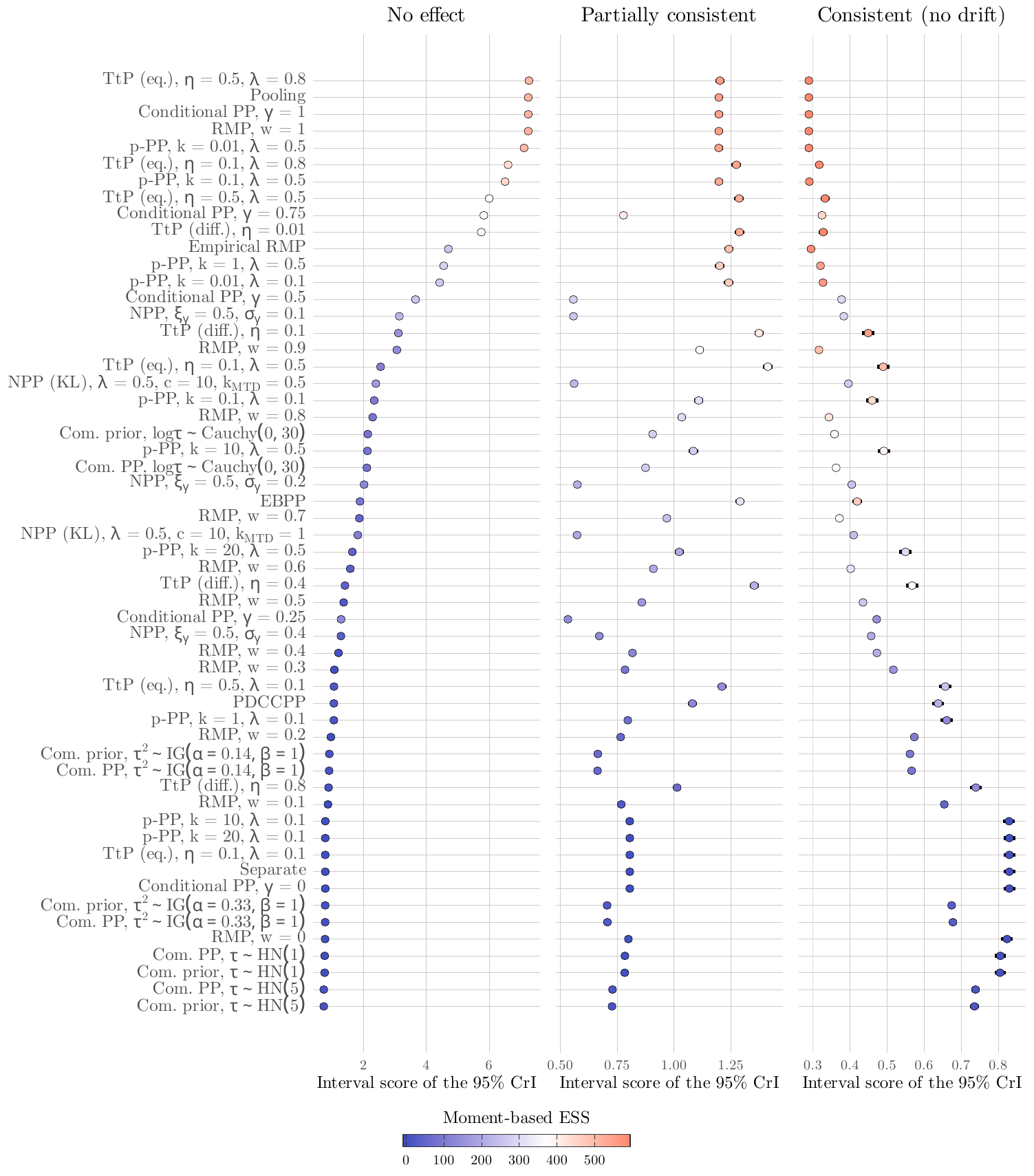}
\caption{Interval score of the 95\% credible interval for the three main treatment effect values considered in the teriflunomide case study ($n_T = 123$). Smaller is better. Error bars correspond to the 95\% CI.}
\label{fig:teriflunomide_interval_score_forest_plot_target_sample_size_per_arm_123}
\end{figure}

\begin{figure}[htbp]
  \centering
  \includegraphics[width=\textwidth]{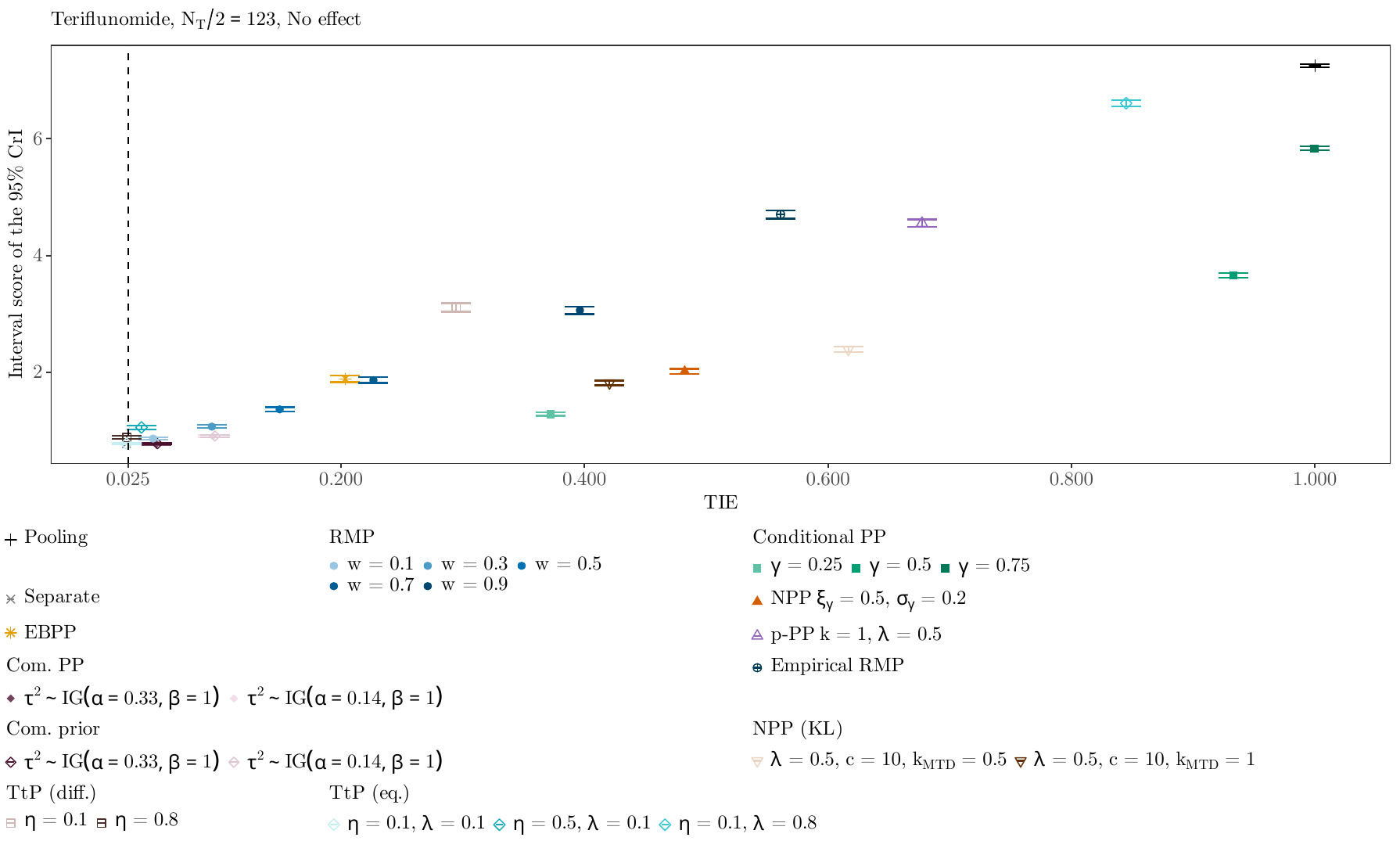}
  \caption{Interval score of the 95\% credible interval as a function of type I error rate. Each point represents one method-parameter combination in the teriflunomide case study with a sample size per arm of 123, without treatment effect. Smaller is better. Error bars correspond to the 95\% CI of the interval score and type I error rate. Dashed vertical line represents the nominal type I error rate of 0.025. } \label{fig:teriflunomide_interval_score_vs_tie_sample_size=123_treatment_effect=no_effect}
\end{figure}

\begin{figure}[htbp]
\centering
\includegraphics[width=0.95\linewidth, height=0.95\textheight,keepaspectratio]{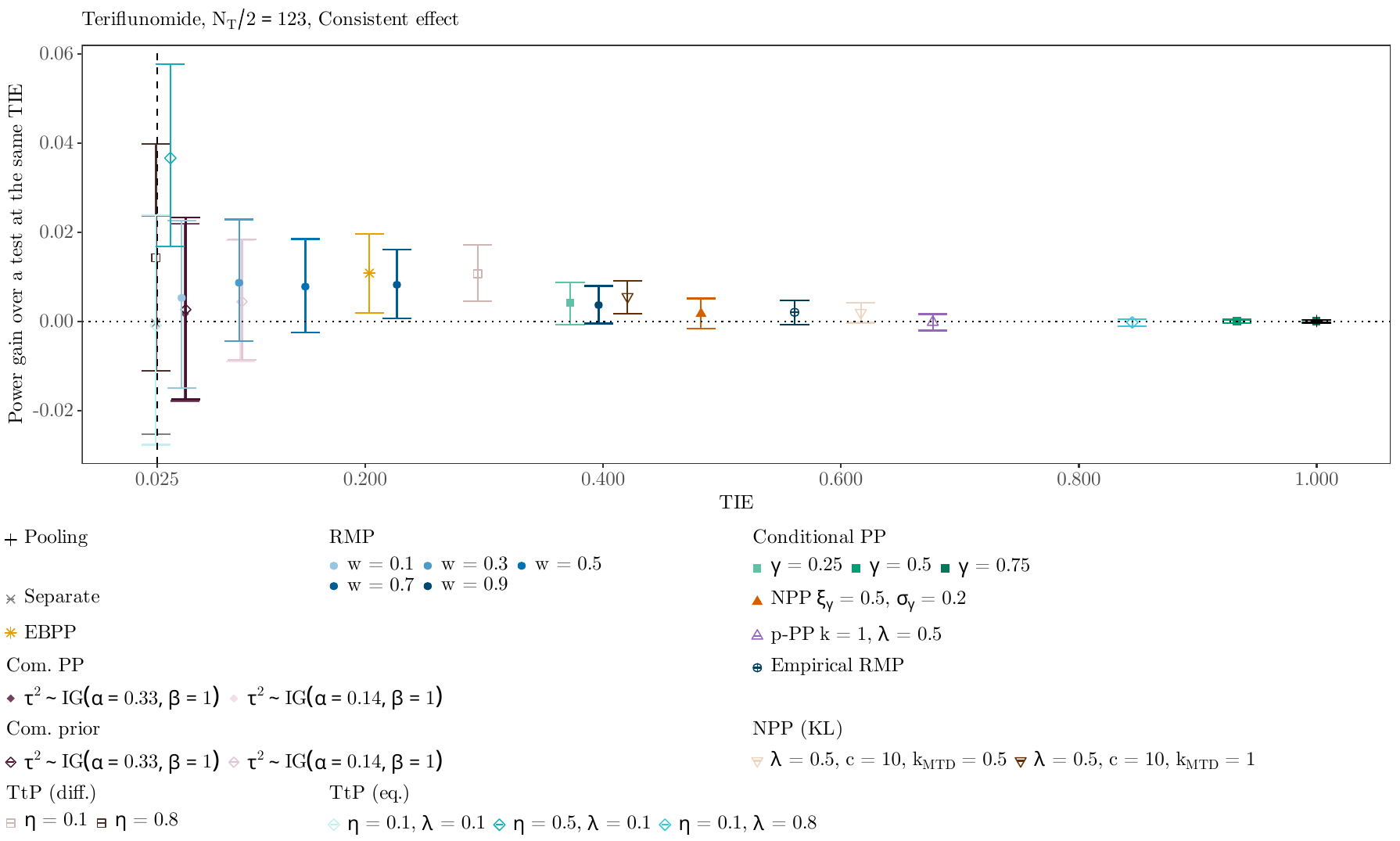}
\caption{Power gain over a frequentist test of the target data alone performed at the same TIE, versus TIE, in the teriflunomide case study, with 123 participants per arm and a consistent treatment effect. Error bars are 95\% intervals for the difference. For readability, only a subset of the parameter settings is shown, and the PDCCPP is omitted.}
\label{fig:teriflunomide_power_gain_vs_tie_sample_size=123_treatment_effect=consistent_source_denominator_change_factor=1}
\end{figure}

\clearpage
\subsection{Supplementary figures for the aprepitant case study (binary endpoint).}

\begin{figure}[htbp]
    \centering
    \includegraphics[width=0.95\linewidth, height=0.95\textheight,keepaspectratio]{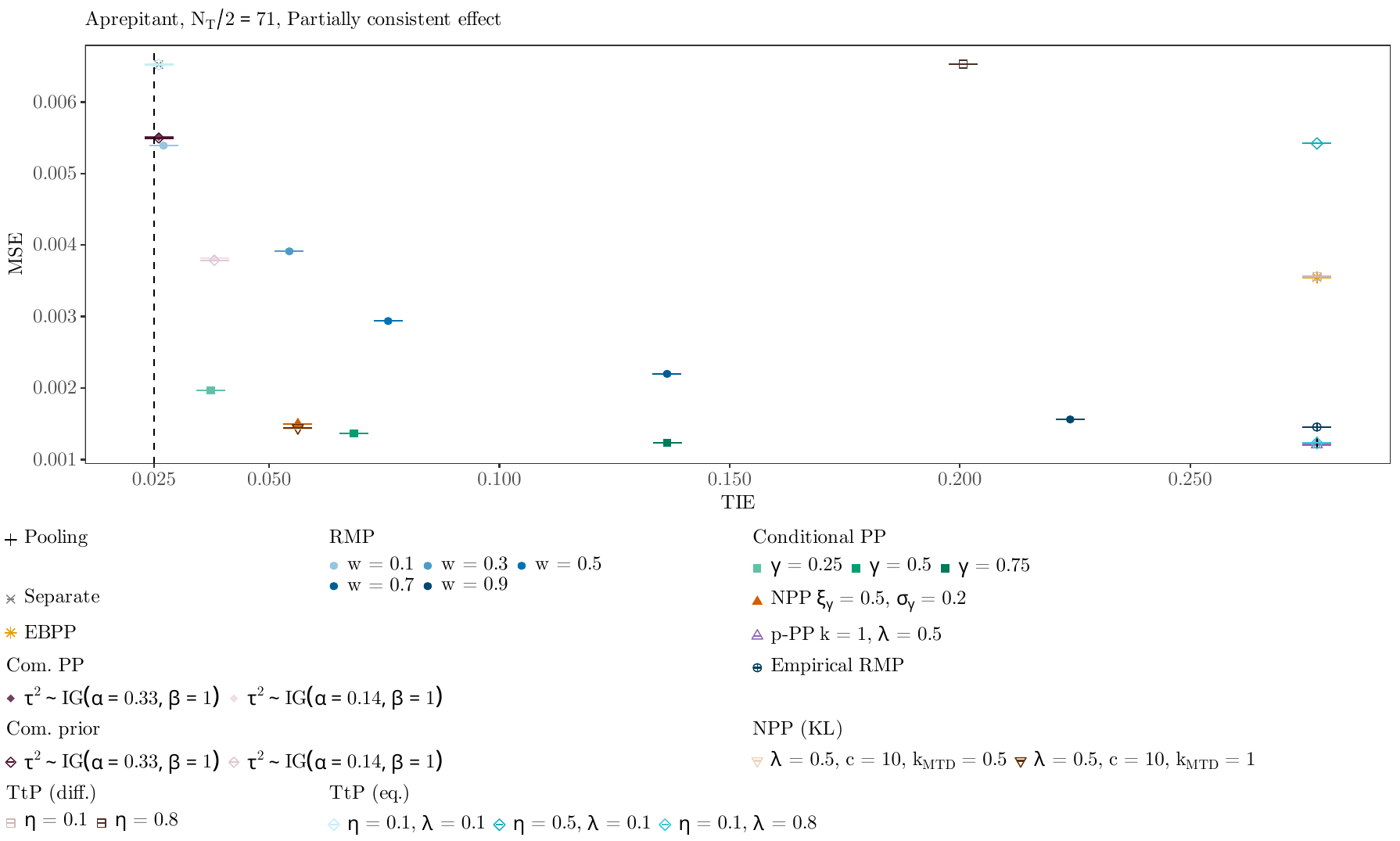}
    \caption{MSE as a function of type I error rate. Each point represents one method-parameter combination in the aprepitant case study ($n_T = 71$), with partially consistent treatment effect. Error bars correspond to the 95\% CI. Dashed vertical line represents the nominal type I error rate of $0.025$. Method abbreviations and parameters as in Table~\ref{tab:abbreviations}.}
    \label{fig:aprepitant_mse_vs_tie_sample_size=71_treatment_effect=partially_consistent}
\end{figure}

\begin{figure}[htpb]
    \centering
    \includegraphics[width=0.95\linewidth, height=0.95\textheight,keepaspectratio]{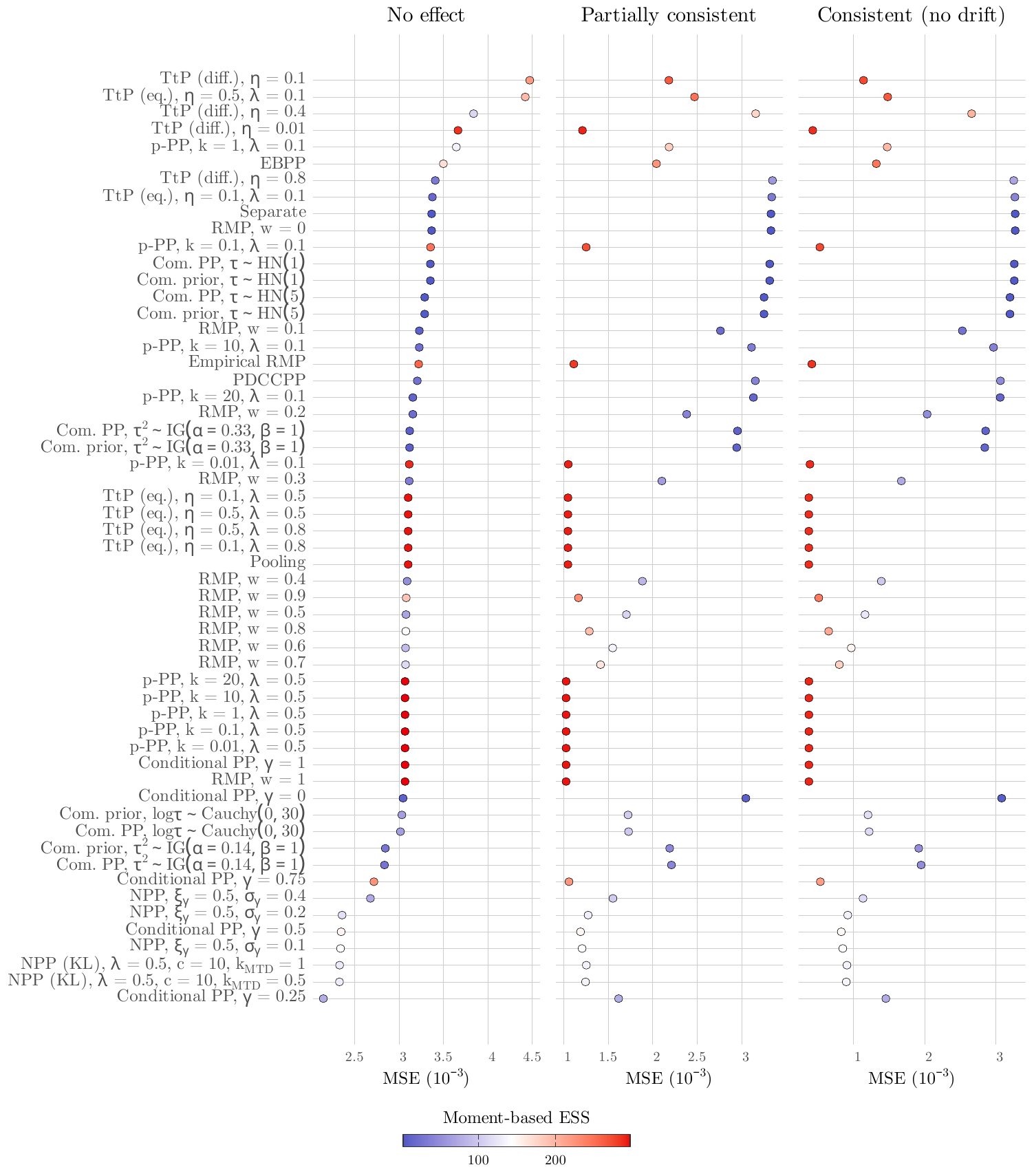}
    \caption{MSE  for the three main treatment effect values considered in the aprepitant case study with a sample size per arm of 143 subjects. Method abbreviations and parameters as in Table~\ref{tab:abbreviations}.} \label{fig:aprepitant_mse_forest_plot_target_sample_size_per_arm_143}
\end{figure}

 \begin{figure}[htpb]
    \centering
    \includegraphics[width=0.95\linewidth, height=0.95\textheight,keepaspectratio]{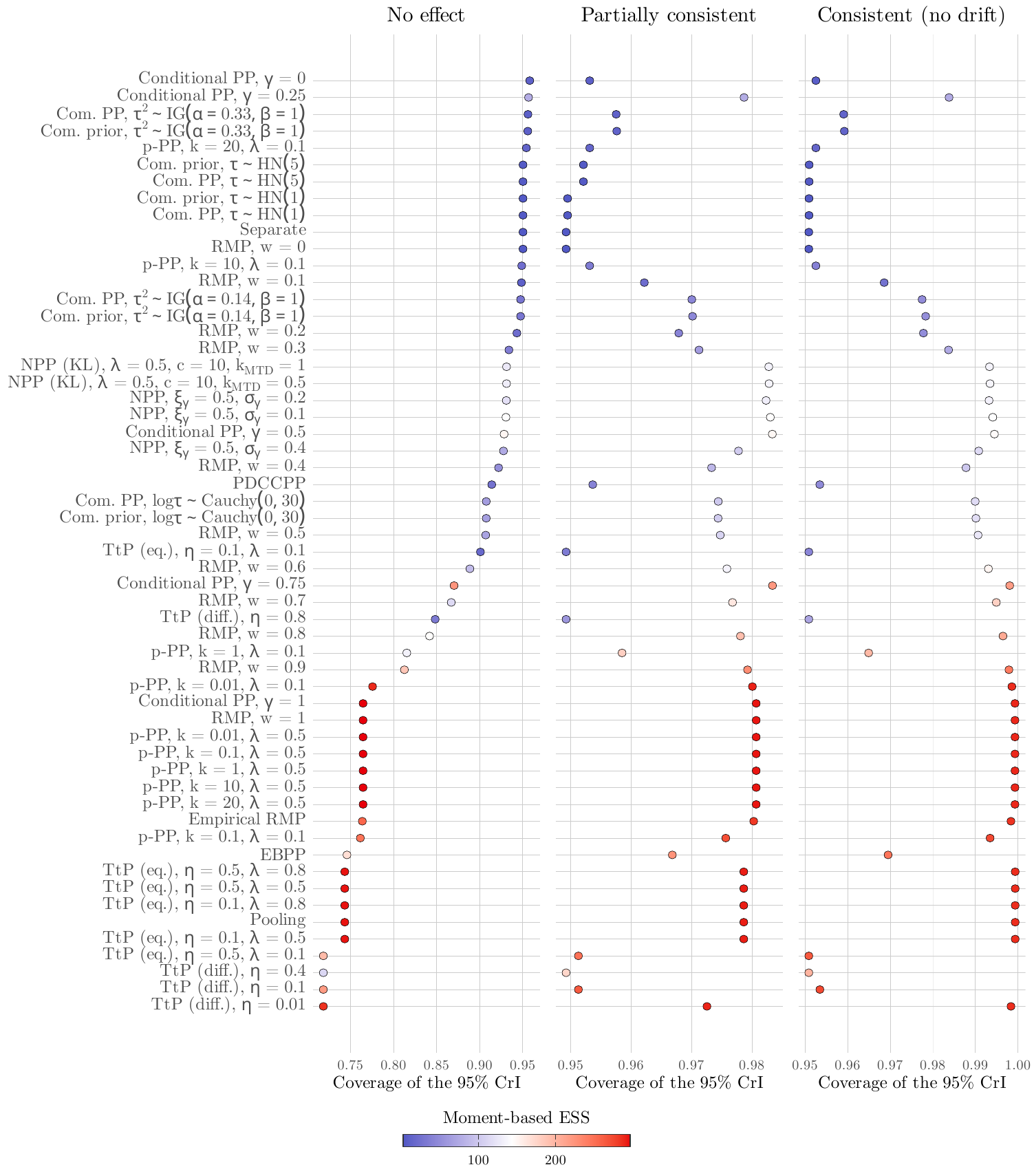}
    \caption{Empirical coverage of the 95\% credible interval for the three main treatment effect values considered in the aprepitant case study with a sample size per arm of 143 subjects. Method abbreviations and parameters as in Table~\ref{tab:abbreviations}.} \label{fig:aprepitant_coverage_forest_plot_target_sample_size_per_arm_143}
\end{figure}

  \begin{figure}[htpb]
     \centering
     \includegraphics[width=0.95\linewidth, height=0.95\textheight,keepaspectratio]{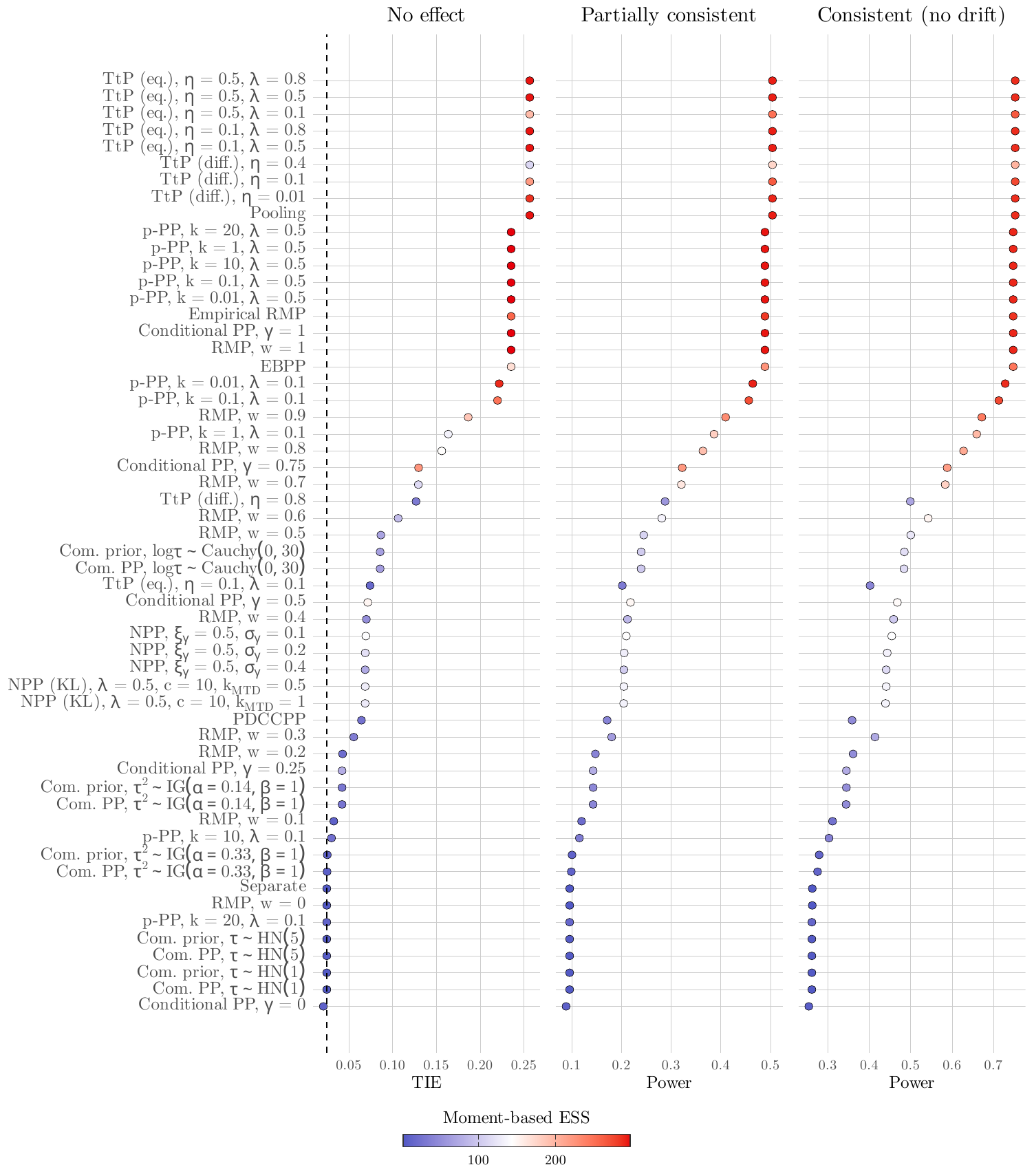}
     \caption{Probability of study success for the three main treatment effect values considered in the aprepitant case study ($n_T = 143$). Method abbreviations and parameters as in Table~\ref{tab:abbreviations}.} \label{fig:aprepitant_success_proba_forest_plot_target_sample_size_per_arm_143}
 \end{figure}

 \begin{figure}[htpb]
     \centering
     \includegraphics[width=0.95\linewidth, height=0.95\textheight,keepaspectratio]{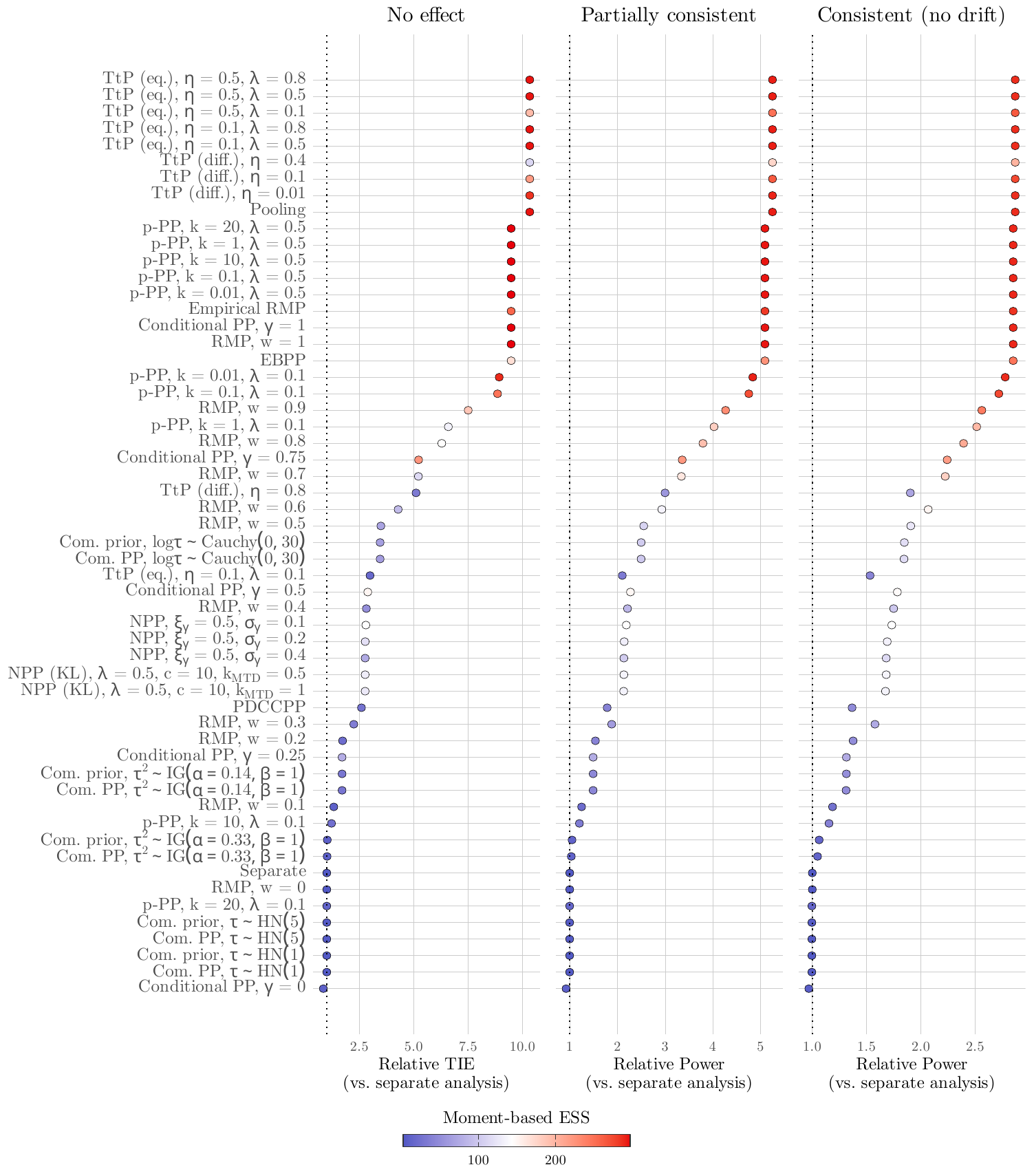}
     \caption{Probability of study success relative to a separate analysis of the target study data for the three main treatment effect values considered in the aprepitant case study ($n_T = 143$). Method abbreviations and parameters as in Table~\ref{tab:abbreviations}.} \label{fig:aprepitant_relative_success_proba_forest_plot_target_sample_size_per_arm_143}
 \end{figure}

\begin{figure}[htbp]
  \centering
  \includegraphics[width=\textwidth]{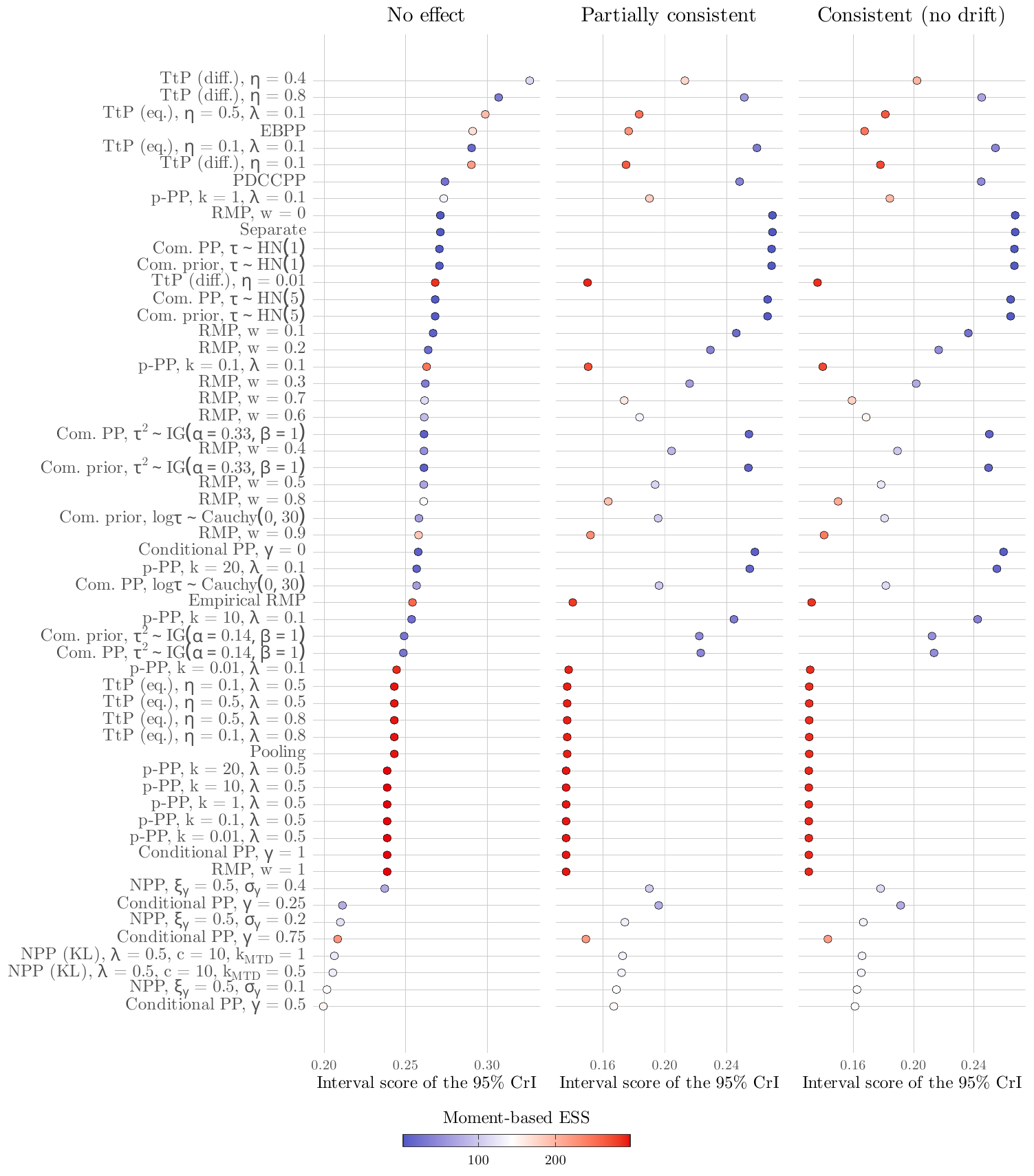}
  \caption{Interval score of the 95\% credible interval for the three main treatment effect values considered in the aprepitant case study with a sample size per arm of 143 subjects. Smaller is better. Error bars correspond to the 95\% CI. }
  \label{fig:aprepitant_interval_score_forest_plot_target_sample_size_per_arm_143}
\end{figure}

\begin{figure}[htbp]
\centering
\includegraphics[width=\textwidth]{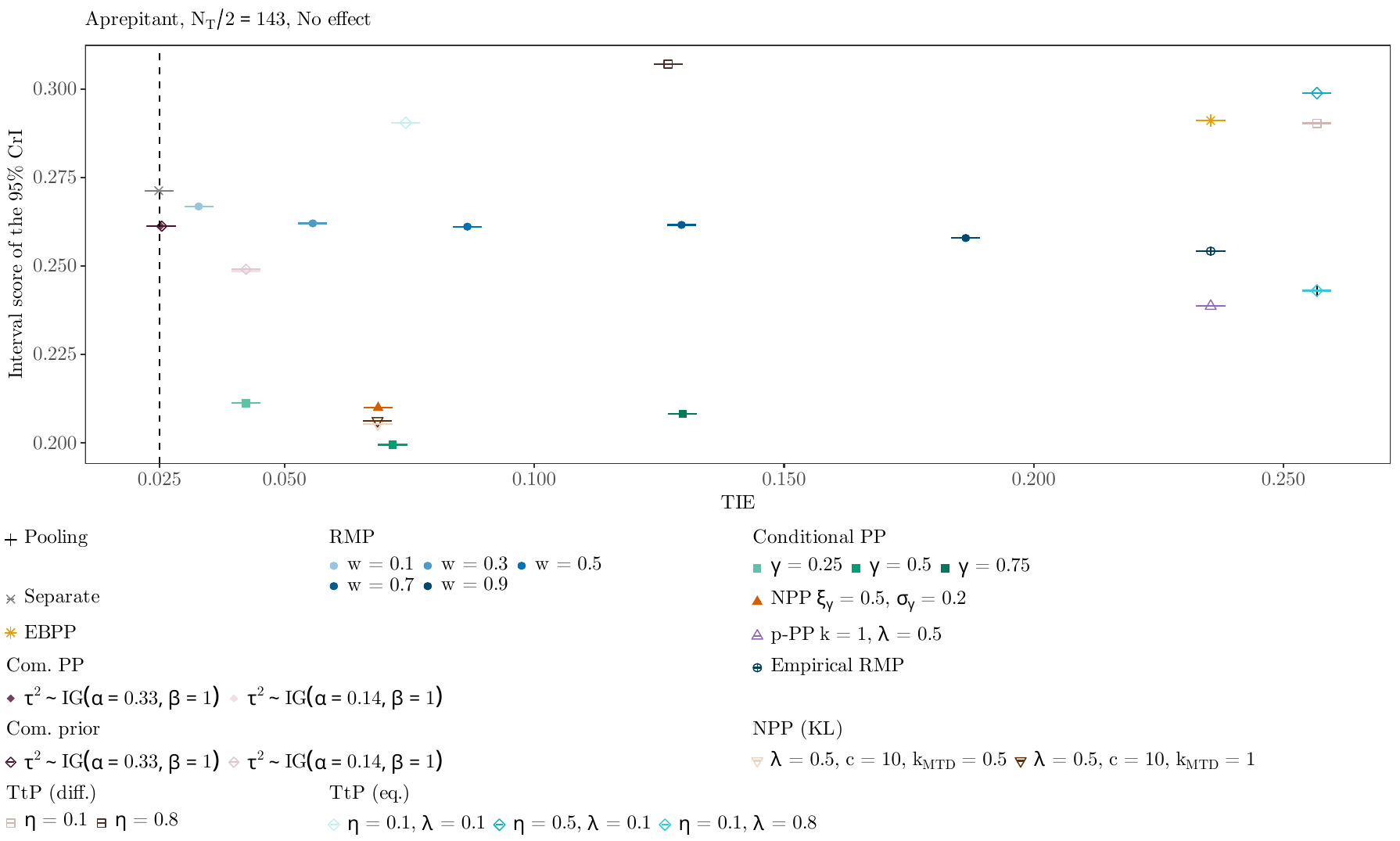}
\caption{Interval score of the 95\% credible interval as a function of type I error rate in the aprepitant case study with a sample size per arm of 143, without treatment effect.  Smaller is better. Error bars correspond to the 95\% CI of the interval score and type I error rate. Dashed vertical line represents the nominal type I error rate of 0.025. } \label{fig:aprepitant_interval_score_vs_tie_sample_size=143_treatment_effect=no_effect}
\end{figure}

\begin{figure}[htbp]
\centering
\includegraphics[width=0.95\linewidth, height=0.95\textheight,keepaspectratio]{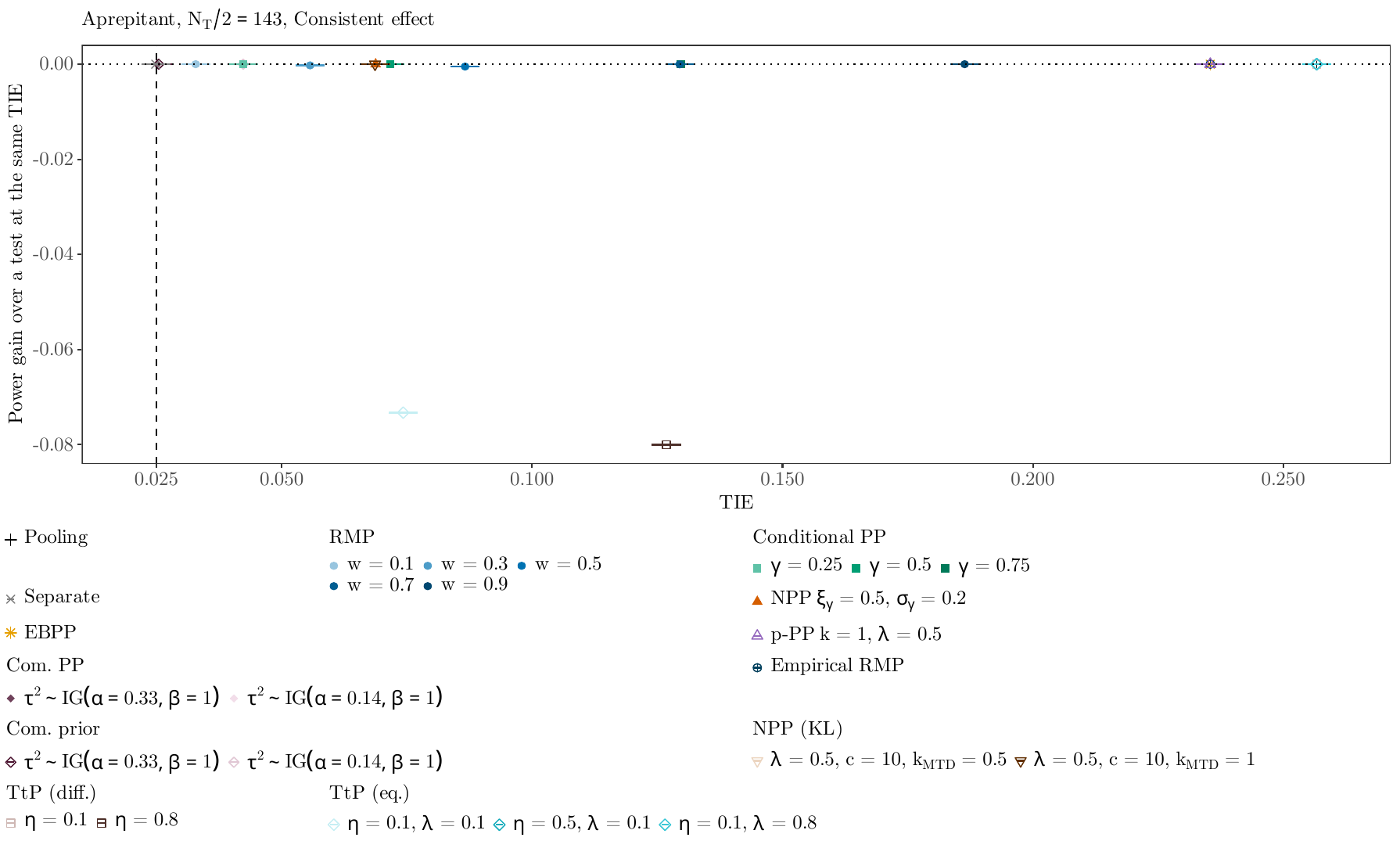}
\caption{Power gain over a frequentist test of the target data alone performed at the same TIE, versus TIE, in the aprepitant case study, with 143 participants per arm and a consistent treatment effect. The operating characteristics are computed exactly, and the reference is the exact randomized test at the method's TIE. For readability, only a subset of the parameter settings is shown, and the PDCCPP is omitted.}
\label{fig:aprepitant_power_gain_vs_tie_sample_size=143_treatment_effect=consistent}
\end{figure}

\begin{figure}[htbp]
\centering
\includegraphics[width=0.95\linewidth, height=0.95\textheight,keepaspectratio]{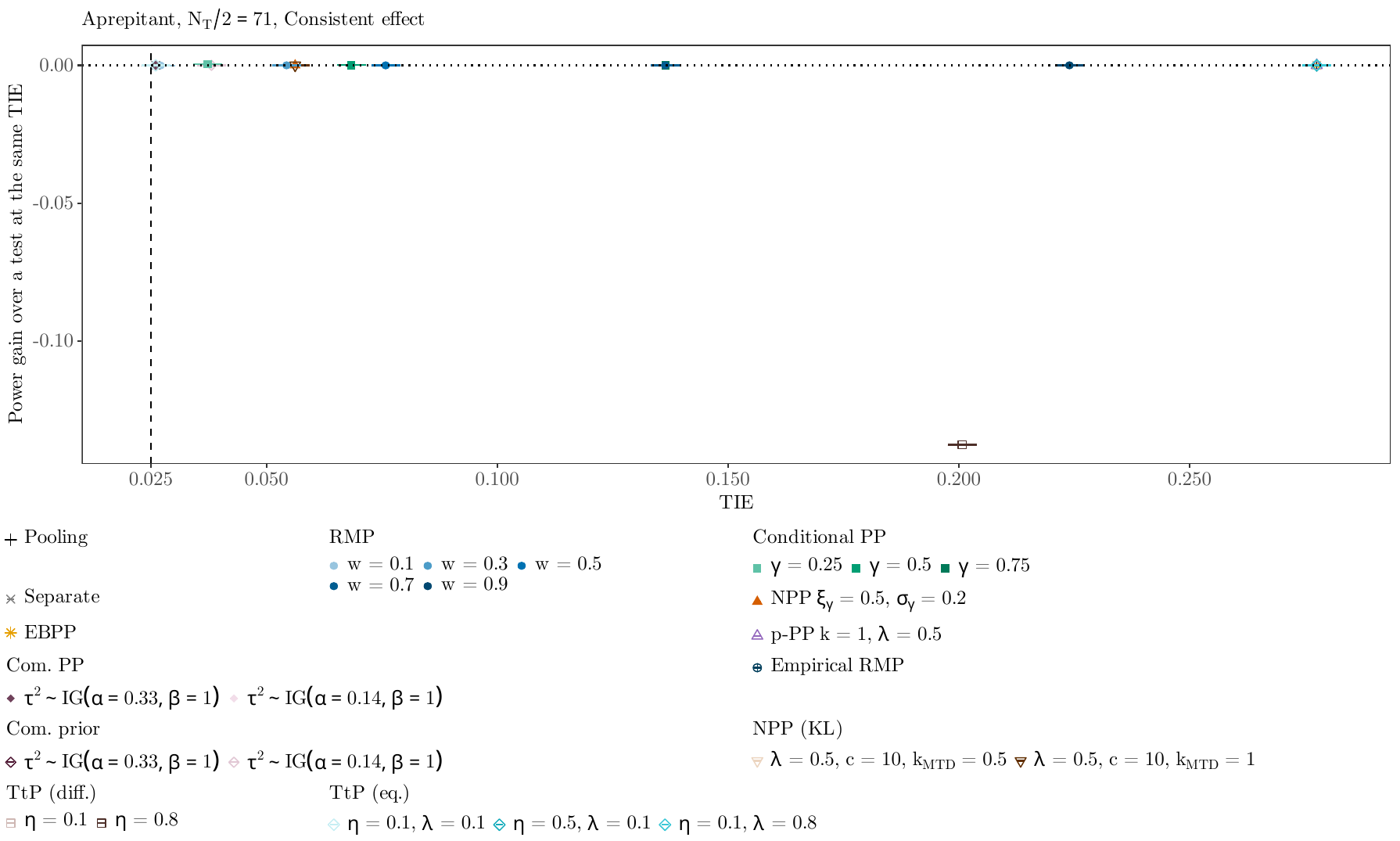}
\caption{Power gain over a frequentist test of the target data alone performed at the same TIE, versus TIE, in the aprepitant case study, with 71 participants per arm and a consistent treatment effect. The operating characteristics are computed exactly, and the reference is the exact randomized test at the method's TIE. For readability, only a subset of the parameter settings is shown, and the PDCCPP is omitted.}
\label{fig:aprepitant_power_gain_vs_tie_sample_size=71_treatment_effect=consistent}
\end{figure}

\stopcontents[appendices]

\end{document}